\newif\ifarxiv     
\newif\ifanonymous 

\arxivtrue
\anonymousfalse

\documentclass[sigplan,10pt
\ifarxiv ,nonacm=true\else\fi\ifanonymous ,anonymous,review\fi]{acmart}

\usepackage[links,references]{agda}
\usepackage{draft}
\usepackage{cleveref}
\usepackage{enumitem}

\newnote{esa}{olive}
\newnote{scw}{purple}

\usepackage[T1]{fontenc}

\usepackage{stmaryrd}
\usepackage{textgreek}
\usepackage{newunicodechar}

\newunicodechar{Γ}{\ensuremath{\mathnormal\Gamma}}
\newunicodechar{∷}{\ensuremath{::}}
\newunicodechar{→}{\ensuremath{\mathnormal\rightarrow}}
\newunicodechar{⇒}{\ensuremath{\mathnormal\Rightarrow}}
\newunicodechar{γ}{\ensuremath{\mathnormal\gamma}}
\newunicodechar{⊨}{\ensuremath{\mathnormal\vDash}}
\newunicodechar{⊢}{\ensuremath{\mathnormal\vdash}}
\newunicodechar{∈}{\ensuremath{\mathnormal\in}}
\newunicodechar{⇓}{\ensuremath{\mathnormal\Downarrow}}
\newunicodechar{δ}{\ensuremath{\mathnormal\delta}}
\newunicodechar{≡}{\ensuremath{\mathnormal\equiv}}
\newunicodechar{↦}{\ensuremath{\mathnormal\mapsto}}
\newunicodechar{ƛ}{\ensuremath{\mathnormal\lambdabar}}
\newunicodechar{⟨}{\ensuremath{\mathnormal\langle}}
\newunicodechar{⟩}{\ensuremath{\mathnormal\rangle}}
\newunicodechar{∈}{\ensuremath{\mathnormal\in}}
\newunicodechar{⟶}{\ensuremath{\mathnormal\longrightarrow}}
\newunicodechar{⇑}{\ensuremath{\mathnormal\Uparrow}}
\newunicodechar{ρ}{\ensuremath{\mathnormal\rho}}
\newunicodechar{ℰ}{\ensuremath{\mathcal{E}}}
\newunicodechar{𝒟}{\ensuremath{\mathcal{D}}}
\newunicodechar{𝒯}{\ensuremath{\mathcal{T}}}
\newunicodechar{𝒢}{\ensuremath{\mathcal{G}}}
\newunicodechar{𝒞}{\ensuremath{\mathcal{C}}}
\newunicodechar{𝔹}{\ensuremath{\mathbb{B}}}
\newunicodechar{𝕋}{\ensuremath{\mathbb{T}}}
\newunicodechar{⊩}{\ensuremath{\mathnormal{\Vdash}}}
\newunicodechar{≢}{\ensuremath{\mathnormal{\nequiv}}}
\newunicodechar{Δ}{\ensuremath{\mathnormal{\Delta}}}
\newunicodechar{⊆}{\ensuremath{\mathnormal{\subseteq}}}
\newunicodechar{∎}{\ensuremath{\mathnormal{\qed}}}
\newunicodechar{₁}{\ensuremath{\mathnormal{_1}}}
\newunicodechar{₂}{\ensuremath{\mathnormal{_2}}}
\newunicodechar{′}{\ensuremath{\mathnormal{^{\prime}}}}
\newunicodechar{⊎}{\ensuremath{\mathnormal{\uplus}}}
\newunicodechar{∅}{\ensuremath{\mathnormal{\varnothing}}}
\newunicodechar{∋}{\ensuremath{\mathnormal{\ni}}}
\newunicodechar{∣}{\ensuremath{\mathnormal{|}}}
\newunicodechar{∀}{\ensuremath{\mathnormal{\forall}}}
\newunicodechar{∃}{\ensuremath{\mathnormal{\exists}}}
\newunicodechar{⟦}{\ensuremath{\mathnormal{\llbracket}}}
\newunicodechar{⟧}{\ensuremath{\mathnormal{\rrbracket}}}
\newunicodechar{⊤}{\ensuremath{\mathnormal{\top}}}
\newunicodechar{⊥}{\ensuremath{\mathnormal{\bot}}}
\newunicodechar{ℕ}{\ensuremath{\mathnormal{\mathbb{N}}}}
\newunicodechar{ⁿ}{\ensuremath{\mathnormal{^{n}}}}
\newunicodechar{ᵉ}{\ensuremath{\mathnormal{^{e}}}}
\newunicodechar{∘}{\ensuremath{\mathnormal{\circ}}}
\newunicodechar{ℓ}{\ensuremath{\mathnormal{\ell}}}

\AtBeginDocument{%
  \providecommand\BibTeX{{%
    \normalfont B\kern-0.5em{\scshape i\kern-0.25em b}\kern-0.8em\TeX}}}

\acmConference[CPP '24]{Certified Programs and Proofs}{January 15--16,
  2024}{London, UK}
\begin{document}

\title[Making Logical Relations More Relatable (Proof Pearl)]{Making Logical Relations More Relatable \\ (Proof Pearl)}

\author{Emmanuel Suárez Acevedo}
\email{emsu@seas.upenn.edu}
\affiliation{%
  \institution{University of Pennsylvania}
  \country{USA}
}

\author{Stephanie Weirich}
\email{sweirich@seas.upenn.edu}
\affiliation{%
  \institution{University of Pennsylvania}
  \country{USA}
}

\begin{abstract}
Mechanical proofs by logical relations often involve tedious reasoning about
substitution. In this paper, we show that this is not necessarily the case, by
developing, in Agda, a proof that all simply typed lambda calculus expressions
evaluate to values. A formalization of the proof is remarkably short ($\sim$40 lines of
code), making for an excellent introduction to the technique of proofs by logical
relations not only on paper but also in a mechanized setting. We then show that
this process extends to more sophisticated reasoning by also proving the totality
of normalization by evaluation. Although these proofs are not new, we believe
presenting them will empower both new and experienced programming language
theorists in their use of logical relations.
\end{abstract}



\maketitle

\begin{code}[hide]%
\>[0]\AgdaKeyword{import}\AgdaSpace{}%
\AgdaModule{Axiom.Extensionality.Propositional}\AgdaSpace{}%
\AgdaSymbol{as}\AgdaSpace{}%
\AgdaModule{Ext}\<%
\\
\>[0]\AgdaKeyword{import}\AgdaSpace{}%
\AgdaModule{Relation.Binary.PropositionalEquality}\AgdaSpace{}%
\AgdaSymbol{as}\AgdaSpace{}%
\AgdaModule{Eq}\<%
\\
\>[0]\AgdaKeyword{open}\AgdaSpace{}%
\AgdaKeyword{import}\AgdaSpace{}%
\AgdaModule{Data.Bool}\AgdaSpace{}%
\AgdaKeyword{using}\AgdaSpace{}%
\AgdaSymbol{(}\AgdaDatatype{Bool}\AgdaSymbol{;}\AgdaSpace{}%
\AgdaInductiveConstructor{true}\AgdaSymbol{;}\AgdaSpace{}%
\AgdaInductiveConstructor{false}\AgdaSymbol{)}\<%
\\
\>[0]\AgdaKeyword{open}\AgdaSpace{}%
\AgdaKeyword{import}\AgdaSpace{}%
\AgdaModule{Data.Empty}\AgdaSpace{}%
\AgdaKeyword{using}\AgdaSpace{}%
\AgdaSymbol{(}\AgdaDatatype{⊥}\AgdaSymbol{)}\<%
\\
\>[0]\AgdaKeyword{open}\AgdaSpace{}%
\AgdaKeyword{import}\AgdaSpace{}%
\AgdaModule{Data.Nat}\AgdaSpace{}%
\AgdaKeyword{using}\AgdaSpace{}%
\AgdaSymbol{(}\AgdaDatatype{ℕ}\AgdaSymbol{;}\AgdaSpace{}%
\AgdaInductiveConstructor{suc}\AgdaSymbol{;}\AgdaSpace{}%
\AgdaInductiveConstructor{zero}\AgdaSymbol{)}\AgdaSpace{}%
\AgdaKeyword{renaming}\AgdaSpace{}%
\AgdaSymbol{(}\AgdaOperator{\AgdaPrimitive{\AgdaUnderscore{}∸\AgdaUnderscore{}}}\AgdaSpace{}%
\AgdaSymbol{to}\AgdaSpace{}%
\AgdaOperator{\AgdaPrimitive{\AgdaUnderscore{}-\AgdaUnderscore{}}}\AgdaSymbol{)}\<%
\\
\>[0]\AgdaKeyword{open}\AgdaSpace{}%
\AgdaKeyword{import}\AgdaSpace{}%
\AgdaModule{Data.Product}\AgdaSpace{}%
\AgdaKeyword{using}\AgdaSpace{}%
\AgdaSymbol{(}\AgdaFunction{∃-syntax}\AgdaSymbol{;}\AgdaSpace{}%
\AgdaOperator{\AgdaFunction{\AgdaUnderscore{}×\AgdaUnderscore{}}}\AgdaSymbol{;}\AgdaSpace{}%
\AgdaOperator{\AgdaInductiveConstructor{\AgdaUnderscore{},\AgdaUnderscore{}}}\AgdaSymbol{)}\<%
\\
\>[0]\AgdaKeyword{open}\AgdaSpace{}%
\AgdaKeyword{import}\AgdaSpace{}%
\AgdaModule{Data.String}\AgdaSpace{}%
\AgdaKeyword{using}\AgdaSpace{}%
\AgdaSymbol{(}\AgdaPostulate{String}\AgdaSymbol{)}\<%
\\
\>[0]\AgdaKeyword{open}\AgdaSpace{}%
\AgdaKeyword{import}\AgdaSpace{}%
\AgdaModule{Data.Sum}\AgdaSpace{}%
\AgdaKeyword{using}\AgdaSpace{}%
\AgdaSymbol{(}\AgdaOperator{\AgdaDatatype{\AgdaUnderscore{}⊎\AgdaUnderscore{}}}\AgdaSymbol{;}\AgdaSpace{}%
\AgdaInductiveConstructor{inj₁}\AgdaSymbol{;}\AgdaSpace{}%
\AgdaInductiveConstructor{inj₂}\AgdaSymbol{)}\<%
\\
\>[0]\AgdaKeyword{open}\AgdaSpace{}%
\AgdaKeyword{import}\AgdaSpace{}%
\AgdaModule{Data.Unit}\AgdaSpace{}%
\AgdaKeyword{using}\AgdaSpace{}%
\AgdaSymbol{(}\AgdaRecord{⊤}\AgdaSymbol{;}\AgdaSpace{}%
\AgdaInductiveConstructor{tt}\AgdaSymbol{)}\<%
\\
\>[0]\AgdaKeyword{open}\AgdaSpace{}%
\AgdaKeyword{import}\AgdaSpace{}%
\AgdaModule{Function}\AgdaSpace{}%
\AgdaKeyword{using}\AgdaSpace{}%
\AgdaSymbol{(}\AgdaOperator{\AgdaFunction{\AgdaUnderscore{}∘\AgdaUnderscore{}}}\AgdaSymbol{)}\<%
\\
\>[0]\AgdaKeyword{open}\AgdaSpace{}%
\AgdaKeyword{import}\AgdaSpace{}%
\AgdaModule{Level}\AgdaSpace{}%
\AgdaKeyword{using}\AgdaSpace{}%
\AgdaSymbol{(}\AgdaFunction{0ℓ}\AgdaSymbol{)}\<%
\\
\>[0]\AgdaKeyword{open}\AgdaSpace{}%
\AgdaKeyword{import}\AgdaSpace{}%
\AgdaModule{Relation.Unary}\AgdaSpace{}%
\AgdaKeyword{using}\AgdaSpace{}%
\AgdaSymbol{(}\AgdaOperator{\AgdaFunction{\AgdaUnderscore{}∈\AgdaUnderscore{}}}\AgdaSymbol{)}\<%
\\
\>[0]\AgdaKeyword{open}\AgdaSpace{}%
\AgdaKeyword{import}\AgdaSpace{}%
\AgdaModule{Relation.Nullary}\AgdaSpace{}%
\AgdaKeyword{using}\AgdaSpace{}%
\AgdaSymbol{(}\AgdaOperator{\AgdaFunction{¬\AgdaUnderscore{}}}\AgdaSymbol{)}\<%
\\
\>[0]\AgdaKeyword{open}\AgdaSpace{}%
\AgdaKeyword{import}\AgdaSpace{}%
\AgdaModule{Relation.Nullary.Negation}\AgdaSpace{}%
\AgdaKeyword{using}\AgdaSpace{}%
\AgdaSymbol{(}\AgdaFunction{contradiction}\AgdaSymbol{)}\<%
\\
\>[0]\AgdaKeyword{open}\AgdaSpace{}%
\AgdaModule{Eq}\AgdaSpace{}%
\AgdaKeyword{using}\AgdaSpace{}%
\AgdaSymbol{(}\AgdaOperator{\AgdaDatatype{\AgdaUnderscore{}≡\AgdaUnderscore{}}}\AgdaSymbol{;}\AgdaSpace{}%
\AgdaInductiveConstructor{refl}\AgdaSymbol{;}\AgdaSpace{}%
\AgdaFunction{cong}\AgdaSymbol{)}\<%
\\
\>[0]\AgdaKeyword{open}\AgdaSpace{}%
\AgdaModule{Eq.≡-Reasoning}\AgdaSpace{}%
\AgdaKeyword{using}\AgdaSpace{}%
\AgdaSymbol{(}\AgdaOperator{\AgdaFunction{begin\AgdaUnderscore{}}}\AgdaSymbol{;}\AgdaSpace{}%
\AgdaFunction{step-≡}\AgdaSymbol{;}\AgdaSpace{}%
\AgdaOperator{\AgdaFunction{\AgdaUnderscore{}∎}}\AgdaSymbol{)}\<%
\\
\>[0]\AgdaKeyword{open}\AgdaSpace{}%
\AgdaModule{Ext}\AgdaSpace{}%
\AgdaKeyword{using}\AgdaSpace{}%
\AgdaSymbol{(}\AgdaFunction{Extensionality}\AgdaSymbol{;}\AgdaSpace{}%
\AgdaFunction{implicit-extensionality}\AgdaSymbol{)}\<%
\\
\\[\AgdaEmptyExtraSkip]%
\>[0]\AgdaKeyword{module}\AgdaSpace{}%
\AgdaModule{paper}\AgdaSpace{}%
\AgdaKeyword{where}\<%
\end{code}

\section{Introduction}
Logical relations are a versatile tool for proving complex properties about
programs and programming languages, making them an integral part of
programming languages research. There are many examples of proofs by logical
relations~\cite[etc.]{tait,girard,plotkin1973lambda,reynolds1983types}, one of
these being a proof of \emph{normalization} for the simply typed lambda
calculus (STLC): if a term is well-typed, then it can be reduced (i.e. computed) to a
``normal form'', a term that cannot be reduced any further.

Typically, we model the semantics of lambda calculi using a set of rules for
reducing terms, featuring a key rule known as the $\beta$-reduction:

$$
(\lambda x. t) s \rightarrow t[s/x]
$$

This rule describes how the application of an abstraction to a term $(\lambda x.
t) s$ reduces to the body of the abstraction $t$ with $s$ substituted for every
instance of the variable $x$ in $t$ (an operation we write as $t[s/x]$).

Normalization, according to these reduction rules, is a well-known result for
the simply-typed lambda calculus~\citep{tait}. This proof cannot be shown by
simple induction on derivations --- instead it requires the use of a proof
technique known as a \emph{logical relation}. One must define, by recursion
over the structure of types, a predicate that holds for all normalizing
terms. Then one shows the fundamental theorem, which states that this
predicate holds for all well-typed terms.

Many tutorials on logical relations start by using this technique to show
normalization for \emph{weak head reduction}
\cite[etc.]{harper2022wn,pierce2002types,skorstengaard2019introduction}.  Weak
head reduction is a limited form of reduction that does not reduce inside the
bodies of abstractions. It corresponds to an evaluation semantics for the
simply typed lambda calculus.  This restriction simplifies the overall
normalization argument making it an appropriate introduction to the general
technique.

However, although there is a growing corpus of programming languages textbooks
that are based on mechanized proofs \cite{plf,plfa} and work in conjunction
with a proof assistant to make sure that students understand each step of the
argument, there are not enough mechanized tutorials of proofs based on logical
relations aimed at beginning researchers.

We believe that the issue is that in a mechanized setting, the structure of 
logical relations proofs can be obscured by the many substitution lemmas that 
are required. While an informal introduction can elide these details, they 
must be included in a mechanized development.
Indeed, \citet{abel2019poplmark} note that substitution lemmas are often the
most cumbersome part of formalizing a proof by logical relations, proposing
the technique as a benchmark for mechanizing metatheory. They quote
\citet{altenkirch1993formalization}, who remarks:
\begin{quote}
``I discovered that the core part of the proof (here proving lemmas about CR)
is fairly straightforward and only requires a good understanding of the paper
version. However, in completing the proof I observed that in certain places I
had to invest much more work than expected, e.g. proving lemmas about
substitution and weakening.''
\end{quote}

But, it doesn't have to be this way. In this paper, we observe that we can
introduce the technique of logical relations without needing a single
substitution lemma. Furthermore, our approach is based on the standard
introductory example: showing that \emph{evaluation of the simply typed
  lambda calculus is total}.

Our key insight is the use of an \emph{environment-based natural
  semantics}~\cite{kahn1987natural}, also called a big-step semantics, to describe the
evaluation of lambda terms. In this setting, our mechanized totality proof is
remarkably concise. In Agda, the proof fits entirely in a single column of
this paper (Figure~\ref{fig:totality-proof}). The overall structure of the
semantics so directly coincides with the structure of the proof that the
fundamental lemma for the logical relation needs no auxiliary lemmas
(substitution or otherwise).\footnote{In contrast, the proof in \citet{plf}
requires nineteen lemmas to show an equivalent result.} We present this proof in
$\S$\ref{sec:totality}, which can be read as a tutorial introduction to logical
relations.

Our proof is not exactly new. In essence, it is a simplification of a much
longer proof related to the \emph{normalization by evaluation} (NbE) algorithm
\cite{berger1991inverse}.  \citet{coquand1997intuitionistic} observe that this
algorithm can be used for a proof of full normalization for STLC---normal forms
can be computed through evaluation. Our simple proof can be viewed as a
restriction of this argument to weak head reduction.

What this means is that we can \emph{also} use a logical relations argument to
show \textbf{full normalization for STLC} without defining or reasoning about
substitution. In $\S$\ref{sec:nbe} we extend the tutorial proof to include this
result. Our proof is based on an informal presentation given in \citet{nbe},
but we find that in Agda, this proof is as nearly as approachable in a
mechanized setting as it would be on paper. This section demonstrates that
this technique is not limited to simple properties that only apply to closed
terms.

Overall, these two examples demonstrate that we can prove complex properties
using a logical relations argument without needing substitutions or
substitution lemmas, a result that we hope empowers researchers in both
mechanizing future proofs and in understanding the proof technique.  As far as
we can tell, this is the first presentation with the explicit goal of
introducing the technique of logical relations in a mechanized setting.

As these results are especially relevant in a mechanized setting, we have
written this paper as a literate Agda script, presenting both proofs in Agda.
Every highlighted line of code rendered here has been checked by Agda, and we
have only omitted imports, precedence declarations for mixfix syntax, and
repeated code. We assume the reader has basic familiarity with Agda, but the
content itself is self-contained and we explain more advanced Agda features as
they are encountered.

We have included the content of the appendix in our supplemental material for
the reviewers, these proofs make up all of the literate Agda code presented in
the main body of the paper.

\section{Totality of evaluation of STLC}
\begin{code}[hide]%
\>[0]\AgdaKeyword{module}\AgdaSpace{}%
\AgdaModule{Totality}\AgdaSpace{}%
\AgdaKeyword{where}\<%
\\
\>[0][@{}l@{\AgdaIndent{0}}]%
\>[2]\AgdaKeyword{infix}\AgdaSpace{}%
\AgdaNumber{5}\AgdaSpace{}%
\AgdaOperator{\AgdaInductiveConstructor{⟨ƛ\AgdaUnderscore{}⟩\AgdaUnderscore{}}}\<%
\\
\>[2]\AgdaKeyword{infix}\AgdaSpace{}%
\AgdaNumber{4}\AgdaSpace{}%
\AgdaOperator{\AgdaDatatype{\AgdaUnderscore{}∣\AgdaUnderscore{}⇓\AgdaUnderscore{}}}\<%
\\
\>[2]\AgdaKeyword{infix}\AgdaSpace{}%
\AgdaNumber{4}\AgdaSpace{}%
\AgdaOperator{\AgdaFunction{\AgdaUnderscore{}⊨\AgdaUnderscore{}}}\<%
\\
\>[2]\AgdaKeyword{infixl}\AgdaSpace{}%
\AgdaNumber{5}\AgdaSpace{}%
\AgdaOperator{\AgdaFunction{\AgdaUnderscore{}++\AgdaUnderscore{}}}\<%
\\
\>[2]\AgdaKeyword{infixr}\AgdaSpace{}%
\AgdaNumber{7}\AgdaSpace{}%
\AgdaOperator{\AgdaInductiveConstructor{\AgdaUnderscore{}⇒\AgdaUnderscore{}}}\<%
\\
\>[2]\AgdaKeyword{infixl}\AgdaSpace{}%
\AgdaNumber{5}\AgdaSpace{}%
\AgdaOperator{\AgdaInductiveConstructor{\AgdaUnderscore{}·:\AgdaUnderscore{}}}\<%
\\
\>[2]\AgdaKeyword{infix}\AgdaSpace{}%
\AgdaNumber{4}\AgdaSpace{}%
\AgdaOperator{\AgdaDatatype{\AgdaUnderscore{}⊢\AgdaUnderscore{}}}\<%
\\
\>[2]\AgdaKeyword{infix}\AgdaSpace{}%
\AgdaNumber{4}\AgdaSpace{}%
\AgdaOperator{\AgdaDatatype{\AgdaUnderscore{}∋\AgdaUnderscore{}}}\<%
\\
\>[2]\AgdaKeyword{infix}\AgdaSpace{}%
\AgdaNumber{5}\AgdaSpace{}%
\AgdaOperator{\AgdaInductiveConstructor{ƛ\AgdaUnderscore{}}}\<%
\\
\>[2]\AgdaKeyword{infixl}\AgdaSpace{}%
\AgdaNumber{7}\AgdaSpace{}%
\AgdaOperator{\AgdaInductiveConstructor{\AgdaUnderscore{}·\AgdaUnderscore{}}}\<%
\\
\>[2]\AgdaKeyword{infix}\AgdaSpace{}%
\AgdaNumber{4}\AgdaSpace{}%
\AgdaFunction{semantic-typing}\<%
\\
\>[2]\AgdaKeyword{infixl}\AgdaSpace{}%
\AgdaNumber{5}\AgdaSpace{}%
\AgdaOperator{\AgdaFunction{\AgdaUnderscore{}\textasciicircum{}\AgdaUnderscore{}}}\<%
\end{code}
\label{sec:totality}
A property that we are generally interested in proving about a program is that
if it can be assigned a type, then its execution is well-defined. Equivalently,
we can prove that its evaluation is total. Proving that evaluation is total
is related to proving normalization with weak head reduction, though the two
properties are distinct and should not be confused. In proving that evaluation is
total, we do not concern ourselves with the normal form that any well-typed term
can be reduced to (i.e. computed), but rather only on showing that the evaluation
of the term (i.e. the computation itself) is well-defined.

We prove that evaluation is total for STLC. To do so, we must first represent
STLC in Agda and model the behavior (the semantics) of an STLC term.

\begin{AgdaAlign}
\begin{figure}
\begin{code}%
\>[2]\AgdaKeyword{data}\AgdaSpace{}%
\AgdaDatatype{Type}\AgdaSpace{}%
\AgdaSymbol{:}\AgdaSpace{}%
\AgdaPrimitive{Set}\AgdaSpace{}%
\AgdaKeyword{where}\<%
\\
\>[2][@{}l@{\AgdaIndent{0}}]%
\>[4]\AgdaInductiveConstructor{bool}\AgdaSpace{}%
\AgdaSymbol{:}\AgdaSpace{}%
\AgdaDatatype{Type}\<%
\\
\>[4]\AgdaOperator{\AgdaInductiveConstructor{\AgdaUnderscore{}⇒\AgdaUnderscore{}}}\AgdaSpace{}%
\AgdaSymbol{:}\AgdaSpace{}%
\AgdaDatatype{Type}\AgdaSpace{}%
\AgdaSymbol{→}\AgdaSpace{}%
\AgdaDatatype{Type}\AgdaSpace{}%
\AgdaSymbol{→}\AgdaSpace{}%
\AgdaDatatype{Type}\<%
\\
\\[\AgdaEmptyExtraSkip]%
\>[2]\AgdaKeyword{variable}\AgdaSpace{}%
\AgdaGeneralizable{S}\AgdaSpace{}%
\AgdaGeneralizable{T}\AgdaSpace{}%
\AgdaSymbol{:}\AgdaSpace{}%
\AgdaDatatype{Type}\<%
\\
\\[\AgdaEmptyExtraSkip]%
\>[2]\AgdaKeyword{data}\AgdaSpace{}%
\AgdaDatatype{Ctx}\AgdaSpace{}%
\AgdaSymbol{:}\AgdaSpace{}%
\AgdaPrimitive{Set}\AgdaSpace{}%
\AgdaKeyword{where}\<%
\\
\>[2][@{}l@{\AgdaIndent{0}}]%
\>[4]\AgdaInductiveConstructor{∅}\AgdaSpace{}%
\AgdaSymbol{:}\AgdaSpace{}%
\AgdaDatatype{Ctx}\<%
\\
\>[4]\AgdaOperator{\AgdaInductiveConstructor{\AgdaUnderscore{}·:\AgdaUnderscore{}}}\AgdaSpace{}%
\AgdaSymbol{:}\AgdaSpace{}%
\AgdaDatatype{Ctx}\AgdaSpace{}%
\AgdaSymbol{→}\AgdaSpace{}%
\AgdaDatatype{Type}\AgdaSpace{}%
\AgdaSymbol{→}\AgdaSpace{}%
\AgdaDatatype{Ctx}\<%
\\
\\[\AgdaEmptyExtraSkip]%
\>[2]\AgdaKeyword{variable}\AgdaSpace{}%
\AgdaGeneralizable{Γ}\AgdaSpace{}%
\AgdaSymbol{:}\AgdaSpace{}%
\AgdaDatatype{Ctx}\<%
\\
\\[\AgdaEmptyExtraSkip]%
\>[2]\AgdaComment{--\ Intrinsically-scoped\ de\ Brujin\ indices}\<%
\\
\>[2]\AgdaKeyword{data}\AgdaSpace{}%
\AgdaOperator{\AgdaDatatype{\AgdaUnderscore{}∋\AgdaUnderscore{}}}\AgdaSpace{}%
\AgdaSymbol{:}\AgdaSpace{}%
\AgdaDatatype{Ctx}\AgdaSpace{}%
\AgdaSymbol{→}\AgdaSpace{}%
\AgdaDatatype{Type}\AgdaSpace{}%
\AgdaSymbol{→}\AgdaSpace{}%
\AgdaPrimitive{Set}\AgdaSpace{}%
\AgdaKeyword{where}\<%
\\
\>[2][@{}l@{\AgdaIndent{0}}]%
\>[4]\AgdaInductiveConstructor{zero}\AgdaSpace{}%
\AgdaSymbol{:}\AgdaSpace{}%
\AgdaGeneralizable{Γ}\AgdaSpace{}%
\AgdaOperator{\AgdaInductiveConstructor{·:}}\AgdaSpace{}%
\AgdaGeneralizable{T}\AgdaSpace{}%
\AgdaOperator{\AgdaDatatype{∋}}\AgdaSpace{}%
\AgdaGeneralizable{T}\<%
\\
\>[4]\AgdaInductiveConstructor{suc}\AgdaSpace{}%
\AgdaSymbol{:}\AgdaSpace{}%
\AgdaGeneralizable{Γ}\AgdaSpace{}%
\AgdaOperator{\AgdaDatatype{∋}}\AgdaSpace{}%
\AgdaGeneralizable{T}\AgdaSpace{}%
\AgdaSymbol{→}\AgdaSpace{}%
\AgdaGeneralizable{Γ}\AgdaSpace{}%
\AgdaOperator{\AgdaInductiveConstructor{·:}}\AgdaSpace{}%
\AgdaGeneralizable{S}\AgdaSpace{}%
\AgdaOperator{\AgdaDatatype{∋}}\AgdaSpace{}%
\AgdaGeneralizable{T}\<%
\\
\\[\AgdaEmptyExtraSkip]%
\>[2]\AgdaKeyword{variable}\AgdaSpace{}%
\AgdaGeneralizable{x}\AgdaSpace{}%
\AgdaSymbol{:}\AgdaSpace{}%
\AgdaGeneralizable{Γ}\AgdaSpace{}%
\AgdaOperator{\AgdaDatatype{∋}}\AgdaSpace{}%
\AgdaGeneralizable{T}\<%
\\
\\[\AgdaEmptyExtraSkip]%
\>[2]\AgdaComment{--\ Intrinsically-typed\ terms}\<%
\\
\>[2]\AgdaKeyword{data}\AgdaSpace{}%
\AgdaOperator{\AgdaDatatype{\AgdaUnderscore{}⊢\AgdaUnderscore{}}}\AgdaSpace{}%
\AgdaSymbol{:}\AgdaSpace{}%
\AgdaDatatype{Ctx}\AgdaSpace{}%
\AgdaSymbol{→}\AgdaSpace{}%
\AgdaDatatype{Type}\AgdaSpace{}%
\AgdaSymbol{→}\AgdaSpace{}%
\AgdaPrimitive{Set}\AgdaSpace{}%
\AgdaKeyword{where}\<%
\\
\>[2][@{}l@{\AgdaIndent{0}}]%
\>[4]\AgdaInductiveConstructor{true}\AgdaSpace{}%
\AgdaInductiveConstructor{false}\AgdaSpace{}%
\AgdaSymbol{:}\AgdaSpace{}%
\AgdaGeneralizable{Γ}\AgdaSpace{}%
\AgdaOperator{\AgdaDatatype{⊢}}\AgdaSpace{}%
\AgdaInductiveConstructor{bool}\<%
\\
\>[4]\AgdaInductiveConstructor{var}\AgdaSpace{}%
\AgdaSymbol{:}\AgdaSpace{}%
\AgdaGeneralizable{Γ}\AgdaSpace{}%
\AgdaOperator{\AgdaDatatype{∋}}\AgdaSpace{}%
\AgdaGeneralizable{T}\AgdaSpace{}%
\AgdaSymbol{→}\AgdaSpace{}%
\AgdaGeneralizable{Γ}\AgdaSpace{}%
\AgdaOperator{\AgdaDatatype{⊢}}\AgdaSpace{}%
\AgdaGeneralizable{T}\<%
\\
\>[4]\AgdaOperator{\AgdaInductiveConstructor{ƛ\AgdaUnderscore{}}}\AgdaSpace{}%
\AgdaSymbol{:}\AgdaSpace{}%
\AgdaGeneralizable{Γ}\AgdaSpace{}%
\AgdaOperator{\AgdaInductiveConstructor{·:}}\AgdaSpace{}%
\AgdaGeneralizable{S}\AgdaSpace{}%
\AgdaOperator{\AgdaDatatype{⊢}}\AgdaSpace{}%
\AgdaGeneralizable{T}\AgdaSpace{}%
\AgdaSymbol{→}\AgdaSpace{}%
\AgdaGeneralizable{Γ}\AgdaSpace{}%
\AgdaOperator{\AgdaDatatype{⊢}}\AgdaSpace{}%
\AgdaGeneralizable{S}\AgdaSpace{}%
\AgdaOperator{\AgdaInductiveConstructor{⇒}}\AgdaSpace{}%
\AgdaGeneralizable{T}\<%
\\
\>[4]\AgdaOperator{\AgdaInductiveConstructor{\AgdaUnderscore{}·\AgdaUnderscore{}}}\AgdaSpace{}%
\AgdaSymbol{:}\AgdaSpace{}%
\AgdaGeneralizable{Γ}\AgdaSpace{}%
\AgdaOperator{\AgdaDatatype{⊢}}\AgdaSpace{}%
\AgdaGeneralizable{S}\AgdaSpace{}%
\AgdaOperator{\AgdaInductiveConstructor{⇒}}\AgdaSpace{}%
\AgdaGeneralizable{T}\AgdaSpace{}%
\AgdaSymbol{→}\AgdaSpace{}%
\AgdaGeneralizable{Γ}\AgdaSpace{}%
\AgdaOperator{\AgdaDatatype{⊢}}\AgdaSpace{}%
\AgdaGeneralizable{S}\AgdaSpace{}%
\AgdaSymbol{→}\AgdaSpace{}%
\AgdaGeneralizable{Γ}\AgdaSpace{}%
\AgdaOperator{\AgdaDatatype{⊢}}\AgdaSpace{}%
\AgdaGeneralizable{T}\<%
\\
\>[4]\AgdaOperator{\AgdaInductiveConstructor{if\AgdaUnderscore{}then\AgdaUnderscore{}else\AgdaUnderscore{}}}\AgdaSpace{}%
\AgdaSymbol{:}\AgdaSpace{}%
\AgdaGeneralizable{Γ}\AgdaSpace{}%
\AgdaOperator{\AgdaDatatype{⊢}}\AgdaSpace{}%
\AgdaInductiveConstructor{bool}\AgdaSpace{}%
\AgdaSymbol{→}\AgdaSpace{}%
\AgdaGeneralizable{Γ}\AgdaSpace{}%
\AgdaOperator{\AgdaDatatype{⊢}}\AgdaSpace{}%
\AgdaGeneralizable{T}\AgdaSpace{}%
\AgdaSymbol{→}\AgdaSpace{}%
\AgdaGeneralizable{Γ}\AgdaSpace{}%
\AgdaOperator{\AgdaDatatype{⊢}}\AgdaSpace{}%
\AgdaGeneralizable{T}\AgdaSpace{}%
\AgdaSymbol{→}\AgdaSpace{}%
\AgdaGeneralizable{Γ}\AgdaSpace{}%
\AgdaOperator{\AgdaDatatype{⊢}}\AgdaSpace{}%
\AgdaGeneralizable{T}\<%
\\
\\[\AgdaEmptyExtraSkip]%
\>[2]\AgdaKeyword{variable}\AgdaSpace{}%
\AgdaGeneralizable{r}\AgdaSpace{}%
\AgdaGeneralizable{s}\AgdaSpace{}%
\AgdaGeneralizable{t}\AgdaSpace{}%
\AgdaSymbol{:}\AgdaSpace{}%
\AgdaGeneralizable{Γ}\AgdaSpace{}%
\AgdaOperator{\AgdaDatatype{⊢}}\AgdaSpace{}%
\AgdaGeneralizable{T}\<%
\end{code}
\caption{STLC}
\label{fig:stlc}
\end{figure}

\subsection{Embedding of STLC in Agda}
We prove that evaluation of STLC is total for a minimal subset of the language
for brevity, shown in Figure \ref{fig:stlc}. The proof has no significant
complexity added when adding more familiar constructs such as sums or products.
We use booleans (\mbox{\AgdaInductiveConstructor{bool}}) for our base type and a
function type (\mbox{\AgdaBound{S} \AgdaInductiveConstructor{⇒} \AgdaBound{T}}).
We consider variables (\mbox{\AgdaInductiveConstructor{var} \AgdaBound{x}}),
abstractions (\mbox{\AgdaInductiveConstructor{ƛ} \AgdaBound{t}}), and
application (\mbox{\AgdaBound{r} \AgdaInductiveConstructor{·} \AgdaBound{s}}).
Additionally, we have the boolean constants \mbox{\AgdaInductiveConstructor{true}}
and \mbox{\AgdaInductiveConstructor{false}}, along with conditional branching
(\mbox{\AgdaInductiveConstructor{if} \AgdaBound{r} \AgdaInductiveConstructor{then}
\AgdaBound{s} \AgdaInductiveConstructor{else} \AgdaBound{t}}).

We represent variable bindings using de Brujin indices \cite{debrujin}, and a
typing context serves as a list of types with each index into the list representing a
variable. We define typing contexts as either the empty context
\AgdaInductiveConstructor{∅} or an extension to a context \mbox{\AgdaBound{Γ}
\AgdaInductiveConstructor{·:} \AgdaBound{T}}. We use a lookup judgement into a
context \mbox{\AgdaBound{Γ} \AgdaFunction{∋} \AgdaBound{T}} for de Brujin
indices so that they may be intrinsically scoped, using the constructors
\AgdaInductiveConstructor{zero} and \AgdaInductiveConstructor{suc} suggestively.

A term \AgdaBound{t} is an intrinsically typed Agda expression; we do not
consider raw terms that may be ill-typed and represent terms by their typing
derivation. We choose an intrinsically typed representation to simplify our
discussion in $\S$\ref{sec:correctness}. The decision is irrelevant for the
proof that evaluation is total, however. We can prove that evaluation is total
with an extrinsic representation of terms with few changes to the formalization
we present here. Surprisingly, this remains the case even if using raw bindings
(e.g. strings) to represent variables.\footnote{We include a formalization of
the totality of evaluation for STLC using an extrinsic representation with raw
bindings in appendix \ref{sec:extrinsic} to make this apparent.} Consequently,
the proof we present here could be introduced as soon as the simply typed lambda
calculus is first taught in material using proof assistants for education, where
it is common to first use an embedding of the STLC using raw bindings for
variables to improve readability (e.g. as done by \citet{plfa} and \citet{plf}).

To make some of our relations and theorems more succinct, we make use of a feature
of Agda that allows us establish metavariables. For example,
\mbox{\AgdaKeyword{variable} \AgdaBound{x} : \AgdaBound{Γ} \AgdaDatatype{∋}
\AgdaBound{T}} indicates the metavariable \AgdaBound{x} will be used for
the de Brujin indices making up variable bindings in our representation of the
STLC. The first such relation that takes advantage of this feature is our
evaluation relation, which we define in the next section.

\subsection{Natural semantics}
\label{sec:natural-semantics}
We can model the behavior of a program with a small-step relation (each ``step''
a program takes in its execution) or a big-step relation (the execution of a
program as a whole).

We often describe the semantics of STLC through the use of substitution, an
operation that replaces every instance of a variable in a term for another term
in a $\beta$-reduction. A $\beta$-reduction closely resembles a single step that
a program may take in its execution, so the rule is often used to describe a
small-step operational semantics.

While a small-step semantics almost always uses substitution to model function
application, a big-step semantics may be defined equivalently using an
environment or with substitutions. Big-step semantics are commonly used to
reason about programs as noted by \citet{chargueraud2013pretty} when they claim
that \emph{big-step is not dead}.

The semantics we use (shown in Figure \ref{fig:semantics}) is called a natural
semantics \cite{kahn1987natural}, modeling the execution of a term through its
evaluation. It is a big-step semantics that uses an environment \AgdaBound{γ} to
evaluate a term \AgdaBound{t} to a result value \AgdaBound{a}, notated
\mbox{\AgdaBound{γ} \AgdaDatatype{∣} \AgdaBound{t} \AgdaDatatype{⇓}
\AgdaBound{a}}. As this semantics does not use substitution, we can avoid having
to reason about the operation.

\begin{figure}[t]
\begin{code}%
\>[2]\AgdaKeyword{mutual}\<%
\\
\>[2][@{}l@{\AgdaIndent{0}}]%
\>[4]\AgdaComment{--\ Environments}\<%
\\
\>[4]\AgdaFunction{Env}\AgdaSpace{}%
\AgdaSymbol{:}\AgdaSpace{}%
\AgdaDatatype{Ctx}\AgdaSpace{}%
\AgdaSymbol{→}\AgdaSpace{}%
\AgdaPrimitive{Set}\<%
\\
\>[4]\AgdaFunction{Env}\AgdaSpace{}%
\AgdaBound{Γ}\AgdaSpace{}%
\AgdaSymbol{=}\AgdaSpace{}%
\AgdaSymbol{∀}\AgdaSpace{}%
\AgdaSymbol{\{}\AgdaBound{T}\AgdaSymbol{\}}\AgdaSpace{}%
\AgdaSymbol{→}\AgdaSpace{}%
\AgdaBound{Γ}\AgdaSpace{}%
\AgdaOperator{\AgdaDatatype{∋}}\AgdaSpace{}%
\AgdaBound{T}\AgdaSpace{}%
\AgdaSymbol{→}\AgdaSpace{}%
\AgdaDatatype{Domain}\AgdaSpace{}%
\AgdaBound{T}\<%
\\
\\[\AgdaEmptyExtraSkip]%
\>[4]\AgdaComment{--\ Domain\ of\ evaluation}\<%
\\
\>[4]\AgdaKeyword{data}\AgdaSpace{}%
\AgdaDatatype{Domain}\AgdaSpace{}%
\AgdaSymbol{:}\AgdaSpace{}%
\AgdaDatatype{Type}\AgdaSpace{}%
\AgdaSymbol{→}\AgdaSpace{}%
\AgdaPrimitive{Set}\AgdaSpace{}%
\AgdaKeyword{where}\<%
\\
\>[4][@{}l@{\AgdaIndent{0}}]%
\>[6]\AgdaComment{--\ Booleans}\<%
\\
\>[6]\AgdaInductiveConstructor{true}\AgdaSpace{}%
\AgdaInductiveConstructor{false}\AgdaSpace{}%
\AgdaSymbol{:}\AgdaSpace{}%
\AgdaDatatype{Domain}\AgdaSpace{}%
\AgdaInductiveConstructor{bool}\<%
\\
\\[\AgdaEmptyExtraSkip]%
\>[6]\AgdaComment{--\ Closures}\<%
\\
\>[6]\AgdaOperator{\AgdaInductiveConstructor{⟨ƛ\AgdaUnderscore{}⟩\AgdaUnderscore{}}}\AgdaSpace{}%
\AgdaSymbol{:}\AgdaSpace{}%
\AgdaGeneralizable{Γ}\AgdaSpace{}%
\AgdaOperator{\AgdaInductiveConstructor{·:}}\AgdaSpace{}%
\AgdaGeneralizable{S}\AgdaSpace{}%
\AgdaOperator{\AgdaDatatype{⊢}}\AgdaSpace{}%
\AgdaGeneralizable{T}\AgdaSpace{}%
\AgdaSymbol{→}\AgdaSpace{}%
\AgdaFunction{Env}\AgdaSpace{}%
\AgdaGeneralizable{Γ}\AgdaSpace{}%
\AgdaSymbol{→}\AgdaSpace{}%
\AgdaDatatype{Domain}\AgdaSpace{}%
\AgdaSymbol{(}\AgdaGeneralizable{S}\AgdaSpace{}%
\AgdaOperator{\AgdaInductiveConstructor{⇒}}\AgdaSpace{}%
\AgdaGeneralizable{T}\AgdaSymbol{)}\<%
\\
\\[\AgdaEmptyExtraSkip]%
\>[2]\AgdaKeyword{variable}\AgdaSpace{}%
\AgdaGeneralizable{γ}\AgdaSpace{}%
\AgdaGeneralizable{δ}\AgdaSpace{}%
\AgdaSymbol{:}\AgdaSpace{}%
\AgdaFunction{Env}\AgdaSpace{}%
\AgdaGeneralizable{Γ}\<%
\\
\>[2]\AgdaKeyword{variable}\AgdaSpace{}%
\AgdaGeneralizable{a}\AgdaSpace{}%
\AgdaGeneralizable{b}\AgdaSpace{}%
\AgdaGeneralizable{d}\AgdaSpace{}%
\AgdaSymbol{:}\AgdaSpace{}%
\AgdaDatatype{Domain}\AgdaSpace{}%
\AgdaGeneralizable{T}\<%
\\
\\[\AgdaEmptyExtraSkip]%
\>[2]\AgdaComment{--\ Extending\ an\ environment}\<%
\\
\>[2]\AgdaOperator{\AgdaFunction{\AgdaUnderscore{}++\AgdaUnderscore{}}}\AgdaSpace{}%
\AgdaSymbol{:}\AgdaSpace{}%
\AgdaFunction{Env}\AgdaSpace{}%
\AgdaGeneralizable{Γ}\AgdaSpace{}%
\AgdaSymbol{→}\AgdaSpace{}%
\AgdaDatatype{Domain}\AgdaSpace{}%
\AgdaGeneralizable{T}\AgdaSpace{}%
\AgdaSymbol{→}\AgdaSpace{}%
\AgdaFunction{Env}\AgdaSpace{}%
\AgdaSymbol{(}\AgdaGeneralizable{Γ}\AgdaSpace{}%
\AgdaOperator{\AgdaInductiveConstructor{·:}}\AgdaSpace{}%
\AgdaGeneralizable{T}\AgdaSymbol{)}\<%
\\
\>[2]\AgdaSymbol{(}\AgdaBound{γ}\AgdaSpace{}%
\AgdaOperator{\AgdaFunction{++}}\AgdaSpace{}%
\AgdaBound{a}\AgdaSymbol{)}\AgdaSpace{}%
\AgdaInductiveConstructor{zero}\AgdaSpace{}%
\AgdaSymbol{=}\AgdaSpace{}%
\AgdaBound{a}\<%
\\
\>[2]\AgdaSymbol{(}\AgdaBound{γ}\AgdaSpace{}%
\AgdaOperator{\AgdaFunction{++}}\AgdaSpace{}%
\AgdaBound{a}\AgdaSymbol{)}\AgdaSpace{}%
\AgdaSymbol{(}\AgdaInductiveConstructor{suc}\AgdaSpace{}%
\AgdaBound{x}\AgdaSymbol{)}\AgdaSpace{}%
\AgdaSymbol{=}\AgdaSpace{}%
\AgdaBound{γ}\AgdaSpace{}%
\AgdaBound{x}\<%
\\
\\[\AgdaEmptyExtraSkip]%
\>[2]\AgdaComment{--\ Evaluation\ of\ terms}\<%
\\
\>[2]\AgdaKeyword{data}\AgdaSpace{}%
\AgdaOperator{\AgdaDatatype{\AgdaUnderscore{}∣\AgdaUnderscore{}⇓\AgdaUnderscore{}}}\AgdaSpace{}%
\AgdaSymbol{:}\AgdaSpace{}%
\AgdaFunction{Env}\AgdaSpace{}%
\AgdaGeneralizable{Γ}\AgdaSpace{}%
\AgdaSymbol{→}\AgdaSpace{}%
\AgdaGeneralizable{Γ}\AgdaSpace{}%
\AgdaOperator{\AgdaDatatype{⊢}}\AgdaSpace{}%
\AgdaGeneralizable{T}\AgdaSpace{}%
\AgdaSymbol{→}\AgdaSpace{}%
\AgdaDatatype{Domain}\AgdaSpace{}%
\AgdaGeneralizable{T}\AgdaSpace{}%
\AgdaSymbol{→}\AgdaSpace{}%
\AgdaPrimitive{Set}\AgdaSpace{}%
\AgdaKeyword{where}\<%
\\
\>[2][@{}l@{\AgdaIndent{0}}]%
\>[4]\AgdaInductiveConstructor{evalTrue}\AgdaSpace{}%
\AgdaSymbol{:}\AgdaSpace{}%
\AgdaGeneralizable{γ}\AgdaSpace{}%
\AgdaOperator{\AgdaDatatype{∣}}\AgdaSpace{}%
\AgdaInductiveConstructor{true}\AgdaSpace{}%
\AgdaOperator{\AgdaDatatype{⇓}}\AgdaSpace{}%
\AgdaInductiveConstructor{true}\<%
\\
\>[4]\AgdaInductiveConstructor{evalFalse}\AgdaSpace{}%
\AgdaSymbol{:}\AgdaSpace{}%
\AgdaGeneralizable{γ}\AgdaSpace{}%
\AgdaOperator{\AgdaDatatype{∣}}\AgdaSpace{}%
\AgdaInductiveConstructor{false}\AgdaSpace{}%
\AgdaOperator{\AgdaDatatype{⇓}}\AgdaSpace{}%
\AgdaInductiveConstructor{false}\<%
\\
\>[4]\AgdaInductiveConstructor{evalVar}\AgdaSpace{}%
\AgdaSymbol{:}\AgdaSpace{}%
\AgdaGeneralizable{γ}\AgdaSpace{}%
\AgdaOperator{\AgdaDatatype{∣}}\AgdaSpace{}%
\AgdaInductiveConstructor{var}\AgdaSpace{}%
\AgdaGeneralizable{x}\AgdaSpace{}%
\AgdaOperator{\AgdaDatatype{⇓}}\AgdaSpace{}%
\AgdaGeneralizable{γ}\AgdaSpace{}%
\AgdaGeneralizable{x}\<%
\\
\>[4]\AgdaInductiveConstructor{evalAbs}\AgdaSpace{}%
\AgdaSymbol{:}\AgdaSpace{}%
\AgdaGeneralizable{γ}\AgdaSpace{}%
\AgdaOperator{\AgdaDatatype{∣}}\AgdaSpace{}%
\AgdaOperator{\AgdaInductiveConstructor{ƛ}}\AgdaSpace{}%
\AgdaGeneralizable{t}\AgdaSpace{}%
\AgdaOperator{\AgdaDatatype{⇓}}\AgdaSpace{}%
\AgdaOperator{\AgdaInductiveConstructor{⟨ƛ}}\AgdaSpace{}%
\AgdaGeneralizable{t}\AgdaSpace{}%
\AgdaOperator{\AgdaInductiveConstructor{⟩}}\AgdaSpace{}%
\AgdaGeneralizable{γ}\<%
\\
\>[4]\AgdaInductiveConstructor{evalApp}\AgdaSpace{}%
\AgdaSymbol{:}\<%
\\
\>[4][@{}l@{\AgdaIndent{0}}]%
\>[8]\AgdaGeneralizable{γ}\AgdaSpace{}%
\AgdaOperator{\AgdaDatatype{∣}}\AgdaSpace{}%
\AgdaGeneralizable{r}\AgdaSpace{}%
\AgdaOperator{\AgdaDatatype{⇓}}\AgdaSpace{}%
\AgdaOperator{\AgdaInductiveConstructor{⟨ƛ}}\AgdaSpace{}%
\AgdaGeneralizable{t}\AgdaSpace{}%
\AgdaOperator{\AgdaInductiveConstructor{⟩}}\AgdaSpace{}%
\AgdaGeneralizable{δ}\<%
\\
\>[4][@{}l@{\AgdaIndent{0}}]%
\>[6]\AgdaSymbol{→}\AgdaSpace{}%
\AgdaGeneralizable{γ}\AgdaSpace{}%
\AgdaOperator{\AgdaDatatype{∣}}\AgdaSpace{}%
\AgdaGeneralizable{s}\AgdaSpace{}%
\AgdaOperator{\AgdaDatatype{⇓}}\AgdaSpace{}%
\AgdaGeneralizable{a}\<%
\\
\>[6]\AgdaSymbol{→}\AgdaSpace{}%
\AgdaGeneralizable{δ}\AgdaSpace{}%
\AgdaOperator{\AgdaFunction{++}}\AgdaSpace{}%
\AgdaGeneralizable{a}\AgdaSpace{}%
\AgdaOperator{\AgdaDatatype{∣}}\AgdaSpace{}%
\AgdaGeneralizable{t}\AgdaSpace{}%
\AgdaOperator{\AgdaDatatype{⇓}}\AgdaSpace{}%
\AgdaGeneralizable{b}\<%
\\
\>[6]\AgdaSymbol{→}\AgdaSpace{}%
\AgdaGeneralizable{γ}\AgdaSpace{}%
\AgdaOperator{\AgdaDatatype{∣}}\AgdaSpace{}%
\AgdaGeneralizable{r}\AgdaSpace{}%
\AgdaOperator{\AgdaInductiveConstructor{·}}\AgdaSpace{}%
\AgdaGeneralizable{s}\AgdaSpace{}%
\AgdaOperator{\AgdaDatatype{⇓}}\AgdaSpace{}%
\AgdaGeneralizable{b}\<%
\\
\>[4]\AgdaInductiveConstructor{evalIfTrue}\AgdaSpace{}%
\AgdaSymbol{:}\<%
\\
\>[4][@{}l@{\AgdaIndent{0}}]%
\>[8]\AgdaGeneralizable{γ}\AgdaSpace{}%
\AgdaOperator{\AgdaDatatype{∣}}\AgdaSpace{}%
\AgdaGeneralizable{r}\AgdaSpace{}%
\AgdaOperator{\AgdaDatatype{⇓}}\AgdaSpace{}%
\AgdaInductiveConstructor{true}\<%
\\
\>[4][@{}l@{\AgdaIndent{0}}]%
\>[6]\AgdaSymbol{→}\AgdaSpace{}%
\AgdaGeneralizable{γ}\AgdaSpace{}%
\AgdaOperator{\AgdaDatatype{∣}}\AgdaSpace{}%
\AgdaGeneralizable{s}\AgdaSpace{}%
\AgdaOperator{\AgdaDatatype{⇓}}\AgdaSpace{}%
\AgdaGeneralizable{a}\<%
\\
\>[6]\AgdaSymbol{→}\AgdaSpace{}%
\AgdaGeneralizable{γ}\AgdaSpace{}%
\AgdaOperator{\AgdaDatatype{∣}}\AgdaSpace{}%
\AgdaOperator{\AgdaInductiveConstructor{if}}\AgdaSpace{}%
\AgdaGeneralizable{r}\AgdaSpace{}%
\AgdaOperator{\AgdaInductiveConstructor{then}}\AgdaSpace{}%
\AgdaGeneralizable{s}\AgdaSpace{}%
\AgdaOperator{\AgdaInductiveConstructor{else}}\AgdaSpace{}%
\AgdaGeneralizable{t}\AgdaSpace{}%
\AgdaOperator{\AgdaDatatype{⇓}}\AgdaSpace{}%
\AgdaGeneralizable{a}\<%
\\
\>[4]\AgdaInductiveConstructor{evalIfFalse}\AgdaSpace{}%
\AgdaSymbol{:}\<%
\\
\>[4][@{}l@{\AgdaIndent{0}}]%
\>[8]\AgdaGeneralizable{γ}\AgdaSpace{}%
\AgdaOperator{\AgdaDatatype{∣}}\AgdaSpace{}%
\AgdaGeneralizable{r}\AgdaSpace{}%
\AgdaOperator{\AgdaDatatype{⇓}}\AgdaSpace{}%
\AgdaInductiveConstructor{false}\<%
\\
\>[4][@{}l@{\AgdaIndent{0}}]%
\>[6]\AgdaSymbol{→}\AgdaSpace{}%
\AgdaGeneralizable{γ}\AgdaSpace{}%
\AgdaOperator{\AgdaDatatype{∣}}\AgdaSpace{}%
\AgdaGeneralizable{t}\AgdaSpace{}%
\AgdaOperator{\AgdaDatatype{⇓}}\AgdaSpace{}%
\AgdaGeneralizable{b}\<%
\\
\>[6]\AgdaSymbol{→}\AgdaSpace{}%
\AgdaGeneralizable{γ}\AgdaSpace{}%
\AgdaOperator{\AgdaDatatype{∣}}\AgdaSpace{}%
\AgdaOperator{\AgdaInductiveConstructor{if}}\AgdaSpace{}%
\AgdaGeneralizable{r}\AgdaSpace{}%
\AgdaOperator{\AgdaInductiveConstructor{then}}\AgdaSpace{}%
\AgdaGeneralizable{s}\AgdaSpace{}%
\AgdaOperator{\AgdaInductiveConstructor{else}}\AgdaSpace{}%
\AgdaGeneralizable{t}\AgdaSpace{}%
\AgdaOperator{\AgdaDatatype{⇓}}\AgdaSpace{}%
\AgdaGeneralizable{b}\<%
\end{code}
\caption{Natural semantics}
\label{fig:semantics}
\end{figure}

We model the evaluation of a term as a relation and not a function because we
cannot write partial functions in Agda, and we do not know if evaluation is
total -- this is the property we want to prove!\footnote{As we are
using an intrinsically-typed representation, Agda can check that such a function
is in fact total, though we do not take advantage of this as we want to present
a proof independent of our use of intrinsic typing.} If we were writing these
semantics as a function in a language such as Haskell, however, they would be
comparable to an interpreter that one might write for the simply typed lambda
calculus (and in fact, the denotational semantics we discuss in
$\S$\ref{sec:correctness} are just that).

Environments are defined mutually with a domain of evaluation that makes up the
set of fully evaluated terms. For our minimal STLC, the only elements in this
domain are \mbox{\AgdaInductiveConstructor{true}}, \mbox{\AgdaInductiveConstructor{false}},
and closures (\mbox{\AgdaInductiveConstructor{⟨ƛ} \AgdaBound{t}
\AgdaInductiveConstructor{⟩} \AgdaBound{δ}}), i.e. an abstraction paired with
a saved environment \cite{landin1964mechanical}.\footnote{We take advantage of
Agda's support for overloading inductive constructors here, so that we can reuse
\mbox{\AgdaInductiveConstructor{false}} and \mbox{\AgdaInductiveConstructor{true}}
instead of using new names.} Both environments and domain elements are well-typed
by construction, an environment \AgdaBound{γ} is typed according to a context
\AgdaBound{Γ} and can only map variables that are present in \AgdaBound{Γ} to
domain elements that are themselves well-typed. Domain elements are typed
independently of contexts, as we close terms with environments.

Definitionally, environments have no real difference from substitutions, with their
distinguishing factor being how they are used. Instead of applying a substitution
to a term to reduce it, we perform a sort of delayed substitution where we
substitute a variable in a term as we evaluate the term itself.

We evaluate terms to a separate domain instead of reducing the term to a value as we
might do in a semantics described using substitutions. This is because in
evaluating an abstraction (\mbox{\AgdaInductiveConstructor{evalAbs}}), we wish
to save the environment that the abstraction is being evaluated under to form a
closure. This allows us to use the environment later on to continue evaluating
the body of the abstraction. We do so in the case of application
(\mbox{\AgdaInductiveConstructor{evalApp}}), where we evaluate the term being
applied to a closure and then extend its environment to continue evaluating its
body. Note that as we evaluate the term being applied to extend the closure's
environment, these are semantics call-by-value.

\subsection{Proof by logical relation}
We now turn to the property we wish to prove, that the execution of a program is
well-defined. With our semantics, this is equivalent to proving that evaluation
of STLC is total. We consider the evaluation of closed terms only as would be
the case with weak head reduction. We do not evaluate inside the bodies of
abstractions, therefore we do not evaluate open terms.

\begin{code}%
\>[2]\AgdaFunction{empty}\AgdaSpace{}%
\AgdaSymbol{:}\AgdaSpace{}%
\AgdaFunction{Env}\AgdaSpace{}%
\AgdaInductiveConstructor{∅}\<%
\\
\>[2]\AgdaFunction{empty}\AgdaSpace{}%
\AgdaSymbol{()}\<%
\\
\\[\AgdaEmptyExtraSkip]%
\>[2]\AgdaFunction{⇓-well-defined}\AgdaSpace{}%
\AgdaSymbol{:}\AgdaSpace{}%
\AgdaInductiveConstructor{∅}\AgdaSpace{}%
\AgdaOperator{\AgdaDatatype{⊢}}\AgdaSpace{}%
\AgdaGeneralizable{T}\AgdaSpace{}%
\AgdaSymbol{→}\AgdaSpace{}%
\AgdaPrimitive{Set}\<%
\\
\>[2]\AgdaFunction{⇓-well-defined}\AgdaSpace{}%
\AgdaBound{t}\AgdaSpace{}%
\AgdaSymbol{=}\AgdaSpace{}%
\AgdaFunction{∃[}\AgdaSpace{}%
\AgdaBound{a}\AgdaSpace{}%
\AgdaFunction{]}\AgdaSpace{}%
\AgdaFunction{empty}\AgdaSpace{}%
\AgdaOperator{\AgdaDatatype{∣}}\AgdaSpace{}%
\AgdaBound{t}\AgdaSpace{}%
\AgdaOperator{\AgdaDatatype{⇓}}\AgdaSpace{}%
\AgdaBound{a}\<%
\end{code}

\begin{figure}
\begin{code}%
\>[2]\AgdaComment{--\ Semantic\ types\ (logical\ predicate)}\<%
\\
\>[2]\AgdaOperator{\AgdaFunction{⟦\AgdaUnderscore{}⟧}}\AgdaSpace{}%
\AgdaSymbol{:}\AgdaSpace{}%
\AgdaSymbol{∀}\AgdaSpace{}%
\AgdaSymbol{(}\AgdaBound{T}\AgdaSpace{}%
\AgdaSymbol{:}\AgdaSpace{}%
\AgdaDatatype{Type}\AgdaSymbol{)}\AgdaSpace{}%
\AgdaSymbol{→}\AgdaSpace{}%
\AgdaSymbol{(}\AgdaDatatype{Domain}\AgdaSpace{}%
\AgdaBound{T}\AgdaSpace{}%
\AgdaSymbol{→}\AgdaSpace{}%
\AgdaPrimitive{Set}\AgdaSymbol{)}\<%
\\
\>[2]\AgdaOperator{\AgdaFunction{⟦}}\AgdaSpace{}%
\AgdaInductiveConstructor{bool}\AgdaSpace{}%
\AgdaOperator{\AgdaFunction{⟧}}\AgdaSpace{}%
\AgdaSymbol{\AgdaUnderscore{}}\AgdaSpace{}%
\AgdaSymbol{=}\AgdaSpace{}%
\AgdaRecord{⊤}\<%
\\
\>[2]\AgdaOperator{\AgdaFunction{⟦}}%
\>[458I]\AgdaBound{S}\AgdaSpace{}%
\AgdaOperator{\AgdaInductiveConstructor{⇒}}\AgdaSpace{}%
\AgdaBound{T}\AgdaSpace{}%
\AgdaOperator{\AgdaFunction{⟧}}\AgdaSpace{}%
\AgdaSymbol{(}\AgdaOperator{\AgdaInductiveConstructor{⟨ƛ}}\AgdaSpace{}%
\AgdaBound{t}\AgdaSpace{}%
\AgdaOperator{\AgdaInductiveConstructor{⟩}}\AgdaSpace{}%
\AgdaBound{δ}\AgdaSymbol{)}\AgdaSpace{}%
\AgdaSymbol{=}\<%
\\
\>[.][@{}l@{}]\<[458I]%
\>[4]\AgdaSymbol{∀}\AgdaSpace{}%
\AgdaSymbol{\{}\AgdaBound{a}\AgdaSymbol{\}}\AgdaSpace{}%
\AgdaSymbol{→}\AgdaSpace{}%
\AgdaBound{a}\AgdaSpace{}%
\AgdaOperator{\AgdaFunction{∈}}\AgdaSpace{}%
\AgdaOperator{\AgdaFunction{⟦}}\AgdaSpace{}%
\AgdaBound{S}\AgdaSpace{}%
\AgdaOperator{\AgdaFunction{⟧}}\<%
\\
\>[4]\AgdaSymbol{→}\AgdaSpace{}%
\AgdaFunction{∃[}\AgdaSpace{}%
\AgdaBound{b}\AgdaSpace{}%
\AgdaFunction{]}\AgdaSpace{}%
\AgdaBound{δ}\AgdaSpace{}%
\AgdaOperator{\AgdaFunction{++}}\AgdaSpace{}%
\AgdaBound{a}\AgdaSpace{}%
\AgdaOperator{\AgdaDatatype{∣}}\AgdaSpace{}%
\AgdaBound{t}\AgdaSpace{}%
\AgdaOperator{\AgdaDatatype{⇓}}\AgdaSpace{}%
\AgdaBound{b}\AgdaSpace{}%
\AgdaOperator{\AgdaFunction{×}}\AgdaSpace{}%
\AgdaBound{b}\AgdaSpace{}%
\AgdaOperator{\AgdaFunction{∈}}\AgdaSpace{}%
\AgdaOperator{\AgdaFunction{⟦}}\AgdaSpace{}%
\AgdaBound{T}\AgdaSpace{}%
\AgdaOperator{\AgdaFunction{⟧}}\<%
\\
\\[\AgdaEmptyExtraSkip]%
\>[2]\AgdaComment{--\ Semantic\ typing\ for\ environments}\<%
\\
\>[2]\AgdaOperator{\AgdaFunction{\AgdaUnderscore{}⊨\AgdaUnderscore{}}}\AgdaSpace{}%
\AgdaSymbol{:}\AgdaSpace{}%
\AgdaSymbol{(}\AgdaBound{Γ}\AgdaSpace{}%
\AgdaSymbol{:}\AgdaSpace{}%
\AgdaDatatype{Ctx}\AgdaSymbol{)}\AgdaSpace{}%
\AgdaSymbol{→}\AgdaSpace{}%
\AgdaFunction{Env}\AgdaSpace{}%
\AgdaBound{Γ}\AgdaSpace{}%
\AgdaSymbol{→}\AgdaSpace{}%
\AgdaPrimitive{Set}\<%
\\
\>[2]\AgdaBound{Γ}\AgdaSpace{}%
\AgdaOperator{\AgdaFunction{⊨}}\AgdaSpace{}%
\AgdaBound{γ}\AgdaSpace{}%
\AgdaSymbol{=}\AgdaSpace{}%
\AgdaSymbol{∀}\AgdaSpace{}%
\AgdaSymbol{\{}\AgdaBound{T}\AgdaSymbol{\}}\AgdaSpace{}%
\AgdaSymbol{→}\AgdaSpace{}%
\AgdaSymbol{(}\AgdaBound{x}\AgdaSpace{}%
\AgdaSymbol{:}\AgdaSpace{}%
\AgdaBound{Γ}\AgdaSpace{}%
\AgdaOperator{\AgdaDatatype{∋}}\AgdaSpace{}%
\AgdaBound{T}\AgdaSymbol{)}\AgdaSpace{}%
\AgdaSymbol{→}\AgdaSpace{}%
\AgdaBound{γ}\AgdaSpace{}%
\AgdaBound{x}\AgdaSpace{}%
\AgdaOperator{\AgdaFunction{∈}}\AgdaSpace{}%
\AgdaOperator{\AgdaFunction{⟦}}\AgdaSpace{}%
\AgdaBound{T}\AgdaSpace{}%
\AgdaOperator{\AgdaFunction{⟧}}\<%
\\
\\[\AgdaEmptyExtraSkip]%
\>[2]\AgdaComment{--\ Extending\ semantically\ typed\ environments}\<%
\\
\>[2]\AgdaOperator{\AgdaFunction{\AgdaUnderscore{}\textasciicircum{}\AgdaUnderscore{}}}\AgdaSpace{}%
\AgdaSymbol{:}\AgdaSpace{}%
\AgdaGeneralizable{Γ}\AgdaSpace{}%
\AgdaOperator{\AgdaFunction{⊨}}\AgdaSpace{}%
\AgdaGeneralizable{γ}\AgdaSpace{}%
\AgdaSymbol{→}\AgdaSpace{}%
\AgdaGeneralizable{a}\AgdaSpace{}%
\AgdaOperator{\AgdaFunction{∈}}\AgdaSpace{}%
\AgdaOperator{\AgdaFunction{⟦}}\AgdaSpace{}%
\AgdaGeneralizable{T}\AgdaSpace{}%
\AgdaOperator{\AgdaFunction{⟧}}\AgdaSpace{}%
\AgdaSymbol{→}\AgdaSpace{}%
\AgdaGeneralizable{Γ}\AgdaSpace{}%
\AgdaOperator{\AgdaInductiveConstructor{·:}}\AgdaSpace{}%
\AgdaGeneralizable{T}\AgdaSpace{}%
\AgdaOperator{\AgdaFunction{⊨}}\AgdaSpace{}%
\AgdaGeneralizable{γ}\AgdaSpace{}%
\AgdaOperator{\AgdaFunction{++}}\AgdaSpace{}%
\AgdaGeneralizable{a}\<%
\\
\>[2]\AgdaSymbol{(}\AgdaBound{⊨γ}\AgdaSpace{}%
\AgdaOperator{\AgdaFunction{\textasciicircum{}}}\AgdaSpace{}%
\AgdaBound{sa}\AgdaSymbol{)}\AgdaSpace{}%
\AgdaInductiveConstructor{zero}\AgdaSpace{}%
\AgdaSymbol{=}\AgdaSpace{}%
\AgdaBound{sa}\<%
\\
\>[2]\AgdaSymbol{(}\AgdaBound{⊨γ}\AgdaSpace{}%
\AgdaOperator{\AgdaFunction{\textasciicircum{}}}\AgdaSpace{}%
\AgdaBound{sa}\AgdaSymbol{)}\AgdaSpace{}%
\AgdaSymbol{(}\AgdaInductiveConstructor{suc}\AgdaSpace{}%
\AgdaBound{x}\AgdaSymbol{)}\AgdaSpace{}%
\AgdaSymbol{=}\AgdaSpace{}%
\AgdaBound{⊨γ}\AgdaSpace{}%
\AgdaBound{x}\<%
\\
\\[\AgdaEmptyExtraSkip]%
\>[2]\AgdaComment{--\ Semantic\ typing\ for\ terms}\<%
\\
\>[2]\AgdaFunction{semantic-typing}\AgdaSpace{}%
\AgdaSymbol{:}\AgdaSpace{}%
\AgdaGeneralizable{Γ}\AgdaSpace{}%
\AgdaOperator{\AgdaDatatype{⊢}}\AgdaSpace{}%
\AgdaGeneralizable{T}\AgdaSpace{}%
\AgdaSymbol{→}\AgdaSpace{}%
\AgdaPrimitive{Set}\<%
\\
\>[2]\AgdaFunction{semantic-typing}\AgdaSpace{}%
\AgdaSymbol{\{}\AgdaBound{Γ}\AgdaSymbol{\}}\AgdaSpace{}%
\AgdaSymbol{\{}\AgdaBound{T}\AgdaSymbol{\}}\AgdaSpace{}%
\AgdaBound{t}\AgdaSpace{}%
\AgdaSymbol{=}\<%
\\
\>[2][@{}l@{\AgdaIndent{0}}]%
\>[4]\AgdaSymbol{∀}\AgdaSpace{}%
\AgdaSymbol{\{}\AgdaBound{γ}\AgdaSpace{}%
\AgdaSymbol{:}\AgdaSpace{}%
\AgdaFunction{Env}\AgdaSpace{}%
\AgdaBound{Γ}\AgdaSymbol{\}}\AgdaSpace{}%
\AgdaSymbol{→}\AgdaSpace{}%
\AgdaBound{Γ}\AgdaSpace{}%
\AgdaOperator{\AgdaFunction{⊨}}\AgdaSpace{}%
\AgdaBound{γ}\AgdaSpace{}%
\AgdaSymbol{→}\AgdaSpace{}%
\AgdaFunction{∃[}\AgdaSpace{}%
\AgdaBound{a}\AgdaSpace{}%
\AgdaFunction{]}\AgdaSpace{}%
\AgdaBound{γ}\AgdaSpace{}%
\AgdaOperator{\AgdaDatatype{∣}}\AgdaSpace{}%
\AgdaBound{t}\AgdaSpace{}%
\AgdaOperator{\AgdaDatatype{⇓}}\AgdaSpace{}%
\AgdaBound{a}\AgdaSpace{}%
\AgdaOperator{\AgdaFunction{×}}\AgdaSpace{}%
\AgdaBound{a}\AgdaSpace{}%
\AgdaOperator{\AgdaFunction{∈}}\AgdaSpace{}%
\AgdaOperator{\AgdaFunction{⟦}}\AgdaSpace{}%
\AgdaBound{T}\AgdaSpace{}%
\AgdaOperator{\AgdaFunction{⟧}}\<%
\\
\\[\AgdaEmptyExtraSkip]%
\>[2]\AgdaKeyword{syntax}\AgdaSpace{}%
\AgdaFunction{semantic-typing}\AgdaSpace{}%
\AgdaSymbol{\{}\AgdaBound{Γ}\AgdaSymbol{\}}\AgdaSpace{}%
\AgdaSymbol{\{}\AgdaBound{T}\AgdaSymbol{\}}\AgdaSpace{}%
\AgdaBound{t}\AgdaSpace{}%
\AgdaSymbol{=}\AgdaSpace{}%
\AgdaBound{Γ}\AgdaSpace{}%
\AgdaFunction{⊨}\AgdaSpace{}%
\AgdaBound{t}\AgdaSpace{}%
\AgdaFunction{∷}\AgdaSpace{}%
\AgdaBound{T}\<%
\\
\\[\AgdaEmptyExtraSkip]%
\>[2]\AgdaComment{--\ Syntactic\ typing\ implies\ semantic\ typing}\<%
\\
\>[2]\AgdaFunction{fundamental-lemma}\AgdaSpace{}%
\AgdaSymbol{:}\AgdaSpace{}%
\AgdaSymbol{∀}\AgdaSpace{}%
\AgdaSymbol{(}\AgdaBound{t}\AgdaSpace{}%
\AgdaSymbol{:}\AgdaSpace{}%
\AgdaGeneralizable{Γ}\AgdaSpace{}%
\AgdaOperator{\AgdaDatatype{⊢}}\AgdaSpace{}%
\AgdaGeneralizable{T}\AgdaSymbol{)}\AgdaSpace{}%
\AgdaSymbol{→}\AgdaSpace{}%
\AgdaGeneralizable{Γ}\AgdaSpace{}%
\AgdaFunction{⊨}\AgdaSpace{}%
\AgdaBound{t}\AgdaSpace{}%
\AgdaFunction{∷}\AgdaSpace{}%
\AgdaGeneralizable{T}\<%
\\
\>[2]\AgdaFunction{fundamental-lemma}\AgdaSpace{}%
\AgdaInductiveConstructor{true}\AgdaSpace{}%
\AgdaSymbol{\AgdaUnderscore{}}%
\>[28]\AgdaSymbol{=}\AgdaSpace{}%
\AgdaInductiveConstructor{true}\AgdaSpace{}%
\AgdaOperator{\AgdaInductiveConstructor{,}}\AgdaSpace{}%
\AgdaInductiveConstructor{evalTrue}\AgdaSpace{}%
\AgdaOperator{\AgdaInductiveConstructor{,}}\AgdaSpace{}%
\AgdaInductiveConstructor{tt}\<%
\\
\>[2]\AgdaFunction{fundamental-lemma}\AgdaSpace{}%
\AgdaInductiveConstructor{false}\AgdaSpace{}%
\AgdaSymbol{\AgdaUnderscore{}}\AgdaSpace{}%
\AgdaSymbol{=}\AgdaSpace{}%
\AgdaInductiveConstructor{false}\AgdaSpace{}%
\AgdaOperator{\AgdaInductiveConstructor{,}}\AgdaSpace{}%
\AgdaInductiveConstructor{evalFalse}\AgdaSpace{}%
\AgdaOperator{\AgdaInductiveConstructor{,}}\AgdaSpace{}%
\AgdaInductiveConstructor{tt}\<%
\\
\>[2]\AgdaFunction{fundamental-lemma}\AgdaSpace{}%
\AgdaSymbol{(}\AgdaInductiveConstructor{var}\AgdaSpace{}%
\AgdaBound{x}\AgdaSymbol{)}\AgdaSpace{}%
\AgdaSymbol{\{}\AgdaBound{γ}\AgdaSymbol{\}}\AgdaSpace{}%
\AgdaBound{⊨γ}\AgdaSpace{}%
\AgdaSymbol{=}\<%
\\
\>[2][@{}l@{\AgdaIndent{0}}]%
\>[4]\AgdaBound{γ}\AgdaSpace{}%
\AgdaBound{x}\AgdaSpace{}%
\AgdaOperator{\AgdaInductiveConstructor{,}}\AgdaSpace{}%
\AgdaInductiveConstructor{evalVar}\AgdaSpace{}%
\AgdaOperator{\AgdaInductiveConstructor{,}}\AgdaSpace{}%
\AgdaBound{⊨γ}\AgdaSpace{}%
\AgdaBound{x}\<%
\\
\>[2]\AgdaFunction{fundamental-lemma}\AgdaSpace{}%
\AgdaSymbol{(}\AgdaOperator{\AgdaInductiveConstructor{ƛ}}\AgdaSpace{}%
\AgdaBound{t}\AgdaSymbol{)}\AgdaSpace{}%
\AgdaSymbol{\{}\AgdaBound{γ}\AgdaSymbol{\}}\AgdaSpace{}%
\AgdaBound{⊨γ}\AgdaSpace{}%
\AgdaSymbol{=}\<%
\\
\>[2][@{}l@{\AgdaIndent{0}}]%
\>[4]\AgdaOperator{\AgdaInductiveConstructor{⟨ƛ}}\AgdaSpace{}%
\AgdaBound{t}\AgdaSpace{}%
\AgdaOperator{\AgdaInductiveConstructor{⟩}}\AgdaSpace{}%
\AgdaBound{γ}\AgdaSpace{}%
\AgdaOperator{\AgdaInductiveConstructor{,}}\AgdaSpace{}%
\AgdaInductiveConstructor{evalAbs}\AgdaSpace{}%
\AgdaOperator{\AgdaInductiveConstructor{,}}\AgdaSpace{}%
\AgdaSymbol{λ}\AgdaSpace{}%
\AgdaBound{sa}\AgdaSpace{}%
\AgdaSymbol{→}\AgdaSpace{}%
\AgdaFunction{fundamental-lemma}\AgdaSpace{}%
\AgdaBound{t}\AgdaSpace{}%
\AgdaSymbol{(}\AgdaBound{⊨γ}\AgdaSpace{}%
\AgdaOperator{\AgdaFunction{\textasciicircum{}}}\AgdaSpace{}%
\AgdaBound{sa}\AgdaSymbol{)}\<%
\\
\>[2]\AgdaFunction{fundamental-lemma}\AgdaSpace{}%
\AgdaSymbol{(}\AgdaBound{r}\AgdaSpace{}%
\AgdaOperator{\AgdaInductiveConstructor{·}}\AgdaSpace{}%
\AgdaBound{s}\AgdaSymbol{)}\AgdaSpace{}%
\AgdaBound{⊨γ}\<%
\\
\>[2][@{}l@{\AgdaIndent{0}}]%
\>[4]\AgdaKeyword{with}\AgdaSpace{}%
\AgdaFunction{fundamental-lemma}\AgdaSpace{}%
\AgdaBound{r}\AgdaSpace{}%
\AgdaBound{⊨γ}\<%
\\
\>[2]\AgdaSymbol{...}\AgdaSpace{}%
\AgdaSymbol{|}\AgdaSpace{}%
\AgdaOperator{\AgdaInductiveConstructor{⟨ƛ}}\AgdaSpace{}%
\AgdaBound{t}\AgdaSpace{}%
\AgdaOperator{\AgdaInductiveConstructor{⟩}}\AgdaSpace{}%
\AgdaBound{δ}\AgdaSpace{}%
\AgdaOperator{\AgdaInductiveConstructor{,}}\AgdaSpace{}%
\AgdaBound{r⇓}\AgdaSpace{}%
\AgdaOperator{\AgdaInductiveConstructor{,}}\AgdaSpace{}%
\AgdaBound{sf}\AgdaSpace{}%
\AgdaSymbol{=}\<%
\\
\>[2][@{}l@{\AgdaIndent{0}}]%
\>[4]\AgdaKeyword{let}\AgdaSpace{}%
\AgdaSymbol{(}\AgdaBound{a}\AgdaSpace{}%
\AgdaOperator{\AgdaInductiveConstructor{,}}\AgdaSpace{}%
\AgdaBound{s⇓}\AgdaSpace{}%
\AgdaOperator{\AgdaInductiveConstructor{,}}\AgdaSpace{}%
\AgdaBound{sa}\AgdaSymbol{)}\AgdaSpace{}%
\AgdaSymbol{=}\AgdaSpace{}%
\AgdaFunction{fundamental-lemma}\AgdaSpace{}%
\AgdaBound{s}\AgdaSpace{}%
\AgdaBound{⊨γ}\AgdaSpace{}%
\AgdaKeyword{in}\<%
\\
\>[4]\AgdaKeyword{let}\AgdaSpace{}%
\AgdaSymbol{(}\AgdaBound{b}\AgdaSpace{}%
\AgdaOperator{\AgdaInductiveConstructor{,}}\AgdaSpace{}%
\AgdaBound{eval-closure}\AgdaSpace{}%
\AgdaOperator{\AgdaInductiveConstructor{,}}\AgdaSpace{}%
\AgdaBound{sb}\AgdaSymbol{)}\AgdaSpace{}%
\AgdaSymbol{=}\AgdaSpace{}%
\AgdaBound{sf}\AgdaSpace{}%
\AgdaBound{sa}\AgdaSpace{}%
\AgdaKeyword{in}\<%
\\
\>[4]\AgdaBound{b}\AgdaSpace{}%
\AgdaOperator{\AgdaInductiveConstructor{,}}\AgdaSpace{}%
\AgdaInductiveConstructor{evalApp}\AgdaSpace{}%
\AgdaBound{r⇓}\AgdaSpace{}%
\AgdaBound{s⇓}\AgdaSpace{}%
\AgdaBound{eval-closure}\AgdaSpace{}%
\AgdaOperator{\AgdaInductiveConstructor{,}}\AgdaSpace{}%
\AgdaBound{sb}\<%
\\
\>[2]\AgdaFunction{fundamental-lemma}\AgdaSpace{}%
\AgdaSymbol{(}\AgdaOperator{\AgdaInductiveConstructor{if}}\AgdaSpace{}%
\AgdaBound{r}\AgdaSpace{}%
\AgdaOperator{\AgdaInductiveConstructor{then}}\AgdaSpace{}%
\AgdaBound{s}\AgdaSpace{}%
\AgdaOperator{\AgdaInductiveConstructor{else}}\AgdaSpace{}%
\AgdaBound{t}\AgdaSymbol{)}\AgdaSpace{}%
\AgdaBound{⊨γ}\<%
\\
\>[2][@{}l@{\AgdaIndent{0}}]%
\>[4]\AgdaKeyword{with}\AgdaSpace{}%
\AgdaFunction{fundamental-lemma}\AgdaSpace{}%
\AgdaBound{r}\AgdaSpace{}%
\AgdaBound{⊨γ}\<%
\\
\>[2]\AgdaSymbol{...}\AgdaSpace{}%
\AgdaSymbol{|}\AgdaSpace{}%
\AgdaInductiveConstructor{true}\AgdaSpace{}%
\AgdaOperator{\AgdaInductiveConstructor{,}}\AgdaSpace{}%
\AgdaBound{r⇓}\AgdaSpace{}%
\AgdaOperator{\AgdaInductiveConstructor{,}}\AgdaSpace{}%
\AgdaSymbol{\AgdaUnderscore{}}\AgdaSpace{}%
\AgdaSymbol{=}\<%
\\
\>[2][@{}l@{\AgdaIndent{0}}]%
\>[4]\AgdaKeyword{let}\AgdaSpace{}%
\AgdaSymbol{(}\AgdaBound{a}\AgdaSpace{}%
\AgdaOperator{\AgdaInductiveConstructor{,}}\AgdaSpace{}%
\AgdaBound{s⇓}\AgdaSpace{}%
\AgdaOperator{\AgdaInductiveConstructor{,}}\AgdaSpace{}%
\AgdaBound{sa}\AgdaSymbol{)}\AgdaSpace{}%
\AgdaSymbol{=}\AgdaSpace{}%
\AgdaFunction{fundamental-lemma}\AgdaSpace{}%
\AgdaBound{s}\AgdaSpace{}%
\AgdaBound{⊨γ}\AgdaSpace{}%
\AgdaKeyword{in}\<%
\\
\>[4]\AgdaBound{a}\AgdaSpace{}%
\AgdaOperator{\AgdaInductiveConstructor{,}}\AgdaSpace{}%
\AgdaInductiveConstructor{evalIfTrue}\AgdaSpace{}%
\AgdaBound{r⇓}\AgdaSpace{}%
\AgdaBound{s⇓}\AgdaSpace{}%
\AgdaOperator{\AgdaInductiveConstructor{,}}\AgdaSpace{}%
\AgdaBound{sa}\<%
\\
\>[2]\AgdaSymbol{...}\AgdaSpace{}%
\AgdaSymbol{|}\AgdaSpace{}%
\AgdaInductiveConstructor{false}\AgdaSpace{}%
\AgdaOperator{\AgdaInductiveConstructor{,}}\AgdaSpace{}%
\AgdaBound{r⇓}\AgdaSpace{}%
\AgdaOperator{\AgdaInductiveConstructor{,}}\AgdaSpace{}%
\AgdaSymbol{\AgdaUnderscore{}}\AgdaSpace{}%
\AgdaSymbol{=}\<%
\\
\>[2][@{}l@{\AgdaIndent{0}}]%
\>[4]\AgdaKeyword{let}\AgdaSpace{}%
\AgdaSymbol{(}\AgdaBound{b}\AgdaSpace{}%
\AgdaOperator{\AgdaInductiveConstructor{,}}\AgdaSpace{}%
\AgdaBound{t⇓}\AgdaSpace{}%
\AgdaOperator{\AgdaInductiveConstructor{,}}\AgdaSpace{}%
\AgdaBound{sb}\AgdaSymbol{)}\AgdaSpace{}%
\AgdaSymbol{=}\AgdaSpace{}%
\AgdaFunction{fundamental-lemma}\AgdaSpace{}%
\AgdaBound{t}\AgdaSpace{}%
\AgdaBound{⊨γ}\AgdaSpace{}%
\AgdaKeyword{in}\<%
\\
\>[4]\AgdaBound{b}\AgdaSpace{}%
\AgdaOperator{\AgdaInductiveConstructor{,}}\AgdaSpace{}%
\AgdaInductiveConstructor{evalIfFalse}\AgdaSpace{}%
\AgdaBound{r⇓}\AgdaSpace{}%
\AgdaBound{t⇓}\AgdaSpace{}%
\AgdaOperator{\AgdaInductiveConstructor{,}}\AgdaSpace{}%
\AgdaBound{sb}\<%
\\
\\[\AgdaEmptyExtraSkip]%
\>[2]\AgdaComment{--\ Evaluation\ is\ total}\<%
\\
\>[2]\AgdaFunction{⇓-total}\AgdaSpace{}%
\AgdaSymbol{:}\AgdaSpace{}%
\AgdaSymbol{∀}\AgdaSpace{}%
\AgdaSymbol{(}\AgdaBound{t}\AgdaSpace{}%
\AgdaSymbol{:}\AgdaSpace{}%
\AgdaInductiveConstructor{∅}\AgdaSpace{}%
\AgdaOperator{\AgdaDatatype{⊢}}\AgdaSpace{}%
\AgdaGeneralizable{T}\AgdaSymbol{)}\AgdaSpace{}%
\AgdaSymbol{→}\AgdaSpace{}%
\AgdaFunction{⇓-well-defined}\AgdaSpace{}%
\AgdaBound{t}\<%
\\
\>[2]\AgdaFunction{⇓-total}\AgdaSpace{}%
\AgdaBound{t}\AgdaSpace{}%
\AgdaSymbol{=}\<%
\\
\>[2][@{}l@{\AgdaIndent{0}}]%
\>[4]\AgdaKeyword{let}\AgdaSpace{}%
\AgdaSymbol{(}\AgdaBound{a}\AgdaSpace{}%
\AgdaOperator{\AgdaInductiveConstructor{,}}\AgdaSpace{}%
\AgdaBound{t⇓a}\AgdaSpace{}%
\AgdaOperator{\AgdaInductiveConstructor{,}}\AgdaSpace{}%
\AgdaSymbol{\AgdaUnderscore{})}\AgdaSpace{}%
\AgdaSymbol{=}\AgdaSpace{}%
\AgdaFunction{fundamental-lemma}\AgdaSpace{}%
\AgdaBound{t}\AgdaSpace{}%
\AgdaSymbol{(λ}\AgdaSpace{}%
\AgdaSymbol{())}\AgdaSpace{}%
\AgdaKeyword{in}\<%
\\
\>[4]\AgdaBound{a}\AgdaSpace{}%
\AgdaOperator{\AgdaInductiveConstructor{,}}\AgdaSpace{}%
\AgdaBound{t⇓a}\<%
\end{code}
\caption{Full proof that evaluation of STLC is total}
\label{fig:totality-proof}
\end{figure}
\end{AgdaAlign}

We cannot prove that evaluation is well-defined for a well-typed term by direct
induction on the typing derivation because our induction hypotheses would not
be strong enough in the case of application \mbox{\AgdaBound{r}
\AgdaInductiveConstructor{·} \AgdaBound{s}}.

With our induction hypotheses, we would have that the evaluation is well-defined
for the term \AgdaBound{r} and that it evaluates to a closure
\mbox{\AgdaInductiveConstructor{⟨ƛ} \AgdaBound{t} \AgdaInductiveConstructor{⟩}
\AgdaBound{δ}}. We would additionally have that evaluation is well-defined for
the term \AgdaBound{s}, evaluating to some domain element \AgdaBound{a}.
However, we want to show that the evaluation of the closure's body
\AgdaBound{t} is itself well-defined when the environment \AgdaBound{δ} is
extended with \AgdaBound{a}, i.e. that there exists some \AgdaBound{b} such that
\mbox{\AgdaBound{δ} \AgdaFunction{++} \AgdaBound{a} \AgdaDatatype{∣}
\AgdaBound{t} \AgdaDatatype{⇓} \AgdaBound{b}}. This is something that is not
given to us by our induction hypothesis.

We turn instead to the use of a logical predicate on the domain of evaluation
(i.e. \mbox{\AgdaDatatype{Domain} \AgdaArgument{→} \AgdaDatatype{Set}}). For any
logical predicate \mbox{\AgdaBound{A}}, we use Agda's set membership notation
\mbox{\AgdaBound{a} \AgdaFunction{∈} \AgdaBound{A}} to refer to a domain element
\AgdaBound{a} that satisfies \AgdaBound{A}.

A logical predicate is a unary logical relation. A logical relation is always
defined recursively on types. At each type, the relation describes the
``logical'' interpretation of that type. In other words, we describe the
behavior that we would \emph{expect} a program to have at that type. In
particular, at function types we generally expect related arguments to give us
related results: if a function is given an argument that satisfies a logical
relation, the result should also satisfy the relation.

We follow this formula to define the logical predicate \mbox{\AgdaFunction{⟦}
\AgdaBound{T} \AgdaFunction{⟧}}, which serves as a ``semantic type'' describing
the well-defined evaluation of terms.

At our boolean type, we have that evaluation is well defined for all domain
elements (\mbox{\AgdaInductiveConstructor{true}} and
\mbox{\AgdaInductiveConstructor{false}}), so we use Agda's unit type
\AgdaDatatype{⊤}.

At our function type, we define the logical predicate on closures such that a
closure \mbox{\AgdaInductiveConstructor{⟨ƛ} \AgdaBound{t}
\AgdaInductiveConstructor{⟩} \AgdaBound{δ}} has a semantic function type
\mbox{\AgdaFunction{⟦} \AgdaBound{S} \AgdaInductiveConstructor{⇒} \AgdaBound{T}
\AgdaFunction{⟧}} if evaluation is well-defined for the body \AgdaBound{t} of
the closure when its environment \AgdaBound{δ} is extended with any arbitrary
\mbox{\AgdaBound{a} \AgdaFunction{∈} \AgdaFunction{⟦} \AgdaBound{S}
\AgdaFunction{⟧}} such that the evaluated body satisfies the logical predicate
described by \mbox{\AgdaFunction{⟦} \AgdaBound{T} \AgdaFunction{⟧}} (i.e.
\mbox{\AgdaBound{δ} \AgdaFunction{++} \AgdaBound{a} \AgdaDatatype{∣}
\AgdaBound{t} \AgdaDatatype{⇓} \AgdaBound{b}} and \mbox{\AgdaBound{b}
\AgdaFunction{∈} \AgdaFunction{⟦} \AgdaBound{T} \AgdaFunction{⟧}}).

This is one the key components mentioned earlier of a proof by logical
relations; where a semantic function type takes related arguments
(\mbox{\AgdaBound{a} \AgdaFunction{∈} \AgdaFunction{⟦} \AgdaBound{S}
\AgdaFunction{⟧}}) to related results (\mbox{\AgdaBound{b} \AgdaFunction{∈}
\AgdaFunction{⟦} \AgdaBound{T} \AgdaFunction{⟧}}). This structure of our logical
relation solves the issue we described in the case of application.

We use this logical predicate to prove a stronger property than just proving
that the evaluation of a term is well-defined, instead we prove that a
well-typed term satisfies the the semantic typing judgement \mbox{\AgdaBound{Γ}
\AgdaFunction{⊨} \AgdaBound{t} \AgdaFunction{∷} \AgdaBound{T}}. Given an
environment that is semantically typed according to the context \AgdaBound{Γ},
we expect the evaluation of the term \AgdaBound{t} in that environment to be
well-defined and for its evaluation \AgdaBound{a} to be semantically typed, i.e.
\mbox{\AgdaBound{a} \AgdaFunction{∈} \AgdaBound{A}}. Proving this property gives
us a strong enough induction hypothesis to prove that evaluation is total.

We define a semantically typed environment as an environment that is made up of
domain elements that are themselves semantically typed, notated \mbox{\AgdaBound{Γ}
\AgdaFunction{⊨} \AgdaBound{γ}}.

Having set up a logical predicate and a semantic typing judgement for terms, we
can prove the fundamental lemma of logical relations: if a term is well-typed,
it is also semantically typed. We prove this lemma by induction on our typing
derivation. The boolean constants \mbox{\AgdaInductiveConstructor{true}} and
\mbox{\AgdaInductiveConstructor{false}} trivially satisfy the logical predicate
described by the semantic boolean type. In the case of a variable, we use the
fact that the environment is semantically typed. In the case of application, we
use the semantic typing of the evaluated subterms to have an appropriately
strong induction hypothesis. For conditional branching, our induction hypothesis
gives us that either branch is semantically typed regardless of whether the
conditional evaluates to \mbox{\AgdaInductiveConstructor{true}} or
\mbox{\AgdaInductiveConstructor{false}}. In our abstraction case, we show that
the closure an abstraction evaluates to is semantically typed by using our
induction hypothesis on an extended semantically typed environment.

This is the familiar structure for a proof of the fundamental lemma of logical
relations. It is our abstraction case that has the most notable difference; our
induction hypothesis no longer has a need for a substitution lemma.

Putting all of these pieces together, we can now prove that evaluation of closed
terms is total in the empty environment as a corollary of the fundamental lemma,
given that an empty environment is vacuously semantically typed according to the
empty context. This is our entire proof by logical relations. We present it all
in Agda (Figure \ref{fig:totality-proof}), without needing to veer too far from
how it might be presented on paper. Now that we have proven that evaluation of
STLC is total, we move on to extending the proof.

\section{Extending the proof: full normalization of STLC}
\label{sec:nbe}
\begin{code}[hide]%
\>[0]\AgdaKeyword{module}\AgdaSpace{}%
\AgdaModule{Normalization}\AgdaSpace{}%
\AgdaKeyword{where}\<%
\\
\>[0][@{}l@{\AgdaIndent{0}}]%
\>[2]\AgdaKeyword{infix}\AgdaSpace{}%
\AgdaNumber{5}\AgdaSpace{}%
\AgdaOperator{\AgdaInductiveConstructor{⟨ƛ\AgdaUnderscore{}⟩\AgdaUnderscore{}}}\<%
\\
\>[2]\AgdaKeyword{infix}\AgdaSpace{}%
\AgdaNumber{7}\AgdaSpace{}%
\AgdaOperator{\AgdaInductiveConstructor{`\AgdaUnderscore{}}}\<%
\\
\>[2]\AgdaKeyword{infix}\AgdaSpace{}%
\AgdaNumber{4}\AgdaSpace{}%
\AgdaOperator{\AgdaDatatype{\AgdaUnderscore{}·\AgdaUnderscore{}⇓\AgdaUnderscore{}}}\<%
\\
\>[2]\AgdaKeyword{infix}\AgdaSpace{}%
\AgdaNumber{4}\AgdaSpace{}%
\AgdaOperator{\AgdaDatatype{\AgdaUnderscore{}∣\AgdaUnderscore{}⇓\AgdaUnderscore{}}}\<%
\\
\>[2]\AgdaKeyword{infix}\AgdaSpace{}%
\AgdaNumber{4}\AgdaSpace{}%
\AgdaOperator{\AgdaDatatype{\AgdaUnderscore{}∣\AgdaUnderscore{}⇑\AgdaUnderscore{}}}\<%
\\
\>[2]\AgdaKeyword{infix}\AgdaSpace{}%
\AgdaNumber{4}\AgdaSpace{}%
\AgdaOperator{\AgdaDatatype{\AgdaUnderscore{}∣\AgdaUnderscore{}⇑ⁿᵉ\AgdaUnderscore{}}}\<%
\\
\>[2]\AgdaKeyword{infixr}\AgdaSpace{}%
\AgdaNumber{7}\AgdaSpace{}%
\AgdaOperator{\AgdaFunction{\AgdaUnderscore{}⟶\AgdaUnderscore{}}}\<%
\\
\>[2]\AgdaKeyword{infix}\AgdaSpace{}%
\AgdaNumber{4}\AgdaSpace{}%
\AgdaOperator{\AgdaFunction{\AgdaUnderscore{}⊨\AgdaUnderscore{}}}\<%
\\
\>[2]\AgdaKeyword{infix}\AgdaSpace{}%
\AgdaNumber{4}\AgdaSpace{}%
\AgdaOperator{\AgdaFunction{\AgdaUnderscore{}⊨\AgdaUnderscore{}∷\AgdaUnderscore{}}}\<%
\\
\>[2]\AgdaKeyword{infixl}\AgdaSpace{}%
\AgdaNumber{5}\AgdaSpace{}%
\AgdaOperator{\AgdaFunction{\AgdaUnderscore{}++\AgdaUnderscore{}}}\<%
\\
\>[2]\AgdaKeyword{infix}\AgdaSpace{}%
\AgdaNumber{5}\AgdaSpace{}%
\AgdaOperator{\AgdaInductiveConstructor{ƛ\AgdaUnderscore{}}}\<%
\\
\>[2]\AgdaKeyword{infixl}\AgdaSpace{}%
\AgdaNumber{7}\AgdaSpace{}%
\AgdaOperator{\AgdaInductiveConstructor{\AgdaUnderscore{}·\AgdaUnderscore{}}}\<%
\\
\>[2]\AgdaKeyword{infix}\AgdaSpace{}%
\AgdaNumber{4}\AgdaSpace{}%
\AgdaOperator{\AgdaDatatype{\AgdaUnderscore{}⊢\AgdaUnderscore{}∷\AgdaUnderscore{}}}\<%
\\
\>[2]\AgdaKeyword{infix}\AgdaSpace{}%
\AgdaNumber{4}\AgdaSpace{}%
\AgdaOperator{\AgdaDatatype{\AgdaUnderscore{}∷\AgdaUnderscore{}∈\AgdaUnderscore{}}}\<%
\\
\>[2]\AgdaKeyword{infixr}\AgdaSpace{}%
\AgdaNumber{7}\AgdaSpace{}%
\AgdaOperator{\AgdaInductiveConstructor{\AgdaUnderscore{}⇒\AgdaUnderscore{}}}\<%
\\
\>[2]\AgdaKeyword{infixl}\AgdaSpace{}%
\AgdaNumber{5}\AgdaSpace{}%
\AgdaOperator{\AgdaInductiveConstructor{\AgdaUnderscore{}·:\AgdaUnderscore{}}}\<%
\end{code}
We extend our proof that evaluation is total to prove normalization of the
simply typed lambda calculus; any well-typed term \AgdaBound{t} reduces to a
normal form \AgdaBound{v}. We can prove normalization with either weak head
reduction or full reduction (reducing under the body of abstractions). In the
previous section, proving that evaluation of the simply typed lambda calculus is
total is very close to proving normalization with weak head reduction. All that
we are missing is determining the normal form of a term from its evaluation.

Determining the normal form of a term from its evaluation is a well-known
technique called normalization by evaluation (NbE). Typically it is described as
an algorithm, though we can also describe it as a proof of normalization. We
prove that evaluation of a well-typed term is well-defined, after which we
prove that we can read back a normal form from its evaluation.\footnote{Though
this does not prove the correctness of the technique, something we discuss in
$\S$\ref{sec:correctness}.}

Extending our proof that evaluation is total to prove normalization with weak
head reduction is not too much additional work: our domain of evaluation is
made up of our normal forms. If a term evaluates to a boolean, then its normal
form is that boolean. If a term evaluates to a closure, all we need to do is
``close'' the term by performing the delayed substitution represented by its
saved environment.

In appendix \ref{sec:intrinsic-full}, we extend our proof that evaluation of
the simply typed lambda calculus is total to prove normalization with weak head
reduction. As we have to perform a delayed substitution to determine a normal
form, reasoning about normalization involves reasoning about substitutions and
related lemmas. For example, we also show that our proof normalization with weak
head reduction is sound, for which we need to show that soundness is closed
under the substitution operation.

If we extend our proof that evaluation is total to prove normalization with full
reductions, the proof resembles the typical algorithm for NbE (which is
described with full reductions). The algorithm for NbE interleaves the
evaluation of terms and reading back normal forms. As a result, it does not need
a substitution operation. This remains the case for a proof by evaluation of
normalization with full reductions.

We extend our proof that evaluation is total and prove normalization with full
reductions (formalizing the proof presented in \citet{nbe}). The proof
illustrates how using the evaluation of programs to prove properties by logical
relations without needing substitutions is not a one-off trick. Indeed, there
are sophisticated logical relations arguments that do not require reasoning
about substitutions. Consequently, the technique is a valuable way to introduce
proofs by logical relations.

The algorithm for NbE can be summarized as follows: first, we evaluate a term in
an environment of the variables in its context, then we read back a normal form
from its evaluation in this environment.

Converting this to a proof of normalization consists of 1) proving that
evaluation in an environment of variables is total for well-typed terms, and 2)
proving that we can determine a normal form from this evaluation.

The first step remains mostly the same as in the previous section. We omit our
proof that evaluation is total, including only the differences from the previous
section.\footnote{We include the full proof, including any omissions in the main
body of the paper, in appendix \ref{sec:nbe-full}.} The most notable difference
from the previous section is that we extend our domain of evaluation to include
blocked evaluations such as variables, so that we can evaluate a term in an
environment of variables.

The second step makes up most of the work in this section. We must extend our
semantics with a relation between the evaluation of a term and the normal form
we read back. Additionally, we have to modify our semantic types so that we can
prove that semantically typed terms can have a normal form read back from their
evaluation.

\begin{AgdaAlign}
\begin{figure}[t]
\begin{code}%
\>[2]\AgdaKeyword{data}\AgdaSpace{}%
\AgdaDatatype{Type}\AgdaSpace{}%
\AgdaSymbol{:}\AgdaSpace{}%
\AgdaPrimitive{Set}\AgdaSpace{}%
\AgdaKeyword{where}\<%
\\
\>[2][@{}l@{\AgdaIndent{0}}]%
\>[4]\AgdaInductiveConstructor{base}\AgdaSpace{}%
\AgdaSymbol{:}\AgdaSpace{}%
\AgdaDatatype{Type}\<%
\\
\>[4]\AgdaOperator{\AgdaInductiveConstructor{\AgdaUnderscore{}⇒\AgdaUnderscore{}}}\AgdaSpace{}%
\AgdaSymbol{:}\AgdaSpace{}%
\AgdaDatatype{Type}\AgdaSpace{}%
\AgdaSymbol{→}\AgdaSpace{}%
\AgdaDatatype{Type}\AgdaSpace{}%
\AgdaSymbol{→}\AgdaSpace{}%
\AgdaDatatype{Type}\<%
\\
\\[\AgdaEmptyExtraSkip]%
\>[2]\AgdaKeyword{variable}\AgdaSpace{}%
\AgdaGeneralizable{S}\AgdaSpace{}%
\AgdaGeneralizable{T}\AgdaSpace{}%
\AgdaSymbol{:}\AgdaSpace{}%
\AgdaDatatype{Type}\<%
\\
\\[\AgdaEmptyExtraSkip]%
\>[2]\AgdaKeyword{data}\AgdaSpace{}%
\AgdaDatatype{Ctx}\AgdaSpace{}%
\AgdaSymbol{:}\AgdaSpace{}%
\AgdaPrimitive{Set}\AgdaSpace{}%
\AgdaKeyword{where}\<%
\\
\>[2][@{}l@{\AgdaIndent{0}}]%
\>[4]\AgdaInductiveConstructor{∅}\AgdaSpace{}%
\AgdaSymbol{:}\AgdaSpace{}%
\AgdaDatatype{Ctx}\<%
\\
\>[4]\AgdaOperator{\AgdaInductiveConstructor{\AgdaUnderscore{}·:\AgdaUnderscore{}}}\AgdaSpace{}%
\AgdaSymbol{:}\AgdaSpace{}%
\AgdaDatatype{Ctx}\AgdaSpace{}%
\AgdaSymbol{→}\AgdaSpace{}%
\AgdaDatatype{Type}\AgdaSpace{}%
\AgdaSymbol{→}\AgdaSpace{}%
\AgdaDatatype{Ctx}\<%
\\
\\[\AgdaEmptyExtraSkip]%
\>[2]\AgdaKeyword{variable}\AgdaSpace{}%
\AgdaGeneralizable{Γ}\AgdaSpace{}%
\AgdaSymbol{:}\AgdaSpace{}%
\AgdaDatatype{Ctx}\<%
\\
\\[\AgdaEmptyExtraSkip]%
\>[2]\AgdaComment{--\ Raw\ terms}\<%
\\
\>[2]\AgdaKeyword{data}\AgdaSpace{}%
\AgdaDatatype{Term}\AgdaSpace{}%
\AgdaSymbol{:}\AgdaSpace{}%
\AgdaPrimitive{Set}\AgdaSpace{}%
\AgdaKeyword{where}\<%
\\
\>[2][@{}l@{\AgdaIndent{0}}]%
\>[4]\AgdaInductiveConstructor{var}\AgdaSpace{}%
\AgdaSymbol{:}\AgdaSpace{}%
\AgdaSymbol{(}\AgdaBound{x}\AgdaSpace{}%
\AgdaSymbol{:}\AgdaSpace{}%
\AgdaDatatype{ℕ}\AgdaSymbol{)}\AgdaSpace{}%
\AgdaSymbol{→}\AgdaSpace{}%
\AgdaDatatype{Term}\<%
\\
\>[4]\AgdaOperator{\AgdaInductiveConstructor{ƛ\AgdaUnderscore{}}}\AgdaSpace{}%
\AgdaSymbol{:}\AgdaSpace{}%
\AgdaDatatype{Term}\AgdaSpace{}%
\AgdaSymbol{→}\AgdaSpace{}%
\AgdaDatatype{Term}\<%
\\
\>[4]\AgdaOperator{\AgdaInductiveConstructor{\AgdaUnderscore{}·\AgdaUnderscore{}}}\AgdaSpace{}%
\AgdaSymbol{:}\AgdaSpace{}%
\AgdaDatatype{Term}\AgdaSpace{}%
\AgdaSymbol{→}\AgdaSpace{}%
\AgdaDatatype{Term}\AgdaSpace{}%
\AgdaSymbol{→}\AgdaSpace{}%
\AgdaDatatype{Term}\<%
\\
\\[\AgdaEmptyExtraSkip]%
\>[2]\AgdaKeyword{variable}\AgdaSpace{}%
\AgdaGeneralizable{x}\AgdaSpace{}%
\AgdaSymbol{:}\AgdaSpace{}%
\AgdaDatatype{ℕ}\<%
\\
\>[2]\AgdaKeyword{variable}\AgdaSpace{}%
\AgdaGeneralizable{r}\AgdaSpace{}%
\AgdaGeneralizable{s}\AgdaSpace{}%
\AgdaGeneralizable{t}\AgdaSpace{}%
\AgdaGeneralizable{u}\AgdaSpace{}%
\AgdaGeneralizable{v}\AgdaSpace{}%
\AgdaSymbol{:}\AgdaSpace{}%
\AgdaDatatype{Term}\<%
\\
\\[\AgdaEmptyExtraSkip]%
\>[2]\AgdaComment{--\ Variable\ lookup}\<%
\\
\>[2]\AgdaKeyword{data}\AgdaSpace{}%
\AgdaOperator{\AgdaDatatype{\AgdaUnderscore{}∷\AgdaUnderscore{}∈\AgdaUnderscore{}}}\AgdaSpace{}%
\AgdaSymbol{:}\AgdaSpace{}%
\AgdaDatatype{ℕ}\AgdaSpace{}%
\AgdaSymbol{→}\AgdaSpace{}%
\AgdaDatatype{Type}\AgdaSpace{}%
\AgdaSymbol{→}\AgdaSpace{}%
\AgdaDatatype{Ctx}\AgdaSpace{}%
\AgdaSymbol{→}\AgdaSpace{}%
\AgdaPrimitive{Set}\AgdaSpace{}%
\AgdaKeyword{where}\<%
\\
\>[2][@{}l@{\AgdaIndent{0}}]%
\>[4]\AgdaInductiveConstructor{here}\AgdaSpace{}%
\AgdaSymbol{:}\AgdaSpace{}%
\AgdaInductiveConstructor{zero}\AgdaSpace{}%
\AgdaOperator{\AgdaDatatype{∷}}\AgdaSpace{}%
\AgdaGeneralizable{T}\AgdaSpace{}%
\AgdaOperator{\AgdaDatatype{∈}}\AgdaSpace{}%
\AgdaGeneralizable{Γ}\AgdaSpace{}%
\AgdaOperator{\AgdaInductiveConstructor{·:}}\AgdaSpace{}%
\AgdaGeneralizable{T}\<%
\\
\>[4]\AgdaInductiveConstructor{there}\AgdaSpace{}%
\AgdaSymbol{:}\AgdaSpace{}%
\AgdaGeneralizable{x}\AgdaSpace{}%
\AgdaOperator{\AgdaDatatype{∷}}\AgdaSpace{}%
\AgdaGeneralizable{T}\AgdaSpace{}%
\AgdaOperator{\AgdaDatatype{∈}}\AgdaSpace{}%
\AgdaGeneralizable{Γ}\AgdaSpace{}%
\AgdaSymbol{→}\AgdaSpace{}%
\AgdaInductiveConstructor{suc}\AgdaSpace{}%
\AgdaGeneralizable{x}\AgdaSpace{}%
\AgdaOperator{\AgdaDatatype{∷}}\AgdaSpace{}%
\AgdaGeneralizable{T}\AgdaSpace{}%
\AgdaOperator{\AgdaDatatype{∈}}\AgdaSpace{}%
\AgdaGeneralizable{Γ}\AgdaSpace{}%
\AgdaOperator{\AgdaInductiveConstructor{·:}}\AgdaSpace{}%
\AgdaGeneralizable{S}\<%
\\
\\[\AgdaEmptyExtraSkip]%
\>[2]\AgdaComment{--\ Typing\ judgement}\<%
\\
\>[2]\AgdaKeyword{data}\AgdaSpace{}%
\AgdaOperator{\AgdaDatatype{\AgdaUnderscore{}⊢\AgdaUnderscore{}∷\AgdaUnderscore{}}}\AgdaSpace{}%
\AgdaSymbol{:}\AgdaSpace{}%
\AgdaDatatype{Ctx}\AgdaSpace{}%
\AgdaSymbol{→}\AgdaSpace{}%
\AgdaDatatype{Term}\AgdaSpace{}%
\AgdaSymbol{→}\AgdaSpace{}%
\AgdaDatatype{Type}\AgdaSpace{}%
\AgdaSymbol{→}\AgdaSpace{}%
\AgdaPrimitive{Set}\AgdaSpace{}%
\AgdaKeyword{where}\<%
\\
\>[2][@{}l@{\AgdaIndent{0}}]%
\>[4]\AgdaInductiveConstructor{⊢var}\AgdaSpace{}%
\AgdaSymbol{:}\AgdaSpace{}%
\AgdaGeneralizable{x}\AgdaSpace{}%
\AgdaOperator{\AgdaDatatype{∷}}\AgdaSpace{}%
\AgdaGeneralizable{T}\AgdaSpace{}%
\AgdaOperator{\AgdaDatatype{∈}}\AgdaSpace{}%
\AgdaGeneralizable{Γ}\AgdaSpace{}%
\AgdaSymbol{→}\AgdaSpace{}%
\AgdaGeneralizable{Γ}\AgdaSpace{}%
\AgdaOperator{\AgdaDatatype{⊢}}\AgdaSpace{}%
\AgdaInductiveConstructor{var}\AgdaSpace{}%
\AgdaGeneralizable{x}\AgdaSpace{}%
\AgdaOperator{\AgdaDatatype{∷}}\AgdaSpace{}%
\AgdaGeneralizable{T}\<%
\\
\>[4]\AgdaInductiveConstructor{⊢abs}\AgdaSpace{}%
\AgdaSymbol{:}\AgdaSpace{}%
\AgdaGeneralizable{Γ}\AgdaSpace{}%
\AgdaOperator{\AgdaInductiveConstructor{·:}}\AgdaSpace{}%
\AgdaGeneralizable{S}\AgdaSpace{}%
\AgdaOperator{\AgdaDatatype{⊢}}\AgdaSpace{}%
\AgdaGeneralizable{t}\AgdaSpace{}%
\AgdaOperator{\AgdaDatatype{∷}}\AgdaSpace{}%
\AgdaGeneralizable{T}\AgdaSpace{}%
\AgdaSymbol{→}\AgdaSpace{}%
\AgdaGeneralizable{Γ}\AgdaSpace{}%
\AgdaOperator{\AgdaDatatype{⊢}}\AgdaSpace{}%
\AgdaOperator{\AgdaInductiveConstructor{ƛ}}\AgdaSpace{}%
\AgdaGeneralizable{t}\AgdaSpace{}%
\AgdaOperator{\AgdaDatatype{∷}}\AgdaSpace{}%
\AgdaGeneralizable{S}\AgdaSpace{}%
\AgdaOperator{\AgdaInductiveConstructor{⇒}}\AgdaSpace{}%
\AgdaGeneralizable{T}\<%
\\
\>[4]\AgdaInductiveConstructor{⊢app}\AgdaSpace{}%
\AgdaSymbol{:}\AgdaSpace{}%
\AgdaGeneralizable{Γ}\AgdaSpace{}%
\AgdaOperator{\AgdaDatatype{⊢}}\AgdaSpace{}%
\AgdaGeneralizable{r}\AgdaSpace{}%
\AgdaOperator{\AgdaDatatype{∷}}\AgdaSpace{}%
\AgdaGeneralizable{S}\AgdaSpace{}%
\AgdaOperator{\AgdaInductiveConstructor{⇒}}\AgdaSpace{}%
\AgdaGeneralizable{T}\AgdaSpace{}%
\AgdaSymbol{→}\AgdaSpace{}%
\AgdaGeneralizable{Γ}\AgdaSpace{}%
\AgdaOperator{\AgdaDatatype{⊢}}\AgdaSpace{}%
\AgdaGeneralizable{s}\AgdaSpace{}%
\AgdaOperator{\AgdaDatatype{∷}}\AgdaSpace{}%
\AgdaGeneralizable{S}\AgdaSpace{}%
\AgdaSymbol{→}\AgdaSpace{}%
\AgdaGeneralizable{Γ}\AgdaSpace{}%
\AgdaOperator{\AgdaDatatype{⊢}}\AgdaSpace{}%
\AgdaGeneralizable{r}\AgdaSpace{}%
\AgdaOperator{\AgdaInductiveConstructor{·}}\AgdaSpace{}%
\AgdaGeneralizable{s}\AgdaSpace{}%
\AgdaOperator{\AgdaDatatype{∷}}\AgdaSpace{}%
\AgdaGeneralizable{T}\<%
\end{code}
\caption{Extrinsic representation of STLC}
\label{fig:extrinsic-stlc}
\end{figure}

\subsection{Switching to an extrinsic representation}
\label{sec:extrinsic-switch}

However, before proceeding to prove normalization, we must first switch to an
extrinsic representation of the simply typed lambda calculus.

An intrinsic representation does not complicate our proof that evaluation is
total, even if we extend it to prove normalization with weak head reduction.
In contrast, an intrinsic representation complicates the proof significantly
if we extend it to prove normalization with full reductions. There are a few
reasons for this, with the most significant of these being that we will now be
including variables in our domain of evaluation (we now evaluate a term in an
environment of variables in the process of normalization). With an intrinsically
typed representation, we need to carry around a context in our domain of evaluation
to represent variables, which results in a need to reason about context extension
and renaming.

For simplicity, we switch to an extrinsic representation of terms (shown in
Figure \ref{fig:extrinsic-stlc}) in this section. We drop booleans and
conditional branching for brevity; adding them back does not affect the overall
complexity of our proof. We only consider variables, abstractions, and
application, changing our boolean base type for an empty base type
\mbox{\AgdaInductiveConstructor{base}} that is inhabited only by variables.
Even with these changes, a fair portion of our proof remains relatively
unchanged from the previous section.

Using this representation, we will prove normalization by proving that any term
that is well-typed has a normal form. We do so by determining the normal term
that a term reduces to by evaluation. We define normal terms mutually with
neutral terms:

\begin{code}%
\>[2]\AgdaKeyword{mutual}\<%
\\
\>[2][@{}l@{\AgdaIndent{0}}]%
\>[4]\AgdaKeyword{data}\AgdaSpace{}%
\AgdaDatatype{Normal}\AgdaSpace{}%
\AgdaSymbol{:}\AgdaSpace{}%
\AgdaDatatype{Term}\AgdaSpace{}%
\AgdaSymbol{→}\AgdaSpace{}%
\AgdaPrimitive{Set}\AgdaSpace{}%
\AgdaKeyword{where}\<%
\\
\>[4][@{}l@{\AgdaIndent{0}}]%
\>[6]\AgdaInductiveConstructor{normalAbs}\AgdaSpace{}%
\AgdaSymbol{:}\AgdaSpace{}%
\AgdaDatatype{Normal}\AgdaSpace{}%
\AgdaGeneralizable{v}\AgdaSpace{}%
\AgdaSymbol{→}\AgdaSpace{}%
\AgdaDatatype{Normal}\AgdaSpace{}%
\AgdaSymbol{(}\AgdaOperator{\AgdaInductiveConstructor{ƛ}}\AgdaSpace{}%
\AgdaGeneralizable{v}\AgdaSymbol{)}\<%
\\
\>[6]\AgdaInductiveConstructor{neutral}\AgdaSpace{}%
\AgdaSymbol{:}\AgdaSpace{}%
\AgdaDatatype{Neutral}\AgdaSpace{}%
\AgdaGeneralizable{u}\AgdaSpace{}%
\AgdaSymbol{→}\AgdaSpace{}%
\AgdaDatatype{Normal}\AgdaSpace{}%
\AgdaGeneralizable{u}\<%
\\
\\[\AgdaEmptyExtraSkip]%
\>[4]\AgdaKeyword{data}\AgdaSpace{}%
\AgdaDatatype{Neutral}\AgdaSpace{}%
\AgdaSymbol{:}\AgdaSpace{}%
\AgdaDatatype{Term}\AgdaSpace{}%
\AgdaSymbol{→}\AgdaSpace{}%
\AgdaPrimitive{Set}\AgdaSpace{}%
\AgdaKeyword{where}\<%
\\
\>[4][@{}l@{\AgdaIndent{0}}]%
\>[6]\AgdaInductiveConstructor{neutralVar}\AgdaSpace{}%
\AgdaSymbol{:}\AgdaSpace{}%
\AgdaDatatype{Neutral}\AgdaSpace{}%
\AgdaSymbol{(}\AgdaInductiveConstructor{var}\AgdaSpace{}%
\AgdaGeneralizable{x}\AgdaSymbol{)}\<%
\\
\>[6]\AgdaInductiveConstructor{neutralApp}\AgdaSpace{}%
\AgdaSymbol{:}\AgdaSpace{}%
\AgdaDatatype{Neutral}\AgdaSpace{}%
\AgdaGeneralizable{u}\AgdaSpace{}%
\AgdaSymbol{→}\AgdaSpace{}%
\AgdaDatatype{Normal}\AgdaSpace{}%
\AgdaGeneralizable{v}\AgdaSpace{}%
\AgdaSymbol{→}\AgdaSpace{}%
\AgdaDatatype{Neutral}\AgdaSpace{}%
\AgdaSymbol{(}\AgdaGeneralizable{u}\AgdaSpace{}%
\AgdaOperator{\AgdaInductiveConstructor{·}}\AgdaSpace{}%
\AgdaGeneralizable{v}\AgdaSymbol{)}\<%
\end{code}

Normal terms are terms in their normal form. They are either abstractions whose
bodies are normal terms, or neutral terms. Neutral terms are either variables or
the application of a neutral term to a normal term.

\subsection{Natural semantics revisited}
\begin{figure}
\begin{code}%
\>[2]\AgdaKeyword{mutual}\<%
\\
\>[2][@{}l@{\AgdaIndent{0}}]%
\>[4]\AgdaComment{--\ Environments}\<%
\\
\>[4]\AgdaFunction{Env}\AgdaSpace{}%
\AgdaSymbol{=}\AgdaSpace{}%
\AgdaDatatype{ℕ}\AgdaSpace{}%
\AgdaSymbol{→}\AgdaSpace{}%
\AgdaDatatype{Domain}\<%
\\
\\[\AgdaEmptyExtraSkip]%
\>[4]\AgdaComment{--\ Domain}\<%
\\
\>[4]\AgdaKeyword{data}\AgdaSpace{}%
\AgdaDatatype{Domain}\AgdaSpace{}%
\AgdaSymbol{:}\AgdaSpace{}%
\AgdaPrimitive{Set}\AgdaSpace{}%
\AgdaKeyword{where}\<%
\\
\>[4][@{}l@{\AgdaIndent{0}}]%
\>[6]\AgdaComment{--\ Closures}\<%
\\
\>[6]\AgdaOperator{\AgdaInductiveConstructor{⟨ƛ\AgdaUnderscore{}⟩\AgdaUnderscore{}}}\AgdaSpace{}%
\AgdaSymbol{:}\AgdaSpace{}%
\AgdaDatatype{Term}\AgdaSpace{}%
\AgdaSymbol{→}\AgdaSpace{}%
\AgdaFunction{Env}\AgdaSpace{}%
\AgdaSymbol{→}\AgdaSpace{}%
\AgdaDatatype{Domain}\<%
\\
\\[\AgdaEmptyExtraSkip]%
\>[6]\AgdaComment{--\ Neutral\ domain\ elements}\<%
\\
\>[6]\AgdaOperator{\AgdaInductiveConstructor{`\AgdaUnderscore{}}}\AgdaSpace{}%
\AgdaSymbol{:}\AgdaSpace{}%
\AgdaDatatype{Domainⁿᵉ}\AgdaSpace{}%
\AgdaSymbol{→}\AgdaSpace{}%
\AgdaDatatype{Domain}\<%
\\
\\[\AgdaEmptyExtraSkip]%
\>[4]\AgdaComment{--\ Neutral\ domain}\<%
\\
\>[4]\AgdaKeyword{data}\AgdaSpace{}%
\AgdaDatatype{Domainⁿᵉ}\AgdaSpace{}%
\AgdaSymbol{:}\AgdaSpace{}%
\AgdaPrimitive{Set}\AgdaSpace{}%
\AgdaKeyword{where}\<%
\\
\>[4][@{}l@{\AgdaIndent{0}}]%
\>[6]\AgdaComment{--\ Variables\ (de\ Brujin\ levels)}\<%
\\
\>[6]\AgdaInductiveConstructor{lvl}\AgdaSpace{}%
\AgdaSymbol{:}\AgdaSpace{}%
\AgdaSymbol{(}\AgdaBound{k}\AgdaSpace{}%
\AgdaSymbol{:}\AgdaSpace{}%
\AgdaDatatype{ℕ}\AgdaSymbol{)}\AgdaSpace{}%
\AgdaSymbol{→}\AgdaSpace{}%
\AgdaDatatype{Domainⁿᵉ}\<%
\\
\\[\AgdaEmptyExtraSkip]%
\>[6]\AgdaComment{--\ Application}\<%
\\
\>[6]\AgdaOperator{\AgdaInductiveConstructor{\AgdaUnderscore{}·\AgdaUnderscore{}}}\AgdaSpace{}%
\AgdaSymbol{:}\AgdaSpace{}%
\AgdaDatatype{Domainⁿᵉ}\AgdaSpace{}%
\AgdaSymbol{→}\AgdaSpace{}%
\AgdaDatatype{Domain}\AgdaSpace{}%
\AgdaSymbol{→}\AgdaSpace{}%
\AgdaDatatype{Domainⁿᵉ}\<%
\\
\\[\AgdaEmptyExtraSkip]%
\>[2]\AgdaKeyword{variable}\AgdaSpace{}%
\AgdaGeneralizable{a}\AgdaSpace{}%
\AgdaGeneralizable{b}\AgdaSpace{}%
\AgdaGeneralizable{d}\AgdaSpace{}%
\AgdaGeneralizable{f}\AgdaSpace{}%
\AgdaSymbol{:}\AgdaSpace{}%
\AgdaDatatype{Domain}\<%
\\
\>[2]\AgdaKeyword{variable}\AgdaSpace{}%
\AgdaGeneralizable{e}\AgdaSpace{}%
\AgdaSymbol{:}\AgdaSpace{}%
\AgdaDatatype{Domainⁿᵉ}\<%
\\
\>[2]\AgdaKeyword{variable}\AgdaSpace{}%
\AgdaGeneralizable{k}\AgdaSpace{}%
\AgdaSymbol{:}\AgdaSpace{}%
\AgdaDatatype{ℕ}\AgdaSpace{}%
\AgdaComment{--\ metavariable\ for\ de\ Brujin\ level}\<%
\\
\>[2]\AgdaKeyword{variable}\AgdaSpace{}%
\AgdaGeneralizable{γ}\AgdaSpace{}%
\AgdaGeneralizable{δ}\AgdaSpace{}%
\AgdaSymbol{:}\AgdaSpace{}%
\AgdaFunction{Env}\<%
\\
\\[\AgdaEmptyExtraSkip]%
\>[2]\AgdaOperator{\AgdaFunction{\AgdaUnderscore{}++\AgdaUnderscore{}}}\AgdaSpace{}%
\AgdaSymbol{:}\AgdaSpace{}%
\AgdaFunction{Env}\AgdaSpace{}%
\AgdaSymbol{→}\AgdaSpace{}%
\AgdaDatatype{Domain}\AgdaSpace{}%
\AgdaSymbol{→}\AgdaSpace{}%
\AgdaFunction{Env}\<%
\\
\>[2]\AgdaSymbol{(\AgdaUnderscore{}}\AgdaSpace{}%
\AgdaOperator{\AgdaFunction{++}}\AgdaSpace{}%
\AgdaBound{a}\AgdaSymbol{)}\AgdaSpace{}%
\AgdaInductiveConstructor{zero}%
\>[19]\AgdaSymbol{=}\AgdaSpace{}%
\AgdaBound{a}\<%
\\
\>[2]\AgdaSymbol{(}\AgdaBound{γ}\AgdaSpace{}%
\AgdaOperator{\AgdaFunction{++}}\AgdaSpace{}%
\AgdaSymbol{\AgdaUnderscore{})}\AgdaSpace{}%
\AgdaSymbol{(}\AgdaInductiveConstructor{suc}\AgdaSpace{}%
\AgdaBound{m}\AgdaSymbol{)}\AgdaSpace{}%
\AgdaSymbol{=}\AgdaSpace{}%
\AgdaBound{γ}\AgdaSpace{}%
\AgdaBound{m}\<%
\end{code}
\caption{Changes to environments and domains}
\label{fig:domain-norm}
\end{figure}

In the previous section, our natural semantics did not evaluate the bodies of
abstractions. As we are now proving normalization with full reductions, this is
no longer enough. However, we do not change our natural semantics to evaluate
inside the bodies of abstractions. Instead, we change our domain of evaluation
in Figure \ref{fig:domain-norm}.

Our domain of evaluation is no longer terms that are fully evaluated, but rather
terms that are in the \emph{process} of being evaluated to have a normal form
read back. Later, we will define a relation for reading back a normal form from
the evaluation of a term, but we focus on the changes to our domain of evaluation
first.

We extend our domain to include ``blocked'' evaluations, which we call the neutral
domain (\mbox{\AgdaDatatype{Domainⁿᵉ}}). A domain element
\mbox{\AgdaInductiveConstructor{`} \AgdaBound{e}} that cannot be evaluated any
further is either a variable (\mbox{\AgdaInductiveConstructor{lvl}
\AgdaBound{k}}) or the application of an element in the neutral domain to any
element in the domain (\mbox{\AgdaBound{e} \AgdaInductiveConstructor{·}
\AgdaBound{d}}). In our domain of evaluation, we use de Brujin levels instead
of de Brujin indices to represent variables. This is because de Brujin levels
act more as constants, allowing us to avoid having to rename variables as we
evaluate terms.

\begin{figure}
\begin{code}%
\>[2]\AgdaKeyword{mutual}\<%
\\
\>[2][@{}l@{\AgdaIndent{0}}]%
\>[4]\AgdaKeyword{data}\AgdaSpace{}%
\AgdaOperator{\AgdaDatatype{\AgdaUnderscore{}∣\AgdaUnderscore{}⇓\AgdaUnderscore{}}}\AgdaSpace{}%
\AgdaSymbol{:}\AgdaSpace{}%
\AgdaFunction{Env}\AgdaSpace{}%
\AgdaSymbol{→}\AgdaSpace{}%
\AgdaDatatype{Term}\AgdaSpace{}%
\AgdaSymbol{→}\AgdaSpace{}%
\AgdaDatatype{Domain}\AgdaSpace{}%
\AgdaSymbol{→}\AgdaSpace{}%
\AgdaPrimitive{Set}\AgdaSpace{}%
\AgdaKeyword{where}\<%
\\
\>[4][@{}l@{\AgdaIndent{0}}]%
\>[6]\AgdaInductiveConstructor{evalVar}\AgdaSpace{}%
\AgdaSymbol{:}\AgdaSpace{}%
\AgdaGeneralizable{γ}\AgdaSpace{}%
\AgdaOperator{\AgdaDatatype{∣}}\AgdaSpace{}%
\AgdaInductiveConstructor{var}\AgdaSpace{}%
\AgdaGeneralizable{x}\AgdaSpace{}%
\AgdaOperator{\AgdaDatatype{⇓}}\AgdaSpace{}%
\AgdaGeneralizable{γ}\AgdaSpace{}%
\AgdaGeneralizable{x}\<%
\\
\>[6]\AgdaInductiveConstructor{evalAbs}\AgdaSpace{}%
\AgdaSymbol{:}\AgdaSpace{}%
\AgdaGeneralizable{γ}\AgdaSpace{}%
\AgdaOperator{\AgdaDatatype{∣}}\AgdaSpace{}%
\AgdaOperator{\AgdaInductiveConstructor{ƛ}}\AgdaSpace{}%
\AgdaGeneralizable{t}\AgdaSpace{}%
\AgdaOperator{\AgdaDatatype{⇓}}\AgdaSpace{}%
\AgdaOperator{\AgdaInductiveConstructor{⟨ƛ}}\AgdaSpace{}%
\AgdaGeneralizable{t}\AgdaSpace{}%
\AgdaOperator{\AgdaInductiveConstructor{⟩}}\AgdaSpace{}%
\AgdaGeneralizable{γ}\<%
\\
\>[6]\AgdaInductiveConstructor{evalApp}\AgdaSpace{}%
\AgdaSymbol{:}\<%
\\
\>[6][@{}l@{\AgdaIndent{0}}]%
\>[10]\AgdaGeneralizable{γ}\AgdaSpace{}%
\AgdaOperator{\AgdaDatatype{∣}}\AgdaSpace{}%
\AgdaGeneralizable{r}\AgdaSpace{}%
\AgdaOperator{\AgdaDatatype{⇓}}\AgdaSpace{}%
\AgdaGeneralizable{f}\<%
\\
\>[6][@{}l@{\AgdaIndent{0}}]%
\>[8]\AgdaSymbol{→}\AgdaSpace{}%
\AgdaGeneralizable{γ}\AgdaSpace{}%
\AgdaOperator{\AgdaDatatype{∣}}\AgdaSpace{}%
\AgdaGeneralizable{s}\AgdaSpace{}%
\AgdaOperator{\AgdaDatatype{⇓}}\AgdaSpace{}%
\AgdaGeneralizable{a}\<%
\\
\>[8]\AgdaSymbol{→}\AgdaSpace{}%
\AgdaGeneralizable{f}\AgdaSpace{}%
\AgdaOperator{\AgdaDatatype{·}}\AgdaSpace{}%
\AgdaGeneralizable{a}\AgdaSpace{}%
\AgdaOperator{\AgdaDatatype{⇓}}\AgdaSpace{}%
\AgdaGeneralizable{b}\<%
\\
\>[8]\AgdaSymbol{→}\AgdaSpace{}%
\AgdaGeneralizable{γ}\AgdaSpace{}%
\AgdaOperator{\AgdaDatatype{∣}}\AgdaSpace{}%
\AgdaGeneralizable{r}\AgdaSpace{}%
\AgdaOperator{\AgdaInductiveConstructor{·}}\AgdaSpace{}%
\AgdaGeneralizable{s}\AgdaSpace{}%
\AgdaOperator{\AgdaDatatype{⇓}}\AgdaSpace{}%
\AgdaGeneralizable{b}\<%
\\
\\[\AgdaEmptyExtraSkip]%
\>[4]\AgdaComment{--\ Well-defined\ application}\<%
\\
\>[4]\AgdaKeyword{data}\AgdaSpace{}%
\AgdaOperator{\AgdaDatatype{\AgdaUnderscore{}·\AgdaUnderscore{}⇓\AgdaUnderscore{}}}\AgdaSpace{}%
\AgdaSymbol{:}\AgdaSpace{}%
\AgdaDatatype{Domain}\AgdaSpace{}%
\AgdaSymbol{→}\AgdaSpace{}%
\AgdaDatatype{Domain}\AgdaSpace{}%
\AgdaSymbol{→}\AgdaSpace{}%
\AgdaDatatype{Domain}\AgdaSpace{}%
\AgdaSymbol{→}\AgdaSpace{}%
\AgdaPrimitive{Set}\AgdaSpace{}%
\AgdaKeyword{where}\<%
\\
\>[4][@{}l@{\AgdaIndent{0}}]%
\>[6]\AgdaInductiveConstructor{appClosure}\AgdaSpace{}%
\AgdaSymbol{:}\AgdaSpace{}%
\AgdaGeneralizable{δ}\AgdaSpace{}%
\AgdaOperator{\AgdaFunction{++}}\AgdaSpace{}%
\AgdaGeneralizable{a}\AgdaSpace{}%
\AgdaOperator{\AgdaDatatype{∣}}\AgdaSpace{}%
\AgdaGeneralizable{t}\AgdaSpace{}%
\AgdaOperator{\AgdaDatatype{⇓}}\AgdaSpace{}%
\AgdaGeneralizable{b}\AgdaSpace{}%
\AgdaSymbol{→}\AgdaSpace{}%
\AgdaOperator{\AgdaInductiveConstructor{⟨ƛ}}\AgdaSpace{}%
\AgdaGeneralizable{t}\AgdaSpace{}%
\AgdaOperator{\AgdaInductiveConstructor{⟩}}\AgdaSpace{}%
\AgdaGeneralizable{δ}\AgdaSpace{}%
\AgdaOperator{\AgdaDatatype{·}}\AgdaSpace{}%
\AgdaGeneralizable{a}\AgdaSpace{}%
\AgdaOperator{\AgdaDatatype{⇓}}\AgdaSpace{}%
\AgdaGeneralizable{b}\<%
\\
\>[6]\AgdaInductiveConstructor{appNeutral}\AgdaSpace{}%
\AgdaSymbol{:}\AgdaSpace{}%
\AgdaOperator{\AgdaInductiveConstructor{`}}\AgdaSpace{}%
\AgdaGeneralizable{e}\AgdaSpace{}%
\AgdaOperator{\AgdaDatatype{·}}\AgdaSpace{}%
\AgdaGeneralizable{d}\AgdaSpace{}%
\AgdaOperator{\AgdaDatatype{⇓}}\AgdaSpace{}%
\AgdaOperator{\AgdaInductiveConstructor{`}}\AgdaSpace{}%
\AgdaSymbol{(}\AgdaGeneralizable{e}\AgdaSpace{}%
\AgdaOperator{\AgdaInductiveConstructor{·}}\AgdaSpace{}%
\AgdaGeneralizable{d}\AgdaSymbol{)}\<%
\end{code}
\caption{Changes to natural semantics}
\label{fig:semantics-norm}
\end{figure}

With these changes to our domain of evaluation, our natural semantics in Figure
\ref{fig:semantics-norm} remain almost the same as before. The main difference
is that we now generalize \mbox{\AgdaInductiveConstructor{evalApp}} with a
condition for well-defined application. This is because there are now two cases
for well-defined application, with \mbox{\AgdaInductiveConstructor{appClosure}}
being the case we had before. Our other case,
\mbox{\AgdaInductiveConstructor{appNeutral}}, is for the application of a domain
element to a neutral domain element. In this case, the evaluation simply remains
blocked.

We now turn to the question of evaluating inside the bodies of abstractions.
Without this, our semantics seem almost incomplete, as we are still evaluating
abstractions to closures. We now describe how to make the semantics feel more
``complete'' by introducing the notion of a reading back a normal form from
an evaluated term.

\subsection{Reading a normal form back from an evaluated term}
\begin{figure}[t]
\begin{code}%
\>[2]\AgdaKeyword{variable}\AgdaSpace{}%
\AgdaGeneralizable{n}\AgdaSpace{}%
\AgdaSymbol{:}\AgdaSpace{}%
\AgdaDatatype{ℕ}\AgdaSpace{}%
\AgdaComment{--\ Scope}\<%
\\
\\[\AgdaEmptyExtraSkip]%
\>[2]\AgdaComment{--\ Converting\ a\ de\ Brujin\ level\ to\ a\ de\ Brujin\ index}\<%
\\
\>[2]\AgdaFunction{lvl→idx}\AgdaSpace{}%
\AgdaSymbol{:}\AgdaSpace{}%
\AgdaDatatype{ℕ}\AgdaSpace{}%
\AgdaSymbol{→}\AgdaSpace{}%
\AgdaDatatype{ℕ}\AgdaSpace{}%
\AgdaSymbol{→}\AgdaSpace{}%
\AgdaDatatype{ℕ}\<%
\\
\>[2]\AgdaFunction{lvl→idx}\AgdaSpace{}%
\AgdaBound{k}\AgdaSpace{}%
\AgdaBound{n}\AgdaSpace{}%
\AgdaSymbol{=}\AgdaSpace{}%
\AgdaBound{n}\AgdaSpace{}%
\AgdaOperator{\AgdaPrimitive{-}}\AgdaSpace{}%
\AgdaInductiveConstructor{suc}\AgdaSpace{}%
\AgdaBound{k}\<%
\\
\\[\AgdaEmptyExtraSkip]%
\>[2]\AgdaKeyword{mutual}\<%
\\
\>[2][@{}l@{\AgdaIndent{0}}]%
\>[4]\AgdaComment{--\ Reading\ back\ a\ normal\ term}\<%
\\
\>[4]\AgdaKeyword{data}\AgdaSpace{}%
\AgdaOperator{\AgdaDatatype{\AgdaUnderscore{}∣\AgdaUnderscore{}⇑\AgdaUnderscore{}}}\AgdaSpace{}%
\AgdaSymbol{:}\AgdaSpace{}%
\AgdaDatatype{ℕ}\AgdaSpace{}%
\AgdaSymbol{→}\AgdaSpace{}%
\AgdaDatatype{Domain}\AgdaSpace{}%
\AgdaSymbol{→}\AgdaSpace{}%
\AgdaDatatype{Term}\AgdaSpace{}%
\AgdaSymbol{→}\AgdaSpace{}%
\AgdaPrimitive{Set}\AgdaSpace{}%
\AgdaKeyword{where}\<%
\\
\>[4][@{}l@{\AgdaIndent{0}}]%
\>[6]\AgdaInductiveConstructor{⇑closure}\AgdaSpace{}%
\AgdaSymbol{:}\<%
\\
\>[6][@{}l@{\AgdaIndent{0}}]%
\>[10]\AgdaGeneralizable{δ}\AgdaSpace{}%
\AgdaOperator{\AgdaFunction{++}}\AgdaSpace{}%
\AgdaOperator{\AgdaInductiveConstructor{`}}\AgdaSpace{}%
\AgdaInductiveConstructor{lvl}\AgdaSpace{}%
\AgdaGeneralizable{n}\AgdaSpace{}%
\AgdaOperator{\AgdaDatatype{∣}}\AgdaSpace{}%
\AgdaGeneralizable{t}\AgdaSpace{}%
\AgdaOperator{\AgdaDatatype{⇓}}\AgdaSpace{}%
\AgdaGeneralizable{a}\<%
\\
\>[6][@{}l@{\AgdaIndent{0}}]%
\>[8]\AgdaSymbol{→}\AgdaSpace{}%
\AgdaGeneralizable{n}\AgdaSpace{}%
\AgdaOperator{\AgdaDatatype{∣}}\AgdaSpace{}%
\AgdaGeneralizable{a}\AgdaSpace{}%
\AgdaOperator{\AgdaDatatype{⇑}}\AgdaSpace{}%
\AgdaGeneralizable{v}\<%
\\
\>[8]\AgdaSymbol{→}\AgdaSpace{}%
\AgdaGeneralizable{n}\AgdaSpace{}%
\AgdaOperator{\AgdaDatatype{∣}}\AgdaSpace{}%
\AgdaOperator{\AgdaInductiveConstructor{⟨ƛ}}\AgdaSpace{}%
\AgdaGeneralizable{t}\AgdaSpace{}%
\AgdaOperator{\AgdaInductiveConstructor{⟩}}\AgdaSpace{}%
\AgdaGeneralizable{δ}\AgdaSpace{}%
\AgdaOperator{\AgdaDatatype{⇑}}\AgdaSpace{}%
\AgdaOperator{\AgdaInductiveConstructor{ƛ}}\AgdaSpace{}%
\AgdaGeneralizable{v}\<%
\\
\\[\AgdaEmptyExtraSkip]%
\>[6]\AgdaInductiveConstructor{⇑neutral}\AgdaSpace{}%
\AgdaSymbol{:}\AgdaSpace{}%
\AgdaGeneralizable{n}\AgdaSpace{}%
\AgdaOperator{\AgdaDatatype{∣}}\AgdaSpace{}%
\AgdaGeneralizable{e}\AgdaSpace{}%
\AgdaOperator{\AgdaDatatype{⇑ⁿᵉ}}\AgdaSpace{}%
\AgdaGeneralizable{u}\AgdaSpace{}%
\AgdaSymbol{→}\AgdaSpace{}%
\AgdaGeneralizable{n}\AgdaSpace{}%
\AgdaOperator{\AgdaDatatype{∣}}\AgdaSpace{}%
\AgdaOperator{\AgdaInductiveConstructor{`}}\AgdaSpace{}%
\AgdaGeneralizable{e}\AgdaSpace{}%
\AgdaOperator{\AgdaDatatype{⇑}}\AgdaSpace{}%
\AgdaGeneralizable{u}\<%
\\
\\[\AgdaEmptyExtraSkip]%
\>[4]\AgdaComment{--\ Reading\ back\ a\ neutral\ term}\<%
\\
\>[4]\AgdaKeyword{data}\AgdaSpace{}%
\AgdaOperator{\AgdaDatatype{\AgdaUnderscore{}∣\AgdaUnderscore{}⇑ⁿᵉ\AgdaUnderscore{}}}\AgdaSpace{}%
\AgdaSymbol{:}\AgdaSpace{}%
\AgdaDatatype{ℕ}\AgdaSpace{}%
\AgdaSymbol{→}\AgdaSpace{}%
\AgdaDatatype{Domainⁿᵉ}\AgdaSpace{}%
\AgdaSymbol{→}\AgdaSpace{}%
\AgdaDatatype{Term}\AgdaSpace{}%
\AgdaSymbol{→}\AgdaSpace{}%
\AgdaPrimitive{Set}\AgdaSpace{}%
\AgdaKeyword{where}\<%
\\
\>[4][@{}l@{\AgdaIndent{0}}]%
\>[6]\AgdaInductiveConstructor{⇑lvl}\AgdaSpace{}%
\AgdaSymbol{:}\AgdaSpace{}%
\AgdaGeneralizable{n}\AgdaSpace{}%
\AgdaOperator{\AgdaDatatype{∣}}\AgdaSpace{}%
\AgdaInductiveConstructor{lvl}\AgdaSpace{}%
\AgdaGeneralizable{k}\AgdaSpace{}%
\AgdaOperator{\AgdaDatatype{⇑ⁿᵉ}}\AgdaSpace{}%
\AgdaInductiveConstructor{var}\AgdaSpace{}%
\AgdaSymbol{(}\AgdaFunction{lvl→idx}\AgdaSpace{}%
\AgdaGeneralizable{k}\AgdaSpace{}%
\AgdaGeneralizable{n}\AgdaSymbol{)}\<%
\\
\\[\AgdaEmptyExtraSkip]%
\>[6]\AgdaInductiveConstructor{⇑app}\AgdaSpace{}%
\AgdaSymbol{:}\<%
\\
\>[6][@{}l@{\AgdaIndent{0}}]%
\>[10]\AgdaGeneralizable{n}\AgdaSpace{}%
\AgdaOperator{\AgdaDatatype{∣}}\AgdaSpace{}%
\AgdaGeneralizable{e}\AgdaSpace{}%
\AgdaOperator{\AgdaDatatype{⇑ⁿᵉ}}\AgdaSpace{}%
\AgdaGeneralizable{u}\<%
\\
\>[6][@{}l@{\AgdaIndent{0}}]%
\>[8]\AgdaSymbol{→}\AgdaSpace{}%
\AgdaGeneralizable{n}\AgdaSpace{}%
\AgdaOperator{\AgdaDatatype{∣}}\AgdaSpace{}%
\AgdaGeneralizable{d}\AgdaSpace{}%
\AgdaOperator{\AgdaDatatype{⇑}}\AgdaSpace{}%
\AgdaGeneralizable{v}\<%
\\
\>[8]\AgdaSymbol{→}\AgdaSpace{}%
\AgdaGeneralizable{n}\AgdaSpace{}%
\AgdaOperator{\AgdaDatatype{∣}}\AgdaSpace{}%
\AgdaGeneralizable{e}\AgdaSpace{}%
\AgdaOperator{\AgdaInductiveConstructor{·}}\AgdaSpace{}%
\AgdaGeneralizable{d}\AgdaSpace{}%
\AgdaOperator{\AgdaDatatype{⇑ⁿᵉ}}\AgdaSpace{}%
\AgdaGeneralizable{u}\AgdaSpace{}%
\AgdaOperator{\AgdaInductiveConstructor{·}}\AgdaSpace{}%
\AgdaGeneralizable{v}\<%
\end{code}
\caption{Reading back a normal form from the evaluation of a term}
\label{fig:readback-nf}
\end{figure}

We now describe the conditions for reading back a normal term \AgdaBound{v} from
the evaluation \AgdaBound{d} of a term \AgdaBound{t}, notated
\mbox{\AgdaBound{n} \AgdaDatatype{∣} \AgdaBound{d} \AgdaDatatype{⇑}
\AgdaBound{v}}. It is defined mutually with the reading back of a neutral term,
notated \mbox{\AgdaBound{n} \AgdaDatatype{∣} \AgdaBound{e} \AgdaDatatype{⇑ⁿᵉ}
\AgdaBound{u}}.

Both relations need to know the overall ``scope'' of the original term
\AgdaBound{t}. This is the total number \AgdaBound{n} of variables in the term's
context. We use this value to convert a de Brujin level to its corresponding de
Brujin index in \mbox{\AgdaInductiveConstructor{⇑lvl}}.

In rule \mbox{\AgdaInductiveConstructor{⇑closure}}, we actually evaluate the
body of the closure to read back a normal term. This was not possible with weak
head reduction, as our domain of evaluation did not include variables. With our
new domain, we can evaluate the body of the closure by extending its environment
with a ``fresh'' variable (the first de Brujin level not in scope). After we
have evaluated the body of the closure, we can read back a normal term. This
rule is why we switched to de Brujin levels in our domain of evaluation: we can
evaluate the body of the closure without needing to introduce a renaming
operation.

With these changes, we can now turn to the property that we want to prove,
normalization. For any term, we wish to show that it can be evaluated in an
environment consisting of the variables in its scope (all converted to their
corresponding de Brujin levels) and then have a normal form read back.

\begin{code}%
\>[2]\AgdaFunction{idx→lvl}\AgdaSpace{}%
\AgdaSymbol{:}\AgdaSpace{}%
\AgdaDatatype{ℕ}\AgdaSpace{}%
\AgdaSymbol{→}\AgdaSpace{}%
\AgdaDatatype{ℕ}\AgdaSpace{}%
\AgdaSymbol{→}\AgdaSpace{}%
\AgdaDatatype{ℕ}\<%
\\
\>[2]\AgdaFunction{idx→lvl}\AgdaSpace{}%
\AgdaBound{i}\AgdaSpace{}%
\AgdaBound{n}\AgdaSpace{}%
\AgdaSymbol{=}\AgdaSpace{}%
\AgdaBound{n}\AgdaSpace{}%
\AgdaOperator{\AgdaPrimitive{-}}\AgdaSpace{}%
\AgdaInductiveConstructor{suc}\AgdaSpace{}%
\AgdaBound{i}\<%
\\
\\[\AgdaEmptyExtraSkip]%
\>[2]\AgdaFunction{env}\AgdaSpace{}%
\AgdaSymbol{:}\AgdaSpace{}%
\AgdaDatatype{ℕ}\AgdaSpace{}%
\AgdaSymbol{→}\AgdaSpace{}%
\AgdaFunction{Env}\<%
\\
\>[2]\AgdaFunction{env}\AgdaSpace{}%
\AgdaBound{n}\AgdaSpace{}%
\AgdaBound{i}\AgdaSpace{}%
\AgdaSymbol{=}\AgdaSpace{}%
\AgdaOperator{\AgdaInductiveConstructor{`}}\AgdaSpace{}%
\AgdaInductiveConstructor{lvl}\AgdaSpace{}%
\AgdaSymbol{(}\AgdaFunction{idx→lvl}\AgdaSpace{}%
\AgdaBound{i}\AgdaSpace{}%
\AgdaBound{n}\AgdaSymbol{)}\<%
\\
\\[\AgdaEmptyExtraSkip]%
\>[2]\AgdaOperator{\AgdaFunction{\AgdaUnderscore{}has-normal-form<\AgdaUnderscore{}>\AgdaUnderscore{}}}\AgdaSpace{}%
\AgdaSymbol{:}\AgdaSpace{}%
\AgdaDatatype{Term}\AgdaSpace{}%
\AgdaSymbol{→}\AgdaSpace{}%
\AgdaDatatype{ℕ}\AgdaSpace{}%
\AgdaSymbol{→}\AgdaSpace{}%
\AgdaDatatype{Term}\AgdaSpace{}%
\AgdaSymbol{→}\AgdaSpace{}%
\AgdaPrimitive{Set}\<%
\\
\>[2]\AgdaBound{t}%
\>[1292I]\AgdaOperator{\AgdaFunction{has-normal-form<}}\AgdaSpace{}%
\AgdaBound{n}\AgdaSpace{}%
\AgdaOperator{\AgdaFunction{>}}\AgdaSpace{}%
\AgdaBound{v}\AgdaSpace{}%
\AgdaSymbol{=}\<%
\\
\>[.][@{}l@{}]\<[1292I]%
\>[4]\AgdaFunction{∃[}\AgdaSpace{}%
\AgdaBound{a}\AgdaSpace{}%
\AgdaFunction{]}\AgdaSpace{}%
\AgdaFunction{env}\AgdaSpace{}%
\AgdaBound{n}\AgdaSpace{}%
\AgdaOperator{\AgdaDatatype{∣}}\AgdaSpace{}%
\AgdaBound{t}\AgdaSpace{}%
\AgdaOperator{\AgdaDatatype{⇓}}\AgdaSpace{}%
\AgdaBound{a}\AgdaSpace{}%
\AgdaOperator{\AgdaFunction{×}}\AgdaSpace{}%
\AgdaBound{n}\AgdaSpace{}%
\AgdaOperator{\AgdaDatatype{∣}}\AgdaSpace{}%
\AgdaBound{a}\AgdaSpace{}%
\AgdaOperator{\AgdaDatatype{⇑}}\AgdaSpace{}%
\AgdaBound{v}\<%
\end{code}

Note that any term that is read back from the domain of evaluation is a normal
term, something that we can prove quickly by induction on the read back relation.
(We include this proof at the end of appendix \ref{sec:nbe-full}). For this
reason, we only need to prove that we can read back a term in the first place
according to the relation.

\subsection{Asking more of semantic types with candidate spaces}

Proving that the evaluation of a well-typed term is total is the same as before.
The only difference is that we are now evaluating the term in an environment of
variables instead of the empty environment. This does not change the fundamental
lemma at all (and we omit its proof), as it is generalized over any environment
that is semantically typed. As a result, all we need to do here is prove that an
environment of variables is semantically typed. This will be a consequence of
proving that if the evaluation of a term is semantically typed, we can read back
a normal form from its evaluation. We begin to address this matter in this
section.

\begin{figure}
\begin{code}%
\>[2]\AgdaFunction{LogPred}\AgdaSpace{}%
\AgdaSymbol{=}\AgdaSpace{}%
\AgdaDatatype{Domain}\AgdaSpace{}%
\AgdaSymbol{→}\AgdaSpace{}%
\AgdaPrimitive{Set}\<%
\\
\\[\AgdaEmptyExtraSkip]%
\>[2]\AgdaComment{--\ Bottom\ of\ candidate\ space}\<%
\\
\>[2]\AgdaComment{--\ (neutral\ term\ can\ be\ read\ back)}\<%
\\
\>[2]\AgdaFunction{𝔹}\AgdaSpace{}%
\AgdaSymbol{:}\AgdaSpace{}%
\AgdaFunction{LogPred}\<%
\\
\>[2]\AgdaFunction{𝔹}\AgdaSpace{}%
\AgdaSymbol{(}\AgdaOperator{\AgdaInductiveConstructor{`}}\AgdaSpace{}%
\AgdaBound{e}\AgdaSymbol{)}\AgdaSpace{}%
\AgdaSymbol{=}\AgdaSpace{}%
\AgdaSymbol{∀}\AgdaSpace{}%
\AgdaBound{n}\AgdaSpace{}%
\AgdaSymbol{→}\AgdaSpace{}%
\AgdaFunction{∃[}\AgdaSpace{}%
\AgdaBound{u}\AgdaSpace{}%
\AgdaFunction{]}\AgdaSpace{}%
\AgdaBound{n}\AgdaSpace{}%
\AgdaOperator{\AgdaDatatype{∣}}\AgdaSpace{}%
\AgdaBound{e}\AgdaSpace{}%
\AgdaOperator{\AgdaDatatype{⇑ⁿᵉ}}\AgdaSpace{}%
\AgdaBound{u}\<%
\\
\>[2]\AgdaCatchallClause{\AgdaFunction{𝔹}}\AgdaSpace{}%
\AgdaCatchallClause{\AgdaSymbol{\AgdaUnderscore{}}}\AgdaSpace{}%
\AgdaSymbol{=}\AgdaSpace{}%
\AgdaDatatype{⊥}\<%
\\
\\[\AgdaEmptyExtraSkip]%
\>[2]\AgdaComment{--\ Top\ of\ candidate\ space}\<%
\\
\>[2]\AgdaComment{--\ (normal\ term\ can\ be\ read\ back)}\<%
\\
\>[2]\AgdaFunction{𝕋}\AgdaSpace{}%
\AgdaSymbol{:}\AgdaSpace{}%
\AgdaFunction{LogPred}\<%
\\
\>[2]\AgdaFunction{𝕋}\AgdaSpace{}%
\AgdaBound{d}\AgdaSpace{}%
\AgdaSymbol{=}\AgdaSpace{}%
\AgdaSymbol{∀}\AgdaSpace{}%
\AgdaBound{n}\AgdaSpace{}%
\AgdaSymbol{→}\AgdaSpace{}%
\AgdaFunction{∃[}\AgdaSpace{}%
\AgdaBound{v}\AgdaSpace{}%
\AgdaFunction{]}\AgdaSpace{}%
\AgdaBound{n}\AgdaSpace{}%
\AgdaOperator{\AgdaDatatype{∣}}\AgdaSpace{}%
\AgdaBound{d}\AgdaSpace{}%
\AgdaOperator{\AgdaDatatype{⇑}}\AgdaSpace{}%
\AgdaBound{v}\<%
\\
\\[\AgdaEmptyExtraSkip]%
\>[2]\AgdaComment{--\ Bottom\ of\ space\ is\ a\ subset\ of\ top}\<%
\\
\>[2]\AgdaFunction{𝔹⊆𝕋}\AgdaSpace{}%
\AgdaSymbol{:}\AgdaSpace{}%
\AgdaGeneralizable{d}\AgdaSpace{}%
\AgdaOperator{\AgdaFunction{∈}}\AgdaSpace{}%
\AgdaFunction{𝔹}\AgdaSpace{}%
\AgdaSymbol{→}\AgdaSpace{}%
\AgdaGeneralizable{d}\AgdaSpace{}%
\AgdaOperator{\AgdaFunction{∈}}\AgdaSpace{}%
\AgdaFunction{𝕋}\<%
\\
\>[2]\AgdaFunction{𝔹⊆𝕋}\AgdaSpace{}%
\AgdaSymbol{\{}\AgdaOperator{\AgdaInductiveConstructor{`}}\AgdaSpace{}%
\AgdaBound{e}\AgdaSymbol{\}}\AgdaSpace{}%
\AgdaBound{eb}\AgdaSpace{}%
\AgdaBound{n}\<%
\\
\>[2][@{}l@{\AgdaIndent{0}}]%
\>[4]\AgdaKeyword{with}\AgdaSpace{}%
\AgdaBound{eb}\AgdaSpace{}%
\AgdaBound{n}\<%
\\
\>[2]\AgdaSymbol{...}\AgdaSpace{}%
\AgdaSymbol{|}\AgdaSpace{}%
\AgdaBound{u}\AgdaSpace{}%
\AgdaOperator{\AgdaInductiveConstructor{,}}\AgdaSpace{}%
\AgdaBound{e⇑u}\AgdaSpace{}%
\AgdaSymbol{=}\<%
\\
\>[2][@{}l@{\AgdaIndent{0}}]%
\>[4]\AgdaBound{u}\AgdaSpace{}%
\AgdaOperator{\AgdaInductiveConstructor{,}}\AgdaSpace{}%
\AgdaInductiveConstructor{⇑neutral}\AgdaSpace{}%
\AgdaBound{e⇑u}\<%
\end{code}
\caption{Candidate space for reading back a normal term}
\label{fig:candidate-space}
\end{figure}

To prove that we can read back a normal form from the evaluation of a term if it
is semantically typed, we must define a ``candidate space'' (Figure
\ref{fig:candidate-space}) of elements in the domain that can actually have a
normal form read back from them. We define a ``bottom'' of the candidate space
\AgdaFunction{𝔹} (the subset of the neutral domain that can have a neutral term
read back) and a ``top'' of the candidate space \AgdaFunction{𝕋} (the subset of
the domain that can have a normal term read back). The bottom of the candidate
space is a subset of the top of the candidate space (as elements in the neutral
domain are a subset of the domain and neutral terms are a subset of normal
terms).

For convenience, we define \mbox{\AgdaFunction{LogPred}} to be
\mbox{\AgdaDatatype{Domain} \AgdaArgument{→} \AgdaDatatype{Set}}, as we will be
using logical predicates more in this section. In particular, the top and bottom
of the candidate space are themselves logical predicates.

We now update our semantic types so that they inhabit this candidate space, this
is referred to as "asking more" of semantic types \cite{girard}.

\subsection{Changes to semantic types}

\begin{figure}
\begin{code}%
\>[2]\AgdaComment{--\ Semantic\ function\ space}\<%
\\
\>[2]\AgdaOperator{\AgdaFunction{\AgdaUnderscore{}⟶\AgdaUnderscore{}}}\AgdaSpace{}%
\AgdaSymbol{:}\AgdaSpace{}%
\AgdaFunction{LogPred}\AgdaSpace{}%
\AgdaSymbol{→}\AgdaSpace{}%
\AgdaFunction{LogPred}\AgdaSpace{}%
\AgdaSymbol{→}\AgdaSpace{}%
\AgdaFunction{LogPred}\<%
\\
\>[2]\AgdaSymbol{(}\AgdaBound{A}\AgdaSpace{}%
\AgdaOperator{\AgdaFunction{⟶}}\AgdaSpace{}%
\AgdaBound{B}\AgdaSymbol{)}\AgdaSpace{}%
\AgdaBound{f}\AgdaSpace{}%
\AgdaSymbol{=}\AgdaSpace{}%
\AgdaSymbol{∀}\AgdaSpace{}%
\AgdaSymbol{\{}\AgdaBound{a}\AgdaSymbol{\}}\AgdaSpace{}%
\AgdaSymbol{→}\AgdaSpace{}%
\AgdaBound{a}\AgdaSpace{}%
\AgdaOperator{\AgdaFunction{∈}}\AgdaSpace{}%
\AgdaBound{A}\AgdaSpace{}%
\AgdaSymbol{→}\AgdaSpace{}%
\AgdaFunction{∃[}\AgdaSpace{}%
\AgdaBound{b}\AgdaSpace{}%
\AgdaFunction{]}\AgdaSpace{}%
\AgdaBound{f}\AgdaSpace{}%
\AgdaOperator{\AgdaDatatype{·}}\AgdaSpace{}%
\AgdaBound{a}\AgdaSpace{}%
\AgdaOperator{\AgdaDatatype{⇓}}\AgdaSpace{}%
\AgdaBound{b}\AgdaSpace{}%
\AgdaOperator{\AgdaFunction{×}}\AgdaSpace{}%
\AgdaBound{b}\AgdaSpace{}%
\AgdaOperator{\AgdaFunction{∈}}\AgdaSpace{}%
\AgdaBound{B}\<%
\\
\\[\AgdaEmptyExtraSkip]%
\>[2]\AgdaComment{--\ Semantic\ types\ (logical\ relation)}\<%
\\
\>[2]\AgdaOperator{\AgdaFunction{⟦\AgdaUnderscore{}⟧}}\AgdaSpace{}%
\AgdaSymbol{:}\AgdaSpace{}%
\AgdaDatatype{Type}\AgdaSpace{}%
\AgdaSymbol{→}\AgdaSpace{}%
\AgdaFunction{LogPred}\<%
\\
\>[2]\AgdaOperator{\AgdaFunction{⟦}}\AgdaSpace{}%
\AgdaInductiveConstructor{base}\AgdaSpace{}%
\AgdaOperator{\AgdaFunction{⟧}}\AgdaSpace{}%
\AgdaSymbol{=}\AgdaSpace{}%
\AgdaFunction{𝔹}\<%
\\
\>[2]\AgdaOperator{\AgdaFunction{⟦}}\AgdaSpace{}%
\AgdaBound{S}\AgdaSpace{}%
\AgdaOperator{\AgdaInductiveConstructor{⇒}}\AgdaSpace{}%
\AgdaBound{T}\AgdaSpace{}%
\AgdaOperator{\AgdaFunction{⟧}}\AgdaSpace{}%
\AgdaSymbol{=}\AgdaSpace{}%
\AgdaOperator{\AgdaFunction{⟦}}\AgdaSpace{}%
\AgdaBound{S}\AgdaSpace{}%
\AgdaOperator{\AgdaFunction{⟧}}\AgdaSpace{}%
\AgdaOperator{\AgdaFunction{⟶}}\AgdaSpace{}%
\AgdaOperator{\AgdaFunction{⟦}}\AgdaSpace{}%
\AgdaBound{T}\AgdaSpace{}%
\AgdaOperator{\AgdaFunction{⟧}}\<%
\end{code}
\caption{Changes to semantic types}
\label{sec:semantic}
\end{figure}

The only domain elements that may inhabit our semantic base type are neutral
domain elements (e.g. variables), we expect closures to inhabit the semantic
function type. If a neutral domain element is semantically typed, we wish to be
able to read back a neutral term from it. Therefore, we insert the bottom of the
candidate space \AgdaFunction{𝔹} at our base semantic type.

As a result of this change to our logical predicate, semantic types are now
restricted so that they inhabit the candidate space, i.e. for any type
\AgdaBound{T}, we have that \mbox{\AgdaFunction{𝔹} $\subseteq$ \AgdaFunction{⟦}
\AgdaBound{T} \AgdaFunction{⟧} $\subseteq$ \AgdaFunction{𝕋}}). This is not
obvious, and we have to prove it explicitly in the next section.

In restricting our semantic types so they inhabit the candidate space, proving
that the evaluation of a term is semantically typed immediately yields that we
can read back a normal term.

Our semantic function type remains unchanged, though we generalize it to use a
semantic function space \mbox{\AgdaBound{A} \AgdaFunction{⟶} \AgdaBound{B}}. We
do so because we will be reusing the semantic function space in our reasoning in
the next section when proving that semantic types inhabit the candidate space.
The semantic function space is equivalent to our former presentation, however,
and we could have defined our semantic function type this way if we had wanted
to.

Note that here, we are now using \mbox{\AgdaBound{f} \AgdaDatatype{·}
\AgdaBound{a} \AgdaDatatype{⇓} \AgdaBound{b}} in our semantic function type
instead of \mbox{\AgdaBound{δ} \AgdaFunction{++} \AgdaBound{a} \AgdaDatatype{∣}
\AgdaBound{t} \AgdaDatatype{⇓} \AgdaBound{b}} as before. This is because we have
extended our semantics to include more cases for well-defined application in the
domain.

\begin{code}[hide]%
\>[2]\AgdaOperator{\AgdaFunction{\AgdaUnderscore{}⊨\AgdaUnderscore{}}}\AgdaSpace{}%
\AgdaSymbol{:}\AgdaSpace{}%
\AgdaDatatype{Ctx}\AgdaSpace{}%
\AgdaSymbol{→}\AgdaSpace{}%
\AgdaFunction{Env}\AgdaSpace{}%
\AgdaSymbol{→}\AgdaSpace{}%
\AgdaPrimitive{Set}\<%
\\
\>[2]\AgdaBound{Γ}\AgdaSpace{}%
\AgdaOperator{\AgdaFunction{⊨}}\AgdaSpace{}%
\AgdaBound{γ}\AgdaSpace{}%
\AgdaSymbol{=}\AgdaSpace{}%
\AgdaSymbol{∀}\AgdaSpace{}%
\AgdaSymbol{\{}\AgdaBound{x}\AgdaSymbol{\}}\AgdaSpace{}%
\AgdaSymbol{\{}\AgdaBound{T}\AgdaSymbol{\}}\AgdaSpace{}%
\AgdaSymbol{→}\AgdaSpace{}%
\AgdaBound{x}\AgdaSpace{}%
\AgdaOperator{\AgdaDatatype{∷}}\AgdaSpace{}%
\AgdaBound{T}\AgdaSpace{}%
\AgdaOperator{\AgdaDatatype{∈}}\AgdaSpace{}%
\AgdaBound{Γ}\AgdaSpace{}%
\AgdaSymbol{→}\AgdaSpace{}%
\AgdaBound{γ}\AgdaSpace{}%
\AgdaBound{x}\AgdaSpace{}%
\AgdaOperator{\AgdaFunction{∈}}\AgdaSpace{}%
\AgdaOperator{\AgdaFunction{⟦}}\AgdaSpace{}%
\AgdaBound{T}\AgdaSpace{}%
\AgdaOperator{\AgdaFunction{⟧}}\<%
\\
\\[\AgdaEmptyExtraSkip]%
\>[2]\AgdaOperator{\AgdaFunction{\AgdaUnderscore{}⊨\AgdaUnderscore{}∷\AgdaUnderscore{}}}\AgdaSpace{}%
\AgdaSymbol{:}\AgdaSpace{}%
\AgdaDatatype{Ctx}\AgdaSpace{}%
\AgdaSymbol{→}\AgdaSpace{}%
\AgdaDatatype{Term}\AgdaSpace{}%
\AgdaSymbol{→}\AgdaSpace{}%
\AgdaDatatype{Type}\AgdaSpace{}%
\AgdaSymbol{→}\AgdaSpace{}%
\AgdaPrimitive{Set}\<%
\\
\>[2]\AgdaBound{Γ}\AgdaSpace{}%
\AgdaOperator{\AgdaFunction{⊨}}\AgdaSpace{}%
\AgdaBound{t}\AgdaSpace{}%
\AgdaOperator{\AgdaFunction{∷}}\AgdaSpace{}%
\AgdaBound{T}\AgdaSpace{}%
\AgdaSymbol{=}\AgdaSpace{}%
\AgdaSymbol{∀}\AgdaSpace{}%
\AgdaSymbol{\{}\AgdaBound{γ}\AgdaSymbol{\}}\AgdaSpace{}%
\AgdaSymbol{→}\AgdaSpace{}%
\AgdaBound{Γ}\AgdaSpace{}%
\AgdaOperator{\AgdaFunction{⊨}}\AgdaSpace{}%
\AgdaBound{γ}\AgdaSpace{}%
\AgdaSymbol{→}\AgdaSpace{}%
\AgdaFunction{∃[}\AgdaSpace{}%
\AgdaBound{a}\AgdaSpace{}%
\AgdaFunction{]}\AgdaSpace{}%
\AgdaBound{γ}\AgdaSpace{}%
\AgdaOperator{\AgdaDatatype{∣}}\AgdaSpace{}%
\AgdaBound{t}\AgdaSpace{}%
\AgdaOperator{\AgdaDatatype{⇓}}\AgdaSpace{}%
\AgdaBound{a}\AgdaSpace{}%
\AgdaOperator{\AgdaFunction{×}}\AgdaSpace{}%
\AgdaBound{a}\AgdaSpace{}%
\AgdaOperator{\AgdaFunction{∈}}\AgdaSpace{}%
\AgdaOperator{\AgdaFunction{⟦}}\AgdaSpace{}%
\AgdaBound{T}\AgdaSpace{}%
\AgdaOperator{\AgdaFunction{⟧}}\<%
\\
\\[\AgdaEmptyExtraSkip]%
\>[2]\AgdaOperator{\AgdaFunction{\AgdaUnderscore{}\textasciicircum{}\AgdaUnderscore{}}}\AgdaSpace{}%
\AgdaSymbol{:}\AgdaSpace{}%
\AgdaGeneralizable{Γ}\AgdaSpace{}%
\AgdaOperator{\AgdaFunction{⊨}}\AgdaSpace{}%
\AgdaGeneralizable{γ}\AgdaSpace{}%
\AgdaSymbol{→}\AgdaSpace{}%
\AgdaGeneralizable{a}\AgdaSpace{}%
\AgdaOperator{\AgdaFunction{∈}}\AgdaSpace{}%
\AgdaOperator{\AgdaFunction{⟦}}\AgdaSpace{}%
\AgdaGeneralizable{S}\AgdaSpace{}%
\AgdaOperator{\AgdaFunction{⟧}}\AgdaSpace{}%
\AgdaSymbol{→}\AgdaSpace{}%
\AgdaGeneralizable{Γ}\AgdaSpace{}%
\AgdaOperator{\AgdaInductiveConstructor{·:}}\AgdaSpace{}%
\AgdaGeneralizable{S}\AgdaSpace{}%
\AgdaOperator{\AgdaFunction{⊨}}\AgdaSpace{}%
\AgdaGeneralizable{γ}\AgdaSpace{}%
\AgdaOperator{\AgdaFunction{++}}\AgdaSpace{}%
\AgdaGeneralizable{a}\<%
\\
\>[2]\AgdaOperator{\AgdaFunction{\AgdaUnderscore{}\textasciicircum{}\AgdaUnderscore{}}}\AgdaSpace{}%
\AgdaSymbol{\AgdaUnderscore{}}\AgdaSpace{}%
\AgdaBound{sa}\AgdaSpace{}%
\AgdaInductiveConstructor{here}\AgdaSpace{}%
\AgdaSymbol{=}\AgdaSpace{}%
\AgdaBound{sa}\<%
\\
\>[2]\AgdaOperator{\AgdaFunction{\AgdaUnderscore{}\textasciicircum{}\AgdaUnderscore{}}}\AgdaSpace{}%
\AgdaBound{⊨γ}\AgdaSpace{}%
\AgdaSymbol{\AgdaUnderscore{}}\AgdaSpace{}%
\AgdaSymbol{(}\AgdaInductiveConstructor{there}\AgdaSpace{}%
\AgdaBound{pf}\AgdaSymbol{)}\AgdaSpace{}%
\AgdaSymbol{=}\AgdaSpace{}%
\AgdaBound{⊨γ}\AgdaSpace{}%
\AgdaBound{pf}\<%
\\
\\[\AgdaEmptyExtraSkip]%
\>[2]\AgdaFunction{fundamental-lemma}\AgdaSpace{}%
\AgdaSymbol{:}\AgdaSpace{}%
\AgdaGeneralizable{Γ}\AgdaSpace{}%
\AgdaOperator{\AgdaDatatype{⊢}}\AgdaSpace{}%
\AgdaGeneralizable{t}\AgdaSpace{}%
\AgdaOperator{\AgdaDatatype{∷}}\AgdaSpace{}%
\AgdaGeneralizable{T}\AgdaSpace{}%
\AgdaSymbol{→}\AgdaSpace{}%
\AgdaGeneralizable{Γ}\AgdaSpace{}%
\AgdaOperator{\AgdaFunction{⊨}}\AgdaSpace{}%
\AgdaGeneralizable{t}\AgdaSpace{}%
\AgdaOperator{\AgdaFunction{∷}}\AgdaSpace{}%
\AgdaGeneralizable{T}\<%
\\
\>[2]\AgdaFunction{fundamental-lemma}\AgdaSpace{}%
\AgdaSymbol{(}\AgdaInductiveConstructor{⊢var}\AgdaSpace{}%
\AgdaSymbol{\{}\AgdaBound{x}\AgdaSymbol{\}}\AgdaSpace{}%
\AgdaBound{pf}\AgdaSymbol{)}\AgdaSpace{}%
\AgdaSymbol{\{}\AgdaBound{γ}\AgdaSymbol{\}}\AgdaSpace{}%
\AgdaBound{⊨γ}\AgdaSpace{}%
\AgdaSymbol{=}\<%
\\
\>[2][@{}l@{\AgdaIndent{0}}]%
\>[4]\AgdaBound{γ}\AgdaSpace{}%
\AgdaBound{x}\AgdaSpace{}%
\AgdaOperator{\AgdaInductiveConstructor{,}}\AgdaSpace{}%
\AgdaInductiveConstructor{evalVar}\AgdaSpace{}%
\AgdaOperator{\AgdaInductiveConstructor{,}}\AgdaSpace{}%
\AgdaBound{⊨γ}\AgdaSpace{}%
\AgdaBound{pf}\<%
\\
\>[2]\AgdaFunction{fundamental-lemma}\AgdaSpace{}%
\AgdaSymbol{\{}\AgdaArgument{Γ}\AgdaSpace{}%
\AgdaSymbol{=}\AgdaSpace{}%
\AgdaBound{Γ}\AgdaSymbol{\}}\AgdaSpace{}%
\AgdaSymbol{\{}\AgdaArgument{t}\AgdaSpace{}%
\AgdaSymbol{=}\AgdaSpace{}%
\AgdaOperator{\AgdaInductiveConstructor{ƛ}}\AgdaSpace{}%
\AgdaBound{t}\AgdaSymbol{\}}\AgdaSpace{}%
\AgdaSymbol{\{}\AgdaBound{S}\AgdaSpace{}%
\AgdaOperator{\AgdaInductiveConstructor{⇒}}\AgdaSpace{}%
\AgdaBound{T}\AgdaSymbol{\}}\AgdaSpace{}%
\AgdaSymbol{(}\AgdaInductiveConstructor{⊢abs}\AgdaSpace{}%
\AgdaBound{⊢t}\AgdaSymbol{)}\AgdaSpace{}%
\AgdaSymbol{\{}\AgdaBound{γ}\AgdaSymbol{\}}\AgdaSpace{}%
\AgdaBound{⊨γ}\AgdaSpace{}%
\AgdaSymbol{=}\<%
\\
\>[2][@{}l@{\AgdaIndent{0}}]%
\>[4]\AgdaOperator{\AgdaInductiveConstructor{⟨ƛ}}\AgdaSpace{}%
\AgdaBound{t}\AgdaSpace{}%
\AgdaOperator{\AgdaInductiveConstructor{⟩}}\AgdaSpace{}%
\AgdaBound{γ}\AgdaSpace{}%
\AgdaOperator{\AgdaInductiveConstructor{,}}\<%
\\
\>[4]\AgdaInductiveConstructor{evalAbs}\AgdaSpace{}%
\AgdaOperator{\AgdaInductiveConstructor{,}}\<%
\\
\>[4]\AgdaSymbol{λ}%
\>[1553I]\AgdaBound{sa}\AgdaSpace{}%
\AgdaSymbol{→}\<%
\\
\>[.][@{}l@{}]\<[1553I]%
\>[6]\AgdaKeyword{let}\AgdaSpace{}%
\AgdaBound{⊨t}\AgdaSpace{}%
\AgdaSymbol{=}\AgdaSpace{}%
\AgdaFunction{fundamental-lemma}\AgdaSpace{}%
\AgdaBound{⊢t}\AgdaSpace{}%
\AgdaKeyword{in}\<%
\\
\>[6]\AgdaKeyword{let}\AgdaSpace{}%
\AgdaSymbol{(}\AgdaBound{b}\AgdaSpace{}%
\AgdaOperator{\AgdaInductiveConstructor{,}}\AgdaSpace{}%
\AgdaBound{eval-closure}\AgdaSpace{}%
\AgdaOperator{\AgdaInductiveConstructor{,}}\AgdaSpace{}%
\AgdaBound{sb}\AgdaSymbol{)}\AgdaSpace{}%
\AgdaSymbol{=}\AgdaSpace{}%
\AgdaBound{⊨t}\AgdaSpace{}%
\AgdaSymbol{(}\AgdaBound{⊨γ}\AgdaSpace{}%
\AgdaOperator{\AgdaFunction{\textasciicircum{}}}\AgdaSpace{}%
\AgdaBound{sa}\AgdaSymbol{)}\AgdaSpace{}%
\AgdaKeyword{in}\<%
\\
\>[6]\AgdaBound{b}\AgdaSpace{}%
\AgdaOperator{\AgdaInductiveConstructor{,}}\AgdaSpace{}%
\AgdaInductiveConstructor{appClosure}\AgdaSpace{}%
\AgdaBound{eval-closure}\AgdaSpace{}%
\AgdaOperator{\AgdaInductiveConstructor{,}}\AgdaSpace{}%
\AgdaBound{sb}\<%
\\
\>[2]\AgdaFunction{fundamental-lemma}\AgdaSpace{}%
\AgdaSymbol{(}\AgdaInductiveConstructor{⊢app}\AgdaSpace{}%
\AgdaBound{⊢r}\AgdaSpace{}%
\AgdaBound{⊢s}\AgdaSymbol{)}\AgdaSpace{}%
\AgdaBound{⊨γ}\AgdaSpace{}%
\AgdaSymbol{=}\<%
\\
\>[2][@{}l@{\AgdaIndent{0}}]%
\>[4]\AgdaKeyword{let}\AgdaSpace{}%
\AgdaSymbol{(}\AgdaBound{f}\AgdaSpace{}%
\AgdaOperator{\AgdaInductiveConstructor{,}}\AgdaSpace{}%
\AgdaBound{r⇓}\AgdaSpace{}%
\AgdaOperator{\AgdaInductiveConstructor{,}}\AgdaSpace{}%
\AgdaBound{sf}\AgdaSymbol{)}\AgdaSpace{}%
\AgdaSymbol{=}\AgdaSpace{}%
\AgdaFunction{fundamental-lemma}\AgdaSpace{}%
\AgdaBound{⊢r}\AgdaSpace{}%
\AgdaBound{⊨γ}\AgdaSpace{}%
\AgdaKeyword{in}\<%
\\
\>[4]\AgdaKeyword{let}\AgdaSpace{}%
\AgdaSymbol{(}\AgdaBound{a}\AgdaSpace{}%
\AgdaOperator{\AgdaInductiveConstructor{,}}\AgdaSpace{}%
\AgdaBound{s⇓}\AgdaSpace{}%
\AgdaOperator{\AgdaInductiveConstructor{,}}\AgdaSpace{}%
\AgdaBound{sa}\AgdaSymbol{)}\AgdaSpace{}%
\AgdaSymbol{=}\AgdaSpace{}%
\AgdaFunction{fundamental-lemma}\AgdaSpace{}%
\AgdaBound{⊢s}\AgdaSpace{}%
\AgdaBound{⊨γ}\AgdaSpace{}%
\AgdaKeyword{in}\<%
\\
\>[4]\AgdaKeyword{let}\AgdaSpace{}%
\AgdaSymbol{(}\AgdaBound{b}\AgdaSpace{}%
\AgdaOperator{\AgdaInductiveConstructor{,}}\AgdaSpace{}%
\AgdaBound{app-eval}\AgdaSpace{}%
\AgdaOperator{\AgdaInductiveConstructor{,}}\AgdaSpace{}%
\AgdaBound{sb}\AgdaSymbol{)}\AgdaSpace{}%
\AgdaSymbol{=}\AgdaSpace{}%
\AgdaBound{sf}\AgdaSpace{}%
\AgdaBound{sa}\AgdaSpace{}%
\AgdaKeyword{in}\<%
\\
\>[4]\AgdaBound{b}\AgdaSpace{}%
\AgdaOperator{\AgdaInductiveConstructor{,}}\AgdaSpace{}%
\AgdaInductiveConstructor{evalApp}\AgdaSpace{}%
\AgdaBound{r⇓}\AgdaSpace{}%
\AgdaBound{s⇓}\AgdaSpace{}%
\AgdaBound{app-eval}\AgdaSpace{}%
\AgdaOperator{\AgdaInductiveConstructor{,}}\AgdaSpace{}%
\AgdaBound{sb}\<%
\end{code}

\subsection{Semantically typed domain elements inhabit the candidate space}

The top and bottom of the candidate space are chosen such that a few key properties
hold, by using these properties we are able to prove that we can read back a normal
form from an evaluated term if it is semantically typed. The first property (Figure \ref{fig:prop1}) is that the
bottom of the candidate space \AgdaFunction{𝔹} is a subset of the semantic function space
\mbox{\AgdaFunction{𝕋} \AgdaFunction{⟶} \AgdaFunction{𝔹}}. Intuitively, this property
holds because for any neutral domain element \mbox{\AgdaBound{e} \AgdaFunction{∈}
\AgdaFunction{𝔹}} we can apply it to any domain element \mbox{\AgdaBound{d}
\AgdaFunction{∈} \AgdaFunction{𝕋}} and have a neutral domain element that can
still have a neutral term read back (the application of each domain element's
corresponding term that is read back).

\begin{figure}
\begin{code}%
\>[2]\AgdaFunction{𝔹⊆𝕋⟶𝔹}\AgdaSpace{}%
\AgdaSymbol{:}\AgdaSpace{}%
\AgdaOperator{\AgdaInductiveConstructor{`}}\AgdaSpace{}%
\AgdaGeneralizable{e}\AgdaSpace{}%
\AgdaOperator{\AgdaFunction{∈}}\AgdaSpace{}%
\AgdaFunction{𝔹}\AgdaSpace{}%
\AgdaSymbol{→}\AgdaSpace{}%
\AgdaOperator{\AgdaInductiveConstructor{`}}\AgdaSpace{}%
\AgdaGeneralizable{e}\AgdaSpace{}%
\AgdaOperator{\AgdaFunction{∈}}\AgdaSpace{}%
\AgdaFunction{𝕋}\AgdaSpace{}%
\AgdaOperator{\AgdaFunction{⟶}}\AgdaSpace{}%
\AgdaFunction{𝔹}\<%
\\
\>[2]\AgdaFunction{𝔹⊆𝕋⟶𝔹}\AgdaSpace{}%
\AgdaSymbol{\{}\AgdaBound{e}\AgdaSymbol{\}}\AgdaSpace{}%
\AgdaBound{eb}\AgdaSpace{}%
\AgdaSymbol{\{}\AgdaBound{d}\AgdaSymbol{\}}\AgdaSpace{}%
\AgdaBound{dt}\AgdaSpace{}%
\AgdaSymbol{=}\<%
\\
\>[2][@{}l@{\AgdaIndent{0}}]%
\>[4]\AgdaOperator{\AgdaInductiveConstructor{`}}\AgdaSpace{}%
\AgdaSymbol{(}\AgdaBound{e}\AgdaSpace{}%
\AgdaOperator{\AgdaInductiveConstructor{·}}\AgdaSpace{}%
\AgdaBound{d}\AgdaSymbol{)}\AgdaSpace{}%
\AgdaOperator{\AgdaInductiveConstructor{,}}\<%
\\
\>[4]\AgdaInductiveConstructor{appNeutral}\AgdaSpace{}%
\AgdaOperator{\AgdaInductiveConstructor{,}}\<%
\\
\>[4]\AgdaSymbol{λ}%
\>[1639I]\AgdaBound{n}\AgdaSpace{}%
\AgdaSymbol{→}\<%
\\
\>[.][@{}l@{}]\<[1639I]%
\>[6]\AgdaKeyword{let}\AgdaSpace{}%
\AgdaSymbol{(}\AgdaBound{u}\AgdaSpace{}%
\AgdaOperator{\AgdaInductiveConstructor{,}}\AgdaSpace{}%
\AgdaBound{e⇑u}\AgdaSymbol{)}\AgdaSpace{}%
\AgdaSymbol{=}\AgdaSpace{}%
\AgdaBound{eb}\AgdaSpace{}%
\AgdaBound{n}\AgdaSpace{}%
\AgdaKeyword{in}\<%
\\
\>[6]\AgdaKeyword{let}\AgdaSpace{}%
\AgdaSymbol{(}\AgdaBound{v}\AgdaSpace{}%
\AgdaOperator{\AgdaInductiveConstructor{,}}\AgdaSpace{}%
\AgdaBound{d⇑v}\AgdaSymbol{)}\AgdaSpace{}%
\AgdaSymbol{=}\AgdaSpace{}%
\AgdaBound{dt}\AgdaSpace{}%
\AgdaBound{n}\AgdaSpace{}%
\AgdaKeyword{in}\<%
\\
\>[6]\AgdaBound{u}\AgdaSpace{}%
\AgdaOperator{\AgdaInductiveConstructor{·}}\AgdaSpace{}%
\AgdaBound{v}\AgdaSpace{}%
\AgdaOperator{\AgdaInductiveConstructor{,}}\AgdaSpace{}%
\AgdaInductiveConstructor{⇑app}\AgdaSpace{}%
\AgdaBound{e⇑u}\AgdaSpace{}%
\AgdaBound{d⇑v}\<%
\end{code}
\caption{First property of candidate space}
\label{fig:prop1}
\end{figure}

For the second property, we first prove a separate lemma that all variables are
in \AgdaFunction{𝔹}. This is because any de Brujin level can be read back to its
corresponding de Brujin index.

\begin{code}%
\>[2]\AgdaFunction{lvl∈𝔹}\AgdaSpace{}%
\AgdaSymbol{:}\AgdaSpace{}%
\AgdaSymbol{∀}\AgdaSpace{}%
\AgdaBound{k}\AgdaSpace{}%
\AgdaSymbol{→}\AgdaSpace{}%
\AgdaOperator{\AgdaInductiveConstructor{`}}\AgdaSpace{}%
\AgdaInductiveConstructor{lvl}\AgdaSpace{}%
\AgdaBound{k}\AgdaSpace{}%
\AgdaOperator{\AgdaFunction{∈}}\AgdaSpace{}%
\AgdaFunction{𝔹}\<%
\\
\>[2]\AgdaFunction{lvl∈𝔹}\AgdaSpace{}%
\AgdaBound{k}\AgdaSpace{}%
\AgdaBound{n}\AgdaSpace{}%
\AgdaSymbol{=}\AgdaSpace{}%
\AgdaInductiveConstructor{var}\AgdaSpace{}%
\AgdaSymbol{(}\AgdaFunction{lvl→idx}\AgdaSpace{}%
\AgdaBound{k}\AgdaSpace{}%
\AgdaBound{n}\AgdaSymbol{)}\AgdaSpace{}%
\AgdaOperator{\AgdaInductiveConstructor{,}}\AgdaSpace{}%
\AgdaInductiveConstructor{⇑lvl}\<%
\end{code}

The second property (Figure \ref{fig:prop2}) is that any object in the semantic function space
\mbox{\AgdaFunction{𝔹} \AgdaFunction{⟶} \AgdaFunction{𝕋}} is a subset of the top
of the candidate space \AgdaFunction{𝕋}. For any domain element \mbox{\AgdaBound{d}
\AgdaFunction{∈} \AgdaFunction{𝔹} \AgdaFunction{⟶} \AgdaFunction{𝕋}}, we have by
the properties of the semantic function space that the evaluation of its application
to the first variable not in scope (which is in \AgdaFunction{𝔹}) can have a normal
form read back. We do a case analysis on the evaluation of this application to
determine the normal form of the domain element \AgdaBound{d}.

\begin{figure}
\begin{code}%
\>[2]\AgdaFunction{𝔹⟶𝕋⊆𝕋}\AgdaSpace{}%
\AgdaSymbol{:}\AgdaSpace{}%
\AgdaGeneralizable{d}\AgdaSpace{}%
\AgdaOperator{\AgdaFunction{∈}}\AgdaSpace{}%
\AgdaFunction{𝔹}\AgdaSpace{}%
\AgdaOperator{\AgdaFunction{⟶}}\AgdaSpace{}%
\AgdaFunction{𝕋}\AgdaSpace{}%
\AgdaSymbol{→}\AgdaSpace{}%
\AgdaGeneralizable{d}\AgdaSpace{}%
\AgdaOperator{\AgdaFunction{∈}}\AgdaSpace{}%
\AgdaFunction{𝕋}\<%
\\
\>[2]\AgdaFunction{𝔹⟶𝕋⊆𝕋}\AgdaSpace{}%
\AgdaSymbol{\{}\AgdaOperator{\AgdaInductiveConstructor{⟨ƛ}}\AgdaSpace{}%
\AgdaBound{t}\AgdaSpace{}%
\AgdaOperator{\AgdaInductiveConstructor{⟩}}\AgdaSpace{}%
\AgdaBound{γ}\AgdaSymbol{\}}\AgdaSpace{}%
\AgdaBound{pf}\AgdaSpace{}%
\AgdaBound{n}\<%
\\
\>[2][@{}l@{\AgdaIndent{0}}]%
\>[4]\AgdaKeyword{with}\AgdaSpace{}%
\AgdaBound{pf}\AgdaSpace{}%
\AgdaSymbol{(}\AgdaFunction{lvl∈𝔹}\AgdaSpace{}%
\AgdaBound{n}\AgdaSymbol{)}\<%
\\
\>[2]\AgdaSymbol{...}\AgdaSpace{}%
\AgdaSymbol{|}\AgdaSpace{}%
\AgdaBound{d}\AgdaSpace{}%
\AgdaOperator{\AgdaInductiveConstructor{,}}\AgdaSpace{}%
\AgdaInductiveConstructor{appClosure}\AgdaSpace{}%
\AgdaBound{eval-closure}\AgdaSpace{}%
\AgdaOperator{\AgdaInductiveConstructor{,}}\AgdaSpace{}%
\AgdaBound{dt}\<%
\\
\>[2][@{}l@{\AgdaIndent{0}}]%
\>[4]\AgdaKeyword{with}\AgdaSpace{}%
\AgdaBound{dt}\AgdaSpace{}%
\AgdaBound{n}\<%
\\
\>[2]\AgdaSymbol{...}\AgdaSpace{}%
\AgdaSymbol{|}\AgdaSpace{}%
\AgdaBound{v}\AgdaSpace{}%
\AgdaOperator{\AgdaInductiveConstructor{,}}\AgdaSpace{}%
\AgdaBound{d⇑v}\AgdaSpace{}%
\AgdaSymbol{=}\<%
\\
\>[2][@{}l@{\AgdaIndent{0}}]%
\>[4]\AgdaOperator{\AgdaInductiveConstructor{ƛ}}\AgdaSpace{}%
\AgdaBound{v}\AgdaSpace{}%
\AgdaOperator{\AgdaInductiveConstructor{,}}\AgdaSpace{}%
\AgdaInductiveConstructor{⇑closure}\AgdaSpace{}%
\AgdaBound{eval-closure}\AgdaSpace{}%
\AgdaBound{d⇑v}\<%
\\
\>[2]\AgdaFunction{𝔹⟶𝕋⊆𝕋}\AgdaSpace{}%
\AgdaSymbol{\{}\AgdaOperator{\AgdaInductiveConstructor{`}}\AgdaSpace{}%
\AgdaBound{e}\AgdaSymbol{\}}\AgdaSpace{}%
\AgdaBound{pf}\AgdaSpace{}%
\AgdaBound{n}\<%
\\
\>[2][@{}l@{\AgdaIndent{0}}]%
\>[4]\AgdaKeyword{with}\AgdaSpace{}%
\AgdaBound{pf}\AgdaSpace{}%
\AgdaSymbol{(}\AgdaFunction{lvl∈𝔹}\AgdaSpace{}%
\AgdaBound{n}\AgdaSymbol{)}\<%
\\
\>[2]\AgdaSymbol{...}\AgdaSpace{}%
\AgdaSymbol{|}\AgdaSpace{}%
\AgdaSymbol{\AgdaUnderscore{}}\AgdaSpace{}%
\AgdaOperator{\AgdaInductiveConstructor{,}}\AgdaSpace{}%
\AgdaInductiveConstructor{appNeutral}\AgdaSpace{}%
\AgdaOperator{\AgdaInductiveConstructor{,}}\AgdaSpace{}%
\AgdaBound{et}\<%
\\
\>[2][@{}l@{\AgdaIndent{0}}]%
\>[4]\AgdaKeyword{with}\AgdaSpace{}%
\AgdaBound{et}\AgdaSpace{}%
\AgdaBound{n}\<%
\\
\>[2]\AgdaSymbol{...}\AgdaSpace{}%
\AgdaSymbol{|}\AgdaSpace{}%
\AgdaBound{u}\AgdaSpace{}%
\AgdaOperator{\AgdaInductiveConstructor{·}}\AgdaSpace{}%
\AgdaBound{v}\AgdaSpace{}%
\AgdaOperator{\AgdaInductiveConstructor{,}}\AgdaSpace{}%
\AgdaInductiveConstructor{⇑neutral}\AgdaSpace{}%
\AgdaSymbol{(}\AgdaInductiveConstructor{⇑app}\AgdaSpace{}%
\AgdaBound{e⇑u}\AgdaSpace{}%
\AgdaSymbol{\AgdaUnderscore{})}\AgdaSpace{}%
\AgdaSymbol{=}\<%
\\
\>[2][@{}l@{\AgdaIndent{0}}]%
\>[4]\AgdaBound{u}\AgdaSpace{}%
\AgdaOperator{\AgdaInductiveConstructor{,}}\AgdaSpace{}%
\AgdaInductiveConstructor{⇑neutral}\AgdaSpace{}%
\AgdaBound{e⇑u}\<%
\end{code}
\caption{Second property of candidate space}
\label{fig:prop2}
\end{figure}

Using these properties, we are now able to prove that semantically typed domain
elements inhabit the candidate space; for any type \AgdaBound{T}, we have that
\mbox{\AgdaFunction{⟦} \AgdaBound{T} \AgdaFunction{⟧}} is a subset of \AgdaFunction{𝕋}.

The result follows in Figure \ref{sec:sem->nf} from the properties we have shown
so far along with two lemmas we prove mutually. The first is that
\mbox{\AgdaFunction{𝕋} \AgdaFunction{⟶} \AgdaFunction{𝔹}} is a subset of
\mbox{\AgdaFunction{⟦} \AgdaBound{S} \AgdaInductiveConstructor{⇒} \AgdaBound{T}
\AgdaFunction{⟧}}, which we use to prove that \AgdaFunction{𝔹} is a subset of
any semantic type. The second is that \mbox{\AgdaFunction{⟦} \AgdaBound{S}
\AgdaInductiveConstructor{⇒} \AgdaBound{T} \AgdaFunction{⟧}} is a subset of
\mbox{\AgdaFunction{𝔹} \AgdaFunction{⟶} \AgdaFunction{𝕋}}, which we use to prove
that any semantic type is a subset of \AgdaFunction{𝕋}.

\begin{figure}
\begin{code}%
\>[2]\AgdaKeyword{mutual}\<%
\\
\>[2][@{}l@{\AgdaIndent{0}}]%
\>[4]\AgdaOperator{\AgdaFunction{𝕋⟶𝔹⊆⟦\AgdaUnderscore{}⇒\AgdaUnderscore{}⟧}}\AgdaSpace{}%
\AgdaSymbol{:}\AgdaSpace{}%
\AgdaSymbol{∀}\AgdaSpace{}%
\AgdaBound{S}\AgdaSpace{}%
\AgdaBound{T}\AgdaSpace{}%
\AgdaSymbol{→}\AgdaSpace{}%
\AgdaGeneralizable{f}\AgdaSpace{}%
\AgdaOperator{\AgdaFunction{∈}}\AgdaSpace{}%
\AgdaFunction{𝕋}\AgdaSpace{}%
\AgdaOperator{\AgdaFunction{⟶}}\AgdaSpace{}%
\AgdaFunction{𝔹}\AgdaSpace{}%
\AgdaSymbol{→}\AgdaSpace{}%
\AgdaGeneralizable{f}\AgdaSpace{}%
\AgdaOperator{\AgdaFunction{∈}}\AgdaSpace{}%
\AgdaOperator{\AgdaFunction{⟦}}\AgdaSpace{}%
\AgdaBound{S}\AgdaSpace{}%
\AgdaOperator{\AgdaInductiveConstructor{⇒}}\AgdaSpace{}%
\AgdaBound{T}\AgdaSpace{}%
\AgdaOperator{\AgdaFunction{⟧}}\<%
\\
\>[4]\AgdaOperator{\AgdaFunction{𝕋⟶𝔹⊆⟦}}\AgdaSpace{}%
\AgdaBound{S}\AgdaSpace{}%
\AgdaOperator{\AgdaFunction{⇒}}\AgdaSpace{}%
\AgdaBound{T}\AgdaSpace{}%
\AgdaOperator{\AgdaFunction{⟧}}\AgdaSpace{}%
\AgdaBound{pf}\AgdaSpace{}%
\AgdaBound{sa}\<%
\\
\>[4][@{}l@{\AgdaIndent{0}}]%
\>[6]\AgdaKeyword{with}\AgdaSpace{}%
\AgdaOperator{\AgdaFunction{⟦}}\AgdaSpace{}%
\AgdaBound{S}\AgdaSpace{}%
\AgdaOperator{\AgdaFunction{⟧⊆𝕋}}\AgdaSpace{}%
\AgdaBound{sa}\<%
\\
\>[4]\AgdaSymbol{...}\AgdaSpace{}%
\AgdaSymbol{|}\AgdaSpace{}%
\AgdaBound{at}\<%
\\
\>[4][@{}l@{\AgdaIndent{0}}]%
\>[6]\AgdaKeyword{with}\AgdaSpace{}%
\AgdaBound{pf}\AgdaSpace{}%
\AgdaBound{at}\<%
\\
\>[4]\AgdaSymbol{...}\AgdaSpace{}%
\AgdaSymbol{|}\AgdaSpace{}%
\AgdaOperator{\AgdaInductiveConstructor{`}}\AgdaSpace{}%
\AgdaBound{e}\AgdaSpace{}%
\AgdaOperator{\AgdaInductiveConstructor{,}}\AgdaSpace{}%
\AgdaBound{app-eval}\AgdaSpace{}%
\AgdaOperator{\AgdaInductiveConstructor{,}}\AgdaSpace{}%
\AgdaBound{eb}\<%
\\
\>[4][@{}l@{\AgdaIndent{0}}]%
\>[6]\AgdaKeyword{with}\AgdaSpace{}%
\AgdaOperator{\AgdaFunction{𝔹⊆⟦}}\AgdaSpace{}%
\AgdaBound{T}\AgdaSpace{}%
\AgdaOperator{\AgdaFunction{⟧}}\AgdaSpace{}%
\AgdaBound{eb}\<%
\\
\>[4]\AgdaSymbol{...}\AgdaSpace{}%
\AgdaSymbol{|}\AgdaSpace{}%
\AgdaBound{se}\AgdaSpace{}%
\AgdaSymbol{=}\<%
\\
\>[4][@{}l@{\AgdaIndent{0}}]%
\>[6]\AgdaOperator{\AgdaInductiveConstructor{`}}\AgdaSpace{}%
\AgdaBound{e}\AgdaSpace{}%
\AgdaOperator{\AgdaInductiveConstructor{,}}\AgdaSpace{}%
\AgdaBound{app-eval}\AgdaSpace{}%
\AgdaOperator{\AgdaInductiveConstructor{,}}\AgdaSpace{}%
\AgdaBound{se}\<%
\\
\\[\AgdaEmptyExtraSkip]%
\>[4]\AgdaOperator{\AgdaFunction{⟦\AgdaUnderscore{}⇒\AgdaUnderscore{}⟧⊆𝔹⟶𝕋}}\AgdaSpace{}%
\AgdaSymbol{:}\AgdaSpace{}%
\AgdaSymbol{∀}\AgdaSpace{}%
\AgdaBound{S}\AgdaSpace{}%
\AgdaBound{T}\AgdaSpace{}%
\AgdaSymbol{→}\AgdaSpace{}%
\AgdaGeneralizable{f}\AgdaSpace{}%
\AgdaOperator{\AgdaFunction{∈}}\AgdaSpace{}%
\AgdaOperator{\AgdaFunction{⟦}}\AgdaSpace{}%
\AgdaBound{S}\AgdaSpace{}%
\AgdaOperator{\AgdaInductiveConstructor{⇒}}\AgdaSpace{}%
\AgdaBound{T}\AgdaSpace{}%
\AgdaOperator{\AgdaFunction{⟧}}\AgdaSpace{}%
\AgdaSymbol{→}\AgdaSpace{}%
\AgdaGeneralizable{f}\AgdaSpace{}%
\AgdaOperator{\AgdaFunction{∈}}\AgdaSpace{}%
\AgdaFunction{𝔹}\AgdaSpace{}%
\AgdaOperator{\AgdaFunction{⟶}}\AgdaSpace{}%
\AgdaFunction{𝕋}\<%
\\
\>[4]\AgdaOperator{\AgdaFunction{⟦}}%
\>[1814I]\AgdaBound{S}\AgdaSpace{}%
\AgdaOperator{\AgdaFunction{⇒}}\AgdaSpace{}%
\AgdaBound{T}\AgdaSpace{}%
\AgdaOperator{\AgdaFunction{⟧⊆𝔹⟶𝕋}}\AgdaSpace{}%
\AgdaBound{sf}\AgdaSpace{}%
\AgdaSymbol{\{}\AgdaOperator{\AgdaInductiveConstructor{`}}\AgdaSpace{}%
\AgdaBound{e}\AgdaSymbol{\}}\AgdaSpace{}%
\AgdaBound{eb}\<%
\\
\>[.][@{}l@{}]\<[1814I]%
\>[6]\AgdaKeyword{with}\AgdaSpace{}%
\AgdaBound{sf}\AgdaSpace{}%
\AgdaSymbol{(}\AgdaOperator{\AgdaFunction{𝔹⊆⟦}}\AgdaSpace{}%
\AgdaBound{S}\AgdaSpace{}%
\AgdaOperator{\AgdaFunction{⟧}}\AgdaSpace{}%
\AgdaBound{eb}\AgdaSymbol{)}\<%
\\
\>[4]\AgdaSymbol{...}\AgdaSpace{}%
\AgdaSymbol{|}\AgdaSpace{}%
\AgdaBound{d}\AgdaSpace{}%
\AgdaOperator{\AgdaInductiveConstructor{,}}\AgdaSpace{}%
\AgdaBound{app-eval}\AgdaSpace{}%
\AgdaOperator{\AgdaInductiveConstructor{,}}\AgdaSpace{}%
\AgdaBound{sd}\AgdaSpace{}%
\AgdaSymbol{=}\<%
\\
\>[4][@{}l@{\AgdaIndent{0}}]%
\>[6]\AgdaBound{d}\AgdaSpace{}%
\AgdaOperator{\AgdaInductiveConstructor{,}}%
\>[11]\AgdaBound{app-eval}\AgdaSpace{}%
\AgdaOperator{\AgdaInductiveConstructor{,}}\AgdaSpace{}%
\AgdaOperator{\AgdaFunction{⟦}}\AgdaSpace{}%
\AgdaBound{T}\AgdaSpace{}%
\AgdaOperator{\AgdaFunction{⟧⊆𝕋}}\AgdaSpace{}%
\AgdaBound{sd}\<%
\\
\\[\AgdaEmptyExtraSkip]%
\>[4]\AgdaComment{--\ Any\ object\ that\ can\ have\ a\ neutral\ form\ read\ back}\<%
\\
\>[4]\AgdaComment{--\ is\ semantically\ typed}\<%
\\
\>[4]\AgdaOperator{\AgdaFunction{𝔹⊆⟦\AgdaUnderscore{}⟧}}\AgdaSpace{}%
\AgdaSymbol{:}\AgdaSpace{}%
\AgdaSymbol{∀}\AgdaSpace{}%
\AgdaBound{T}\AgdaSpace{}%
\AgdaSymbol{→}\AgdaSpace{}%
\AgdaOperator{\AgdaInductiveConstructor{`}}\AgdaSpace{}%
\AgdaGeneralizable{e}\AgdaSpace{}%
\AgdaOperator{\AgdaFunction{∈}}\AgdaSpace{}%
\AgdaFunction{𝔹}\AgdaSpace{}%
\AgdaSymbol{→}\AgdaSpace{}%
\AgdaOperator{\AgdaInductiveConstructor{`}}\AgdaSpace{}%
\AgdaGeneralizable{e}\AgdaSpace{}%
\AgdaOperator{\AgdaFunction{∈}}\AgdaSpace{}%
\AgdaOperator{\AgdaFunction{⟦}}\AgdaSpace{}%
\AgdaBound{T}\AgdaSpace{}%
\AgdaOperator{\AgdaFunction{⟧}}\<%
\\
\>[4]\AgdaOperator{\AgdaFunction{𝔹⊆⟦}}\AgdaSpace{}%
\AgdaInductiveConstructor{base}\AgdaSpace{}%
\AgdaOperator{\AgdaFunction{⟧}}\AgdaSpace{}%
\AgdaBound{eb}\AgdaSpace{}%
\AgdaSymbol{=}\AgdaSpace{}%
\AgdaBound{eb}\<%
\\
\>[4]\AgdaOperator{\AgdaFunction{𝔹⊆⟦}}\AgdaSpace{}%
\AgdaBound{S}\AgdaSpace{}%
\AgdaOperator{\AgdaInductiveConstructor{⇒}}\AgdaSpace{}%
\AgdaBound{T}\AgdaSpace{}%
\AgdaOperator{\AgdaFunction{⟧}}\AgdaSpace{}%
\AgdaSymbol{=}\AgdaSpace{}%
\AgdaOperator{\AgdaFunction{𝕋⟶𝔹⊆⟦}}\AgdaSpace{}%
\AgdaBound{S}\AgdaSpace{}%
\AgdaOperator{\AgdaFunction{⇒}}\AgdaSpace{}%
\AgdaBound{T}\AgdaSpace{}%
\AgdaOperator{\AgdaFunction{⟧}}\AgdaSpace{}%
\AgdaOperator{\AgdaFunction{∘}}\AgdaSpace{}%
\AgdaFunction{𝔹⊆𝕋⟶𝔹}\<%
\\
\\[\AgdaEmptyExtraSkip]%
\>[4]\AgdaComment{--\ Semantic\ typing\ implies\ a\ normal\ form\ can\ be\ read}\<%
\\
\>[4]\AgdaComment{--\ back}\<%
\\
\>[4]\AgdaOperator{\AgdaFunction{⟦\AgdaUnderscore{}⟧⊆𝕋}}\AgdaSpace{}%
\AgdaSymbol{:}\AgdaSpace{}%
\AgdaSymbol{∀}\AgdaSpace{}%
\AgdaBound{T}\AgdaSpace{}%
\AgdaSymbol{→}\AgdaSpace{}%
\AgdaGeneralizable{d}\AgdaSpace{}%
\AgdaOperator{\AgdaFunction{∈}}\AgdaSpace{}%
\AgdaOperator{\AgdaFunction{⟦}}\AgdaSpace{}%
\AgdaBound{T}\AgdaSpace{}%
\AgdaOperator{\AgdaFunction{⟧}}\AgdaSpace{}%
\AgdaSymbol{→}\AgdaSpace{}%
\AgdaGeneralizable{d}\AgdaSpace{}%
\AgdaOperator{\AgdaFunction{∈}}\AgdaSpace{}%
\AgdaFunction{𝕋}\<%
\\
\>[4]\AgdaOperator{\AgdaFunction{⟦}}\AgdaSpace{}%
\AgdaInductiveConstructor{base}\AgdaSpace{}%
\AgdaOperator{\AgdaFunction{⟧⊆𝕋}}\AgdaSpace{}%
\AgdaSymbol{=}\AgdaSpace{}%
\AgdaFunction{𝔹⊆𝕋}\<%
\\
\>[4]\AgdaOperator{\AgdaFunction{⟦}}\AgdaSpace{}%
\AgdaBound{S}\AgdaSpace{}%
\AgdaOperator{\AgdaInductiveConstructor{⇒}}\AgdaSpace{}%
\AgdaBound{T}\AgdaSpace{}%
\AgdaOperator{\AgdaFunction{⟧⊆𝕋}}\AgdaSpace{}%
\AgdaSymbol{=}\AgdaSpace{}%
\AgdaFunction{𝔹⟶𝕋⊆𝕋}\AgdaSpace{}%
\AgdaOperator{\AgdaFunction{∘}}\AgdaSpace{}%
\AgdaOperator{\AgdaFunction{⟦}}\AgdaSpace{}%
\AgdaBound{S}\AgdaSpace{}%
\AgdaOperator{\AgdaFunction{⇒}}\AgdaSpace{}%
\AgdaBound{T}\AgdaSpace{}%
\AgdaOperator{\AgdaFunction{⟧⊆𝔹⟶𝕋}}\<%
\end{code}
\caption{Semantic types inhabit the candidate space}
\label{sec:sem->nf}
\end{figure}

\subsection{Establishing normalization}
\begin{figure}
\begin{code}%
\>[2]\AgdaOperator{\AgdaFunction{∣\AgdaUnderscore{}∣}}\AgdaSpace{}%
\AgdaSymbol{:}\AgdaSpace{}%
\AgdaDatatype{Ctx}\AgdaSpace{}%
\AgdaSymbol{→}\AgdaSpace{}%
\AgdaDatatype{ℕ}\<%
\\
\>[2]\AgdaOperator{\AgdaFunction{∣}}\AgdaSpace{}%
\AgdaInductiveConstructor{∅}\AgdaSpace{}%
\AgdaOperator{\AgdaFunction{∣}}\AgdaSpace{}%
\AgdaSymbol{=}\AgdaSpace{}%
\AgdaInductiveConstructor{zero}\<%
\\
\>[2]\AgdaOperator{\AgdaFunction{∣}}\AgdaSpace{}%
\AgdaBound{Γ}\AgdaSpace{}%
\AgdaOperator{\AgdaInductiveConstructor{·:}}\AgdaSpace{}%
\AgdaSymbol{\AgdaUnderscore{}}\AgdaSpace{}%
\AgdaOperator{\AgdaFunction{∣}}\AgdaSpace{}%
\AgdaSymbol{=}\AgdaSpace{}%
\AgdaInductiveConstructor{suc}\AgdaSpace{}%
\AgdaOperator{\AgdaFunction{∣}}\AgdaSpace{}%
\AgdaBound{Γ}\AgdaSpace{}%
\AgdaOperator{\AgdaFunction{∣}}\<%
\\
\\[\AgdaEmptyExtraSkip]%
\>[2]\AgdaOperator{\AgdaFunction{⊨env∣\AgdaUnderscore{}∣}}\AgdaSpace{}%
\AgdaSymbol{:}\AgdaSpace{}%
\AgdaSymbol{∀}\AgdaSpace{}%
\AgdaBound{Γ}\AgdaSpace{}%
\AgdaSymbol{→}\AgdaSpace{}%
\AgdaBound{Γ}\AgdaSpace{}%
\AgdaOperator{\AgdaFunction{⊨}}\AgdaSpace{}%
\AgdaFunction{env}\AgdaSpace{}%
\AgdaOperator{\AgdaFunction{∣}}\AgdaSpace{}%
\AgdaBound{Γ}\AgdaSpace{}%
\AgdaOperator{\AgdaFunction{∣}}\<%
\\
\>[2]\AgdaOperator{\AgdaFunction{⊨env∣\AgdaUnderscore{}∣}}\AgdaSpace{}%
\AgdaBound{Γ}\AgdaSpace{}%
\AgdaSymbol{\{}\AgdaBound{x}\AgdaSymbol{\}}\AgdaSpace{}%
\AgdaSymbol{\{}\AgdaBound{T}\AgdaSymbol{\}}\AgdaSpace{}%
\AgdaSymbol{\AgdaUnderscore{}}\AgdaSpace{}%
\AgdaSymbol{=}\<%
\\
\>[2][@{}l@{\AgdaIndent{0}}]%
\>[4]\AgdaOperator{\AgdaFunction{𝔹⊆⟦}}\AgdaSpace{}%
\AgdaBound{T}\AgdaSpace{}%
\AgdaOperator{\AgdaFunction{⟧}}\AgdaSpace{}%
\AgdaSymbol{(}\AgdaFunction{lvl∈𝔹}\AgdaSpace{}%
\AgdaSymbol{(}\AgdaFunction{idx→lvl}\AgdaSpace{}%
\AgdaBound{x}\AgdaSpace{}%
\AgdaOperator{\AgdaFunction{∣}}\AgdaSpace{}%
\AgdaBound{Γ}\AgdaSpace{}%
\AgdaOperator{\AgdaFunction{∣}}\AgdaSymbol{))}\<%
\\
\\[\AgdaEmptyExtraSkip]%
\>[2]\AgdaFunction{normalization}\AgdaSpace{}%
\AgdaSymbol{:}\AgdaSpace{}%
\AgdaGeneralizable{Γ}\AgdaSpace{}%
\AgdaOperator{\AgdaDatatype{⊢}}\AgdaSpace{}%
\AgdaGeneralizable{t}\AgdaSpace{}%
\AgdaOperator{\AgdaDatatype{∷}}\AgdaSpace{}%
\AgdaGeneralizable{T}\AgdaSpace{}%
\AgdaSymbol{→}\AgdaSpace{}%
\AgdaFunction{∃[}\AgdaSpace{}%
\AgdaBound{v}\AgdaSpace{}%
\AgdaFunction{]}\AgdaSpace{}%
\AgdaGeneralizable{t}\AgdaSpace{}%
\AgdaOperator{\AgdaFunction{has-normal-form<}}\AgdaSpace{}%
\AgdaOperator{\AgdaFunction{∣}}\AgdaSpace{}%
\AgdaGeneralizable{Γ}\AgdaSpace{}%
\AgdaOperator{\AgdaFunction{∣}}\AgdaSpace{}%
\AgdaOperator{\AgdaFunction{>}}\AgdaSpace{}%
\AgdaBound{v}\<%
\\
\>[2]\AgdaFunction{normalization}\AgdaSpace{}%
\AgdaSymbol{\{}\AgdaBound{Γ}\AgdaSymbol{\}}\AgdaSpace{}%
\AgdaSymbol{\{}\AgdaArgument{T}\AgdaSpace{}%
\AgdaSymbol{=}\AgdaSpace{}%
\AgdaBound{T}\AgdaSymbol{\}}\AgdaSpace{}%
\AgdaBound{⊢t}\<%
\\
\>[2][@{}l@{\AgdaIndent{0}}]%
\>[4]\AgdaKeyword{with}\AgdaSpace{}%
\AgdaFunction{fundamental-lemma}\AgdaSpace{}%
\AgdaBound{⊢t}\AgdaSpace{}%
\AgdaOperator{\AgdaFunction{⊨env∣}}\AgdaSpace{}%
\AgdaBound{Γ}\AgdaSpace{}%
\AgdaOperator{\AgdaFunction{∣}}\<%
\\
\>[2]\AgdaSymbol{...}\AgdaSpace{}%
\AgdaSymbol{|}\AgdaSpace{}%
\AgdaBound{a}\AgdaSpace{}%
\AgdaOperator{\AgdaInductiveConstructor{,}}\AgdaSpace{}%
\AgdaBound{t⇓a}\AgdaSpace{}%
\AgdaOperator{\AgdaInductiveConstructor{,}}\AgdaSpace{}%
\AgdaBound{st}\<%
\\
\>[2][@{}l@{\AgdaIndent{0}}]%
\>[4]\AgdaKeyword{with}\AgdaSpace{}%
\AgdaOperator{\AgdaFunction{⟦}}\AgdaSpace{}%
\AgdaBound{T}\AgdaSpace{}%
\AgdaOperator{\AgdaFunction{⟧⊆𝕋}}\AgdaSpace{}%
\AgdaBound{st}\AgdaSpace{}%
\AgdaOperator{\AgdaFunction{∣}}\AgdaSpace{}%
\AgdaBound{Γ}\AgdaSpace{}%
\AgdaOperator{\AgdaFunction{∣}}\<%
\\
\>[2]\AgdaSymbol{...}\AgdaSpace{}%
\AgdaSymbol{|}\AgdaSpace{}%
\AgdaBound{v}\AgdaSpace{}%
\AgdaOperator{\AgdaInductiveConstructor{,}}\AgdaSpace{}%
\AgdaBound{a⇑v}\AgdaSpace{}%
\AgdaSymbol{=}\<%
\\
\>[2][@{}l@{\AgdaIndent{0}}]%
\>[4]\AgdaBound{v}\AgdaSpace{}%
\AgdaOperator{\AgdaInductiveConstructor{,}}\AgdaSpace{}%
\AgdaBound{a}\AgdaSpace{}%
\AgdaOperator{\AgdaInductiveConstructor{,}}\AgdaSpace{}%
\AgdaBound{t⇓a}\AgdaSpace{}%
\AgdaOperator{\AgdaInductiveConstructor{,}}\AgdaSpace{}%
\AgdaBound{a⇑v}\<%
\end{code}
\caption{Normalization of STLC}
\label{fig:nbe}
\end{figure}
\end{AgdaAlign}
We prove normalization in Figure \ref{fig:nbe}. We use the fundamental lemma to
prove that the evaluation of a well-typed term in an environment of variables is
well-defined.

We prove that the environment of variables is semantically typed by using the
fact that all elements of \AgdaFunction{𝔹} are semantically typed, composed with
the fact all variables are elements of \AgdaFunction{𝔹}.

By the fundamental lemma, we also have that the evaluation of a well-typed term
is semantically typed. As we have just shown, this implies that we can read back
a normal form from the evaluation of the term.

We therefore prove normalization of STLC, as for any term that is well-typed its
evaluation is well-defined and we can read back a normal form from its
evaluation.

\section{Correctness}
\label{sec:correctness}
Our formalization has almost a direct correspondence to how it may be presented
in an informal mathematical presentation. For our proof that evaluation is total,
the main difference is that we use an intrinsic representation. The formalization
becomes more complicated when extended to prove normalization with full reduction,
though remains nearly as approachable as it would if presented informally. This
brings us to the question: is it okay to avoid substitutions as we have done here?

Natural semantics are a well established model for the semantics of programming languages,
so it makes sense to use a natural semantics to prove that evaluation of the simply
typed lambda calculus is total. However, in the case of 
full reduction, we use the evaluation relation more as a tool
to obtain the normal form than a semantic definition. In this sense, our proof
is more algorithmic, and indeed there are parts of the algorithm that may be
incorrect, such as the conversion between de Brujin indices and de Brujin
levels. The proof does not  verify that this conversion is correct, only that it
terminates. Showing that normalization is correct requires a separate proof.

For this reason, in future work we would like to show how we can additionally
prove that the technique of normalization by evaluation is both complete and
sound with respect to an equivalence model for the simply typed lambda calculus.
The completeness property states that if two terms are equivalent, they have the
same normal form. Soundness is that a term should be equivalent to its normal
form. Proving these two properties shows us the correctness of of normalization
by evaluation.

Abel proves that the technique is both complete and sound with respect to
$\beta\eta$-equivalence.  Doing so requires extending normalization by
evaluation to include explicit substitutions and $\eta$-expansions (for unique
$\eta$-long forms). After adding these extensions, the proof employs a Kripke
logical relation, used to reasons about future extended contexts.

In the case of our version, we would want to show that using evaluation to prove
normalization with full reduction is complete and sound with respect to
$\beta$-equivalence (as we are not computing $\eta$-long forms). Doing so,
however, would naturally involve reasoning about substitutions, which we have
explicitly avoided so far.

\begin{code}[hide]%
\>[0]\AgdaKeyword{module}\AgdaSpace{}%
\AgdaModule{Correctness}\AgdaSpace{}%
\AgdaKeyword{where}\<%
\\
\>[0][@{}l@{\AgdaIndent{0}}]%
\>[2]\AgdaKeyword{open}\AgdaSpace{}%
\AgdaModule{Totality}\<%
\end{code}

\begin{figure}
\begin{code}%
\>[2]\AgdaComment{--\ Denotation\ of\ types}\<%
\\
\>[2]\AgdaOperator{\AgdaFunction{𝒯⟦\AgdaUnderscore{}⟧}}\AgdaSpace{}%
\AgdaSymbol{:}\AgdaSpace{}%
\AgdaDatatype{Type}\AgdaSpace{}%
\AgdaSymbol{→}\AgdaSpace{}%
\AgdaPrimitive{Set}\<%
\\
\>[2]\AgdaOperator{\AgdaFunction{𝒯⟦}}\AgdaSpace{}%
\AgdaInductiveConstructor{bool}\AgdaSpace{}%
\AgdaOperator{\AgdaFunction{⟧}}\AgdaSpace{}%
\AgdaSymbol{=}\AgdaSpace{}%
\AgdaDatatype{Bool}%
\>[31]\AgdaComment{--\ Agda's\ boolean\ type}\<%
\\
\>[2]\AgdaOperator{\AgdaFunction{𝒯⟦}}\AgdaSpace{}%
\AgdaBound{S}\AgdaSpace{}%
\AgdaOperator{\AgdaInductiveConstructor{⇒}}\AgdaSpace{}%
\AgdaBound{T}\AgdaSpace{}%
\AgdaOperator{\AgdaFunction{⟧}}\AgdaSpace{}%
\AgdaSymbol{=}\AgdaSpace{}%
\AgdaOperator{\AgdaFunction{𝒯⟦}}\AgdaSpace{}%
\AgdaBound{S}\AgdaSpace{}%
\AgdaOperator{\AgdaFunction{⟧}}\AgdaSpace{}%
\AgdaSymbol{→}\AgdaSpace{}%
\AgdaOperator{\AgdaFunction{𝒯⟦}}\AgdaSpace{}%
\AgdaBound{T}\AgdaSpace{}%
\AgdaOperator{\AgdaFunction{⟧}}\<%
\\
\\[\AgdaEmptyExtraSkip]%
\>[2]\AgdaComment{--\ Denotation\ of\ contexts}\<%
\\
\>[2]\AgdaOperator{\AgdaFunction{𝒞⟦\AgdaUnderscore{}⟧}}\AgdaSpace{}%
\AgdaSymbol{:}\AgdaSpace{}%
\AgdaDatatype{Ctx}\AgdaSpace{}%
\AgdaSymbol{→}\AgdaSpace{}%
\AgdaPrimitive{Set}\<%
\\
\>[2]\AgdaOperator{\AgdaFunction{𝒞⟦}}\AgdaSpace{}%
\AgdaBound{Γ}\AgdaSpace{}%
\AgdaOperator{\AgdaFunction{⟧}}\AgdaSpace{}%
\AgdaSymbol{=}\AgdaSpace{}%
\AgdaSymbol{∀}\AgdaSpace{}%
\AgdaSymbol{\{}\AgdaBound{T}\AgdaSymbol{\}}\AgdaSpace{}%
\AgdaSymbol{→}\AgdaSpace{}%
\AgdaBound{Γ}\AgdaSpace{}%
\AgdaOperator{\AgdaDatatype{∋}}\AgdaSpace{}%
\AgdaBound{T}\AgdaSpace{}%
\AgdaSymbol{→}\AgdaSpace{}%
\AgdaOperator{\AgdaFunction{𝒯⟦}}\AgdaSpace{}%
\AgdaBound{T}\AgdaSpace{}%
\AgdaOperator{\AgdaFunction{⟧}}\<%
\\
\\[\AgdaEmptyExtraSkip]%
\>[2]\AgdaComment{--\ Extending\ context\ denotations}\<%
\\
\>[2]\AgdaOperator{\AgdaFunction{\AgdaUnderscore{}\&\AgdaUnderscore{}}}\AgdaSpace{}%
\AgdaSymbol{:}\AgdaSpace{}%
\AgdaOperator{\AgdaFunction{𝒞⟦}}\AgdaSpace{}%
\AgdaGeneralizable{Γ}\AgdaSpace{}%
\AgdaOperator{\AgdaFunction{⟧}}\AgdaSpace{}%
\AgdaSymbol{→}\AgdaSpace{}%
\AgdaOperator{\AgdaFunction{𝒯⟦}}\AgdaSpace{}%
\AgdaGeneralizable{T}\AgdaSpace{}%
\AgdaOperator{\AgdaFunction{⟧}}\AgdaSpace{}%
\AgdaSymbol{→}\AgdaSpace{}%
\AgdaOperator{\AgdaFunction{𝒞⟦}}\AgdaSpace{}%
\AgdaGeneralizable{Γ}\AgdaSpace{}%
\AgdaOperator{\AgdaInductiveConstructor{·:}}\AgdaSpace{}%
\AgdaGeneralizable{T}\AgdaSpace{}%
\AgdaOperator{\AgdaFunction{⟧}}\<%
\\
\>[2]\AgdaSymbol{(}\AgdaBound{ρ}\AgdaSpace{}%
\AgdaOperator{\AgdaFunction{\&}}\AgdaSpace{}%
\AgdaBound{z}\AgdaSymbol{)}\AgdaSpace{}%
\AgdaInductiveConstructor{zero}\AgdaSpace{}%
\AgdaSymbol{=}\AgdaSpace{}%
\AgdaBound{z}\<%
\\
\>[2]\AgdaSymbol{(}\AgdaBound{ρ}\AgdaSpace{}%
\AgdaOperator{\AgdaFunction{\&}}\AgdaSpace{}%
\AgdaBound{z}\AgdaSymbol{)}\AgdaSpace{}%
\AgdaSymbol{(}\AgdaInductiveConstructor{suc}\AgdaSpace{}%
\AgdaBound{x}\AgdaSymbol{)}\AgdaSpace{}%
\AgdaSymbol{=}\AgdaSpace{}%
\AgdaBound{ρ}\AgdaSpace{}%
\AgdaBound{x}\<%
\\
\\[\AgdaEmptyExtraSkip]%
\>[2]\AgdaComment{--\ Denotation\ of\ terms}\<%
\\
\>[2]\AgdaOperator{\AgdaFunction{ℰ⟦\AgdaUnderscore{}⟧}}\AgdaSpace{}%
\AgdaSymbol{:}\AgdaSpace{}%
\AgdaGeneralizable{Γ}\AgdaSpace{}%
\AgdaOperator{\AgdaDatatype{⊢}}\AgdaSpace{}%
\AgdaGeneralizable{T}\AgdaSpace{}%
\AgdaSymbol{→}\AgdaSpace{}%
\AgdaOperator{\AgdaFunction{𝒞⟦}}\AgdaSpace{}%
\AgdaGeneralizable{Γ}\AgdaSpace{}%
\AgdaOperator{\AgdaFunction{⟧}}\AgdaSpace{}%
\AgdaSymbol{→}\AgdaSpace{}%
\AgdaOperator{\AgdaFunction{𝒯⟦}}\AgdaSpace{}%
\AgdaGeneralizable{T}\AgdaSpace{}%
\AgdaOperator{\AgdaFunction{⟧}}\<%
\\
\>[2]\AgdaOperator{\AgdaFunction{ℰ⟦}}\AgdaSpace{}%
\AgdaInductiveConstructor{true}\AgdaSpace{}%
\AgdaOperator{\AgdaFunction{⟧}}\AgdaSpace{}%
\AgdaBound{ρ}\AgdaSpace{}%
\AgdaSymbol{=}\AgdaSpace{}%
\AgdaInductiveConstructor{true}\<%
\\
\>[2]\AgdaOperator{\AgdaFunction{ℰ⟦}}\AgdaSpace{}%
\AgdaInductiveConstructor{false}\AgdaSpace{}%
\AgdaOperator{\AgdaFunction{⟧}}\AgdaSpace{}%
\AgdaBound{ρ}\AgdaSpace{}%
\AgdaSymbol{=}\AgdaSpace{}%
\AgdaInductiveConstructor{false}\<%
\\
\>[2]\AgdaOperator{\AgdaFunction{ℰ⟦}}\AgdaSpace{}%
\AgdaInductiveConstructor{var}\AgdaSpace{}%
\AgdaBound{x}\AgdaSpace{}%
\AgdaOperator{\AgdaFunction{⟧}}\AgdaSpace{}%
\AgdaBound{ρ}\AgdaSpace{}%
\AgdaSymbol{=}\AgdaSpace{}%
\AgdaBound{ρ}\AgdaSpace{}%
\AgdaBound{x}\<%
\\
\>[2]\AgdaOperator{\AgdaFunction{ℰ⟦}}\AgdaSpace{}%
\AgdaOperator{\AgdaInductiveConstructor{ƛ}}\AgdaSpace{}%
\AgdaBound{t}\AgdaSpace{}%
\AgdaOperator{\AgdaFunction{⟧}}\AgdaSpace{}%
\AgdaBound{ρ}\AgdaSpace{}%
\AgdaSymbol{=}\AgdaSpace{}%
\AgdaSymbol{λ}\AgdaSpace{}%
\AgdaBound{z}\AgdaSpace{}%
\AgdaSymbol{→}\AgdaSpace{}%
\AgdaOperator{\AgdaFunction{ℰ⟦}}\AgdaSpace{}%
\AgdaBound{t}\AgdaSpace{}%
\AgdaOperator{\AgdaFunction{⟧}}\AgdaSpace{}%
\AgdaSymbol{(}\AgdaBound{ρ}\AgdaSpace{}%
\AgdaOperator{\AgdaFunction{\&}}\AgdaSpace{}%
\AgdaBound{z}\AgdaSymbol{)}\<%
\\
\>[2]\AgdaOperator{\AgdaFunction{ℰ⟦}}\AgdaSpace{}%
\AgdaBound{r}\AgdaSpace{}%
\AgdaOperator{\AgdaInductiveConstructor{·}}\AgdaSpace{}%
\AgdaBound{s}\AgdaSpace{}%
\AgdaOperator{\AgdaFunction{⟧}}\AgdaSpace{}%
\AgdaBound{ρ}\AgdaSpace{}%
\AgdaSymbol{=}\AgdaSpace{}%
\AgdaOperator{\AgdaFunction{ℰ⟦}}\AgdaSpace{}%
\AgdaBound{r}\AgdaSpace{}%
\AgdaOperator{\AgdaFunction{⟧}}\AgdaSpace{}%
\AgdaBound{ρ}\AgdaSpace{}%
\AgdaSymbol{(}\AgdaOperator{\AgdaFunction{ℰ⟦}}\AgdaSpace{}%
\AgdaBound{s}\AgdaSpace{}%
\AgdaOperator{\AgdaFunction{⟧}}\AgdaSpace{}%
\AgdaBound{ρ}\AgdaSymbol{)}\<%
\\
\>[2]\AgdaOperator{\AgdaFunction{ℰ⟦}}\AgdaSpace{}%
\AgdaOperator{\AgdaInductiveConstructor{if}}\AgdaSpace{}%
\AgdaBound{r}\AgdaSpace{}%
\AgdaOperator{\AgdaInductiveConstructor{then}}\AgdaSpace{}%
\AgdaBound{s}\AgdaSpace{}%
\AgdaOperator{\AgdaInductiveConstructor{else}}\AgdaSpace{}%
\AgdaBound{t}\AgdaSpace{}%
\AgdaOperator{\AgdaFunction{⟧}}\AgdaSpace{}%
\AgdaBound{ρ}\<%
\\
\>[2][@{}l@{\AgdaIndent{0}}]%
\>[4]\AgdaKeyword{with}\AgdaSpace{}%
\AgdaOperator{\AgdaFunction{ℰ⟦}}\AgdaSpace{}%
\AgdaBound{r}\AgdaSpace{}%
\AgdaOperator{\AgdaFunction{⟧}}\AgdaSpace{}%
\AgdaBound{ρ}\<%
\\
\>[2]\AgdaSymbol{...}\AgdaSpace{}%
\AgdaSymbol{|}\AgdaSpace{}%
\AgdaInductiveConstructor{true}\AgdaSpace{}%
\AgdaSymbol{=}\AgdaSpace{}%
\AgdaOperator{\AgdaFunction{ℰ⟦}}\AgdaSpace{}%
\AgdaBound{s}\AgdaSpace{}%
\AgdaOperator{\AgdaFunction{⟧}}\AgdaSpace{}%
\AgdaBound{ρ}\<%
\\
\>[2]\AgdaSymbol{...}\AgdaSpace{}%
\AgdaSymbol{|}\AgdaSpace{}%
\AgdaInductiveConstructor{false}\AgdaSpace{}%
\AgdaSymbol{=}\AgdaSpace{}%
\AgdaOperator{\AgdaFunction{ℰ⟦}}\AgdaSpace{}%
\AgdaBound{t}\AgdaSpace{}%
\AgdaOperator{\AgdaFunction{⟧}}\AgdaSpace{}%
\AgdaBound{ρ}\<%
\\
\\[\AgdaEmptyExtraSkip]%
\>[2]\AgdaKeyword{mutual}\<%
\\
\>[2][@{}l@{\AgdaIndent{0}}]%
\>[4]\AgdaComment{--\ Denotation\ of\ environments}\<%
\\
\>[4]\AgdaOperator{\AgdaFunction{𝒢⟦\AgdaUnderscore{}⟧}}\AgdaSpace{}%
\AgdaSymbol{:}\AgdaSpace{}%
\AgdaFunction{Env}\AgdaSpace{}%
\AgdaGeneralizable{Γ}\AgdaSpace{}%
\AgdaSymbol{→}\AgdaSpace{}%
\AgdaOperator{\AgdaFunction{𝒞⟦}}\AgdaSpace{}%
\AgdaGeneralizable{Γ}\AgdaSpace{}%
\AgdaOperator{\AgdaFunction{⟧}}\<%
\\
\>[4]\AgdaOperator{\AgdaFunction{𝒢⟦}}\AgdaSpace{}%
\AgdaBound{γ}\AgdaSpace{}%
\AgdaOperator{\AgdaFunction{⟧}}\AgdaSpace{}%
\AgdaBound{x}\AgdaSpace{}%
\AgdaSymbol{=}\AgdaSpace{}%
\AgdaOperator{\AgdaFunction{𝒟⟦}}\AgdaSpace{}%
\AgdaBound{γ}\AgdaSpace{}%
\AgdaBound{x}\AgdaSpace{}%
\AgdaOperator{\AgdaFunction{⟧}}\<%
\\
\\[\AgdaEmptyExtraSkip]%
\>[4]\AgdaComment{--\ Denotation\ of\ domain\ elements}\<%
\\
\>[4]\AgdaOperator{\AgdaFunction{𝒟⟦\AgdaUnderscore{}⟧}}\AgdaSpace{}%
\AgdaSymbol{:}\AgdaSpace{}%
\AgdaDatatype{Domain}\AgdaSpace{}%
\AgdaGeneralizable{T}\AgdaSpace{}%
\AgdaSymbol{→}\AgdaSpace{}%
\AgdaOperator{\AgdaFunction{𝒯⟦}}\AgdaSpace{}%
\AgdaGeneralizable{T}\AgdaSpace{}%
\AgdaOperator{\AgdaFunction{⟧}}\<%
\\
\>[4]\AgdaOperator{\AgdaFunction{𝒟⟦}}\AgdaSpace{}%
\AgdaInductiveConstructor{true}\AgdaSpace{}%
\AgdaOperator{\AgdaFunction{⟧}}\AgdaSpace{}%
\AgdaSymbol{=}\AgdaSpace{}%
\AgdaInductiveConstructor{true}\<%
\\
\>[4]\AgdaOperator{\AgdaFunction{𝒟⟦}}\AgdaSpace{}%
\AgdaInductiveConstructor{false}\AgdaSpace{}%
\AgdaOperator{\AgdaFunction{⟧}}\AgdaSpace{}%
\AgdaSymbol{=}\AgdaSpace{}%
\AgdaInductiveConstructor{false}\<%
\\
\>[4]\AgdaOperator{\AgdaFunction{𝒟⟦}}\AgdaSpace{}%
\AgdaOperator{\AgdaInductiveConstructor{⟨ƛ}}\AgdaSpace{}%
\AgdaBound{t}\AgdaSpace{}%
\AgdaOperator{\AgdaInductiveConstructor{⟩}}\AgdaSpace{}%
\AgdaBound{δ}\AgdaSpace{}%
\AgdaOperator{\AgdaFunction{⟧}}\AgdaSpace{}%
\AgdaSymbol{=}\AgdaSpace{}%
\AgdaSymbol{λ}\AgdaSpace{}%
\AgdaBound{z}\AgdaSpace{}%
\AgdaSymbol{→}\AgdaSpace{}%
\AgdaOperator{\AgdaFunction{ℰ⟦}}\AgdaSpace{}%
\AgdaBound{t}\AgdaSpace{}%
\AgdaOperator{\AgdaFunction{⟧}}\AgdaSpace{}%
\AgdaSymbol{(}\AgdaOperator{\AgdaFunction{𝒢⟦}}\AgdaSpace{}%
\AgdaBound{δ}\AgdaSpace{}%
\AgdaOperator{\AgdaFunction{⟧}}\AgdaSpace{}%
\AgdaOperator{\AgdaFunction{\&}}\AgdaSpace{}%
\AgdaBound{z}\AgdaSymbol{)}\<%
\\
\\[\AgdaEmptyExtraSkip]%
\>[2]\AgdaComment{--\ Denotational\ equivalence}\<%
\\
\>[2]\AgdaOperator{\AgdaFunction{\AgdaUnderscore{}ℰ≡\AgdaUnderscore{}}}\AgdaSpace{}%
\AgdaSymbol{:}\AgdaSpace{}%
\AgdaSymbol{∀}\AgdaSpace{}%
\AgdaSymbol{(}\AgdaBound{t}\AgdaSpace{}%
\AgdaBound{v}\AgdaSpace{}%
\AgdaSymbol{:}\AgdaSpace{}%
\AgdaGeneralizable{Γ}\AgdaSpace{}%
\AgdaOperator{\AgdaDatatype{⊢}}\AgdaSpace{}%
\AgdaGeneralizable{T}\AgdaSymbol{)}\AgdaSpace{}%
\AgdaSymbol{→}\AgdaSpace{}%
\AgdaPrimitive{Set}\<%
\\
\>[2]\AgdaOperator{\AgdaFunction{\AgdaUnderscore{}ℰ≡\AgdaUnderscore{}}}\AgdaSpace{}%
\AgdaSymbol{\{}\AgdaBound{Γ}\AgdaSymbol{\}}\AgdaSpace{}%
\AgdaBound{t}\AgdaSpace{}%
\AgdaBound{v}\AgdaSpace{}%
\AgdaSymbol{=}\AgdaSpace{}%
\AgdaSymbol{∀}\AgdaSpace{}%
\AgdaSymbol{\{}\AgdaBound{ρ}\AgdaSpace{}%
\AgdaSymbol{:}\AgdaSpace{}%
\AgdaOperator{\AgdaFunction{𝒞⟦}}\AgdaSpace{}%
\AgdaBound{Γ}\AgdaSpace{}%
\AgdaOperator{\AgdaFunction{⟧}}\AgdaSymbol{\}}\AgdaSpace{}%
\AgdaSymbol{→}\AgdaSpace{}%
\AgdaOperator{\AgdaFunction{ℰ⟦}}\AgdaSpace{}%
\AgdaBound{t}\AgdaSpace{}%
\AgdaOperator{\AgdaFunction{⟧}}\AgdaSpace{}%
\AgdaBound{ρ}\AgdaSpace{}%
\AgdaOperator{\AgdaDatatype{≡}}\AgdaSpace{}%
\AgdaOperator{\AgdaFunction{ℰ⟦}}\AgdaSpace{}%
\AgdaBound{v}\AgdaSpace{}%
\AgdaOperator{\AgdaFunction{⟧}}\AgdaSpace{}%
\AgdaBound{ρ}\<%
\end{code}
\caption{Denotational semantics}
\label{fig:denotational}
\end{figure}

Instead, we would consider a different model of equivalence: denotational
equivalence. Figure~\ref{fig:denotational} shows a denotational semantics of
STLC terms that interprets them as Agda values. Using this model, we wish to
show the soundness of normalization by proving that a term \AgdaBound{t} is
denotationally equivalent to its normal form \AgdaBound{v}:
\mbox{\AgdaBound{t} \AgdaFunction{ℰ≡} \AgdaBound{v}}.

We have proven that normalization using weak head reduction is sound. This
proof appears in appendix \ref{sec:intrinsic-full}. In this proof, we must
reason about substitution as we essentially implement the operation (which we
name \mbox{\AgdaFunction{close}}) to perform the delayed substitution
represented by a closure's saved environment.

However, we cannot prove completeness as weak head reduction does not reduce
under abstractions. Consider the terms $\lambda x. x$ and
$\lambda x. (\lambda y. y) x$, which are denotationally equivalent but do not
have the same normal form; both terms are already in their normal form.

In future work, we would like to extend this proof to show the soundness of
proving normalization under full reduction by evaluation. We conjecture that we
may not need to reason about substitutions to prove this result, as in going from
weak head reduction to full reduction we were able to avoid having to implement the
operation. This would make for another valuable proof by logical relations that
did not require substitution lemmas.

\section{Related Work}
\emph{Teaching proofs by logical relations}. There are many  tutorials that
introduce the proof technique of logical relations. \citet{ahmed2013logical}
provides a general overview, discussing their use in proving several properties
of programs. \citet{harper2022wn} shows how Tait's method can can be used to
prove normalization with weak head reduction of the simply typed lambda calculus
and later extends the technique to prove normalization with full reductions
\cite{harper2022sn}. \citet{skorstengaard2019introduction} gives an in-depth
introduction, going on to show how to prove properties more complex than
normalization. Additionally, \citet{abel2019poplmark} provide a general tutorial
for proving strong normalization by logical relations in their proposal. These
tutorials continue to be excellent sources to familiarize oneself with the
technique of proofs by logical relations. Despite this, they are tricky to adapt
to a mechanized setting, as the technique of logical relations becomes obscured
by the many substitution lemmas required. We provide an introduction to logical
relations that is appropriate even when presented in a formalization.
\citet{cave2013first} also provide a mechanization of a proof of normalization
with weak head reduction that does not suffer from substitution lemmas, though
this is is because they mechanize the proof using Beluga \cite{beluga}. Beluga
is a programming language with direct support for reasoning about the binding of
variables using higher-order abstract syntax \cite{hoas}, and substitution
lemmas are handled automatically.

\emph{Normalization by evaluation}. \citet{martin1975intuitionistic}
first demonstrates normalization by a semantic argument, later coined
normalization by evaluation. \citet{coquand1997intuitionistic} expand on this
idea, observing that an algorithm for normalization by evaluation can be used
for a proof of normalization in a more traditional sense. They prove
normalization of STLC and additionally prove soundness of the technique.
Although they formalize their results in ALF, the proof is presented in plain text
for readability. With the capabilities of Agda for literate programming, and the
increased popularity of mechanized reasoning, we are able to present our proofs in
a formalized setting. \citet{wieczorek2018coq} formalize Abel's complete proof of
normalization by evaluation. They extend STLC with explicit substitutions to prove
the soundness and completeness with respect to $\beta\eta$-equivalence. We do not
use explicit substitutions, enabling us to draw a more direct comparison with our
initial proof that evaluation is total. \citet{sestini2019normalization} shows how
the technique of proving normalization with full reductions by evaluation can be
modified to prove normalization with weak $\lambda$-reduction. Weak
$\lambda$-reduction is an extension of weak head reduction that allows for
limited reductions in the bodies of abstractions, making the system confluent.
Formalized in Agda, their result is related to our proof of normalization with
weak head reduction in appendix \ref{sec:intrinsic-full}. We focus on weak head
reduction here as our goal is to show a full proof by logical relations that can
be used to introduce the technique.

\emph{Formalizing substitution lemmas}. There are many options for alleviating the
burden that is posed by substitution lemmas in a formalization. These range from
the automation of such proofs \cite[etc.]{autosubst,lngen} to having these
properties hold automatically, as \citet{cave2018mechanizing} show is possible
with Beluga. Specifically for Agda, \citet{allais2018type} introduce the
\texttt{generic-syntax} library for representing syntax that can prove generic
lemmas such as the fundamental lemma of logical relations. Although there has
been significant progress in aiding the mechanization of proofs requiring
substitution lemmas, it continues to be a topic of interest, with proofs by
logical relations recently being proposed as a benchmark for mechanizing
metatheory \cite{abel2019poplmark}.

\section{Conclusion}
We have demonstrated that proofs by logical relations do not always need substitution
lemmas. Consequently, the technique can be taught in a mechanized setting without
being obscured by the many lemmas typically required. We prove that evaluation is
total for the simply typed lambda calculus. Formalized in Agda, the proof by
logical relations is remarkably short and remains as readable as it might in an
informal setting. The result is a proof that makes for an appropriate introduction
to the technique of logical relations, even when teaching programming language
foundations through the use of a proof assistant as has become common.

We extend our proof to prove the more common property of normalization of the
simply typed lambda calculus. We prove normalization with full reductions by
formalizing the work of \citet{nbe}. The proof shows how we can establish richer
properties of programs by logical relations while still avoiding substitution
lemmas. We do not formally show that our proof of normalization with full
reductions is correct, though refer to \citet{wieczorek2018coq}, who formalize
Abel's full proof in Coq.

We hope that in presenting both proofs we are making logical relations more
relatable in mechanized settings, and that future work can expand on how they
can be used to prove other properties of programs with logical relations without
needing substitution lemmas.

\ifanonymous
\else
\begin{acks}
  This work has been partially supported by the National Science Foundation
  under grant NSF 2006535.
\end{acks}
\fi

\bibliographystyle{ACM-Reference-Format}
\bibliography{references}

\appendix

\section{Extending $\S$\ref{sec:totality}: Normalization With Weak Head Reduction}
\label{sec:intrinsic-full}

We show in this section how to extend our proof of the totality of evaluation
for the simply typed lambda calculus in \ref{sec:totality} to prove normalization
with weak head reduction. Most of the content is identical to the figures presented
in the paper, we only extend it to show that we can read back a normal form from a
term that is well-typed.

We additionally show that the normal form that we read back is denotationally
equivalent to the original term, thereby proving the soundness of the technique.

\begin{code}[hide]%
\>[0]\AgdaKeyword{module}\AgdaSpace{}%
\AgdaModule{IntrinsicWeakHeadNormalization}\AgdaSpace{}%
\AgdaKeyword{where}\<%
\\
\>[0][@{}l@{\AgdaIndent{0}}]%
\>[2]\AgdaKeyword{infixr}\AgdaSpace{}%
\AgdaNumber{7}\AgdaSpace{}%
\AgdaOperator{\AgdaInductiveConstructor{\AgdaUnderscore{}⇒\AgdaUnderscore{}}}\<%
\\
\>[2]\AgdaKeyword{infixl}\AgdaSpace{}%
\AgdaNumber{5}\AgdaSpace{}%
\AgdaOperator{\AgdaInductiveConstructor{\AgdaUnderscore{}·:\AgdaUnderscore{}}}\<%
\\
\>[2]\AgdaKeyword{infix}\AgdaSpace{}%
\AgdaNumber{4}\AgdaSpace{}%
\AgdaOperator{\AgdaDatatype{\AgdaUnderscore{}∋\AgdaUnderscore{}}}\<%
\\
\>[2]\AgdaKeyword{infix}\AgdaSpace{}%
\AgdaNumber{4}\AgdaSpace{}%
\AgdaOperator{\AgdaDatatype{\AgdaUnderscore{}⊢\AgdaUnderscore{}}}\<%
\\
\>[2]\AgdaKeyword{infixl}\AgdaSpace{}%
\AgdaNumber{7}\AgdaSpace{}%
\AgdaOperator{\AgdaInductiveConstructor{\AgdaUnderscore{}·\AgdaUnderscore{}}}\<%
\\
\>[2]\AgdaKeyword{infix}\AgdaSpace{}%
\AgdaNumber{5}\AgdaSpace{}%
\AgdaOperator{\AgdaInductiveConstructor{ƛ\AgdaUnderscore{}}}\<%
\\
\>[2]\AgdaKeyword{infix}\AgdaSpace{}%
\AgdaNumber{5}\AgdaSpace{}%
\AgdaOperator{\AgdaInductiveConstructor{⟨ƛ\AgdaUnderscore{}⟩\AgdaUnderscore{}}}\<%
\\
\>[2]\AgdaKeyword{infixl}\AgdaSpace{}%
\AgdaNumber{5}\AgdaSpace{}%
\AgdaOperator{\AgdaFunction{\AgdaUnderscore{}++\AgdaUnderscore{}}}\<%
\\
\>[2]\AgdaKeyword{infix}\AgdaSpace{}%
\AgdaNumber{4}\AgdaSpace{}%
\AgdaOperator{\AgdaDatatype{\AgdaUnderscore{}∣\AgdaUnderscore{}⇓\AgdaUnderscore{}}}\<%
\\
\>[2]\AgdaKeyword{infix}\AgdaSpace{}%
\AgdaNumber{4}\AgdaSpace{}%
\AgdaOperator{\AgdaFunction{\AgdaUnderscore{}⊨\AgdaUnderscore{}}}\<%
\\
\>[2]\AgdaKeyword{infix}\AgdaSpace{}%
\AgdaNumber{4}\AgdaSpace{}%
\AgdaFunction{semantic-type}\<%
\\
\>[2]\AgdaKeyword{infixl}\AgdaSpace{}%
\AgdaNumber{5}\AgdaSpace{}%
\AgdaOperator{\AgdaFunction{\AgdaUnderscore{}<>\AgdaUnderscore{}}}\<%
\\
\>[2]\AgdaKeyword{infix}\AgdaSpace{}%
\AgdaNumber{4}\AgdaSpace{}%
\AgdaOperator{\AgdaFunction{\AgdaUnderscore{}⇑}}\<%
\\
\>[2]\AgdaKeyword{infixl}\AgdaSpace{}%
\AgdaNumber{5}\AgdaSpace{}%
\AgdaOperator{\AgdaFunction{\AgdaUnderscore{}\&\AgdaUnderscore{}}}\<%
\\
\>[2]\AgdaKeyword{infixl}\AgdaSpace{}%
\AgdaNumber{5}\AgdaSpace{}%
\AgdaOperator{\AgdaFunction{\AgdaUnderscore{}\&\&\AgdaUnderscore{}}}\<%
\end{code}

\subsection{Embedding of STLC in Agda (Figure \ref{fig:stlc})}

\begin{AgdaAlign}
\begin{code}%
\>[2]\AgdaComment{--\ Types}\<%
\\
\>[2]\AgdaKeyword{data}\AgdaSpace{}%
\AgdaDatatype{Type}\AgdaSpace{}%
\AgdaSymbol{:}\AgdaSpace{}%
\AgdaPrimitive{Set}\AgdaSpace{}%
\AgdaKeyword{where}\<%
\\
\>[2][@{}l@{\AgdaIndent{0}}]%
\>[4]\AgdaInductiveConstructor{bool}\AgdaSpace{}%
\AgdaSymbol{:}\AgdaSpace{}%
\AgdaDatatype{Type}\<%
\\
\>[4]\AgdaOperator{\AgdaInductiveConstructor{\AgdaUnderscore{}⇒\AgdaUnderscore{}}}\AgdaSpace{}%
\AgdaSymbol{:}\AgdaSpace{}%
\AgdaDatatype{Type}\AgdaSpace{}%
\AgdaSymbol{→}\AgdaSpace{}%
\AgdaDatatype{Type}\AgdaSpace{}%
\AgdaSymbol{→}\AgdaSpace{}%
\AgdaDatatype{Type}\<%
\\
\\[\AgdaEmptyExtraSkip]%
\>[2]\AgdaKeyword{variable}\AgdaSpace{}%
\AgdaGeneralizable{S}\AgdaSpace{}%
\AgdaGeneralizable{T}\AgdaSpace{}%
\AgdaSymbol{:}\AgdaSpace{}%
\AgdaDatatype{Type}\<%
\\
\\[\AgdaEmptyExtraSkip]%
\>[2]\AgdaComment{--\ Typing\ contexts}\<%
\\
\>[2]\AgdaKeyword{data}\AgdaSpace{}%
\AgdaDatatype{Ctx}\AgdaSpace{}%
\AgdaSymbol{:}\AgdaSpace{}%
\AgdaPrimitive{Set}\AgdaSpace{}%
\AgdaKeyword{where}\<%
\\
\>[2][@{}l@{\AgdaIndent{0}}]%
\>[4]\AgdaInductiveConstructor{∅}\AgdaSpace{}%
\AgdaSymbol{:}\AgdaSpace{}%
\AgdaDatatype{Ctx}\<%
\\
\>[4]\AgdaOperator{\AgdaInductiveConstructor{\AgdaUnderscore{}·:\AgdaUnderscore{}}}\AgdaSpace{}%
\AgdaSymbol{:}\AgdaSpace{}%
\AgdaDatatype{Ctx}\AgdaSpace{}%
\AgdaSymbol{→}\AgdaSpace{}%
\AgdaDatatype{Type}\AgdaSpace{}%
\AgdaSymbol{→}\AgdaSpace{}%
\AgdaDatatype{Ctx}\<%
\\
\\[\AgdaEmptyExtraSkip]%
\>[2]\AgdaKeyword{variable}\AgdaSpace{}%
\AgdaGeneralizable{Γ}\AgdaSpace{}%
\AgdaGeneralizable{Δ}\AgdaSpace{}%
\AgdaSymbol{:}\AgdaSpace{}%
\AgdaDatatype{Ctx}\<%
\\
\\[\AgdaEmptyExtraSkip]%
\>[2]\AgdaComment{--\ Intrinsically\ scoped\ de\ Brujin\ indices}\<%
\\
\>[2]\AgdaKeyword{data}\AgdaSpace{}%
\AgdaOperator{\AgdaDatatype{\AgdaUnderscore{}∋\AgdaUnderscore{}}}\AgdaSpace{}%
\AgdaSymbol{:}\AgdaSpace{}%
\AgdaDatatype{Ctx}\AgdaSpace{}%
\AgdaSymbol{→}\AgdaSpace{}%
\AgdaDatatype{Type}\AgdaSpace{}%
\AgdaSymbol{→}\AgdaSpace{}%
\AgdaPrimitive{Set}\AgdaSpace{}%
\AgdaKeyword{where}\<%
\\
\>[2][@{}l@{\AgdaIndent{0}}]%
\>[4]\AgdaInductiveConstructor{zero}\AgdaSpace{}%
\AgdaSymbol{:}\AgdaSpace{}%
\AgdaGeneralizable{Γ}\AgdaSpace{}%
\AgdaOperator{\AgdaInductiveConstructor{·:}}\AgdaSpace{}%
\AgdaGeneralizable{T}\AgdaSpace{}%
\AgdaOperator{\AgdaDatatype{∋}}\AgdaSpace{}%
\AgdaGeneralizable{T}\<%
\\
\>[4]\AgdaInductiveConstructor{suc}\AgdaSpace{}%
\AgdaSymbol{:}\AgdaSpace{}%
\AgdaGeneralizable{Γ}\AgdaSpace{}%
\AgdaOperator{\AgdaDatatype{∋}}\AgdaSpace{}%
\AgdaGeneralizable{T}\AgdaSpace{}%
\AgdaSymbol{→}\AgdaSpace{}%
\AgdaGeneralizable{Γ}\AgdaSpace{}%
\AgdaOperator{\AgdaInductiveConstructor{·:}}\AgdaSpace{}%
\AgdaGeneralizable{S}\AgdaSpace{}%
\AgdaOperator{\AgdaDatatype{∋}}\AgdaSpace{}%
\AgdaGeneralizable{T}\<%
\\
\\[\AgdaEmptyExtraSkip]%
\>[2]\AgdaKeyword{variable}\AgdaSpace{}%
\AgdaGeneralizable{x}\AgdaSpace{}%
\AgdaSymbol{:}\AgdaSpace{}%
\AgdaGeneralizable{Γ}\AgdaSpace{}%
\AgdaOperator{\AgdaDatatype{∋}}\AgdaSpace{}%
\AgdaGeneralizable{T}\<%
\\
\\[\AgdaEmptyExtraSkip]%
\>[2]\AgdaComment{--\ Intrinsically\ typed\ terms}\<%
\\
\>[2]\AgdaKeyword{data}\AgdaSpace{}%
\AgdaOperator{\AgdaDatatype{\AgdaUnderscore{}⊢\AgdaUnderscore{}}}\AgdaSpace{}%
\AgdaSymbol{:}\AgdaSpace{}%
\AgdaDatatype{Ctx}\AgdaSpace{}%
\AgdaSymbol{→}\AgdaSpace{}%
\AgdaDatatype{Type}\AgdaSpace{}%
\AgdaSymbol{→}\AgdaSpace{}%
\AgdaPrimitive{Set}\AgdaSpace{}%
\AgdaKeyword{where}\<%
\\
\>[2][@{}l@{\AgdaIndent{0}}]%
\>[4]\AgdaInductiveConstructor{true}\AgdaSpace{}%
\AgdaInductiveConstructor{false}\AgdaSpace{}%
\AgdaSymbol{:}\AgdaSpace{}%
\AgdaGeneralizable{Γ}\AgdaSpace{}%
\AgdaOperator{\AgdaDatatype{⊢}}\AgdaSpace{}%
\AgdaInductiveConstructor{bool}\<%
\\
\>[4]\AgdaOperator{\AgdaInductiveConstructor{if\AgdaUnderscore{}then\AgdaUnderscore{}else\AgdaUnderscore{}}}\AgdaSpace{}%
\AgdaSymbol{:}\AgdaSpace{}%
\AgdaGeneralizable{Γ}\AgdaSpace{}%
\AgdaOperator{\AgdaDatatype{⊢}}\AgdaSpace{}%
\AgdaInductiveConstructor{bool}\AgdaSpace{}%
\AgdaSymbol{→}\AgdaSpace{}%
\AgdaGeneralizable{Γ}\AgdaSpace{}%
\AgdaOperator{\AgdaDatatype{⊢}}\AgdaSpace{}%
\AgdaGeneralizable{T}\AgdaSpace{}%
\AgdaSymbol{→}\AgdaSpace{}%
\AgdaGeneralizable{Γ}\AgdaSpace{}%
\AgdaOperator{\AgdaDatatype{⊢}}\AgdaSpace{}%
\AgdaGeneralizable{T}\AgdaSpace{}%
\AgdaSymbol{→}\AgdaSpace{}%
\AgdaGeneralizable{Γ}\AgdaSpace{}%
\AgdaOperator{\AgdaDatatype{⊢}}\AgdaSpace{}%
\AgdaGeneralizable{T}\<%
\\
\>[4]\AgdaInductiveConstructor{var}\AgdaSpace{}%
\AgdaSymbol{:}\AgdaSpace{}%
\AgdaGeneralizable{Γ}\AgdaSpace{}%
\AgdaOperator{\AgdaDatatype{∋}}\AgdaSpace{}%
\AgdaGeneralizable{T}\AgdaSpace{}%
\AgdaSymbol{→}\AgdaSpace{}%
\AgdaGeneralizable{Γ}\AgdaSpace{}%
\AgdaOperator{\AgdaDatatype{⊢}}\AgdaSpace{}%
\AgdaGeneralizable{T}\<%
\\
\>[4]\AgdaOperator{\AgdaInductiveConstructor{ƛ\AgdaUnderscore{}}}\AgdaSpace{}%
\AgdaSymbol{:}\AgdaSpace{}%
\AgdaGeneralizable{Γ}\AgdaSpace{}%
\AgdaOperator{\AgdaInductiveConstructor{·:}}\AgdaSpace{}%
\AgdaGeneralizable{S}\AgdaSpace{}%
\AgdaOperator{\AgdaDatatype{⊢}}\AgdaSpace{}%
\AgdaGeneralizable{T}\AgdaSpace{}%
\AgdaSymbol{→}\AgdaSpace{}%
\AgdaGeneralizable{Γ}\AgdaSpace{}%
\AgdaOperator{\AgdaDatatype{⊢}}\AgdaSpace{}%
\AgdaGeneralizable{S}\AgdaSpace{}%
\AgdaOperator{\AgdaInductiveConstructor{⇒}}\AgdaSpace{}%
\AgdaGeneralizable{T}\<%
\\
\>[4]\AgdaOperator{\AgdaInductiveConstructor{\AgdaUnderscore{}·\AgdaUnderscore{}}}\AgdaSpace{}%
\AgdaSymbol{:}\AgdaSpace{}%
\AgdaGeneralizable{Γ}\AgdaSpace{}%
\AgdaOperator{\AgdaDatatype{⊢}}\AgdaSpace{}%
\AgdaGeneralizable{S}\AgdaSpace{}%
\AgdaOperator{\AgdaInductiveConstructor{⇒}}\AgdaSpace{}%
\AgdaGeneralizable{T}\AgdaSpace{}%
\AgdaSymbol{→}\AgdaSpace{}%
\AgdaGeneralizable{Γ}\AgdaSpace{}%
\AgdaOperator{\AgdaDatatype{⊢}}\AgdaSpace{}%
\AgdaGeneralizable{S}\AgdaSpace{}%
\AgdaSymbol{→}\AgdaSpace{}%
\AgdaGeneralizable{Γ}\AgdaSpace{}%
\AgdaOperator{\AgdaDatatype{⊢}}\AgdaSpace{}%
\AgdaGeneralizable{T}\<%
\\
\\[\AgdaEmptyExtraSkip]%
\>[2]\AgdaKeyword{variable}\AgdaSpace{}%
\AgdaGeneralizable{r}\AgdaSpace{}%
\AgdaGeneralizable{s}\AgdaSpace{}%
\AgdaGeneralizable{t}\AgdaSpace{}%
\AgdaGeneralizable{u}\AgdaSpace{}%
\AgdaGeneralizable{v}\AgdaSpace{}%
\AgdaSymbol{:}\AgdaSpace{}%
\AgdaGeneralizable{Γ}\AgdaSpace{}%
\AgdaOperator{\AgdaDatatype{⊢}}\AgdaSpace{}%
\AgdaGeneralizable{T}\<%
\end{code}

\subsection{Natural semantics (Figure \ref{fig:semantics})}

\begin{code}%
\>[2]\AgdaComment{--\ Environments\ and\ domain}\<%
\\
\>[2]\AgdaKeyword{mutual}\<%
\\
\>[2][@{}l@{\AgdaIndent{0}}]%
\>[4]\AgdaFunction{Env}\AgdaSpace{}%
\AgdaSymbol{:}\AgdaSpace{}%
\AgdaDatatype{Ctx}\AgdaSpace{}%
\AgdaSymbol{→}\AgdaSpace{}%
\AgdaPrimitive{Set}\<%
\\
\>[4]\AgdaFunction{Env}\AgdaSpace{}%
\AgdaBound{Γ}\AgdaSpace{}%
\AgdaSymbol{=}\AgdaSpace{}%
\AgdaSymbol{∀}\AgdaSpace{}%
\AgdaSymbol{\{}\AgdaBound{T}\AgdaSymbol{\}}\AgdaSpace{}%
\AgdaSymbol{→}\AgdaSpace{}%
\AgdaBound{Γ}\AgdaSpace{}%
\AgdaOperator{\AgdaDatatype{∋}}\AgdaSpace{}%
\AgdaBound{T}\AgdaSpace{}%
\AgdaSymbol{→}\AgdaSpace{}%
\AgdaDatatype{Domain}\AgdaSpace{}%
\AgdaBound{T}\<%
\\
\\[\AgdaEmptyExtraSkip]%
\>[4]\AgdaKeyword{data}\AgdaSpace{}%
\AgdaDatatype{Domain}\AgdaSpace{}%
\AgdaSymbol{:}\AgdaSpace{}%
\AgdaDatatype{Type}\AgdaSpace{}%
\AgdaSymbol{→}\AgdaSpace{}%
\AgdaPrimitive{Set}\AgdaSpace{}%
\AgdaKeyword{where}\<%
\\
\>[4][@{}l@{\AgdaIndent{0}}]%
\>[6]\AgdaInductiveConstructor{true}\AgdaSpace{}%
\AgdaInductiveConstructor{false}\AgdaSpace{}%
\AgdaSymbol{:}\AgdaSpace{}%
\AgdaDatatype{Domain}\AgdaSpace{}%
\AgdaInductiveConstructor{bool}\<%
\\
\>[6]\AgdaComment{--\ Closures}\<%
\\
\>[6]\AgdaOperator{\AgdaInductiveConstructor{⟨ƛ\AgdaUnderscore{}⟩\AgdaUnderscore{}}}\AgdaSpace{}%
\AgdaSymbol{:}\AgdaSpace{}%
\AgdaGeneralizable{Γ}\AgdaSpace{}%
\AgdaOperator{\AgdaInductiveConstructor{·:}}\AgdaSpace{}%
\AgdaGeneralizable{S}\AgdaSpace{}%
\AgdaOperator{\AgdaDatatype{⊢}}\AgdaSpace{}%
\AgdaGeneralizable{T}\AgdaSpace{}%
\AgdaSymbol{→}\AgdaSpace{}%
\AgdaFunction{Env}\AgdaSpace{}%
\AgdaGeneralizable{Γ}\AgdaSpace{}%
\AgdaSymbol{→}\AgdaSpace{}%
\AgdaDatatype{Domain}\AgdaSpace{}%
\AgdaSymbol{(}\AgdaGeneralizable{S}\AgdaSpace{}%
\AgdaOperator{\AgdaInductiveConstructor{⇒}}\AgdaSpace{}%
\AgdaGeneralizable{T}\AgdaSymbol{)}\<%
\\
\\[\AgdaEmptyExtraSkip]%
\\[\AgdaEmptyExtraSkip]%
\>[2]\AgdaKeyword{variable}\AgdaSpace{}%
\AgdaGeneralizable{γ}\AgdaSpace{}%
\AgdaGeneralizable{δ}\AgdaSpace{}%
\AgdaSymbol{:}\AgdaSpace{}%
\AgdaFunction{Env}\AgdaSpace{}%
\AgdaGeneralizable{Γ}\<%
\\
\>[2]\AgdaKeyword{variable}\AgdaSpace{}%
\AgdaGeneralizable{a}\AgdaSpace{}%
\AgdaGeneralizable{b}\AgdaSpace{}%
\AgdaGeneralizable{d}\AgdaSpace{}%
\AgdaGeneralizable{f}\AgdaSpace{}%
\AgdaSymbol{:}\AgdaSpace{}%
\AgdaDatatype{Domain}\AgdaSpace{}%
\AgdaGeneralizable{T}\<%
\\
\\[\AgdaEmptyExtraSkip]%
\>[2]\AgdaOperator{\AgdaFunction{\AgdaUnderscore{}++\AgdaUnderscore{}}}\AgdaSpace{}%
\AgdaSymbol{:}\AgdaSpace{}%
\AgdaFunction{Env}\AgdaSpace{}%
\AgdaGeneralizable{Γ}\AgdaSpace{}%
\AgdaSymbol{→}\AgdaSpace{}%
\AgdaDatatype{Domain}\AgdaSpace{}%
\AgdaGeneralizable{T}\AgdaSpace{}%
\AgdaSymbol{→}\AgdaSpace{}%
\AgdaFunction{Env}\AgdaSpace{}%
\AgdaSymbol{(}\AgdaGeneralizable{Γ}\AgdaSpace{}%
\AgdaOperator{\AgdaInductiveConstructor{·:}}\AgdaSpace{}%
\AgdaGeneralizable{T}\AgdaSymbol{)}\<%
\\
\>[2]\AgdaSymbol{(\AgdaUnderscore{}}\AgdaSpace{}%
\AgdaOperator{\AgdaFunction{++}}\AgdaSpace{}%
\AgdaBound{a}\AgdaSymbol{)}\AgdaSpace{}%
\AgdaInductiveConstructor{zero}\AgdaSpace{}%
\AgdaSymbol{=}\AgdaSpace{}%
\AgdaBound{a}\<%
\\
\>[2]\AgdaSymbol{(}\AgdaBound{γ}\AgdaSpace{}%
\AgdaOperator{\AgdaFunction{++}}\AgdaSpace{}%
\AgdaSymbol{\AgdaUnderscore{})}\AgdaSpace{}%
\AgdaSymbol{(}\AgdaInductiveConstructor{suc}\AgdaSpace{}%
\AgdaBound{x}\AgdaSymbol{)}\AgdaSpace{}%
\AgdaSymbol{=}\AgdaSpace{}%
\AgdaBound{γ}\AgdaSpace{}%
\AgdaBound{x}\<%
\\
\\[\AgdaEmptyExtraSkip]%
\>[2]\AgdaKeyword{data}\AgdaSpace{}%
\AgdaOperator{\AgdaDatatype{\AgdaUnderscore{}∣\AgdaUnderscore{}⇓\AgdaUnderscore{}}}\AgdaSpace{}%
\AgdaSymbol{:}\AgdaSpace{}%
\AgdaFunction{Env}\AgdaSpace{}%
\AgdaGeneralizable{Γ}\AgdaSpace{}%
\AgdaSymbol{→}\AgdaSpace{}%
\AgdaGeneralizable{Γ}\AgdaSpace{}%
\AgdaOperator{\AgdaDatatype{⊢}}\AgdaSpace{}%
\AgdaGeneralizable{T}\AgdaSpace{}%
\AgdaSymbol{→}\AgdaSpace{}%
\AgdaDatatype{Domain}\AgdaSpace{}%
\AgdaGeneralizable{T}\AgdaSpace{}%
\AgdaSymbol{→}\AgdaSpace{}%
\AgdaPrimitive{Set}\AgdaSpace{}%
\AgdaKeyword{where}\<%
\\
\>[2][@{}l@{\AgdaIndent{0}}]%
\>[4]\AgdaInductiveConstructor{evalTrue}\AgdaSpace{}%
\AgdaSymbol{:}\AgdaSpace{}%
\AgdaGeneralizable{γ}\AgdaSpace{}%
\AgdaOperator{\AgdaDatatype{∣}}\AgdaSpace{}%
\AgdaInductiveConstructor{true}\AgdaSpace{}%
\AgdaOperator{\AgdaDatatype{⇓}}\AgdaSpace{}%
\AgdaInductiveConstructor{true}\<%
\\
\>[4]\AgdaInductiveConstructor{evalFalse}\AgdaSpace{}%
\AgdaSymbol{:}\AgdaSpace{}%
\AgdaGeneralizable{γ}\AgdaSpace{}%
\AgdaOperator{\AgdaDatatype{∣}}\AgdaSpace{}%
\AgdaInductiveConstructor{false}\AgdaSpace{}%
\AgdaOperator{\AgdaDatatype{⇓}}\AgdaSpace{}%
\AgdaInductiveConstructor{false}\<%
\\
\>[4]\AgdaInductiveConstructor{evalVar}\AgdaSpace{}%
\AgdaSymbol{:}\AgdaSpace{}%
\AgdaGeneralizable{γ}\AgdaSpace{}%
\AgdaOperator{\AgdaDatatype{∣}}\AgdaSpace{}%
\AgdaInductiveConstructor{var}\AgdaSpace{}%
\AgdaGeneralizable{x}\AgdaSpace{}%
\AgdaOperator{\AgdaDatatype{⇓}}\AgdaSpace{}%
\AgdaGeneralizable{γ}\AgdaSpace{}%
\AgdaGeneralizable{x}\<%
\\
\>[4]\AgdaInductiveConstructor{evalAbs}\AgdaSpace{}%
\AgdaSymbol{:}\AgdaSpace{}%
\AgdaGeneralizable{γ}\AgdaSpace{}%
\AgdaOperator{\AgdaDatatype{∣}}\AgdaSpace{}%
\AgdaOperator{\AgdaInductiveConstructor{ƛ}}\AgdaSpace{}%
\AgdaGeneralizable{t}\AgdaSpace{}%
\AgdaOperator{\AgdaDatatype{⇓}}\AgdaSpace{}%
\AgdaOperator{\AgdaInductiveConstructor{⟨ƛ}}\AgdaSpace{}%
\AgdaGeneralizable{t}\AgdaSpace{}%
\AgdaOperator{\AgdaInductiveConstructor{⟩}}\AgdaSpace{}%
\AgdaGeneralizable{γ}\<%
\\
\>[4]\AgdaInductiveConstructor{evalApp}\AgdaSpace{}%
\AgdaSymbol{:}\<%
\\
\>[4][@{}l@{\AgdaIndent{0}}]%
\>[8]\AgdaGeneralizable{γ}\AgdaSpace{}%
\AgdaOperator{\AgdaDatatype{∣}}\AgdaSpace{}%
\AgdaGeneralizable{r}\AgdaSpace{}%
\AgdaOperator{\AgdaDatatype{⇓}}\AgdaSpace{}%
\AgdaOperator{\AgdaInductiveConstructor{⟨ƛ}}\AgdaSpace{}%
\AgdaGeneralizable{t}\AgdaSpace{}%
\AgdaOperator{\AgdaInductiveConstructor{⟩}}\AgdaSpace{}%
\AgdaGeneralizable{δ}\<%
\\
\>[4][@{}l@{\AgdaIndent{0}}]%
\>[6]\AgdaSymbol{→}\AgdaSpace{}%
\AgdaGeneralizable{γ}\AgdaSpace{}%
\AgdaOperator{\AgdaDatatype{∣}}\AgdaSpace{}%
\AgdaGeneralizable{s}\AgdaSpace{}%
\AgdaOperator{\AgdaDatatype{⇓}}\AgdaSpace{}%
\AgdaGeneralizable{a}\<%
\\
\>[6]\AgdaSymbol{→}\AgdaSpace{}%
\AgdaGeneralizable{δ}\AgdaSpace{}%
\AgdaOperator{\AgdaFunction{++}}\AgdaSpace{}%
\AgdaGeneralizable{a}\AgdaSpace{}%
\AgdaOperator{\AgdaDatatype{∣}}\AgdaSpace{}%
\AgdaGeneralizable{t}\AgdaSpace{}%
\AgdaOperator{\AgdaDatatype{⇓}}\AgdaSpace{}%
\AgdaGeneralizable{b}\<%
\\
\>[6]\AgdaSymbol{→}\AgdaSpace{}%
\AgdaGeneralizable{γ}\AgdaSpace{}%
\AgdaOperator{\AgdaDatatype{∣}}\AgdaSpace{}%
\AgdaGeneralizable{r}\AgdaSpace{}%
\AgdaOperator{\AgdaInductiveConstructor{·}}\AgdaSpace{}%
\AgdaGeneralizable{s}\AgdaSpace{}%
\AgdaOperator{\AgdaDatatype{⇓}}\AgdaSpace{}%
\AgdaGeneralizable{b}\<%
\\
\>[4]\AgdaInductiveConstructor{evalIfTrue}\AgdaSpace{}%
\AgdaSymbol{:}\<%
\\
\>[4][@{}l@{\AgdaIndent{0}}]%
\>[8]\AgdaGeneralizable{γ}\AgdaSpace{}%
\AgdaOperator{\AgdaDatatype{∣}}\AgdaSpace{}%
\AgdaGeneralizable{r}\AgdaSpace{}%
\AgdaOperator{\AgdaDatatype{⇓}}\AgdaSpace{}%
\AgdaInductiveConstructor{true}\<%
\\
\>[4][@{}l@{\AgdaIndent{0}}]%
\>[6]\AgdaSymbol{→}\AgdaSpace{}%
\AgdaGeneralizable{γ}\AgdaSpace{}%
\AgdaOperator{\AgdaDatatype{∣}}\AgdaSpace{}%
\AgdaGeneralizable{s}\AgdaSpace{}%
\AgdaOperator{\AgdaDatatype{⇓}}\AgdaSpace{}%
\AgdaGeneralizable{a}\<%
\\
\>[6]\AgdaSymbol{→}\AgdaSpace{}%
\AgdaGeneralizable{γ}\AgdaSpace{}%
\AgdaOperator{\AgdaDatatype{∣}}\AgdaSpace{}%
\AgdaOperator{\AgdaInductiveConstructor{if}}\AgdaSpace{}%
\AgdaGeneralizable{r}\AgdaSpace{}%
\AgdaOperator{\AgdaInductiveConstructor{then}}\AgdaSpace{}%
\AgdaGeneralizable{s}\AgdaSpace{}%
\AgdaOperator{\AgdaInductiveConstructor{else}}\AgdaSpace{}%
\AgdaGeneralizable{t}\AgdaSpace{}%
\AgdaOperator{\AgdaDatatype{⇓}}\AgdaSpace{}%
\AgdaGeneralizable{a}\<%
\\
\>[4]\AgdaInductiveConstructor{evalIfFalse}\AgdaSpace{}%
\AgdaSymbol{:}\<%
\\
\>[4][@{}l@{\AgdaIndent{0}}]%
\>[8]\AgdaGeneralizable{γ}\AgdaSpace{}%
\AgdaOperator{\AgdaDatatype{∣}}\AgdaSpace{}%
\AgdaGeneralizable{r}\AgdaSpace{}%
\AgdaOperator{\AgdaDatatype{⇓}}\AgdaSpace{}%
\AgdaInductiveConstructor{false}\<%
\\
\>[4][@{}l@{\AgdaIndent{0}}]%
\>[6]\AgdaSymbol{→}\AgdaSpace{}%
\AgdaGeneralizable{γ}\AgdaSpace{}%
\AgdaOperator{\AgdaDatatype{∣}}\AgdaSpace{}%
\AgdaGeneralizable{t}\AgdaSpace{}%
\AgdaOperator{\AgdaDatatype{⇓}}\AgdaSpace{}%
\AgdaGeneralizable{b}\<%
\\
\>[6]\AgdaSymbol{→}\AgdaSpace{}%
\AgdaGeneralizable{γ}\AgdaSpace{}%
\AgdaOperator{\AgdaDatatype{∣}}\AgdaSpace{}%
\AgdaOperator{\AgdaInductiveConstructor{if}}\AgdaSpace{}%
\AgdaGeneralizable{r}\AgdaSpace{}%
\AgdaOperator{\AgdaInductiveConstructor{then}}\AgdaSpace{}%
\AgdaGeneralizable{s}\AgdaSpace{}%
\AgdaOperator{\AgdaInductiveConstructor{else}}\AgdaSpace{}%
\AgdaGeneralizable{t}\AgdaSpace{}%
\AgdaOperator{\AgdaDatatype{⇓}}\AgdaSpace{}%
\AgdaGeneralizable{b}\<%
\end{code}

\subsection{Proof by logical relations (Figure \ref{fig:totality-proof})}

This is the property we wish to prove, that evaluation is well-defined:

\begin{code}%
\>[2]\AgdaFunction{empty}\AgdaSpace{}%
\AgdaSymbol{:}\AgdaSpace{}%
\AgdaFunction{Env}\AgdaSpace{}%
\AgdaInductiveConstructor{∅}\<%
\\
\>[2]\AgdaFunction{empty}\AgdaSpace{}%
\AgdaSymbol{()}\<%
\\
\\[\AgdaEmptyExtraSkip]%
\>[2]\AgdaComment{--\ Evaluation\ is\ well-defined}\<%
\\
\>[2]\AgdaFunction{⇓-well-defined}\AgdaSpace{}%
\AgdaSymbol{:}\AgdaSpace{}%
\AgdaInductiveConstructor{∅}\AgdaSpace{}%
\AgdaOperator{\AgdaDatatype{⊢}}\AgdaSpace{}%
\AgdaGeneralizable{T}\AgdaSpace{}%
\AgdaSymbol{→}\AgdaSpace{}%
\AgdaPrimitive{Set}\<%
\\
\>[2]\AgdaFunction{⇓-well-defined}\AgdaSpace{}%
\AgdaBound{t}\AgdaSpace{}%
\AgdaSymbol{=}\AgdaSpace{}%
\AgdaFunction{∃[}\AgdaSpace{}%
\AgdaBound{a}\AgdaSpace{}%
\AgdaFunction{]}\AgdaSpace{}%
\AgdaFunction{empty}\AgdaSpace{}%
\AgdaOperator{\AgdaDatatype{∣}}\AgdaSpace{}%
\AgdaBound{t}\AgdaSpace{}%
\AgdaOperator{\AgdaDatatype{⇓}}\AgdaSpace{}%
\AgdaBound{a}\<%
\end{code}

Here is the full proof by logical relations:

\begin{code}%
\>[2]\AgdaOperator{\AgdaFunction{⟦\AgdaUnderscore{}⟧}}\AgdaSpace{}%
\AgdaSymbol{:}\AgdaSpace{}%
\AgdaSymbol{∀}\AgdaSpace{}%
\AgdaSymbol{(}\AgdaBound{T}\AgdaSpace{}%
\AgdaSymbol{:}\AgdaSpace{}%
\AgdaDatatype{Type}\AgdaSymbol{)}\AgdaSpace{}%
\AgdaSymbol{→}\AgdaSpace{}%
\AgdaSymbol{(}\AgdaDatatype{Domain}\AgdaSpace{}%
\AgdaBound{T}\AgdaSpace{}%
\AgdaSymbol{→}\AgdaSpace{}%
\AgdaPrimitive{Set}\AgdaSymbol{)}\<%
\\
\>[2]\AgdaOperator{\AgdaFunction{⟦}}\AgdaSpace{}%
\AgdaInductiveConstructor{bool}\AgdaSpace{}%
\AgdaOperator{\AgdaFunction{⟧}}\AgdaSpace{}%
\AgdaSymbol{\AgdaUnderscore{}}\AgdaSpace{}%
\AgdaSymbol{=}\AgdaSpace{}%
\AgdaRecord{⊤}\<%
\\
\>[2]\AgdaOperator{\AgdaFunction{⟦}}%
\>[2603I]\AgdaBound{S}\AgdaSpace{}%
\AgdaOperator{\AgdaInductiveConstructor{⇒}}\AgdaSpace{}%
\AgdaBound{T}\AgdaSpace{}%
\AgdaOperator{\AgdaFunction{⟧}}\AgdaSpace{}%
\AgdaSymbol{(}\AgdaOperator{\AgdaInductiveConstructor{⟨ƛ}}\AgdaSpace{}%
\AgdaBound{t}\AgdaSpace{}%
\AgdaOperator{\AgdaInductiveConstructor{⟩}}\AgdaSpace{}%
\AgdaBound{δ}\AgdaSymbol{)}\AgdaSpace{}%
\AgdaSymbol{=}\<%
\\
\>[.][@{}l@{}]\<[2603I]%
\>[4]\AgdaSymbol{∀}\AgdaSpace{}%
\AgdaSymbol{\{}\AgdaBound{a}\AgdaSymbol{\}}\AgdaSpace{}%
\AgdaSymbol{→}\AgdaSpace{}%
\AgdaBound{a}\AgdaSpace{}%
\AgdaOperator{\AgdaFunction{∈}}\AgdaSpace{}%
\AgdaOperator{\AgdaFunction{⟦}}\AgdaSpace{}%
\AgdaBound{S}\AgdaSpace{}%
\AgdaOperator{\AgdaFunction{⟧}}\<%
\\
\>[4]\AgdaSymbol{→}\AgdaSpace{}%
\AgdaFunction{∃[}\AgdaSpace{}%
\AgdaBound{b}\AgdaSpace{}%
\AgdaFunction{]}\AgdaSpace{}%
\AgdaBound{δ}\AgdaSpace{}%
\AgdaOperator{\AgdaFunction{++}}\AgdaSpace{}%
\AgdaBound{a}\AgdaSpace{}%
\AgdaOperator{\AgdaDatatype{∣}}\AgdaSpace{}%
\AgdaBound{t}\AgdaSpace{}%
\AgdaOperator{\AgdaDatatype{⇓}}\AgdaSpace{}%
\AgdaBound{b}\AgdaSpace{}%
\AgdaOperator{\AgdaFunction{×}}\AgdaSpace{}%
\AgdaBound{b}\AgdaSpace{}%
\AgdaOperator{\AgdaFunction{∈}}\AgdaSpace{}%
\AgdaOperator{\AgdaFunction{⟦}}\AgdaSpace{}%
\AgdaBound{T}\AgdaSpace{}%
\AgdaOperator{\AgdaFunction{⟧}}\<%
\\
\\[\AgdaEmptyExtraSkip]%
\>[2]\AgdaOperator{\AgdaFunction{\AgdaUnderscore{}⊨\AgdaUnderscore{}}}\AgdaSpace{}%
\AgdaSymbol{:}\AgdaSpace{}%
\AgdaSymbol{(}\AgdaBound{Γ}\AgdaSpace{}%
\AgdaSymbol{:}\AgdaSpace{}%
\AgdaDatatype{Ctx}\AgdaSymbol{)}\AgdaSpace{}%
\AgdaSymbol{→}\AgdaSpace{}%
\AgdaFunction{Env}\AgdaSpace{}%
\AgdaBound{Γ}\AgdaSpace{}%
\AgdaSymbol{→}\AgdaSpace{}%
\AgdaPrimitive{Set}\<%
\\
\>[2]\AgdaBound{Γ}\AgdaSpace{}%
\AgdaOperator{\AgdaFunction{⊨}}\AgdaSpace{}%
\AgdaBound{γ}\AgdaSpace{}%
\AgdaSymbol{=}\AgdaSpace{}%
\AgdaSymbol{∀}\AgdaSpace{}%
\AgdaSymbol{\{}\AgdaBound{T}\AgdaSymbol{\}}\AgdaSpace{}%
\AgdaSymbol{→}\AgdaSpace{}%
\AgdaSymbol{(}\AgdaBound{x}\AgdaSpace{}%
\AgdaSymbol{:}\AgdaSpace{}%
\AgdaBound{Γ}\AgdaSpace{}%
\AgdaOperator{\AgdaDatatype{∋}}\AgdaSpace{}%
\AgdaBound{T}\AgdaSymbol{)}\AgdaSpace{}%
\AgdaSymbol{→}\AgdaSpace{}%
\AgdaBound{γ}\AgdaSpace{}%
\AgdaBound{x}\AgdaSpace{}%
\AgdaOperator{\AgdaFunction{∈}}\AgdaSpace{}%
\AgdaOperator{\AgdaFunction{⟦}}\AgdaSpace{}%
\AgdaBound{T}\AgdaSpace{}%
\AgdaOperator{\AgdaFunction{⟧}}\<%
\\
\\[\AgdaEmptyExtraSkip]%
\>[2]\AgdaOperator{\AgdaFunction{\AgdaUnderscore{}\textasciicircum{}\AgdaUnderscore{}}}\AgdaSpace{}%
\AgdaSymbol{:}\AgdaSpace{}%
\AgdaGeneralizable{Γ}\AgdaSpace{}%
\AgdaOperator{\AgdaFunction{⊨}}\AgdaSpace{}%
\AgdaGeneralizable{γ}\AgdaSpace{}%
\AgdaSymbol{→}\AgdaSpace{}%
\AgdaGeneralizable{a}\AgdaSpace{}%
\AgdaOperator{\AgdaFunction{∈}}\AgdaSpace{}%
\AgdaOperator{\AgdaFunction{⟦}}\AgdaSpace{}%
\AgdaGeneralizable{T}\AgdaSpace{}%
\AgdaOperator{\AgdaFunction{⟧}}\AgdaSpace{}%
\AgdaSymbol{→}\AgdaSpace{}%
\AgdaGeneralizable{Γ}\AgdaSpace{}%
\AgdaOperator{\AgdaInductiveConstructor{·:}}\AgdaSpace{}%
\AgdaGeneralizable{T}\AgdaSpace{}%
\AgdaOperator{\AgdaFunction{⊨}}\AgdaSpace{}%
\AgdaGeneralizable{γ}\AgdaSpace{}%
\AgdaOperator{\AgdaFunction{++}}\AgdaSpace{}%
\AgdaGeneralizable{a}\<%
\\
\>[2]\AgdaSymbol{(}\AgdaBound{⊨γ}\AgdaSpace{}%
\AgdaOperator{\AgdaFunction{\textasciicircum{}}}\AgdaSpace{}%
\AgdaBound{sa}\AgdaSymbol{)}\AgdaSpace{}%
\AgdaInductiveConstructor{zero}\AgdaSpace{}%
\AgdaSymbol{=}\AgdaSpace{}%
\AgdaBound{sa}\<%
\\
\>[2]\AgdaSymbol{(}\AgdaBound{⊨γ}\AgdaSpace{}%
\AgdaOperator{\AgdaFunction{\textasciicircum{}}}\AgdaSpace{}%
\AgdaBound{sa}\AgdaSymbol{)}\AgdaSpace{}%
\AgdaSymbol{(}\AgdaInductiveConstructor{suc}\AgdaSpace{}%
\AgdaBound{x}\AgdaSymbol{)}\AgdaSpace{}%
\AgdaSymbol{=}\AgdaSpace{}%
\AgdaBound{⊨γ}\AgdaSpace{}%
\AgdaBound{x}\<%
\\
\\[\AgdaEmptyExtraSkip]%
\>[2]\AgdaFunction{semantic-type}\AgdaSpace{}%
\AgdaSymbol{:}\AgdaSpace{}%
\AgdaGeneralizable{Γ}\AgdaSpace{}%
\AgdaOperator{\AgdaDatatype{⊢}}\AgdaSpace{}%
\AgdaGeneralizable{T}\AgdaSpace{}%
\AgdaSymbol{→}\AgdaSpace{}%
\AgdaPrimitive{Set}\<%
\\
\>[2]\AgdaFunction{semantic-type}\AgdaSpace{}%
\AgdaSymbol{\{}\AgdaBound{Γ}\AgdaSymbol{\}}\AgdaSpace{}%
\AgdaSymbol{\{}\AgdaBound{T}\AgdaSymbol{\}}\AgdaSpace{}%
\AgdaBound{t}\AgdaSpace{}%
\AgdaSymbol{=}\<%
\\
\>[2][@{}l@{\AgdaIndent{0}}]%
\>[5]\AgdaSymbol{∀}\AgdaSpace{}%
\AgdaSymbol{\{}\AgdaBound{γ}\AgdaSpace{}%
\AgdaSymbol{:}\AgdaSpace{}%
\AgdaFunction{Env}\AgdaSpace{}%
\AgdaBound{Γ}\AgdaSymbol{\}}\AgdaSpace{}%
\AgdaSymbol{→}\AgdaSpace{}%
\AgdaBound{Γ}\AgdaSpace{}%
\AgdaOperator{\AgdaFunction{⊨}}\AgdaSpace{}%
\AgdaBound{γ}\<%
\\
\>[5]\AgdaSymbol{→}\AgdaSpace{}%
\AgdaFunction{∃[}\AgdaSpace{}%
\AgdaBound{a}\AgdaSpace{}%
\AgdaFunction{]}\AgdaSpace{}%
\AgdaBound{γ}\AgdaSpace{}%
\AgdaOperator{\AgdaDatatype{∣}}\AgdaSpace{}%
\AgdaBound{t}\AgdaSpace{}%
\AgdaOperator{\AgdaDatatype{⇓}}\AgdaSpace{}%
\AgdaBound{a}\AgdaSpace{}%
\AgdaOperator{\AgdaFunction{×}}\AgdaSpace{}%
\AgdaBound{a}\AgdaSpace{}%
\AgdaOperator{\AgdaFunction{∈}}\AgdaSpace{}%
\AgdaOperator{\AgdaFunction{⟦}}\AgdaSpace{}%
\AgdaBound{T}\AgdaSpace{}%
\AgdaOperator{\AgdaFunction{⟧}}\<%
\\
\\[\AgdaEmptyExtraSkip]%
\>[2]\AgdaKeyword{syntax}\AgdaSpace{}%
\AgdaFunction{semantic-type}\AgdaSpace{}%
\AgdaSymbol{\{}\AgdaBound{Γ}\AgdaSymbol{\}}\AgdaSpace{}%
\AgdaSymbol{\{}\AgdaBound{T}\AgdaSymbol{\}}\AgdaSpace{}%
\AgdaBound{t}\AgdaSpace{}%
\AgdaSymbol{=}\AgdaSpace{}%
\AgdaBound{Γ}\AgdaSpace{}%
\AgdaFunction{⊨}\AgdaSpace{}%
\AgdaBound{t}\AgdaSpace{}%
\AgdaFunction{∷}\AgdaSpace{}%
\AgdaBound{T}\<%
\\
\\[\AgdaEmptyExtraSkip]%
\>[2]\AgdaFunction{fundamental-lemma}\AgdaSpace{}%
\AgdaSymbol{:}\AgdaSpace{}%
\AgdaSymbol{∀}\AgdaSpace{}%
\AgdaSymbol{(}\AgdaBound{t}\AgdaSpace{}%
\AgdaSymbol{:}\AgdaSpace{}%
\AgdaGeneralizable{Γ}\AgdaSpace{}%
\AgdaOperator{\AgdaDatatype{⊢}}\AgdaSpace{}%
\AgdaGeneralizable{T}\AgdaSymbol{)}\AgdaSpace{}%
\AgdaSymbol{→}\AgdaSpace{}%
\AgdaGeneralizable{Γ}\AgdaSpace{}%
\AgdaFunction{⊨}\AgdaSpace{}%
\AgdaBound{t}\AgdaSpace{}%
\AgdaFunction{∷}\AgdaSpace{}%
\AgdaGeneralizable{T}\<%
\\
\>[2]\AgdaFunction{fundamental-lemma}\AgdaSpace{}%
\AgdaInductiveConstructor{true}\AgdaSpace{}%
\AgdaBound{⊨γ}\AgdaSpace{}%
\AgdaSymbol{=}\AgdaSpace{}%
\AgdaInductiveConstructor{true}\AgdaSpace{}%
\AgdaOperator{\AgdaInductiveConstructor{,}}\AgdaSpace{}%
\AgdaInductiveConstructor{evalTrue}\AgdaSpace{}%
\AgdaOperator{\AgdaInductiveConstructor{,}}\AgdaSpace{}%
\AgdaInductiveConstructor{tt}\<%
\\
\>[2]\AgdaFunction{fundamental-lemma}\AgdaSpace{}%
\AgdaInductiveConstructor{false}\AgdaSpace{}%
\AgdaBound{⊨γ}\AgdaSpace{}%
\AgdaSymbol{=}\AgdaSpace{}%
\AgdaInductiveConstructor{false}\AgdaSpace{}%
\AgdaOperator{\AgdaInductiveConstructor{,}}\AgdaSpace{}%
\AgdaInductiveConstructor{evalFalse}\AgdaSpace{}%
\AgdaOperator{\AgdaInductiveConstructor{,}}\AgdaSpace{}%
\AgdaInductiveConstructor{tt}\<%
\\
\>[2]\AgdaFunction{fundamental-lemma}\AgdaSpace{}%
\AgdaSymbol{(}\AgdaInductiveConstructor{var}\AgdaSpace{}%
\AgdaBound{x}\AgdaSymbol{)}\AgdaSpace{}%
\AgdaSymbol{\{}\AgdaBound{γ}\AgdaSymbol{\}}\AgdaSpace{}%
\AgdaBound{⊨γ}\AgdaSpace{}%
\AgdaSymbol{=}\<%
\\
\>[2][@{}l@{\AgdaIndent{0}}]%
\>[4]\AgdaBound{γ}\AgdaSpace{}%
\AgdaBound{x}\AgdaSpace{}%
\AgdaOperator{\AgdaInductiveConstructor{,}}\AgdaSpace{}%
\AgdaInductiveConstructor{evalVar}\AgdaSpace{}%
\AgdaOperator{\AgdaInductiveConstructor{,}}\AgdaSpace{}%
\AgdaBound{⊨γ}\AgdaSpace{}%
\AgdaBound{x}\<%
\\
\>[2]\AgdaFunction{fundamental-lemma}\AgdaSpace{}%
\AgdaSymbol{(}\AgdaOperator{\AgdaInductiveConstructor{ƛ}}\AgdaSpace{}%
\AgdaBound{t}\AgdaSymbol{)}\AgdaSpace{}%
\AgdaSymbol{\{}\AgdaBound{γ}\AgdaSymbol{\}}\AgdaSpace{}%
\AgdaBound{⊨γ}\AgdaSpace{}%
\AgdaSymbol{=}\<%
\\
\>[2][@{}l@{\AgdaIndent{0}}]%
\>[5]\AgdaOperator{\AgdaInductiveConstructor{⟨ƛ}}\AgdaSpace{}%
\AgdaBound{t}\AgdaSpace{}%
\AgdaOperator{\AgdaInductiveConstructor{⟩}}\AgdaSpace{}%
\AgdaBound{γ}\AgdaSpace{}%
\AgdaOperator{\AgdaInductiveConstructor{,}}\<%
\\
\>[5]\AgdaInductiveConstructor{evalAbs}\AgdaSpace{}%
\AgdaOperator{\AgdaInductiveConstructor{,}}\<%
\\
\>[5]\AgdaSymbol{λ}\AgdaSpace{}%
\AgdaBound{sa}\AgdaSpace{}%
\AgdaSymbol{→}\AgdaSpace{}%
\AgdaFunction{fundamental-lemma}\AgdaSpace{}%
\AgdaBound{t}\AgdaSpace{}%
\AgdaSymbol{(}\AgdaBound{⊨γ}\AgdaSpace{}%
\AgdaOperator{\AgdaFunction{\textasciicircum{}}}\AgdaSpace{}%
\AgdaBound{sa}\AgdaSymbol{)}\<%
\\
\>[2]\AgdaFunction{fundamental-lemma}\AgdaSpace{}%
\AgdaSymbol{(}\AgdaBound{r}\AgdaSpace{}%
\AgdaOperator{\AgdaInductiveConstructor{·}}\AgdaSpace{}%
\AgdaBound{s}\AgdaSymbol{)}\AgdaSpace{}%
\AgdaBound{⊨γ}\<%
\\
\>[2][@{}l@{\AgdaIndent{0}}]%
\>[4]\AgdaKeyword{with}\AgdaSpace{}%
\AgdaFunction{fundamental-lemma}\AgdaSpace{}%
\AgdaBound{r}\AgdaSpace{}%
\AgdaBound{⊨γ}\<%
\\
\>[2]\AgdaSymbol{...}\AgdaSpace{}%
\AgdaSymbol{|}\AgdaSpace{}%
\AgdaOperator{\AgdaInductiveConstructor{⟨ƛ}}\AgdaSpace{}%
\AgdaBound{t}\AgdaSpace{}%
\AgdaOperator{\AgdaInductiveConstructor{⟩}}\AgdaSpace{}%
\AgdaBound{δ}\AgdaSpace{}%
\AgdaOperator{\AgdaInductiveConstructor{,}}\AgdaSpace{}%
\AgdaBound{r⇓}\AgdaSpace{}%
\AgdaOperator{\AgdaInductiveConstructor{,}}\AgdaSpace{}%
\AgdaBound{sf}%
\>[28]\AgdaSymbol{=}\<%
\\
\>[2][@{}l@{\AgdaIndent{0}}]%
\>[4]\AgdaKeyword{let}\AgdaSpace{}%
\AgdaSymbol{(}\AgdaBound{a}\AgdaSpace{}%
\AgdaOperator{\AgdaInductiveConstructor{,}}\AgdaSpace{}%
\AgdaBound{s⇓}\AgdaSpace{}%
\AgdaOperator{\AgdaInductiveConstructor{,}}\AgdaSpace{}%
\AgdaBound{sa}\AgdaSymbol{)}\AgdaSpace{}%
\AgdaSymbol{=}\AgdaSpace{}%
\AgdaFunction{fundamental-lemma}\AgdaSpace{}%
\AgdaBound{s}\AgdaSpace{}%
\AgdaBound{⊨γ}\AgdaSpace{}%
\AgdaKeyword{in}\<%
\\
\>[4]\AgdaKeyword{let}\AgdaSpace{}%
\AgdaSymbol{(}\AgdaBound{b}\AgdaSpace{}%
\AgdaOperator{\AgdaInductiveConstructor{,}}\AgdaSpace{}%
\AgdaBound{t⇓}\AgdaSpace{}%
\AgdaOperator{\AgdaInductiveConstructor{,}}\AgdaSpace{}%
\AgdaBound{sb}\AgdaSymbol{)}\AgdaSpace{}%
\AgdaSymbol{=}\AgdaSpace{}%
\AgdaBound{sf}\AgdaSpace{}%
\AgdaBound{sa}\AgdaSpace{}%
\AgdaKeyword{in}\<%
\\
\>[4]\AgdaBound{b}\AgdaSpace{}%
\AgdaOperator{\AgdaInductiveConstructor{,}}\AgdaSpace{}%
\AgdaInductiveConstructor{evalApp}\AgdaSpace{}%
\AgdaBound{r⇓}\AgdaSpace{}%
\AgdaBound{s⇓}\AgdaSpace{}%
\AgdaBound{t⇓}\AgdaSpace{}%
\AgdaOperator{\AgdaInductiveConstructor{,}}\AgdaSpace{}%
\AgdaBound{sb}\<%
\\
\>[2]\AgdaFunction{fundamental-lemma}\AgdaSpace{}%
\AgdaSymbol{(}\AgdaOperator{\AgdaInductiveConstructor{if}}\AgdaSpace{}%
\AgdaBound{r}\AgdaSpace{}%
\AgdaOperator{\AgdaInductiveConstructor{then}}\AgdaSpace{}%
\AgdaBound{s}\AgdaSpace{}%
\AgdaOperator{\AgdaInductiveConstructor{else}}\AgdaSpace{}%
\AgdaBound{t}\AgdaSymbol{)}\AgdaSpace{}%
\AgdaBound{⊨γ}\<%
\\
\>[2][@{}l@{\AgdaIndent{0}}]%
\>[4]\AgdaKeyword{with}\AgdaSpace{}%
\AgdaFunction{fundamental-lemma}\AgdaSpace{}%
\AgdaBound{r}\AgdaSpace{}%
\AgdaBound{⊨γ}\<%
\\
\>[2]\AgdaSymbol{...}\AgdaSpace{}%
\AgdaSymbol{|}\AgdaSpace{}%
\AgdaInductiveConstructor{true}\AgdaSpace{}%
\AgdaOperator{\AgdaInductiveConstructor{,}}\AgdaSpace{}%
\AgdaBound{r⇓}\AgdaSpace{}%
\AgdaOperator{\AgdaInductiveConstructor{,}}\AgdaSpace{}%
\AgdaSymbol{\AgdaUnderscore{}}\AgdaSpace{}%
\AgdaSymbol{=}\<%
\\
\>[2][@{}l@{\AgdaIndent{0}}]%
\>[4]\AgdaKeyword{let}\AgdaSpace{}%
\AgdaSymbol{(}\AgdaBound{a}\AgdaSpace{}%
\AgdaOperator{\AgdaInductiveConstructor{,}}\AgdaSpace{}%
\AgdaBound{s⇓}\AgdaSpace{}%
\AgdaOperator{\AgdaInductiveConstructor{,}}\AgdaSpace{}%
\AgdaBound{sa}\AgdaSymbol{)}\AgdaSpace{}%
\AgdaSymbol{=}\AgdaSpace{}%
\AgdaFunction{fundamental-lemma}\AgdaSpace{}%
\AgdaBound{s}\AgdaSpace{}%
\AgdaBound{⊨γ}\AgdaSpace{}%
\AgdaKeyword{in}\<%
\\
\>[4]\AgdaBound{a}\AgdaSpace{}%
\AgdaOperator{\AgdaInductiveConstructor{,}}\AgdaSpace{}%
\AgdaInductiveConstructor{evalIfTrue}\AgdaSpace{}%
\AgdaBound{r⇓}\AgdaSpace{}%
\AgdaBound{s⇓}\AgdaSpace{}%
\AgdaOperator{\AgdaInductiveConstructor{,}}\AgdaSpace{}%
\AgdaBound{sa}\<%
\\
\>[2]\AgdaSymbol{...}\AgdaSpace{}%
\AgdaSymbol{|}\AgdaSpace{}%
\AgdaInductiveConstructor{false}\AgdaSpace{}%
\AgdaOperator{\AgdaInductiveConstructor{,}}\AgdaSpace{}%
\AgdaBound{r⇓}\AgdaSpace{}%
\AgdaOperator{\AgdaInductiveConstructor{,}}\AgdaSpace{}%
\AgdaSymbol{\AgdaUnderscore{}}\AgdaSpace{}%
\AgdaSymbol{=}\<%
\\
\>[2][@{}l@{\AgdaIndent{0}}]%
\>[4]\AgdaKeyword{let}\AgdaSpace{}%
\AgdaSymbol{(}\AgdaBound{b}\AgdaSpace{}%
\AgdaOperator{\AgdaInductiveConstructor{,}}\AgdaSpace{}%
\AgdaBound{t⇓}\AgdaSpace{}%
\AgdaOperator{\AgdaInductiveConstructor{,}}\AgdaSpace{}%
\AgdaBound{sb}\AgdaSymbol{)}\AgdaSpace{}%
\AgdaSymbol{=}\AgdaSpace{}%
\AgdaFunction{fundamental-lemma}\AgdaSpace{}%
\AgdaBound{t}\AgdaSpace{}%
\AgdaBound{⊨γ}\AgdaSpace{}%
\AgdaKeyword{in}\<%
\\
\>[4]\AgdaBound{b}\AgdaSpace{}%
\AgdaOperator{\AgdaInductiveConstructor{,}}\AgdaSpace{}%
\AgdaInductiveConstructor{evalIfFalse}\AgdaSpace{}%
\AgdaBound{r⇓}\AgdaSpace{}%
\AgdaBound{t⇓}\AgdaSpace{}%
\AgdaOperator{\AgdaInductiveConstructor{,}}\AgdaSpace{}%
\AgdaBound{sb}\<%
\\
\\[\AgdaEmptyExtraSkip]%
\>[2]\AgdaComment{--\ Evaluation\ is\ total}\<%
\\
\>[2]\AgdaFunction{⇓-total}\AgdaSpace{}%
\AgdaSymbol{:}\AgdaSpace{}%
\AgdaSymbol{∀}\AgdaSpace{}%
\AgdaSymbol{(}\AgdaBound{t}\AgdaSpace{}%
\AgdaSymbol{:}\AgdaSpace{}%
\AgdaInductiveConstructor{∅}\AgdaSpace{}%
\AgdaOperator{\AgdaDatatype{⊢}}\AgdaSpace{}%
\AgdaGeneralizable{T}\AgdaSymbol{)}\AgdaSpace{}%
\AgdaSymbol{→}\AgdaSpace{}%
\AgdaFunction{⇓-well-defined}\AgdaSpace{}%
\AgdaBound{t}\<%
\\
\>[2]\AgdaFunction{⇓-total}\AgdaSpace{}%
\AgdaBound{t}\AgdaSpace{}%
\AgdaSymbol{=}\<%
\\
\>[2][@{}l@{\AgdaIndent{0}}]%
\>[4]\AgdaKeyword{let}\AgdaSpace{}%
\AgdaSymbol{(}\AgdaBound{a}\AgdaSpace{}%
\AgdaOperator{\AgdaInductiveConstructor{,}}\AgdaSpace{}%
\AgdaBound{t⇓a}\AgdaSpace{}%
\AgdaOperator{\AgdaInductiveConstructor{,}}\AgdaSpace{}%
\AgdaSymbol{\AgdaUnderscore{})}\AgdaSpace{}%
\AgdaSymbol{=}\AgdaSpace{}%
\AgdaFunction{fundamental-lemma}\AgdaSpace{}%
\AgdaBound{t}\AgdaSpace{}%
\AgdaSymbol{(λ}\AgdaSpace{}%
\AgdaSymbol{())}\AgdaSpace{}%
\AgdaKeyword{in}\<%
\\
\>[4]\AgdaBound{a}\AgdaSpace{}%
\AgdaOperator{\AgdaInductiveConstructor{,}}\AgdaSpace{}%
\AgdaBound{t⇓a}\<%
\end{code}

\subsection{Denotational semantics (Figure \ref{fig:denotational})}

\begin{code}%
\>[2]\AgdaComment{--\ Denotation\ of\ types}\<%
\\
\>[2]\AgdaOperator{\AgdaFunction{𝒯⟦\AgdaUnderscore{}⟧}}\AgdaSpace{}%
\AgdaSymbol{:}\AgdaSpace{}%
\AgdaDatatype{Type}\AgdaSpace{}%
\AgdaSymbol{→}\AgdaSpace{}%
\AgdaPrimitive{Set}\<%
\\
\>[2]\AgdaOperator{\AgdaFunction{𝒯⟦}}\AgdaSpace{}%
\AgdaInductiveConstructor{bool}\AgdaSpace{}%
\AgdaOperator{\AgdaFunction{⟧}}\AgdaSpace{}%
\AgdaSymbol{=}\AgdaSpace{}%
\AgdaDatatype{Bool}\<%
\\
\>[2]\AgdaOperator{\AgdaFunction{𝒯⟦}}\AgdaSpace{}%
\AgdaBound{S}\AgdaSpace{}%
\AgdaOperator{\AgdaInductiveConstructor{⇒}}\AgdaSpace{}%
\AgdaBound{T}\AgdaSpace{}%
\AgdaOperator{\AgdaFunction{⟧}}\AgdaSpace{}%
\AgdaSymbol{=}\AgdaSpace{}%
\AgdaOperator{\AgdaFunction{𝒯⟦}}\AgdaSpace{}%
\AgdaBound{S}\AgdaSpace{}%
\AgdaOperator{\AgdaFunction{⟧}}\AgdaSpace{}%
\AgdaSymbol{→}\AgdaSpace{}%
\AgdaOperator{\AgdaFunction{𝒯⟦}}\AgdaSpace{}%
\AgdaBound{T}\AgdaSpace{}%
\AgdaOperator{\AgdaFunction{⟧}}\<%
\\
\\[\AgdaEmptyExtraSkip]%
\>[2]\AgdaComment{--\ Denotation\ of\ contexts}\<%
\\
\>[2]\AgdaOperator{\AgdaFunction{𝒞⟦\AgdaUnderscore{}⟧}}\AgdaSpace{}%
\AgdaSymbol{:}\AgdaSpace{}%
\AgdaDatatype{Ctx}\AgdaSpace{}%
\AgdaSymbol{→}\AgdaSpace{}%
\AgdaPrimitive{Set}\<%
\\
\>[2]\AgdaOperator{\AgdaFunction{𝒞⟦}}\AgdaSpace{}%
\AgdaBound{Γ}\AgdaSpace{}%
\AgdaOperator{\AgdaFunction{⟧}}\AgdaSpace{}%
\AgdaSymbol{=}\AgdaSpace{}%
\AgdaSymbol{∀}\AgdaSpace{}%
\AgdaSymbol{\{}\AgdaBound{T}\AgdaSymbol{\}}\AgdaSpace{}%
\AgdaSymbol{→}\AgdaSpace{}%
\AgdaBound{Γ}\AgdaSpace{}%
\AgdaOperator{\AgdaDatatype{∋}}\AgdaSpace{}%
\AgdaBound{T}\AgdaSpace{}%
\AgdaSymbol{→}\AgdaSpace{}%
\AgdaOperator{\AgdaFunction{𝒯⟦}}\AgdaSpace{}%
\AgdaBound{T}\AgdaSpace{}%
\AgdaOperator{\AgdaFunction{⟧}}\<%
\\
\\[\AgdaEmptyExtraSkip]%
\>[2]\AgdaComment{--\ Extending\ denoted\ contexts}\<%
\\
\>[2]\AgdaOperator{\AgdaFunction{\AgdaUnderscore{}\&\AgdaUnderscore{}}}\AgdaSpace{}%
\AgdaSymbol{:}\AgdaSpace{}%
\AgdaOperator{\AgdaFunction{𝒞⟦}}\AgdaSpace{}%
\AgdaGeneralizable{Γ}\AgdaSpace{}%
\AgdaOperator{\AgdaFunction{⟧}}\AgdaSpace{}%
\AgdaSymbol{→}\AgdaSpace{}%
\AgdaOperator{\AgdaFunction{𝒯⟦}}\AgdaSpace{}%
\AgdaGeneralizable{T}\AgdaSpace{}%
\AgdaOperator{\AgdaFunction{⟧}}\AgdaSpace{}%
\AgdaSymbol{→}\AgdaSpace{}%
\AgdaOperator{\AgdaFunction{𝒞⟦}}\AgdaSpace{}%
\AgdaGeneralizable{Γ}\AgdaSpace{}%
\AgdaOperator{\AgdaInductiveConstructor{·:}}\AgdaSpace{}%
\AgdaGeneralizable{T}\AgdaSpace{}%
\AgdaOperator{\AgdaFunction{⟧}}\<%
\\
\>[2]\AgdaSymbol{(\AgdaUnderscore{}}\AgdaSpace{}%
\AgdaOperator{\AgdaFunction{\&}}\AgdaSpace{}%
\AgdaBound{a}\AgdaSymbol{)}\AgdaSpace{}%
\AgdaInductiveConstructor{zero}\AgdaSpace{}%
\AgdaSymbol{=}\AgdaSpace{}%
\AgdaBound{a}\<%
\\
\>[2]\AgdaSymbol{(}\AgdaBound{ρ}\AgdaSpace{}%
\AgdaOperator{\AgdaFunction{\&}}\AgdaSpace{}%
\AgdaSymbol{\AgdaUnderscore{})}\AgdaSpace{}%
\AgdaSymbol{(}\AgdaInductiveConstructor{suc}\AgdaSpace{}%
\AgdaBound{x}\AgdaSymbol{)}\AgdaSpace{}%
\AgdaSymbol{=}\AgdaSpace{}%
\AgdaBound{ρ}\AgdaSpace{}%
\AgdaBound{x}\<%
\\
\\[\AgdaEmptyExtraSkip]%
\>[2]\AgdaComment{--\ Denotation\ of\ terms}\<%
\\
\>[2]\AgdaOperator{\AgdaFunction{ℰ⟦\AgdaUnderscore{}⟧}}\AgdaSpace{}%
\AgdaSymbol{:}\AgdaSpace{}%
\AgdaGeneralizable{Γ}\AgdaSpace{}%
\AgdaOperator{\AgdaDatatype{⊢}}\AgdaSpace{}%
\AgdaGeneralizable{T}\AgdaSpace{}%
\AgdaSymbol{→}\AgdaSpace{}%
\AgdaOperator{\AgdaFunction{𝒞⟦}}\AgdaSpace{}%
\AgdaGeneralizable{Γ}\AgdaSpace{}%
\AgdaOperator{\AgdaFunction{⟧}}\AgdaSpace{}%
\AgdaSymbol{→}\AgdaSpace{}%
\AgdaOperator{\AgdaFunction{𝒯⟦}}\AgdaSpace{}%
\AgdaGeneralizable{T}\AgdaSpace{}%
\AgdaOperator{\AgdaFunction{⟧}}\<%
\\
\>[2]\AgdaOperator{\AgdaFunction{ℰ⟦}}\AgdaSpace{}%
\AgdaInductiveConstructor{true}\AgdaSpace{}%
\AgdaOperator{\AgdaFunction{⟧}}\AgdaSpace{}%
\AgdaBound{ρ}\AgdaSpace{}%
\AgdaSymbol{=}\AgdaSpace{}%
\AgdaInductiveConstructor{true}\<%
\\
\>[2]\AgdaOperator{\AgdaFunction{ℰ⟦}}\AgdaSpace{}%
\AgdaInductiveConstructor{false}\AgdaSpace{}%
\AgdaOperator{\AgdaFunction{⟧}}\AgdaSpace{}%
\AgdaBound{ρ}\AgdaSpace{}%
\AgdaSymbol{=}\AgdaSpace{}%
\AgdaInductiveConstructor{false}\<%
\\
\>[2]\AgdaOperator{\AgdaFunction{ℰ⟦}}\AgdaSpace{}%
\AgdaInductiveConstructor{var}\AgdaSpace{}%
\AgdaBound{x}\AgdaSpace{}%
\AgdaOperator{\AgdaFunction{⟧}}\AgdaSpace{}%
\AgdaBound{ρ}\AgdaSpace{}%
\AgdaSymbol{=}\AgdaSpace{}%
\AgdaBound{ρ}\AgdaSpace{}%
\AgdaBound{x}\<%
\\
\>[2]\AgdaOperator{\AgdaFunction{ℰ⟦}}\AgdaSpace{}%
\AgdaOperator{\AgdaInductiveConstructor{ƛ}}\AgdaSpace{}%
\AgdaBound{t}\AgdaSpace{}%
\AgdaOperator{\AgdaFunction{⟧}}\AgdaSpace{}%
\AgdaBound{ρ}\AgdaSpace{}%
\AgdaSymbol{=}\AgdaSpace{}%
\AgdaSymbol{λ}\AgdaSpace{}%
\AgdaBound{a}\AgdaSpace{}%
\AgdaSymbol{→}\AgdaSpace{}%
\AgdaOperator{\AgdaFunction{ℰ⟦}}\AgdaSpace{}%
\AgdaBound{t}\AgdaSpace{}%
\AgdaOperator{\AgdaFunction{⟧}}\AgdaSpace{}%
\AgdaSymbol{(}\AgdaBound{ρ}\AgdaSpace{}%
\AgdaOperator{\AgdaFunction{\&}}\AgdaSpace{}%
\AgdaBound{a}\AgdaSymbol{)}\<%
\\
\>[2]\AgdaOperator{\AgdaFunction{ℰ⟦}}\AgdaSpace{}%
\AgdaBound{r}\AgdaSpace{}%
\AgdaOperator{\AgdaInductiveConstructor{·}}\AgdaSpace{}%
\AgdaBound{s}\AgdaSpace{}%
\AgdaOperator{\AgdaFunction{⟧}}\AgdaSpace{}%
\AgdaBound{ρ}\AgdaSpace{}%
\AgdaSymbol{=}\AgdaSpace{}%
\AgdaOperator{\AgdaFunction{ℰ⟦}}\AgdaSpace{}%
\AgdaBound{r}\AgdaSpace{}%
\AgdaOperator{\AgdaFunction{⟧}}\AgdaSpace{}%
\AgdaBound{ρ}\AgdaSpace{}%
\AgdaSymbol{(}\AgdaOperator{\AgdaFunction{ℰ⟦}}\AgdaSpace{}%
\AgdaBound{s}\AgdaSpace{}%
\AgdaOperator{\AgdaFunction{⟧}}\AgdaSpace{}%
\AgdaBound{ρ}\AgdaSymbol{)}\<%
\\
\>[2]\AgdaOperator{\AgdaFunction{ℰ⟦}}\AgdaSpace{}%
\AgdaOperator{\AgdaInductiveConstructor{if}}\AgdaSpace{}%
\AgdaBound{r}\AgdaSpace{}%
\AgdaOperator{\AgdaInductiveConstructor{then}}\AgdaSpace{}%
\AgdaBound{s}\AgdaSpace{}%
\AgdaOperator{\AgdaInductiveConstructor{else}}\AgdaSpace{}%
\AgdaBound{t}\AgdaSpace{}%
\AgdaOperator{\AgdaFunction{⟧}}\AgdaSpace{}%
\AgdaBound{ρ}\<%
\\
\>[2][@{}l@{\AgdaIndent{0}}]%
\>[4]\AgdaKeyword{with}\AgdaSpace{}%
\AgdaOperator{\AgdaFunction{ℰ⟦}}\AgdaSpace{}%
\AgdaBound{r}\AgdaSpace{}%
\AgdaOperator{\AgdaFunction{⟧}}\AgdaSpace{}%
\AgdaBound{ρ}\<%
\\
\>[2]\AgdaSymbol{...}\AgdaSpace{}%
\AgdaSymbol{|}\AgdaSpace{}%
\AgdaInductiveConstructor{true}\AgdaSpace{}%
\AgdaSymbol{=}\AgdaSpace{}%
\AgdaOperator{\AgdaFunction{ℰ⟦}}\AgdaSpace{}%
\AgdaBound{s}\AgdaSpace{}%
\AgdaOperator{\AgdaFunction{⟧}}\AgdaSpace{}%
\AgdaBound{ρ}\<%
\\
\>[2]\AgdaSymbol{...}\AgdaSpace{}%
\AgdaSymbol{|}\AgdaSpace{}%
\AgdaInductiveConstructor{false}\AgdaSpace{}%
\AgdaSymbol{=}\AgdaSpace{}%
\AgdaOperator{\AgdaFunction{ℰ⟦}}\AgdaSpace{}%
\AgdaBound{t}\AgdaSpace{}%
\AgdaOperator{\AgdaFunction{⟧}}\AgdaSpace{}%
\AgdaBound{ρ}\<%
\\
\\[\AgdaEmptyExtraSkip]%
\>[2]\AgdaKeyword{mutual}\<%
\\
\>[2][@{}l@{\AgdaIndent{0}}]%
\>[4]\AgdaComment{--\ Denotation\ of\ environments}\<%
\\
\>[4]\AgdaOperator{\AgdaFunction{𝒢⟦\AgdaUnderscore{}⟧}}\AgdaSpace{}%
\AgdaSymbol{:}\AgdaSpace{}%
\AgdaFunction{Env}\AgdaSpace{}%
\AgdaGeneralizable{Γ}\AgdaSpace{}%
\AgdaSymbol{→}\AgdaSpace{}%
\AgdaOperator{\AgdaFunction{𝒞⟦}}\AgdaSpace{}%
\AgdaGeneralizable{Γ}\AgdaSpace{}%
\AgdaOperator{\AgdaFunction{⟧}}\<%
\\
\>[4]\AgdaOperator{\AgdaFunction{𝒢⟦}}\AgdaSpace{}%
\AgdaBound{γ}\AgdaSpace{}%
\AgdaOperator{\AgdaFunction{⟧}}\AgdaSpace{}%
\AgdaBound{x}\AgdaSpace{}%
\AgdaSymbol{=}\AgdaSpace{}%
\AgdaOperator{\AgdaFunction{𝒟⟦}}\AgdaSpace{}%
\AgdaBound{γ}\AgdaSpace{}%
\AgdaBound{x}\AgdaSpace{}%
\AgdaOperator{\AgdaFunction{⟧}}\<%
\\
\\[\AgdaEmptyExtraSkip]%
\>[4]\AgdaComment{--\ Denotation\ of\ domain\ elements}\<%
\\
\>[4]\AgdaOperator{\AgdaFunction{𝒟⟦\AgdaUnderscore{}⟧}}\AgdaSpace{}%
\AgdaSymbol{:}\AgdaSpace{}%
\AgdaDatatype{Domain}\AgdaSpace{}%
\AgdaGeneralizable{T}\AgdaSpace{}%
\AgdaSymbol{→}\AgdaSpace{}%
\AgdaOperator{\AgdaFunction{𝒯⟦}}\AgdaSpace{}%
\AgdaGeneralizable{T}\AgdaSpace{}%
\AgdaOperator{\AgdaFunction{⟧}}\<%
\\
\>[4]\AgdaOperator{\AgdaFunction{𝒟⟦}}\AgdaSpace{}%
\AgdaInductiveConstructor{true}\AgdaSpace{}%
\AgdaOperator{\AgdaFunction{⟧}}\AgdaSpace{}%
\AgdaSymbol{=}\AgdaSpace{}%
\AgdaInductiveConstructor{true}\<%
\\
\>[4]\AgdaOperator{\AgdaFunction{𝒟⟦}}\AgdaSpace{}%
\AgdaInductiveConstructor{false}\AgdaSpace{}%
\AgdaOperator{\AgdaFunction{⟧}}\AgdaSpace{}%
\AgdaSymbol{=}\AgdaSpace{}%
\AgdaInductiveConstructor{false}\<%
\\
\>[4]\AgdaOperator{\AgdaFunction{𝒟⟦}}\AgdaSpace{}%
\AgdaOperator{\AgdaInductiveConstructor{⟨ƛ}}\AgdaSpace{}%
\AgdaBound{t}\AgdaSpace{}%
\AgdaOperator{\AgdaInductiveConstructor{⟩}}\AgdaSpace{}%
\AgdaBound{γ}\AgdaSpace{}%
\AgdaOperator{\AgdaFunction{⟧}}\AgdaSpace{}%
\AgdaSymbol{=}\AgdaSpace{}%
\AgdaSymbol{λ}\AgdaSpace{}%
\AgdaBound{a}\AgdaSpace{}%
\AgdaSymbol{→}\AgdaSpace{}%
\AgdaOperator{\AgdaFunction{ℰ⟦}}\AgdaSpace{}%
\AgdaBound{t}\AgdaSpace{}%
\AgdaOperator{\AgdaFunction{⟧}}\AgdaSpace{}%
\AgdaSymbol{(}\AgdaOperator{\AgdaFunction{𝒢⟦}}\AgdaSpace{}%
\AgdaBound{γ}\AgdaSpace{}%
\AgdaOperator{\AgdaFunction{⟧}}\AgdaSpace{}%
\AgdaOperator{\AgdaFunction{\&}}\AgdaSpace{}%
\AgdaBound{a}\AgdaSymbol{)}\<%
\\
\\[\AgdaEmptyExtraSkip]%
\>[2]\AgdaComment{--\ Denotational\ equivalence}\<%
\\
\>[2]\AgdaOperator{\AgdaFunction{\AgdaUnderscore{}ℰ≡\AgdaUnderscore{}}}\AgdaSpace{}%
\AgdaSymbol{:}\AgdaSpace{}%
\AgdaSymbol{∀}\AgdaSpace{}%
\AgdaSymbol{(}\AgdaBound{t}\AgdaSpace{}%
\AgdaBound{v}\AgdaSpace{}%
\AgdaSymbol{:}\AgdaSpace{}%
\AgdaGeneralizable{Γ}\AgdaSpace{}%
\AgdaOperator{\AgdaDatatype{⊢}}\AgdaSpace{}%
\AgdaGeneralizable{T}\AgdaSymbol{)}\AgdaSpace{}%
\AgdaSymbol{→}\AgdaSpace{}%
\AgdaPrimitive{Set}\<%
\\
\>[2]\AgdaOperator{\AgdaFunction{\AgdaUnderscore{}ℰ≡\AgdaUnderscore{}}}\AgdaSpace{}%
\AgdaSymbol{\{}\AgdaBound{Γ}\AgdaSymbol{\}}\AgdaSpace{}%
\AgdaBound{t}\AgdaSpace{}%
\AgdaBound{v}\AgdaSpace{}%
\AgdaSymbol{=}\AgdaSpace{}%
\AgdaSymbol{∀}\AgdaSpace{}%
\AgdaSymbol{\{}\AgdaBound{ρ}\AgdaSpace{}%
\AgdaSymbol{:}\AgdaSpace{}%
\AgdaOperator{\AgdaFunction{𝒞⟦}}\AgdaSpace{}%
\AgdaBound{Γ}\AgdaSpace{}%
\AgdaOperator{\AgdaFunction{⟧}}\AgdaSymbol{\}}\AgdaSpace{}%
\AgdaSymbol{→}\AgdaSpace{}%
\AgdaOperator{\AgdaFunction{ℰ⟦}}\AgdaSpace{}%
\AgdaBound{t}\AgdaSpace{}%
\AgdaOperator{\AgdaFunction{⟧}}\AgdaSpace{}%
\AgdaBound{ρ}\AgdaSpace{}%
\AgdaOperator{\AgdaDatatype{≡}}\AgdaSpace{}%
\AgdaOperator{\AgdaFunction{ℰ⟦}}\AgdaSpace{}%
\AgdaBound{v}\AgdaSpace{}%
\AgdaOperator{\AgdaFunction{⟧}}\AgdaSpace{}%
\AgdaBound{ρ}\<%
\end{code}

\subsection{Reading back a value from the evaluation of a term}

So far, everything is the same as we have presented in the paper. Now, on top of
proving that evaluation is total, we prove normalization with weak head
reduction. To do so, we read back a value from the evaluation of a term.
For this, we have to truly ``close'' a closure to produce a closed term that
is a value.

\begin{code}%
\>[2]\AgdaComment{--\ Context\ append}\<%
\\
\>[2]\AgdaOperator{\AgdaFunction{\AgdaUnderscore{}<>\AgdaUnderscore{}}}\AgdaSpace{}%
\AgdaSymbol{:}\AgdaSpace{}%
\AgdaDatatype{Ctx}\AgdaSpace{}%
\AgdaSymbol{→}\AgdaSpace{}%
\AgdaDatatype{Ctx}\AgdaSpace{}%
\AgdaSymbol{→}\AgdaSpace{}%
\AgdaDatatype{Ctx}\<%
\\
\>[2]\AgdaBound{Δ}\AgdaSpace{}%
\AgdaOperator{\AgdaFunction{<>}}\AgdaSpace{}%
\AgdaInductiveConstructor{∅}\AgdaSpace{}%
\AgdaSymbol{=}\AgdaSpace{}%
\AgdaBound{Δ}\<%
\\
\>[2]\AgdaBound{Δ}\AgdaSpace{}%
\AgdaOperator{\AgdaFunction{<>}}\AgdaSpace{}%
\AgdaSymbol{(}\AgdaBound{Γ}\AgdaSpace{}%
\AgdaOperator{\AgdaInductiveConstructor{·:}}\AgdaSpace{}%
\AgdaBound{S}\AgdaSymbol{)}\AgdaSpace{}%
\AgdaSymbol{=}\AgdaSpace{}%
\AgdaSymbol{(}\AgdaBound{Δ}\AgdaSpace{}%
\AgdaOperator{\AgdaFunction{<>}}\AgdaSpace{}%
\AgdaBound{Γ}\AgdaSymbol{)}\AgdaSpace{}%
\AgdaOperator{\AgdaInductiveConstructor{·:}}\AgdaSpace{}%
\AgdaBound{S}\<%
\\
\\[\AgdaEmptyExtraSkip]%
\>[2]\AgdaComment{--\ Inject\ de\ Brujin\ index\ into\ larger\ context}\<%
\\
\>[2]\AgdaFunction{inject-var}\AgdaSpace{}%
\AgdaSymbol{:}\AgdaSpace{}%
\AgdaGeneralizable{Γ}\AgdaSpace{}%
\AgdaOperator{\AgdaDatatype{∋}}\AgdaSpace{}%
\AgdaGeneralizable{T}\AgdaSpace{}%
\AgdaSymbol{→}\AgdaSpace{}%
\AgdaGeneralizable{Δ}\AgdaSpace{}%
\AgdaOperator{\AgdaFunction{<>}}\AgdaSpace{}%
\AgdaGeneralizable{Γ}\AgdaSpace{}%
\AgdaOperator{\AgdaDatatype{∋}}\AgdaSpace{}%
\AgdaGeneralizable{T}\<%
\\
\>[2]\AgdaFunction{inject-var}\AgdaSpace{}%
\AgdaInductiveConstructor{zero}\AgdaSpace{}%
\AgdaSymbol{=}\AgdaSpace{}%
\AgdaInductiveConstructor{zero}\<%
\\
\>[2]\AgdaFunction{inject-var}\AgdaSpace{}%
\AgdaSymbol{(}\AgdaInductiveConstructor{suc}\AgdaSpace{}%
\AgdaBound{x}\AgdaSymbol{)}\AgdaSpace{}%
\AgdaSymbol{=}\AgdaSpace{}%
\AgdaInductiveConstructor{suc}\AgdaSpace{}%
\AgdaSymbol{(}\AgdaFunction{inject-var}\AgdaSpace{}%
\AgdaBound{x}\AgdaSymbol{)}\<%
\\
\\[\AgdaEmptyExtraSkip]%
\>[2]\AgdaComment{--\ Inject\ term\ into\ larger\ context\ (similar\ to\ weakening)}\<%
\\
\>[2]\AgdaFunction{inject}\AgdaSpace{}%
\AgdaSymbol{:}\AgdaSpace{}%
\AgdaGeneralizable{Γ}\AgdaSpace{}%
\AgdaOperator{\AgdaDatatype{⊢}}\AgdaSpace{}%
\AgdaGeneralizable{T}\AgdaSpace{}%
\AgdaSymbol{→}\AgdaSpace{}%
\AgdaGeneralizable{Δ}\AgdaSpace{}%
\AgdaOperator{\AgdaFunction{<>}}\AgdaSpace{}%
\AgdaGeneralizable{Γ}\AgdaSpace{}%
\AgdaOperator{\AgdaDatatype{⊢}}\AgdaSpace{}%
\AgdaGeneralizable{T}\<%
\\
\>[2]\AgdaFunction{inject}\AgdaSpace{}%
\AgdaSymbol{(}\AgdaInductiveConstructor{var}\AgdaSpace{}%
\AgdaBound{x}\AgdaSymbol{)}\AgdaSpace{}%
\AgdaSymbol{=}\AgdaSpace{}%
\AgdaInductiveConstructor{var}\AgdaSpace{}%
\AgdaSymbol{(}\AgdaFunction{inject-var}\AgdaSpace{}%
\AgdaBound{x}\AgdaSymbol{)}\<%
\\
\>[2]\AgdaFunction{inject}\AgdaSpace{}%
\AgdaSymbol{(}\AgdaOperator{\AgdaInductiveConstructor{ƛ}}\AgdaSpace{}%
\AgdaBound{t}\AgdaSymbol{)}\AgdaSpace{}%
\AgdaSymbol{=}\AgdaSpace{}%
\AgdaOperator{\AgdaInductiveConstructor{ƛ}}\AgdaSpace{}%
\AgdaFunction{inject}\AgdaSpace{}%
\AgdaBound{t}\<%
\\
\>[2]\AgdaFunction{inject}\AgdaSpace{}%
\AgdaSymbol{(}\AgdaBound{r}\AgdaSpace{}%
\AgdaOperator{\AgdaInductiveConstructor{·}}\AgdaSpace{}%
\AgdaBound{s}\AgdaSymbol{)}\AgdaSpace{}%
\AgdaSymbol{=}\AgdaSpace{}%
\AgdaFunction{inject}\AgdaSpace{}%
\AgdaBound{r}\AgdaSpace{}%
\AgdaOperator{\AgdaInductiveConstructor{·}}\AgdaSpace{}%
\AgdaFunction{inject}\AgdaSpace{}%
\AgdaBound{s}\<%
\\
\>[2]\AgdaFunction{inject}\AgdaSpace{}%
\AgdaInductiveConstructor{true}\AgdaSpace{}%
\AgdaSymbol{=}\AgdaSpace{}%
\AgdaInductiveConstructor{true}\<%
\\
\>[2]\AgdaFunction{inject}\AgdaSpace{}%
\AgdaInductiveConstructor{false}\AgdaSpace{}%
\AgdaSymbol{=}\AgdaSpace{}%
\AgdaInductiveConstructor{false}\<%
\\
\>[2]\AgdaFunction{inject}\AgdaSpace{}%
\AgdaSymbol{(}\AgdaOperator{\AgdaInductiveConstructor{if}}\AgdaSpace{}%
\AgdaBound{t}\AgdaSpace{}%
\AgdaOperator{\AgdaInductiveConstructor{then}}\AgdaSpace{}%
\AgdaBound{u}\AgdaSpace{}%
\AgdaOperator{\AgdaInductiveConstructor{else}}\AgdaSpace{}%
\AgdaBound{v}\AgdaSymbol{)}\AgdaSpace{}%
\AgdaSymbol{=}\<%
\\
\>[2][@{}l@{\AgdaIndent{0}}]%
\>[4]\AgdaOperator{\AgdaInductiveConstructor{if}}\AgdaSpace{}%
\AgdaFunction{inject}\AgdaSpace{}%
\AgdaBound{t}\AgdaSpace{}%
\AgdaOperator{\AgdaInductiveConstructor{then}}\AgdaSpace{}%
\AgdaFunction{inject}\AgdaSpace{}%
\AgdaBound{u}\AgdaSpace{}%
\AgdaOperator{\AgdaInductiveConstructor{else}}\AgdaSpace{}%
\AgdaFunction{inject}\AgdaSpace{}%
\AgdaBound{v}\<%
\\
\\[\AgdaEmptyExtraSkip]%
\>[2]\AgdaComment{--\ If\ we\ have\ a\ variable\ in\ a\ injected\ context}\<%
\\
\>[2]\AgdaComment{--\ we\ can\ determine\ where\ it\ came\ from}\<%
\\
\>[2]\AgdaFunction{split}\AgdaSpace{}%
\AgdaSymbol{:}\AgdaSpace{}%
\AgdaGeneralizable{Γ}\AgdaSpace{}%
\AgdaOperator{\AgdaFunction{<>}}\AgdaSpace{}%
\AgdaGeneralizable{Δ}\AgdaSpace{}%
\AgdaOperator{\AgdaDatatype{∋}}\AgdaSpace{}%
\AgdaGeneralizable{T}\AgdaSpace{}%
\AgdaSymbol{→}\AgdaSpace{}%
\AgdaSymbol{(}\AgdaGeneralizable{Γ}\AgdaSpace{}%
\AgdaOperator{\AgdaDatatype{∋}}\AgdaSpace{}%
\AgdaGeneralizable{T}\AgdaSymbol{)}\AgdaSpace{}%
\AgdaOperator{\AgdaDatatype{⊎}}\AgdaSpace{}%
\AgdaSymbol{(}\AgdaGeneralizable{Δ}\AgdaSpace{}%
\AgdaOperator{\AgdaDatatype{∋}}\AgdaSpace{}%
\AgdaGeneralizable{T}\AgdaSymbol{)}\<%
\\
\>[2]\AgdaFunction{split}\AgdaSpace{}%
\AgdaSymbol{\{}\AgdaArgument{Δ}\AgdaSpace{}%
\AgdaSymbol{=}\AgdaSpace{}%
\AgdaInductiveConstructor{∅}\AgdaSymbol{\}}\AgdaSpace{}%
\AgdaBound{x}\AgdaSpace{}%
\AgdaSymbol{=}\AgdaSpace{}%
\AgdaInductiveConstructor{inj₁}\AgdaSpace{}%
\AgdaBound{x}\<%
\\
\>[2]\AgdaFunction{split}\AgdaSpace{}%
\AgdaSymbol{\{}\AgdaArgument{Δ}\AgdaSpace{}%
\AgdaSymbol{=}\AgdaSpace{}%
\AgdaSymbol{\AgdaUnderscore{}}\AgdaSpace{}%
\AgdaOperator{\AgdaInductiveConstructor{·:}}\AgdaSpace{}%
\AgdaSymbol{\AgdaUnderscore{}\}}\AgdaSpace{}%
\AgdaInductiveConstructor{zero}\AgdaSpace{}%
\AgdaSymbol{=}\AgdaSpace{}%
\AgdaInductiveConstructor{inj₂}\AgdaSpace{}%
\AgdaInductiveConstructor{zero}\<%
\\
\>[2]\AgdaFunction{split}\AgdaSpace{}%
\AgdaSymbol{\{}\AgdaArgument{Δ}\AgdaSpace{}%
\AgdaSymbol{=}\AgdaSpace{}%
\AgdaBound{Δ}\AgdaSpace{}%
\AgdaOperator{\AgdaInductiveConstructor{·:}}\AgdaSpace{}%
\AgdaSymbol{\AgdaUnderscore{}\}}\AgdaSpace{}%
\AgdaSymbol{(}\AgdaInductiveConstructor{suc}\AgdaSpace{}%
\AgdaBound{x}\AgdaSymbol{)}\<%
\\
\>[2][@{}l@{\AgdaIndent{0}}]%
\>[4]\AgdaKeyword{with}\AgdaSpace{}%
\AgdaFunction{split}\AgdaSpace{}%
\AgdaSymbol{\{}\AgdaArgument{Δ}\AgdaSpace{}%
\AgdaSymbol{=}\AgdaSpace{}%
\AgdaBound{Δ}\AgdaSymbol{\}}\AgdaSpace{}%
\AgdaBound{x}\<%
\\
\>[2]\AgdaSymbol{...}\AgdaSpace{}%
\AgdaSymbol{|}\AgdaSpace{}%
\AgdaInductiveConstructor{inj₁}\AgdaSpace{}%
\AgdaBound{x}\AgdaSpace{}%
\AgdaSymbol{=}\AgdaSpace{}%
\AgdaInductiveConstructor{inj₁}\AgdaSpace{}%
\AgdaBound{x}\<%
\\
\>[2]\AgdaSymbol{...}\AgdaSpace{}%
\AgdaSymbol{|}\AgdaSpace{}%
\AgdaInductiveConstructor{inj₂}\AgdaSpace{}%
\AgdaBound{y}\AgdaSpace{}%
\AgdaSymbol{=}\AgdaSpace{}%
\AgdaInductiveConstructor{inj₂}\AgdaSpace{}%
\AgdaSymbol{(}\AgdaInductiveConstructor{suc}\AgdaSpace{}%
\AgdaBound{y}\AgdaSymbol{)}\<%
\\
\\[\AgdaEmptyExtraSkip]%
\>[2]\AgdaComment{--\ Reading\ back\ a\ normal\ form\ from\ the\ evaluation\ of}\<%
\\
\>[2]\AgdaComment{--\ a\ term.\ We\ truly\ "close"\ a\ closure\ in\ reading\ it}\<%
\\
\>[2]\AgdaComment{--\ back\ to\ a\ normal\ form\ by\ replacing\ every\ variable}\<%
\\
\>[2]\AgdaComment{--\ in\ its\ context\ using\ its\ environment}\<%
\\
\>[2]\AgdaKeyword{mutual}\<%
\\
\>[2][@{}l@{\AgdaIndent{0}}]%
\>[4]\AgdaOperator{\AgdaFunction{\AgdaUnderscore{}⇑}}\AgdaSpace{}%
\AgdaSymbol{:}\AgdaSpace{}%
\AgdaDatatype{Domain}\AgdaSpace{}%
\AgdaGeneralizable{T}\AgdaSpace{}%
\AgdaSymbol{→}\AgdaSpace{}%
\AgdaInductiveConstructor{∅}\AgdaSpace{}%
\AgdaOperator{\AgdaDatatype{⊢}}\AgdaSpace{}%
\AgdaGeneralizable{T}\<%
\\
\>[4]\AgdaSymbol{(}\AgdaOperator{\AgdaInductiveConstructor{⟨ƛ}}\AgdaSpace{}%
\AgdaBound{u}\AgdaSpace{}%
\AgdaOperator{\AgdaInductiveConstructor{⟩}}\AgdaSpace{}%
\AgdaBound{γ}\AgdaSymbol{)}\AgdaSpace{}%
\AgdaOperator{\AgdaFunction{⇑}}\AgdaSpace{}%
\AgdaSymbol{=}\AgdaSpace{}%
\AgdaSymbol{(}\AgdaOperator{\AgdaInductiveConstructor{ƛ}}\AgdaSpace{}%
\AgdaSymbol{(}\AgdaFunction{close}\AgdaSpace{}%
\AgdaBound{γ}\AgdaSpace{}%
\AgdaBound{u}\AgdaSymbol{))}\<%
\\
\>[4]\AgdaInductiveConstructor{true}\AgdaSpace{}%
\AgdaOperator{\AgdaFunction{⇑}}\AgdaSpace{}%
\AgdaSymbol{=}\AgdaSpace{}%
\AgdaInductiveConstructor{true}\<%
\\
\>[4]\AgdaInductiveConstructor{false}\AgdaSpace{}%
\AgdaOperator{\AgdaFunction{⇑}}\AgdaSpace{}%
\AgdaSymbol{=}\AgdaSpace{}%
\AgdaInductiveConstructor{false}\<%
\\
\\[\AgdaEmptyExtraSkip]%
\>[4]\AgdaComment{--\ Note\ that\ this\ operation\ is\ essentially\ a}\<%
\\
\>[4]\AgdaComment{--\ substitution}\<%
\\
\>[4]\AgdaFunction{close}\AgdaSpace{}%
\AgdaSymbol{:}\AgdaSpace{}%
\AgdaSymbol{(}\AgdaFunction{Env}\AgdaSpace{}%
\AgdaGeneralizable{Γ}\AgdaSymbol{)}\AgdaSpace{}%
\AgdaSymbol{→}\AgdaSpace{}%
\AgdaGeneralizable{Γ}\AgdaSpace{}%
\AgdaOperator{\AgdaFunction{<>}}\AgdaSpace{}%
\AgdaGeneralizable{Δ}\AgdaSpace{}%
\AgdaOperator{\AgdaDatatype{⊢}}\AgdaSpace{}%
\AgdaGeneralizable{T}\AgdaSpace{}%
\AgdaSymbol{→}\AgdaSpace{}%
\AgdaGeneralizable{Δ}\AgdaSpace{}%
\AgdaOperator{\AgdaDatatype{⊢}}\AgdaSpace{}%
\AgdaGeneralizable{T}\<%
\\
\>[4]\AgdaFunction{close}\AgdaSpace{}%
\AgdaBound{γ}\AgdaSpace{}%
\AgdaInductiveConstructor{true}\AgdaSpace{}%
\AgdaSymbol{=}\AgdaSpace{}%
\AgdaInductiveConstructor{true}\<%
\\
\>[4]\AgdaFunction{close}\AgdaSpace{}%
\AgdaBound{γ}\AgdaSpace{}%
\AgdaInductiveConstructor{false}\AgdaSpace{}%
\AgdaSymbol{=}\AgdaSpace{}%
\AgdaInductiveConstructor{false}\<%
\\
\>[4]\AgdaFunction{close}\AgdaSpace{}%
\AgdaSymbol{\{}\AgdaArgument{Γ}\AgdaSpace{}%
\AgdaSymbol{=}\AgdaSpace{}%
\AgdaBound{Γ}\AgdaSymbol{\}}\AgdaSpace{}%
\AgdaBound{γ}\AgdaSpace{}%
\AgdaSymbol{(}\AgdaInductiveConstructor{var}\AgdaSpace{}%
\AgdaBound{x}\AgdaSymbol{)}\<%
\\
\>[4][@{}l@{\AgdaIndent{0}}]%
\>[6]\AgdaKeyword{with}\AgdaSpace{}%
\AgdaFunction{split}\AgdaSpace{}%
\AgdaSymbol{\{}\AgdaBound{Γ}\AgdaSymbol{\}}\AgdaSpace{}%
\AgdaBound{x}\<%
\\
\>[4]\AgdaSymbol{...}\AgdaSpace{}%
\AgdaSymbol{|}\AgdaSpace{}%
\AgdaInductiveConstructor{inj₂}\AgdaSpace{}%
\AgdaBound{y}\AgdaSpace{}%
\AgdaSymbol{=}\AgdaSpace{}%
\AgdaInductiveConstructor{var}\AgdaSpace{}%
\AgdaBound{y}\<%
\\
\>[4]\AgdaSymbol{...}\AgdaSpace{}%
\AgdaSymbol{|}\AgdaSpace{}%
\AgdaInductiveConstructor{inj₁}\AgdaSpace{}%
\AgdaBound{x}\AgdaSpace{}%
\AgdaKeyword{with}\AgdaSpace{}%
\AgdaSymbol{(}\AgdaBound{γ}\AgdaSpace{}%
\AgdaBound{x}\AgdaSymbol{)}\AgdaSpace{}%
\AgdaOperator{\AgdaFunction{⇑}}\<%
\\
\>[4]\AgdaSymbol{...}\AgdaSpace{}%
\AgdaSymbol{|}\AgdaSpace{}%
\AgdaBound{p}\AgdaSpace{}%
\AgdaSymbol{=}\AgdaSpace{}%
\AgdaFunction{inject}\AgdaSpace{}%
\AgdaBound{p}\<%
\\
\>[4]\AgdaFunction{close}\AgdaSpace{}%
\AgdaBound{γ}\AgdaSpace{}%
\AgdaSymbol{(}\AgdaOperator{\AgdaInductiveConstructor{ƛ}}\AgdaSpace{}%
\AgdaBound{t}\AgdaSymbol{)}\AgdaSpace{}%
\AgdaSymbol{=}\AgdaSpace{}%
\AgdaOperator{\AgdaInductiveConstructor{ƛ}}\AgdaSpace{}%
\AgdaSymbol{(}\AgdaFunction{close}\AgdaSpace{}%
\AgdaBound{γ}\AgdaSpace{}%
\AgdaBound{t}\AgdaSymbol{)}\<%
\\
\>[4]\AgdaFunction{close}\AgdaSpace{}%
\AgdaBound{γ}\AgdaSpace{}%
\AgdaSymbol{(}\AgdaBound{r}\AgdaSpace{}%
\AgdaOperator{\AgdaInductiveConstructor{·}}\AgdaSpace{}%
\AgdaBound{s}\AgdaSymbol{)}\AgdaSpace{}%
\AgdaSymbol{=}%
\>[23]\AgdaFunction{close}\AgdaSpace{}%
\AgdaBound{γ}\AgdaSpace{}%
\AgdaBound{r}\AgdaSpace{}%
\AgdaOperator{\AgdaInductiveConstructor{·}}\AgdaSpace{}%
\AgdaFunction{close}\AgdaSpace{}%
\AgdaBound{γ}\AgdaSpace{}%
\AgdaBound{s}\<%
\\
\>[4]\AgdaFunction{close}\AgdaSpace{}%
\AgdaBound{γ}\AgdaSpace{}%
\AgdaSymbol{(}\AgdaOperator{\AgdaInductiveConstructor{if}}\AgdaSpace{}%
\AgdaBound{t}\AgdaSpace{}%
\AgdaOperator{\AgdaInductiveConstructor{then}}\AgdaSpace{}%
\AgdaBound{u}\AgdaSpace{}%
\AgdaOperator{\AgdaInductiveConstructor{else}}\AgdaSpace{}%
\AgdaBound{v}\AgdaSymbol{)}\AgdaSpace{}%
\AgdaSymbol{=}\<%
\\
\>[4][@{}l@{\AgdaIndent{0}}]%
\>[6]\AgdaOperator{\AgdaInductiveConstructor{if}}\AgdaSpace{}%
\AgdaFunction{close}\AgdaSpace{}%
\AgdaBound{γ}\AgdaSpace{}%
\AgdaBound{t}\AgdaSpace{}%
\AgdaOperator{\AgdaInductiveConstructor{then}}\AgdaSpace{}%
\AgdaFunction{close}\AgdaSpace{}%
\AgdaBound{γ}\AgdaSpace{}%
\AgdaBound{u}\AgdaSpace{}%
\AgdaOperator{\AgdaInductiveConstructor{else}}\AgdaSpace{}%
\AgdaFunction{close}\AgdaSpace{}%
\AgdaBound{γ}\AgdaSpace{}%
\AgdaBound{v}\<%
\end{code}

\subsection{Normalization with weak head reduction}

\begin{code}%
\>[2]\AgdaOperator{\AgdaFunction{\AgdaUnderscore{}normalizes-to\AgdaUnderscore{}}}\AgdaSpace{}%
\AgdaSymbol{:}\AgdaSpace{}%
\AgdaInductiveConstructor{∅}\AgdaSpace{}%
\AgdaOperator{\AgdaDatatype{⊢}}\AgdaSpace{}%
\AgdaGeneralizable{T}\AgdaSpace{}%
\AgdaSymbol{→}\AgdaSpace{}%
\AgdaInductiveConstructor{∅}\AgdaSpace{}%
\AgdaOperator{\AgdaDatatype{⊢}}\AgdaSpace{}%
\AgdaGeneralizable{T}\AgdaSpace{}%
\AgdaSymbol{→}\AgdaSpace{}%
\AgdaPrimitive{Set}\<%
\\
\>[2]\AgdaBound{t}%
\>[3401I]\AgdaOperator{\AgdaFunction{normalizes-to}}\AgdaSpace{}%
\AgdaBound{v}%
\>[21]\AgdaSymbol{=}\<%
\\
\>[.][@{}l@{}]\<[3401I]%
\>[4]\AgdaFunction{∃[}\AgdaSpace{}%
\AgdaBound{a}\AgdaSpace{}%
\AgdaFunction{]}\AgdaSpace{}%
\AgdaFunction{empty}\AgdaSpace{}%
\AgdaOperator{\AgdaDatatype{∣}}\AgdaSpace{}%
\AgdaBound{t}\AgdaSpace{}%
\AgdaOperator{\AgdaDatatype{⇓}}\AgdaSpace{}%
\AgdaBound{a}\AgdaSpace{}%
\AgdaOperator{\AgdaFunction{×}}\AgdaSpace{}%
\AgdaSymbol{(}\AgdaBound{a}\AgdaSpace{}%
\AgdaOperator{\AgdaFunction{⇑}}\AgdaSymbol{)}\AgdaSpace{}%
\AgdaOperator{\AgdaDatatype{≡}}\AgdaSpace{}%
\AgdaBound{v}\<%
\\
\\[\AgdaEmptyExtraSkip]%
\>[2]\AgdaFunction{normalization}\AgdaSpace{}%
\AgdaSymbol{:}\AgdaSpace{}%
\AgdaSymbol{(}\AgdaBound{t}\AgdaSpace{}%
\AgdaSymbol{:}\AgdaSpace{}%
\AgdaInductiveConstructor{∅}\AgdaSpace{}%
\AgdaOperator{\AgdaDatatype{⊢}}\AgdaSpace{}%
\AgdaGeneralizable{T}\AgdaSymbol{)}\AgdaSpace{}%
\AgdaSymbol{→}\AgdaSpace{}%
\AgdaFunction{∃[}\AgdaSpace{}%
\AgdaBound{v}\AgdaSpace{}%
\AgdaFunction{]}\AgdaSpace{}%
\AgdaBound{t}\AgdaSpace{}%
\AgdaOperator{\AgdaFunction{normalizes-to}}\AgdaSpace{}%
\AgdaBound{v}\<%
\\
\>[2]\AgdaFunction{normalization}\AgdaSpace{}%
\AgdaBound{t}\<%
\\
\>[2][@{}l@{\AgdaIndent{0}}]%
\>[4]\AgdaKeyword{with}\AgdaSpace{}%
\AgdaFunction{fundamental-lemma}\AgdaSpace{}%
\AgdaBound{t}\AgdaSpace{}%
\AgdaSymbol{(λ}\AgdaSpace{}%
\AgdaSymbol{())}\<%
\\
\>[2]\AgdaSymbol{...}\AgdaSpace{}%
\AgdaSymbol{|}\AgdaSpace{}%
\AgdaBound{a}\AgdaSpace{}%
\AgdaOperator{\AgdaInductiveConstructor{,}}\AgdaSpace{}%
\AgdaBound{t⇓}\AgdaSpace{}%
\AgdaOperator{\AgdaInductiveConstructor{,}}\AgdaSpace{}%
\AgdaSymbol{\AgdaUnderscore{}}%
\>[20]\AgdaSymbol{=}\AgdaSpace{}%
\AgdaSymbol{(}\AgdaBound{a}\AgdaSpace{}%
\AgdaOperator{\AgdaFunction{⇑}}\AgdaSymbol{)}\AgdaSpace{}%
\AgdaOperator{\AgdaInductiveConstructor{,}}\AgdaSpace{}%
\AgdaBound{a}\AgdaSpace{}%
\AgdaOperator{\AgdaInductiveConstructor{,}}\AgdaSpace{}%
\AgdaBound{t⇓}\AgdaSpace{}%
\AgdaOperator{\AgdaInductiveConstructor{,}}\AgdaSpace{}%
\AgdaInductiveConstructor{refl}\<%
\\
\\[\AgdaEmptyExtraSkip]%
\>[2]\AgdaComment{--\ Normal\ form\ of\ a\ term\ is\ indeed\ a\ normal\ term}\<%
\\
\>[2]\AgdaKeyword{data}\AgdaSpace{}%
\AgdaDatatype{Value}\AgdaSpace{}%
\AgdaSymbol{:}\AgdaSpace{}%
\AgdaInductiveConstructor{∅}\AgdaSpace{}%
\AgdaOperator{\AgdaDatatype{⊢}}\AgdaSpace{}%
\AgdaGeneralizable{T}\AgdaSpace{}%
\AgdaSymbol{→}\AgdaSpace{}%
\AgdaPrimitive{Set}\AgdaSpace{}%
\AgdaKeyword{where}\<%
\\
\>[2][@{}l@{\AgdaIndent{0}}]%
\>[4]\AgdaInductiveConstructor{valueTrue}\AgdaSpace{}%
\AgdaSymbol{:}\AgdaSpace{}%
\AgdaDatatype{Value}\AgdaSpace{}%
\AgdaInductiveConstructor{true}\<%
\\
\>[4]\AgdaInductiveConstructor{valueFalse}\AgdaSpace{}%
\AgdaSymbol{:}\AgdaSpace{}%
\AgdaDatatype{Value}\AgdaSpace{}%
\AgdaInductiveConstructor{false}\<%
\\
\>[4]\AgdaInductiveConstructor{valueAbs}\AgdaSpace{}%
\AgdaSymbol{:}\AgdaSpace{}%
\AgdaSymbol{(}\AgdaBound{t}\AgdaSpace{}%
\AgdaSymbol{:}\AgdaSpace{}%
\AgdaInductiveConstructor{∅}\AgdaSpace{}%
\AgdaOperator{\AgdaInductiveConstructor{·:}}\AgdaSpace{}%
\AgdaGeneralizable{S}\AgdaSpace{}%
\AgdaOperator{\AgdaDatatype{⊢}}\AgdaSpace{}%
\AgdaGeneralizable{T}\AgdaSymbol{)}\AgdaSpace{}%
\AgdaSymbol{→}\AgdaSpace{}%
\AgdaDatatype{Value}\AgdaSpace{}%
\AgdaSymbol{(}\AgdaOperator{\AgdaInductiveConstructor{ƛ}}\AgdaSpace{}%
\AgdaBound{t}\AgdaSymbol{)}\<%
\\
\\[\AgdaEmptyExtraSkip]%
\>[2]\AgdaFunction{⇑-Value}\AgdaSpace{}%
\AgdaSymbol{:}\AgdaSpace{}%
\AgdaSymbol{(}\AgdaBound{a}\AgdaSpace{}%
\AgdaSymbol{:}\AgdaSpace{}%
\AgdaDatatype{Domain}\AgdaSpace{}%
\AgdaGeneralizable{T}\AgdaSymbol{)}\AgdaSpace{}%
\AgdaSymbol{→}\AgdaSpace{}%
\AgdaDatatype{Value}\AgdaSpace{}%
\AgdaSymbol{(}\AgdaBound{a}\AgdaSpace{}%
\AgdaOperator{\AgdaFunction{⇑}}\AgdaSymbol{)}\<%
\\
\>[2]\AgdaFunction{⇑-Value}\AgdaSpace{}%
\AgdaSymbol{(}\AgdaOperator{\AgdaInductiveConstructor{⟨ƛ}}\AgdaSpace{}%
\AgdaBound{t}\AgdaSpace{}%
\AgdaOperator{\AgdaInductiveConstructor{⟩}}\AgdaSpace{}%
\AgdaBound{γ}\AgdaSymbol{)}\AgdaSpace{}%
\AgdaSymbol{=}\AgdaSpace{}%
\AgdaInductiveConstructor{valueAbs}\AgdaSpace{}%
\AgdaSymbol{(}\AgdaFunction{close}\AgdaSpace{}%
\AgdaBound{γ}\AgdaSpace{}%
\AgdaBound{t}\AgdaSymbol{)}\<%
\\
\>[2]\AgdaFunction{⇑-Value}\AgdaSpace{}%
\AgdaInductiveConstructor{true}\AgdaSpace{}%
\AgdaSymbol{=}\AgdaSpace{}%
\AgdaInductiveConstructor{valueTrue}\<%
\\
\>[2]\AgdaFunction{⇑-Value}\AgdaSpace{}%
\AgdaInductiveConstructor{false}\AgdaSpace{}%
\AgdaSymbol{=}\AgdaSpace{}%
\AgdaInductiveConstructor{valueFalse}\<%
\end{code}

\subsection{Soundness}

Now, we show that a term is in fact denotationally equivalent to its normal form
as determined by our proof. To do so, we need to show that the normal form we
read back is equivalent to the evaluation of the term (which we have already
proven is equivalent to the term itself).

\begin{code}%
\>[2]\AgdaKeyword{postulate}\<%
\\
\>[2][@{}l@{\AgdaIndent{0}}]%
\>[4]\AgdaPostulate{fext}\AgdaSpace{}%
\AgdaSymbol{:}\AgdaSpace{}%
\AgdaFunction{Extensionality}\AgdaSpace{}%
\AgdaFunction{0ℓ}\AgdaSpace{}%
\AgdaFunction{0ℓ}\<%
\\
\>[2]\AgdaFunction{fext-implicit}\AgdaSpace{}%
\AgdaSymbol{=}\AgdaSpace{}%
\AgdaFunction{implicit-extensionality}\AgdaSpace{}%
\AgdaPostulate{fext}\<%
\\
\\[\AgdaEmptyExtraSkip]%
\>[2]\AgdaFunction{denote-\&-++}\AgdaSpace{}%
\AgdaSymbol{:}\AgdaSpace{}%
\AgdaOperator{\AgdaFunction{𝒢⟦}}\AgdaSpace{}%
\AgdaGeneralizable{δ}\AgdaSpace{}%
\AgdaOperator{\AgdaFunction{⟧}}\AgdaSpace{}%
\AgdaOperator{\AgdaFunction{\&}}\AgdaSpace{}%
\AgdaOperator{\AgdaFunction{𝒟⟦}}\AgdaSpace{}%
\AgdaGeneralizable{a}\AgdaSpace{}%
\AgdaOperator{\AgdaFunction{⟧}}\AgdaSpace{}%
\AgdaOperator{\AgdaDatatype{≡}}\AgdaSpace{}%
\AgdaOperator{\AgdaFunction{𝒢⟦}}\AgdaSpace{}%
\AgdaGeneralizable{δ}\AgdaSpace{}%
\AgdaOperator{\AgdaFunction{++}}\AgdaSpace{}%
\AgdaGeneralizable{a}\AgdaSpace{}%
\AgdaOperator{\AgdaFunction{⟧}}\AgdaSpace{}%
\AgdaSymbol{\{}\AgdaGeneralizable{T}\AgdaSymbol{\}}\<%
\\
\>[2]\AgdaFunction{denote-\&-++}\AgdaSpace{}%
\AgdaSymbol{=}\<%
\\
\>[2][@{}l@{\AgdaIndent{0}}]%
\>[4]\AgdaPostulate{fext}\AgdaSpace{}%
\AgdaSymbol{λ}\AgdaSpace{}%
\AgdaKeyword{where}\AgdaSpace{}%
\AgdaInductiveConstructor{zero}\AgdaSpace{}%
\AgdaSymbol{→}\AgdaSpace{}%
\AgdaInductiveConstructor{refl}\AgdaSpace{}%
\AgdaSymbol{;}\AgdaSpace{}%
\AgdaSymbol{(}\AgdaInductiveConstructor{suc}\AgdaSpace{}%
\AgdaSymbol{\AgdaUnderscore{})}\AgdaSpace{}%
\AgdaSymbol{→}\AgdaSpace{}%
\AgdaInductiveConstructor{refl}\<%
\\
\\[\AgdaEmptyExtraSkip]%
\>[2]\AgdaFunction{⇓-sound}\AgdaSpace{}%
\AgdaSymbol{:}\AgdaSpace{}%
\AgdaGeneralizable{γ}\AgdaSpace{}%
\AgdaOperator{\AgdaDatatype{∣}}\AgdaSpace{}%
\AgdaGeneralizable{t}\AgdaSpace{}%
\AgdaOperator{\AgdaDatatype{⇓}}\AgdaSpace{}%
\AgdaGeneralizable{a}\AgdaSpace{}%
\AgdaSymbol{→}\AgdaSpace{}%
\AgdaOperator{\AgdaFunction{ℰ⟦}}\AgdaSpace{}%
\AgdaGeneralizable{t}\AgdaSpace{}%
\AgdaOperator{\AgdaFunction{⟧}}\AgdaSpace{}%
\AgdaOperator{\AgdaFunction{𝒢⟦}}\AgdaSpace{}%
\AgdaGeneralizable{γ}\AgdaSpace{}%
\AgdaOperator{\AgdaFunction{⟧}}\AgdaSpace{}%
\AgdaOperator{\AgdaDatatype{≡}}\AgdaSpace{}%
\AgdaOperator{\AgdaFunction{𝒟⟦}}\AgdaSpace{}%
\AgdaGeneralizable{a}\AgdaSpace{}%
\AgdaOperator{\AgdaFunction{⟧}}\<%
\\
\>[2]\AgdaFunction{⇓-sound}\AgdaSpace{}%
\AgdaInductiveConstructor{evalTrue}\AgdaSpace{}%
\AgdaSymbol{=}\AgdaSpace{}%
\AgdaInductiveConstructor{refl}\<%
\\
\>[2]\AgdaFunction{⇓-sound}\AgdaSpace{}%
\AgdaInductiveConstructor{evalFalse}\AgdaSpace{}%
\AgdaSymbol{=}\AgdaSpace{}%
\AgdaInductiveConstructor{refl}\<%
\\
\>[2]\AgdaFunction{⇓-sound}\AgdaSpace{}%
\AgdaInductiveConstructor{evalVar}\AgdaSpace{}%
\AgdaSymbol{=}\AgdaSpace{}%
\AgdaInductiveConstructor{refl}\<%
\\
\>[2]\AgdaFunction{⇓-sound}\AgdaSpace{}%
\AgdaInductiveConstructor{evalAbs}\AgdaSpace{}%
\AgdaSymbol{=}\AgdaSpace{}%
\AgdaInductiveConstructor{refl}\<%
\\
\>[2]\AgdaFunction{⇓-sound}\AgdaSpace{}%
\AgdaSymbol{(}\AgdaInductiveConstructor{evalApp}\AgdaSpace{}%
\AgdaSymbol{\{}\AgdaArgument{t}\AgdaSpace{}%
\AgdaSymbol{=}\AgdaSpace{}%
\AgdaBound{t}\AgdaSymbol{\}}\AgdaSpace{}%
\AgdaSymbol{\{}\AgdaBound{δ}\AgdaSymbol{\}}\AgdaSpace{}%
\AgdaSymbol{\{}\AgdaArgument{a}\AgdaSpace{}%
\AgdaSymbol{=}\AgdaSpace{}%
\AgdaBound{a}\AgdaSymbol{\}}\AgdaSpace{}%
\AgdaSymbol{\{}\AgdaBound{b}\AgdaSymbol{\}}\AgdaSpace{}%
\AgdaBound{r⇓}\AgdaSpace{}%
\AgdaBound{s⇓}\AgdaSpace{}%
\AgdaBound{t⇓}\AgdaSymbol{)}\<%
\\
\>[2][@{}l@{\AgdaIndent{0}}]%
\>[4]\AgdaKeyword{rewrite}\AgdaSpace{}%
\AgdaFunction{⇓-sound}\AgdaSpace{}%
\AgdaBound{r⇓}\AgdaSpace{}%
\AgdaSymbol{|}\AgdaSpace{}%
\AgdaFunction{⇓-sound}\AgdaSpace{}%
\AgdaBound{s⇓}\AgdaSpace{}%
\AgdaSymbol{=}\<%
\\
\>[4]\AgdaOperator{\AgdaFunction{begin}}\<%
\\
\>[4][@{}l@{\AgdaIndent{0}}]%
\>[6]\AgdaOperator{\AgdaFunction{ℰ⟦}}\AgdaSpace{}%
\AgdaBound{t}\AgdaSpace{}%
\AgdaOperator{\AgdaFunction{⟧}}\AgdaSpace{}%
\AgdaSymbol{(}\AgdaOperator{\AgdaFunction{𝒢⟦}}\AgdaSpace{}%
\AgdaBound{δ}\AgdaSpace{}%
\AgdaOperator{\AgdaFunction{⟧}}\AgdaSpace{}%
\AgdaOperator{\AgdaFunction{\&}}\AgdaSpace{}%
\AgdaOperator{\AgdaFunction{𝒟⟦}}\AgdaSpace{}%
\AgdaBound{a}\AgdaSpace{}%
\AgdaOperator{\AgdaFunction{⟧}}\AgdaSymbol{)}\<%
\\
\>[4]\AgdaFunction{≡⟨}\AgdaSpace{}%
\AgdaFunction{cong}\AgdaSpace{}%
\AgdaOperator{\AgdaFunction{ℰ⟦}}\AgdaSpace{}%
\AgdaBound{t}\AgdaSpace{}%
\AgdaOperator{\AgdaFunction{⟧}}\AgdaSpace{}%
\AgdaSymbol{(}\AgdaFunction{fext-implicit}\AgdaSpace{}%
\AgdaFunction{denote-\&-++}\AgdaSymbol{)}\AgdaSpace{}%
\AgdaFunction{⟩}\<%
\\
\>[4][@{}l@{\AgdaIndent{0}}]%
\>[6]\AgdaOperator{\AgdaFunction{ℰ⟦}}\AgdaSpace{}%
\AgdaBound{t}\AgdaSpace{}%
\AgdaOperator{\AgdaFunction{⟧}}\AgdaSpace{}%
\AgdaOperator{\AgdaFunction{𝒢⟦}}\AgdaSpace{}%
\AgdaBound{δ}\AgdaSpace{}%
\AgdaOperator{\AgdaFunction{++}}\AgdaSpace{}%
\AgdaBound{a}\AgdaSpace{}%
\AgdaOperator{\AgdaFunction{⟧}}\<%
\\
\>[4]\AgdaFunction{≡⟨}\AgdaSpace{}%
\AgdaFunction{⇓-sound}\AgdaSpace{}%
\AgdaBound{t⇓}\AgdaSpace{}%
\AgdaFunction{⟩}\<%
\\
\>[4][@{}l@{\AgdaIndent{0}}]%
\>[6]\AgdaOperator{\AgdaFunction{𝒟⟦}}\AgdaSpace{}%
\AgdaBound{b}\AgdaSpace{}%
\AgdaOperator{\AgdaFunction{⟧}}\<%
\\
\>[4]\AgdaOperator{\AgdaFunction{∎}}\<%
\\
\>[2]\AgdaFunction{⇓-sound}\AgdaSpace{}%
\AgdaSymbol{(}\AgdaInductiveConstructor{evalIfTrue}\AgdaSpace{}%
\AgdaBound{r⇓}\AgdaSpace{}%
\AgdaBound{s⇓}\AgdaSymbol{)}\<%
\\
\>[2][@{}l@{\AgdaIndent{0}}]%
\>[4]\AgdaKeyword{rewrite}\AgdaSpace{}%
\AgdaFunction{⇓-sound}\AgdaSpace{}%
\AgdaBound{r⇓}\AgdaSpace{}%
\AgdaSymbol{|}\AgdaSpace{}%
\AgdaFunction{⇓-sound}\AgdaSpace{}%
\AgdaBound{s⇓}\AgdaSpace{}%
\AgdaSymbol{=}\AgdaSpace{}%
\AgdaInductiveConstructor{refl}\<%
\\
\>[2]\AgdaFunction{⇓-sound}\AgdaSpace{}%
\AgdaSymbol{(}\AgdaInductiveConstructor{evalIfFalse}\AgdaSpace{}%
\AgdaBound{r⇓}\AgdaSpace{}%
\AgdaBound{t⇓}\AgdaSymbol{)}\<%
\\
\>[2][@{}l@{\AgdaIndent{0}}]%
\>[4]\AgdaKeyword{rewrite}\AgdaSpace{}%
\AgdaFunction{⇓-sound}\AgdaSpace{}%
\AgdaBound{r⇓}\AgdaSpace{}%
\AgdaSymbol{|}\AgdaSpace{}%
\AgdaFunction{⇓-sound}\AgdaSpace{}%
\AgdaBound{t⇓}\AgdaSpace{}%
\AgdaSymbol{=}\AgdaSpace{}%
\AgdaInductiveConstructor{refl}\<%
\\
\\[\AgdaEmptyExtraSkip]%
\>[2]\AgdaComment{--\ Natural\ semantics\ is\ deterministic}\<%
\\
\>[2]\AgdaFunction{⇓-uniq}\AgdaSpace{}%
\AgdaSymbol{:}\AgdaSpace{}%
\AgdaGeneralizable{γ}\AgdaSpace{}%
\AgdaOperator{\AgdaDatatype{∣}}\AgdaSpace{}%
\AgdaGeneralizable{t}\AgdaSpace{}%
\AgdaOperator{\AgdaDatatype{⇓}}\AgdaSpace{}%
\AgdaGeneralizable{a}\AgdaSpace{}%
\AgdaSymbol{→}\AgdaSpace{}%
\AgdaGeneralizable{γ}\AgdaSpace{}%
\AgdaOperator{\AgdaDatatype{∣}}\AgdaSpace{}%
\AgdaGeneralizable{t}\AgdaSpace{}%
\AgdaOperator{\AgdaDatatype{⇓}}\AgdaSpace{}%
\AgdaGeneralizable{b}\AgdaSpace{}%
\AgdaSymbol{→}\AgdaSpace{}%
\AgdaGeneralizable{a}\AgdaSpace{}%
\AgdaOperator{\AgdaDatatype{≡}}\AgdaSpace{}%
\AgdaGeneralizable{b}\<%
\\
\>[2]\AgdaFunction{⇓-uniq}\AgdaSpace{}%
\AgdaInductiveConstructor{evalTrue}\AgdaSpace{}%
\AgdaInductiveConstructor{evalTrue}\AgdaSpace{}%
\AgdaSymbol{=}\AgdaSpace{}%
\AgdaInductiveConstructor{refl}\<%
\\
\>[2]\AgdaFunction{⇓-uniq}\AgdaSpace{}%
\AgdaInductiveConstructor{evalFalse}\AgdaSpace{}%
\AgdaInductiveConstructor{evalFalse}\AgdaSpace{}%
\AgdaSymbol{=}\AgdaSpace{}%
\AgdaInductiveConstructor{refl}\<%
\\
\>[2]\AgdaFunction{⇓-uniq}\AgdaSpace{}%
\AgdaInductiveConstructor{evalVar}\AgdaSpace{}%
\AgdaInductiveConstructor{evalVar}\AgdaSpace{}%
\AgdaSymbol{=}\AgdaSpace{}%
\AgdaInductiveConstructor{refl}\<%
\\
\>[2]\AgdaFunction{⇓-uniq}\AgdaSpace{}%
\AgdaInductiveConstructor{evalAbs}\AgdaSpace{}%
\AgdaInductiveConstructor{evalAbs}\AgdaSpace{}%
\AgdaSymbol{=}\AgdaSpace{}%
\AgdaInductiveConstructor{refl}\<%
\\
\>[2]\AgdaFunction{⇓-uniq}\AgdaSpace{}%
\AgdaSymbol{(}\AgdaInductiveConstructor{evalApp}\AgdaSpace{}%
\AgdaBound{r⇓₁}\AgdaSpace{}%
\AgdaBound{s⇓₁}\AgdaSpace{}%
\AgdaBound{t⇓₁}\AgdaSymbol{)}\AgdaSpace{}%
\AgdaSymbol{(}\AgdaInductiveConstructor{evalApp}\AgdaSpace{}%
\AgdaBound{r⇓₂}\AgdaSpace{}%
\AgdaBound{s⇓₂}\AgdaSpace{}%
\AgdaBound{t⇓₂}\AgdaSymbol{)}\<%
\\
\>[2][@{}l@{\AgdaIndent{0}}]%
\>[4]\AgdaKeyword{with}\AgdaSpace{}%
\AgdaFunction{⇓-uniq}\AgdaSpace{}%
\AgdaBound{r⇓₁}\AgdaSpace{}%
\AgdaBound{r⇓₂}\<%
\\
\>[2]\AgdaSymbol{...}\AgdaSpace{}%
\AgdaSymbol{|}\AgdaSpace{}%
\AgdaInductiveConstructor{refl}\<%
\\
\>[2][@{}l@{\AgdaIndent{0}}]%
\>[4]\AgdaKeyword{rewrite}\AgdaSpace{}%
\AgdaFunction{⇓-uniq}\AgdaSpace{}%
\AgdaBound{s⇓₁}\AgdaSpace{}%
\AgdaBound{s⇓₂}\AgdaSpace{}%
\AgdaSymbol{|}\AgdaSpace{}%
\AgdaFunction{⇓-uniq}\AgdaSpace{}%
\AgdaBound{t⇓₁}\AgdaSpace{}%
\AgdaBound{t⇓₂}\AgdaSpace{}%
\AgdaSymbol{=}\AgdaSpace{}%
\AgdaInductiveConstructor{refl}\<%
\\
\>[2]\AgdaFunction{⇓-uniq}\AgdaSpace{}%
\AgdaSymbol{(}\AgdaInductiveConstructor{evalIfTrue}\AgdaSpace{}%
\AgdaBound{r⇓₁}\AgdaSpace{}%
\AgdaBound{s⇓₁}\AgdaSymbol{)}\AgdaSpace{}%
\AgdaBound{evalIf}\<%
\\
\>[2][@{}l@{\AgdaIndent{0}}]%
\>[4]\AgdaKeyword{with}\AgdaSpace{}%
\AgdaBound{evalIf}\<%
\\
\>[2]\AgdaSymbol{...}\AgdaSpace{}%
\AgdaSymbol{|}\AgdaSpace{}%
\AgdaInductiveConstructor{evalIfTrue}\AgdaSpace{}%
\AgdaBound{r⇓₂}\AgdaSpace{}%
\AgdaBound{s⇓₂}\<%
\\
\>[2][@{}l@{\AgdaIndent{0}}]%
\>[4]\AgdaKeyword{rewrite}\AgdaSpace{}%
\AgdaFunction{⇓-uniq}\AgdaSpace{}%
\AgdaBound{r⇓₁}\AgdaSpace{}%
\AgdaBound{r⇓₂}\AgdaSpace{}%
\AgdaSymbol{|}\AgdaSpace{}%
\AgdaFunction{⇓-uniq}\AgdaSpace{}%
\AgdaBound{s⇓₁}\AgdaSpace{}%
\AgdaBound{s⇓₂}\AgdaSpace{}%
\AgdaSymbol{=}\AgdaSpace{}%
\AgdaInductiveConstructor{refl}\<%
\\
\>[2]\AgdaSymbol{...}\AgdaSpace{}%
\AgdaSymbol{|}\AgdaSpace{}%
\AgdaInductiveConstructor{evalIfFalse}\AgdaSpace{}%
\AgdaBound{r⇓₂}\AgdaSpace{}%
\AgdaSymbol{\AgdaUnderscore{}}\AgdaSpace{}%
\AgdaSymbol{=}\<%
\\
\>[2][@{}l@{\AgdaIndent{0}}]%
\>[4]\AgdaFunction{contradiction}\AgdaSpace{}%
\AgdaSymbol{(}\AgdaFunction{⇓-uniq}\AgdaSpace{}%
\AgdaBound{r⇓₁}\AgdaSpace{}%
\AgdaBound{r⇓₂}\AgdaSymbol{)}\AgdaSpace{}%
\AgdaSymbol{(λ}\AgdaSpace{}%
\AgdaSymbol{())}\<%
\\
\>[2]\AgdaFunction{⇓-uniq}\AgdaSpace{}%
\AgdaSymbol{(}\AgdaInductiveConstructor{evalIfFalse}\AgdaSpace{}%
\AgdaBound{r⇓₁}\AgdaSpace{}%
\AgdaBound{t⇓₁}\AgdaSymbol{)}\AgdaSpace{}%
\AgdaBound{evalIf}\<%
\\
\>[2][@{}l@{\AgdaIndent{0}}]%
\>[4]\AgdaKeyword{with}\AgdaSpace{}%
\AgdaBound{evalIf}\<%
\\
\>[2]\AgdaSymbol{...}\AgdaSpace{}%
\AgdaSymbol{|}\AgdaSpace{}%
\AgdaInductiveConstructor{evalIfFalse}\AgdaSpace{}%
\AgdaBound{r⇓₂}\AgdaSpace{}%
\AgdaBound{t⇓₂}\<%
\\
\>[2][@{}l@{\AgdaIndent{0}}]%
\>[4]\AgdaKeyword{rewrite}\AgdaSpace{}%
\AgdaFunction{⇓-uniq}\AgdaSpace{}%
\AgdaBound{r⇓₁}\AgdaSpace{}%
\AgdaBound{r⇓₂}\AgdaSpace{}%
\AgdaSymbol{|}\AgdaSpace{}%
\AgdaFunction{⇓-uniq}\AgdaSpace{}%
\AgdaBound{t⇓₁}\AgdaSpace{}%
\AgdaBound{t⇓₂}\AgdaSpace{}%
\AgdaSymbol{=}\AgdaSpace{}%
\AgdaInductiveConstructor{refl}\<%
\\
\>[2]\AgdaSymbol{...}\AgdaSpace{}%
\AgdaSymbol{|}\AgdaSpace{}%
\AgdaInductiveConstructor{evalIfTrue}\AgdaSpace{}%
\AgdaBound{r⇓₂}\AgdaSpace{}%
\AgdaSymbol{\AgdaUnderscore{}}\AgdaSpace{}%
\AgdaSymbol{=}\<%
\\
\>[2][@{}l@{\AgdaIndent{0}}]%
\>[4]\AgdaFunction{contradiction}\AgdaSpace{}%
\AgdaSymbol{(}\AgdaFunction{⇓-uniq}\AgdaSpace{}%
\AgdaBound{r⇓₁}\AgdaSpace{}%
\AgdaBound{r⇓₂}\AgdaSymbol{)}\AgdaSpace{}%
\AgdaSymbol{(λ}\AgdaSpace{}%
\AgdaSymbol{())}\<%
\\
\\[\AgdaEmptyExtraSkip]%
\>[2]\AgdaComment{--\ Appending\ denoted\ contexts}\<%
\\
\>[2]\AgdaOperator{\AgdaFunction{\AgdaUnderscore{}\&\&\AgdaUnderscore{}}}\AgdaSpace{}%
\AgdaSymbol{:}\AgdaSpace{}%
\AgdaOperator{\AgdaFunction{𝒞⟦}}\AgdaSpace{}%
\AgdaGeneralizable{Γ}\AgdaSpace{}%
\AgdaOperator{\AgdaFunction{⟧}}\AgdaSpace{}%
\AgdaSymbol{→}\AgdaSpace{}%
\AgdaOperator{\AgdaFunction{𝒞⟦}}\AgdaSpace{}%
\AgdaGeneralizable{Δ}\AgdaSpace{}%
\AgdaOperator{\AgdaFunction{⟧}}\AgdaSpace{}%
\AgdaSymbol{→}\AgdaSpace{}%
\AgdaOperator{\AgdaFunction{𝒞⟦}}\AgdaSpace{}%
\AgdaGeneralizable{Γ}\AgdaSpace{}%
\AgdaOperator{\AgdaFunction{<>}}\AgdaSpace{}%
\AgdaGeneralizable{Δ}\AgdaSpace{}%
\AgdaOperator{\AgdaFunction{⟧}}\<%
\\
\>[2]\AgdaOperator{\AgdaFunction{\AgdaUnderscore{}\&\&\AgdaUnderscore{}}}\AgdaSpace{}%
\AgdaSymbol{\{}\AgdaArgument{Δ}\AgdaSpace{}%
\AgdaSymbol{=}\AgdaSpace{}%
\AgdaInductiveConstructor{∅}\AgdaSymbol{\}}\AgdaSpace{}%
\AgdaBound{p}\AgdaSpace{}%
\AgdaBound{q}\AgdaSpace{}%
\AgdaSymbol{=}\AgdaSpace{}%
\AgdaBound{p}\<%
\\
\>[2]\AgdaOperator{\AgdaFunction{\AgdaUnderscore{}\&\&\AgdaUnderscore{}}}\AgdaSpace{}%
\AgdaSymbol{\{}\AgdaArgument{Δ}\AgdaSpace{}%
\AgdaSymbol{=}\AgdaSpace{}%
\AgdaBound{Δ}\AgdaSpace{}%
\AgdaOperator{\AgdaInductiveConstructor{·:}}\AgdaSpace{}%
\AgdaBound{T}\AgdaSymbol{\}}\AgdaSpace{}%
\AgdaBound{p}\AgdaSpace{}%
\AgdaBound{q}\AgdaSpace{}%
\AgdaSymbol{=}\AgdaSpace{}%
\AgdaSymbol{(}\AgdaBound{p}\AgdaSpace{}%
\AgdaOperator{\AgdaFunction{\&\&}}\AgdaSpace{}%
\AgdaBound{q}\AgdaSpace{}%
\AgdaOperator{\AgdaFunction{∘}}\AgdaSpace{}%
\AgdaInductiveConstructor{suc}\AgdaSymbol{)}\AgdaSpace{}%
\AgdaOperator{\AgdaFunction{\&}}\AgdaSpace{}%
\AgdaBound{q}\AgdaSpace{}%
\AgdaInductiveConstructor{zero}\<%
\\
\\[\AgdaEmptyExtraSkip]%
\>[2]\AgdaComment{--\ Evaluation\ of\ a\ term\ injected\ to\ an\ extended\ context}\<%
\\
\>[2]\AgdaComment{--\ is\ equivalent}\<%
\\
\>[2]\AgdaFunction{ℰ-inject}\AgdaSpace{}%
\AgdaSymbol{:}\<%
\\
\>[2][@{}l@{\AgdaIndent{0}}]%
\>[4]\AgdaSymbol{∀}\AgdaSpace{}%
\AgdaSymbol{(}\AgdaBound{t}\AgdaSpace{}%
\AgdaSymbol{:}\AgdaSpace{}%
\AgdaGeneralizable{Γ}\AgdaSpace{}%
\AgdaOperator{\AgdaDatatype{⊢}}\AgdaSpace{}%
\AgdaGeneralizable{T}\AgdaSymbol{)}\AgdaSpace{}%
\AgdaSymbol{(}\AgdaBound{p}\AgdaSpace{}%
\AgdaSymbol{:}\AgdaSpace{}%
\AgdaOperator{\AgdaFunction{𝒞⟦}}\AgdaSpace{}%
\AgdaGeneralizable{Γ}\AgdaSpace{}%
\AgdaOperator{\AgdaFunction{⟧}}\AgdaSymbol{)}\AgdaSpace{}%
\AgdaSymbol{(}\AgdaBound{q}\AgdaSpace{}%
\AgdaSymbol{:}\AgdaSpace{}%
\AgdaOperator{\AgdaFunction{𝒞⟦}}\AgdaSpace{}%
\AgdaGeneralizable{Δ}\AgdaSpace{}%
\AgdaOperator{\AgdaFunction{⟧}}\AgdaSymbol{)}\<%
\\
\>[4]\AgdaSymbol{→}\AgdaSpace{}%
\AgdaOperator{\AgdaFunction{ℰ⟦}}\AgdaSpace{}%
\AgdaBound{t}\AgdaSpace{}%
\AgdaOperator{\AgdaFunction{⟧}}\AgdaSpace{}%
\AgdaBound{p}\AgdaSpace{}%
\AgdaOperator{\AgdaDatatype{≡}}\AgdaSpace{}%
\AgdaOperator{\AgdaFunction{ℰ⟦}}\AgdaSpace{}%
\AgdaFunction{inject}\AgdaSpace{}%
\AgdaBound{t}\AgdaSpace{}%
\AgdaOperator{\AgdaFunction{⟧}}\AgdaSpace{}%
\AgdaSymbol{(}\AgdaBound{q}\AgdaSpace{}%
\AgdaOperator{\AgdaFunction{\&\&}}\AgdaSpace{}%
\AgdaBound{p}\AgdaSymbol{)}\<%
\\
\>[2]\AgdaFunction{ℰ-inject}\AgdaSpace{}%
\AgdaInductiveConstructor{true}\AgdaSpace{}%
\AgdaBound{p}\AgdaSpace{}%
\AgdaBound{q}\AgdaSpace{}%
\AgdaSymbol{=}\AgdaSpace{}%
\AgdaInductiveConstructor{refl}\<%
\\
\>[2]\AgdaFunction{ℰ-inject}\AgdaSpace{}%
\AgdaInductiveConstructor{false}\AgdaSpace{}%
\AgdaBound{p}\AgdaSpace{}%
\AgdaBound{q}\AgdaSpace{}%
\AgdaSymbol{=}\AgdaSpace{}%
\AgdaInductiveConstructor{refl}\<%
\\
\>[2]\AgdaFunction{ℰ-inject}\AgdaSpace{}%
\AgdaSymbol{(}\AgdaInductiveConstructor{var}\AgdaSpace{}%
\AgdaInductiveConstructor{zero}\AgdaSymbol{)}\AgdaSpace{}%
\AgdaBound{p}\AgdaSpace{}%
\AgdaBound{q}\AgdaSpace{}%
\AgdaSymbol{=}\AgdaSpace{}%
\AgdaInductiveConstructor{refl}\<%
\\
\>[2]\AgdaFunction{ℰ-inject}\AgdaSpace{}%
\AgdaSymbol{(}\AgdaInductiveConstructor{var}\AgdaSpace{}%
\AgdaSymbol{(}\AgdaInductiveConstructor{suc}\AgdaSpace{}%
\AgdaBound{x}\AgdaSymbol{))}\AgdaSpace{}%
\AgdaBound{p}\AgdaSpace{}%
\AgdaBound{q}\<%
\\
\>[2][@{}l@{\AgdaIndent{0}}]%
\>[4]\AgdaKeyword{rewrite}\AgdaSpace{}%
\AgdaFunction{ℰ-inject}\AgdaSpace{}%
\AgdaSymbol{(}\AgdaInductiveConstructor{var}\AgdaSpace{}%
\AgdaBound{x}\AgdaSymbol{)}\AgdaSpace{}%
\AgdaSymbol{(}\AgdaBound{p}\AgdaSpace{}%
\AgdaOperator{\AgdaFunction{∘}}\AgdaSpace{}%
\AgdaInductiveConstructor{suc}\AgdaSymbol{)}\AgdaSpace{}%
\AgdaBound{q}\AgdaSpace{}%
\AgdaSymbol{=}\AgdaSpace{}%
\AgdaInductiveConstructor{refl}\<%
\\
\>[2]\AgdaFunction{ℰ-inject}\AgdaSpace{}%
\AgdaSymbol{(}\AgdaOperator{\AgdaInductiveConstructor{ƛ}}\AgdaSpace{}%
\AgdaBound{t}\AgdaSymbol{)}\AgdaSpace{}%
\AgdaBound{p}\AgdaSpace{}%
\AgdaBound{q}\AgdaSpace{}%
\AgdaSymbol{=}\<%
\\
\>[2][@{}l@{\AgdaIndent{0}}]%
\>[4]\AgdaPostulate{fext}\AgdaSpace{}%
\AgdaSymbol{(λ}\AgdaSpace{}%
\AgdaBound{x}\AgdaSpace{}%
\AgdaSymbol{→}\AgdaSpace{}%
\AgdaFunction{ℰ-inject}\AgdaSpace{}%
\AgdaBound{t}\AgdaSpace{}%
\AgdaSymbol{(}\AgdaBound{p}\AgdaSpace{}%
\AgdaOperator{\AgdaFunction{\&}}\AgdaSpace{}%
\AgdaBound{x}\AgdaSymbol{)}\AgdaSpace{}%
\AgdaBound{q}\AgdaSymbol{)}\<%
\\
\>[2]\AgdaFunction{ℰ-inject}\AgdaSpace{}%
\AgdaSymbol{(}\AgdaBound{r}\AgdaSpace{}%
\AgdaOperator{\AgdaInductiveConstructor{·}}\AgdaSpace{}%
\AgdaBound{s}\AgdaSymbol{)}\AgdaSpace{}%
\AgdaBound{p}\AgdaSpace{}%
\AgdaBound{q}\<%
\\
\>[2][@{}l@{\AgdaIndent{0}}]%
\>[4]\AgdaKeyword{rewrite}\AgdaSpace{}%
\AgdaFunction{ℰ-inject}\AgdaSpace{}%
\AgdaBound{r}\AgdaSpace{}%
\AgdaBound{p}\AgdaSpace{}%
\AgdaBound{q}\AgdaSpace{}%
\AgdaSymbol{|}\AgdaSpace{}%
\AgdaFunction{ℰ-inject}\AgdaSpace{}%
\AgdaBound{s}\AgdaSpace{}%
\AgdaBound{p}\AgdaSpace{}%
\AgdaBound{q}\AgdaSpace{}%
\AgdaSymbol{=}\AgdaSpace{}%
\AgdaInductiveConstructor{refl}\<%
\\
\>[2]\AgdaFunction{ℰ-inject}\AgdaSpace{}%
\AgdaSymbol{(}\AgdaOperator{\AgdaInductiveConstructor{if}}\AgdaSpace{}%
\AgdaBound{r}\AgdaSpace{}%
\AgdaOperator{\AgdaInductiveConstructor{then}}\AgdaSpace{}%
\AgdaBound{s}\AgdaSpace{}%
\AgdaOperator{\AgdaInductiveConstructor{else}}\AgdaSpace{}%
\AgdaBound{t}\AgdaSymbol{)}\AgdaSpace{}%
\AgdaBound{p}\AgdaSpace{}%
\AgdaBound{q}\<%
\\
\>[2][@{}l@{\AgdaIndent{0}}]%
\>[4]\AgdaKeyword{with}\AgdaSpace{}%
\AgdaOperator{\AgdaFunction{ℰ⟦}}\AgdaSpace{}%
\AgdaBound{r}\AgdaSpace{}%
\AgdaOperator{\AgdaFunction{⟧}}\AgdaSpace{}%
\AgdaBound{p}\AgdaSpace{}%
\AgdaSymbol{|}\AgdaSpace{}%
\AgdaOperator{\AgdaFunction{ℰ⟦}}\AgdaSpace{}%
\AgdaFunction{inject}\AgdaSpace{}%
\AgdaBound{r}\AgdaSpace{}%
\AgdaOperator{\AgdaFunction{⟧}}\AgdaSpace{}%
\AgdaSymbol{(}\AgdaBound{q}\AgdaSpace{}%
\AgdaOperator{\AgdaFunction{\&\&}}\AgdaSpace{}%
\AgdaBound{p}\AgdaSymbol{)}\AgdaSpace{}%
\AgdaSymbol{|}\AgdaSpace{}%
\AgdaFunction{ℰ-inject}\AgdaSpace{}%
\AgdaBound{r}\AgdaSpace{}%
\AgdaBound{p}\AgdaSpace{}%
\AgdaBound{q}\<%
\\
\>[2]\AgdaSymbol{...}\AgdaSpace{}%
\AgdaSymbol{|}\AgdaSpace{}%
\AgdaInductiveConstructor{true}\AgdaSpace{}%
\AgdaSymbol{|}\AgdaSpace{}%
\AgdaInductiveConstructor{true}\AgdaSpace{}%
\AgdaSymbol{|}\AgdaSpace{}%
\AgdaInductiveConstructor{refl}\AgdaSpace{}%
\AgdaSymbol{=}\AgdaSpace{}%
\AgdaFunction{ℰ-inject}\AgdaSpace{}%
\AgdaBound{s}\AgdaSpace{}%
\AgdaBound{p}\AgdaSpace{}%
\AgdaBound{q}\<%
\\
\>[2]\AgdaSymbol{...}\AgdaSpace{}%
\AgdaSymbol{|}\AgdaSpace{}%
\AgdaInductiveConstructor{false}\AgdaSpace{}%
\AgdaSymbol{|}\AgdaSpace{}%
\AgdaInductiveConstructor{false}\AgdaSpace{}%
\AgdaSymbol{|}\AgdaSpace{}%
\AgdaInductiveConstructor{refl}\AgdaSpace{}%
\AgdaSymbol{=}\AgdaSpace{}%
\AgdaFunction{ℰ-inject}\AgdaSpace{}%
\AgdaBound{t}\AgdaSpace{}%
\AgdaBound{p}\AgdaSpace{}%
\AgdaBound{q}\<%
\\
\\[\AgdaEmptyExtraSkip]%
\>[2]\AgdaComment{--\ Splitting\ variable\ into\ left\ context}\<%
\\
\>[2]\AgdaFunction{split-left}\AgdaSpace{}%
\AgdaSymbol{:}\<%
\\
\>[2][@{}l@{\AgdaIndent{0}}]%
\>[4]\AgdaSymbol{∀}\AgdaSpace{}%
\AgdaSymbol{\{}\AgdaBound{x}\AgdaSpace{}%
\AgdaSymbol{:}\AgdaSpace{}%
\AgdaGeneralizable{Γ}\AgdaSpace{}%
\AgdaOperator{\AgdaFunction{<>}}\AgdaSpace{}%
\AgdaGeneralizable{Δ}\AgdaSpace{}%
\AgdaOperator{\AgdaDatatype{∋}}\AgdaSpace{}%
\AgdaGeneralizable{T}\AgdaSymbol{\}}\AgdaSpace{}%
\AgdaSymbol{\{}\AgdaBound{y}\AgdaSpace{}%
\AgdaSymbol{:}\AgdaSpace{}%
\AgdaGeneralizable{Γ}\AgdaSpace{}%
\AgdaOperator{\AgdaDatatype{∋}}\AgdaSpace{}%
\AgdaGeneralizable{T}\AgdaSymbol{\}}\AgdaSpace{}%
\AgdaSymbol{\{}\AgdaBound{p}\AgdaSpace{}%
\AgdaSymbol{:}\AgdaSpace{}%
\AgdaOperator{\AgdaFunction{𝒞⟦}}\AgdaSpace{}%
\AgdaGeneralizable{Γ}\AgdaSpace{}%
\AgdaOperator{\AgdaFunction{⟧}}\AgdaSpace{}%
\AgdaSymbol{\}}\AgdaSpace{}%
\AgdaSymbol{\{}\AgdaBound{q}\AgdaSpace{}%
\AgdaSymbol{:}\AgdaSpace{}%
\AgdaOperator{\AgdaFunction{𝒞⟦}}\AgdaSpace{}%
\AgdaGeneralizable{Δ}\AgdaSpace{}%
\AgdaOperator{\AgdaFunction{⟧}}\AgdaSymbol{\}}\<%
\\
\>[4]\AgdaSymbol{→}\AgdaSpace{}%
\AgdaFunction{split}\AgdaSpace{}%
\AgdaBound{x}\AgdaSpace{}%
\AgdaOperator{\AgdaDatatype{≡}}\AgdaSpace{}%
\AgdaInductiveConstructor{inj₁}\AgdaSpace{}%
\AgdaBound{y}\<%
\\
\>[4]\AgdaSymbol{→}\AgdaSpace{}%
\AgdaSymbol{(}\AgdaBound{p}\AgdaSpace{}%
\AgdaOperator{\AgdaFunction{\&\&}}\AgdaSpace{}%
\AgdaBound{q}\AgdaSymbol{)}\AgdaSpace{}%
\AgdaBound{x}\AgdaSpace{}%
\AgdaOperator{\AgdaDatatype{≡}}\AgdaSpace{}%
\AgdaBound{p}\AgdaSpace{}%
\AgdaBound{y}\<%
\\
\>[2]\AgdaFunction{split-left}\AgdaSpace{}%
\AgdaSymbol{\{}\AgdaArgument{Δ}\AgdaSpace{}%
\AgdaSymbol{=}\AgdaSpace{}%
\AgdaInductiveConstructor{∅}\AgdaSymbol{\}}\AgdaSpace{}%
\AgdaSymbol{\{}\AgdaArgument{x}\AgdaSpace{}%
\AgdaSymbol{=}\AgdaSpace{}%
\AgdaInductiveConstructor{zero}\AgdaSymbol{\}}\AgdaSpace{}%
\AgdaInductiveConstructor{refl}\AgdaSpace{}%
\AgdaSymbol{=}\AgdaSpace{}%
\AgdaInductiveConstructor{refl}\<%
\\
\>[2]\AgdaFunction{split-left}\AgdaSpace{}%
\AgdaSymbol{\{}\AgdaArgument{Δ}\AgdaSpace{}%
\AgdaSymbol{=}\AgdaSpace{}%
\AgdaInductiveConstructor{∅}\AgdaSymbol{\}}\AgdaSpace{}%
\AgdaSymbol{\{}\AgdaArgument{x}\AgdaSpace{}%
\AgdaSymbol{=}\AgdaSpace{}%
\AgdaInductiveConstructor{suc}\AgdaSpace{}%
\AgdaBound{x}\AgdaSymbol{\}}\AgdaSpace{}%
\AgdaInductiveConstructor{refl}\AgdaSpace{}%
\AgdaSymbol{=}\AgdaSpace{}%
\AgdaInductiveConstructor{refl}\<%
\\
\>[2]\AgdaFunction{split-left}\AgdaSpace{}%
\AgdaSymbol{\{}\AgdaArgument{Δ}\AgdaSpace{}%
\AgdaSymbol{=}\AgdaSpace{}%
\AgdaBound{Δ}\AgdaSpace{}%
\AgdaOperator{\AgdaInductiveConstructor{·:}}\AgdaSpace{}%
\AgdaBound{S}\AgdaSymbol{\}}\AgdaSpace{}%
\AgdaSymbol{\{}\AgdaArgument{x}\AgdaSpace{}%
\AgdaSymbol{=}\AgdaSpace{}%
\AgdaInductiveConstructor{suc}\AgdaSpace{}%
\AgdaBound{x}\AgdaSymbol{\}}\AgdaSpace{}%
\AgdaBound{eq}\<%
\\
\>[2][@{}l@{\AgdaIndent{0}}]%
\>[4]\AgdaKeyword{with}\AgdaSpace{}%
\AgdaFunction{split}\AgdaSpace{}%
\AgdaSymbol{\{}\AgdaArgument{Δ}\AgdaSpace{}%
\AgdaSymbol{=}\AgdaSpace{}%
\AgdaBound{Δ}\AgdaSymbol{\}}\AgdaSpace{}%
\AgdaBound{x}\AgdaSpace{}%
\AgdaKeyword{in}\AgdaSpace{}%
\AgdaArgument{eq′}\AgdaSpace{}%
\AgdaSymbol{|}\AgdaSpace{}%
\AgdaBound{eq}\<%
\\
\>[2]\AgdaSymbol{...}\AgdaSpace{}%
\AgdaSymbol{|}\AgdaSpace{}%
\AgdaInductiveConstructor{inj₁}\AgdaSpace{}%
\AgdaSymbol{\AgdaUnderscore{}}%
\>[32]\AgdaSymbol{|}\AgdaSpace{}%
\AgdaInductiveConstructor{refl}\AgdaSpace{}%
\AgdaSymbol{=}\<%
\\
\>[2][@{}l@{\AgdaIndent{0}}]%
\>[4]\AgdaFunction{split-left}\AgdaSpace{}%
\AgdaSymbol{\{}\AgdaArgument{Δ}\AgdaSpace{}%
\AgdaSymbol{=}\AgdaSpace{}%
\AgdaBound{Δ}\AgdaSymbol{\}}\AgdaSpace{}%
\AgdaBound{eq′}\<%
\\
\\[\AgdaEmptyExtraSkip]%
\>[2]\AgdaComment{--\ Splitting\ variable\ into\ right\ context}\<%
\\
\>[2]\AgdaFunction{split-right}\AgdaSpace{}%
\AgdaSymbol{:}\<%
\\
\>[2][@{}l@{\AgdaIndent{0}}]%
\>[4]\AgdaSymbol{∀}\AgdaSpace{}%
\AgdaSymbol{\{}\AgdaBound{x}\AgdaSpace{}%
\AgdaSymbol{:}\AgdaSpace{}%
\AgdaGeneralizable{Γ}\AgdaSpace{}%
\AgdaOperator{\AgdaFunction{<>}}\AgdaSpace{}%
\AgdaGeneralizable{Δ}\AgdaSpace{}%
\AgdaOperator{\AgdaDatatype{∋}}\AgdaSpace{}%
\AgdaGeneralizable{T}\AgdaSymbol{\}}\AgdaSpace{}%
\AgdaSymbol{\{}\AgdaBound{y}\AgdaSpace{}%
\AgdaSymbol{:}\AgdaSpace{}%
\AgdaGeneralizable{Δ}\AgdaSpace{}%
\AgdaOperator{\AgdaDatatype{∋}}\AgdaSpace{}%
\AgdaGeneralizable{T}\AgdaSymbol{\}}\AgdaSpace{}%
\AgdaSymbol{\{}\AgdaBound{p}\AgdaSpace{}%
\AgdaSymbol{:}\AgdaSpace{}%
\AgdaOperator{\AgdaFunction{𝒞⟦}}\AgdaSpace{}%
\AgdaGeneralizable{Γ}\AgdaSpace{}%
\AgdaOperator{\AgdaFunction{⟧}}\AgdaSymbol{\}}\AgdaSpace{}%
\AgdaSymbol{\{}\AgdaBound{q}\AgdaSpace{}%
\AgdaSymbol{:}\AgdaSpace{}%
\AgdaOperator{\AgdaFunction{𝒞⟦}}\AgdaSpace{}%
\AgdaGeneralizable{Δ}\AgdaSpace{}%
\AgdaOperator{\AgdaFunction{⟧}}\AgdaSymbol{\}}\<%
\\
\>[4]\AgdaSymbol{→}\AgdaSpace{}%
\AgdaFunction{split}\AgdaSpace{}%
\AgdaBound{x}\AgdaSpace{}%
\AgdaOperator{\AgdaDatatype{≡}}\AgdaSpace{}%
\AgdaInductiveConstructor{inj₂}\AgdaSpace{}%
\AgdaBound{y}\AgdaSpace{}%
\AgdaSymbol{→}\AgdaSpace{}%
\AgdaSymbol{(}\AgdaBound{p}\AgdaSpace{}%
\AgdaOperator{\AgdaFunction{\&\&}}\AgdaSpace{}%
\AgdaBound{q}\AgdaSymbol{)}\AgdaSpace{}%
\AgdaBound{x}\AgdaSpace{}%
\AgdaOperator{\AgdaDatatype{≡}}\AgdaSpace{}%
\AgdaBound{q}\AgdaSpace{}%
\AgdaBound{y}\<%
\\
\>[2]\AgdaFunction{split-right}\AgdaSpace{}%
\AgdaSymbol{\{}\AgdaArgument{Δ}\AgdaSpace{}%
\AgdaSymbol{=}\AgdaSpace{}%
\AgdaBound{Δ}\AgdaSpace{}%
\AgdaOperator{\AgdaInductiveConstructor{·:}}\AgdaSpace{}%
\AgdaSymbol{\AgdaUnderscore{}\}}\AgdaSpace{}%
\AgdaSymbol{\{}\AgdaArgument{x}\AgdaSpace{}%
\AgdaSymbol{=}\AgdaSpace{}%
\AgdaInductiveConstructor{zero}\AgdaSymbol{\}}\AgdaSpace{}%
\AgdaInductiveConstructor{refl}\AgdaSpace{}%
\AgdaSymbol{=}\AgdaSpace{}%
\AgdaInductiveConstructor{refl}\<%
\\
\>[2]\AgdaFunction{split-right}\AgdaSpace{}%
\AgdaSymbol{\{}\AgdaArgument{Δ}\AgdaSpace{}%
\AgdaSymbol{=}\AgdaSpace{}%
\AgdaBound{Δ}\AgdaSpace{}%
\AgdaOperator{\AgdaInductiveConstructor{·:}}\AgdaSpace{}%
\AgdaSymbol{\AgdaUnderscore{}\}}\AgdaSpace{}%
\AgdaSymbol{\{}\AgdaArgument{x}\AgdaSpace{}%
\AgdaSymbol{=}\AgdaSpace{}%
\AgdaInductiveConstructor{suc}\AgdaSpace{}%
\AgdaBound{x}\AgdaSymbol{\}}\AgdaSpace{}%
\AgdaBound{eq}\<%
\\
\>[2][@{}l@{\AgdaIndent{0}}]%
\>[4]\AgdaKeyword{with}\AgdaSpace{}%
\AgdaFunction{split}\AgdaSpace{}%
\AgdaSymbol{\{}\AgdaArgument{Δ}\AgdaSpace{}%
\AgdaSymbol{=}\AgdaSpace{}%
\AgdaBound{Δ}\AgdaSymbol{\}}\AgdaSpace{}%
\AgdaBound{x}\AgdaSpace{}%
\AgdaKeyword{in}\AgdaSpace{}%
\AgdaArgument{eq′}\AgdaSpace{}%
\AgdaSymbol{|}\AgdaSpace{}%
\AgdaBound{eq}\<%
\\
\>[2]\AgdaSymbol{...}\AgdaSpace{}%
\AgdaSymbol{|}\AgdaSpace{}%
\AgdaInductiveConstructor{inj₂}\AgdaSpace{}%
\AgdaSymbol{\AgdaUnderscore{}}%
\>[32]\AgdaSymbol{|}\AgdaSpace{}%
\AgdaInductiveConstructor{refl}\AgdaSpace{}%
\AgdaSymbol{=}\<%
\\
\>[2][@{}l@{\AgdaIndent{0}}]%
\>[4]\AgdaFunction{split-right}\AgdaSpace{}%
\AgdaSymbol{\{}\AgdaArgument{Δ}\AgdaSpace{}%
\AgdaSymbol{=}\AgdaSpace{}%
\AgdaBound{Δ}\AgdaSymbol{\}}\AgdaSpace{}%
\AgdaBound{eq′}\<%
\\
\\[\AgdaEmptyExtraSkip]%
\>[2]\AgdaKeyword{mutual}\<%
\\
\>[2][@{}l@{\AgdaIndent{0}}]%
\>[4]\AgdaComment{--\ Evaluation\ of\ a\ term\ is\ equivalent\ to\ evaluation}\<%
\\
\>[4]\AgdaComment{--\ of\ its\ closed\ self}\<%
\\
\>[4]\AgdaFunction{ℰ-close}\AgdaSpace{}%
\AgdaSymbol{:}\<%
\\
\>[4][@{}l@{\AgdaIndent{0}}]%
\>[6]\AgdaSymbol{∀}\AgdaSpace{}%
\AgdaSymbol{(}\AgdaBound{t}\AgdaSpace{}%
\AgdaSymbol{:}\AgdaSpace{}%
\AgdaGeneralizable{Γ}\AgdaSpace{}%
\AgdaOperator{\AgdaFunction{<>}}\AgdaSpace{}%
\AgdaGeneralizable{Δ}\AgdaSpace{}%
\AgdaOperator{\AgdaDatatype{⊢}}\AgdaSpace{}%
\AgdaGeneralizable{T}\AgdaSymbol{)}\AgdaSpace{}%
\AgdaSymbol{(}\AgdaBound{γ}\AgdaSpace{}%
\AgdaSymbol{:}\AgdaSpace{}%
\AgdaFunction{Env}\AgdaSpace{}%
\AgdaGeneralizable{Γ}\AgdaSymbol{)}\AgdaSpace{}%
\AgdaSymbol{(}\AgdaBound{q}\AgdaSpace{}%
\AgdaSymbol{:}\AgdaSpace{}%
\AgdaOperator{\AgdaFunction{𝒞⟦}}\AgdaSpace{}%
\AgdaGeneralizable{Δ}\AgdaSpace{}%
\AgdaOperator{\AgdaFunction{⟧}}\AgdaSymbol{)}\<%
\\
\>[6]\AgdaSymbol{→}\AgdaSpace{}%
\AgdaOperator{\AgdaFunction{ℰ⟦}}\AgdaSpace{}%
\AgdaBound{t}\AgdaSpace{}%
\AgdaOperator{\AgdaFunction{⟧}}\AgdaSpace{}%
\AgdaSymbol{(}\AgdaOperator{\AgdaFunction{𝒢⟦}}\AgdaSpace{}%
\AgdaBound{γ}\AgdaSpace{}%
\AgdaOperator{\AgdaFunction{⟧}}\AgdaSpace{}%
\AgdaOperator{\AgdaFunction{\&\&}}\AgdaSpace{}%
\AgdaBound{q}\AgdaSymbol{)}\AgdaSpace{}%
\AgdaOperator{\AgdaDatatype{≡}}\AgdaSpace{}%
\AgdaOperator{\AgdaFunction{ℰ⟦}}\AgdaSpace{}%
\AgdaFunction{close}\AgdaSpace{}%
\AgdaBound{γ}\AgdaSpace{}%
\AgdaBound{t}\AgdaSpace{}%
\AgdaOperator{\AgdaFunction{⟧}}\AgdaSpace{}%
\AgdaBound{q}\<%
\\
\>[4]\AgdaFunction{ℰ-close}\AgdaSpace{}%
\AgdaInductiveConstructor{true}\AgdaSpace{}%
\AgdaSymbol{\AgdaUnderscore{}}\AgdaSpace{}%
\AgdaSymbol{\AgdaUnderscore{}}\AgdaSpace{}%
\AgdaSymbol{=}\AgdaSpace{}%
\AgdaInductiveConstructor{refl}\<%
\\
\>[4]\AgdaFunction{ℰ-close}\AgdaSpace{}%
\AgdaInductiveConstructor{false}\AgdaSpace{}%
\AgdaSymbol{\AgdaUnderscore{}}\AgdaSpace{}%
\AgdaSymbol{\AgdaUnderscore{}}\AgdaSpace{}%
\AgdaSymbol{=}\AgdaSpace{}%
\AgdaInductiveConstructor{refl}\<%
\\
\>[4]\AgdaFunction{ℰ-close}\AgdaSpace{}%
\AgdaSymbol{\{}\AgdaArgument{Δ}\AgdaSpace{}%
\AgdaSymbol{=}\AgdaSpace{}%
\AgdaBound{Δ}\AgdaSymbol{\}}\AgdaSpace{}%
\AgdaSymbol{(}\AgdaInductiveConstructor{var}\AgdaSpace{}%
\AgdaBound{x}\AgdaSymbol{)}\AgdaSpace{}%
\AgdaBound{γ}\AgdaSpace{}%
\AgdaBound{q}\<%
\\
\>[4][@{}l@{\AgdaIndent{0}}]%
\>[6]\AgdaKeyword{with}\AgdaSpace{}%
\AgdaFunction{split}\AgdaSpace{}%
\AgdaSymbol{\{}\AgdaArgument{Δ}\AgdaSpace{}%
\AgdaSymbol{=}\AgdaSpace{}%
\AgdaBound{Δ}\AgdaSymbol{\}}\AgdaSpace{}%
\AgdaBound{x}\AgdaSpace{}%
\AgdaKeyword{in}\AgdaSpace{}%
\AgdaArgument{eq}\<%
\\
\>[4]\AgdaSymbol{...}\AgdaSpace{}%
\AgdaSymbol{|}\AgdaSpace{}%
\AgdaInductiveConstructor{inj₂}\AgdaSpace{}%
\AgdaBound{y}\AgdaSpace{}%
\AgdaSymbol{=}\AgdaSpace{}%
\AgdaFunction{split-right}\AgdaSpace{}%
\AgdaSymbol{\{}\AgdaArgument{p}\AgdaSpace{}%
\AgdaSymbol{=}\AgdaSpace{}%
\AgdaOperator{\AgdaFunction{𝒢⟦}}\AgdaSpace{}%
\AgdaBound{γ}\AgdaSpace{}%
\AgdaOperator{\AgdaFunction{⟧}}\AgdaSymbol{\}}\AgdaSpace{}%
\AgdaBound{eq}\<%
\\
\>[4]\AgdaSymbol{...}\AgdaSpace{}%
\AgdaSymbol{|}\AgdaSpace{}%
\AgdaInductiveConstructor{inj₁}\AgdaSpace{}%
\AgdaBound{x′}\<%
\\
\>[4][@{}l@{\AgdaIndent{0}}]%
\>[6]\AgdaKeyword{rewrite}\AgdaSpace{}%
\AgdaFunction{split-left}\AgdaSpace{}%
\AgdaSymbol{\{}\AgdaArgument{p}\AgdaSpace{}%
\AgdaSymbol{=}\AgdaSpace{}%
\AgdaOperator{\AgdaFunction{𝒢⟦}}\AgdaSpace{}%
\AgdaBound{γ}\AgdaSpace{}%
\AgdaOperator{\AgdaFunction{⟧}}\AgdaSymbol{\}}\AgdaSpace{}%
\AgdaSymbol{\{}\AgdaArgument{q}\AgdaSpace{}%
\AgdaSymbol{=}\AgdaSpace{}%
\AgdaBound{q}\AgdaSymbol{\}}\AgdaSpace{}%
\AgdaBound{eq}\<%
\\
\>[6]\AgdaKeyword{rewrite}\AgdaSpace{}%
\AgdaFunction{⇑-sound}\AgdaSpace{}%
\AgdaSymbol{\{}\AgdaArgument{ρ}\AgdaSpace{}%
\AgdaSymbol{=}\AgdaSpace{}%
\AgdaSymbol{λ}\AgdaSpace{}%
\AgdaSymbol{()\}}\AgdaSpace{}%
\AgdaSymbol{(}\AgdaBound{γ}\AgdaSpace{}%
\AgdaBound{x′}\AgdaSymbol{)}\AgdaSpace{}%
\AgdaSymbol{=}\<%
\\
\>[6]\AgdaFunction{ℰ-inject}\AgdaSpace{}%
\AgdaSymbol{((}\AgdaBound{γ}\AgdaSpace{}%
\AgdaBound{x′}\AgdaSymbol{)}\AgdaSpace{}%
\AgdaOperator{\AgdaFunction{⇑}}\AgdaSymbol{)}\AgdaSpace{}%
\AgdaSymbol{(λ}\AgdaSpace{}%
\AgdaSymbol{())}\AgdaSpace{}%
\AgdaBound{q}\<%
\\
\>[4]\AgdaFunction{ℰ-close}\AgdaSpace{}%
\AgdaSymbol{(}\AgdaOperator{\AgdaInductiveConstructor{ƛ}}\AgdaSpace{}%
\AgdaBound{t}\AgdaSymbol{)}\AgdaSpace{}%
\AgdaBound{γ}\AgdaSpace{}%
\AgdaBound{q}\AgdaSpace{}%
\AgdaSymbol{=}\<%
\\
\>[4][@{}l@{\AgdaIndent{0}}]%
\>[6]\AgdaPostulate{fext}\AgdaSpace{}%
\AgdaSymbol{(λ}\AgdaSpace{}%
\AgdaBound{x}\AgdaSpace{}%
\AgdaSymbol{→}\AgdaSpace{}%
\AgdaFunction{ℰ-close}\AgdaSpace{}%
\AgdaBound{t}\AgdaSpace{}%
\AgdaBound{γ}\AgdaSpace{}%
\AgdaSymbol{(}\AgdaBound{q}\AgdaSpace{}%
\AgdaOperator{\AgdaFunction{\&}}\AgdaSpace{}%
\AgdaBound{x}\AgdaSymbol{))}\<%
\\
\>[4]\AgdaFunction{ℰ-close}\AgdaSpace{}%
\AgdaSymbol{(}\AgdaBound{r}\AgdaSpace{}%
\AgdaOperator{\AgdaInductiveConstructor{·}}\AgdaSpace{}%
\AgdaBound{s}\AgdaSymbol{)}\AgdaSpace{}%
\AgdaBound{γ}\AgdaSpace{}%
\AgdaBound{q}\<%
\\
\>[4][@{}l@{\AgdaIndent{0}}]%
\>[6]\AgdaKeyword{rewrite}\AgdaSpace{}%
\AgdaFunction{ℰ-close}\AgdaSpace{}%
\AgdaBound{r}\AgdaSpace{}%
\AgdaBound{γ}\AgdaSpace{}%
\AgdaBound{q}\<%
\\
\>[6][@{}l@{\AgdaIndent{0}}]%
\>[12]\AgdaSymbol{|}\AgdaSpace{}%
\AgdaFunction{ℰ-close}\AgdaSpace{}%
\AgdaBound{s}\AgdaSpace{}%
\AgdaBound{γ}\AgdaSpace{}%
\AgdaBound{q}\AgdaSpace{}%
\AgdaSymbol{=}\AgdaSpace{}%
\AgdaInductiveConstructor{refl}\<%
\\
\>[4]\AgdaFunction{ℰ-close}\AgdaSpace{}%
\AgdaSymbol{(}\AgdaOperator{\AgdaInductiveConstructor{if}}\AgdaSpace{}%
\AgdaBound{r}\AgdaSpace{}%
\AgdaOperator{\AgdaInductiveConstructor{then}}\AgdaSpace{}%
\AgdaBound{s}\AgdaSpace{}%
\AgdaOperator{\AgdaInductiveConstructor{else}}\AgdaSpace{}%
\AgdaBound{t}\AgdaSymbol{)}\AgdaSpace{}%
\AgdaBound{γ}\AgdaSpace{}%
\AgdaBound{q}\<%
\\
\>[4][@{}l@{\AgdaIndent{0}}]%
\>[6]\AgdaKeyword{with}\AgdaSpace{}%
\AgdaOperator{\AgdaFunction{ℰ⟦}}\AgdaSpace{}%
\AgdaBound{r}\AgdaSpace{}%
\AgdaOperator{\AgdaFunction{⟧}}\AgdaSpace{}%
\AgdaSymbol{(}\AgdaOperator{\AgdaFunction{𝒢⟦}}\AgdaSpace{}%
\AgdaBound{γ}\AgdaSpace{}%
\AgdaOperator{\AgdaFunction{⟧}}\AgdaSpace{}%
\AgdaOperator{\AgdaFunction{\&\&}}\AgdaSpace{}%
\AgdaBound{q}\AgdaSymbol{)}\AgdaSpace{}%
\AgdaSymbol{|}\AgdaSpace{}%
\AgdaOperator{\AgdaFunction{ℰ⟦}}\AgdaSpace{}%
\AgdaFunction{close}\AgdaSpace{}%
\AgdaBound{γ}\AgdaSpace{}%
\AgdaBound{r}\AgdaSpace{}%
\AgdaOperator{\AgdaFunction{⟧}}\AgdaSpace{}%
\AgdaBound{q}\AgdaSpace{}%
\AgdaSymbol{|}\AgdaSpace{}%
\AgdaFunction{ℰ-close}\AgdaSpace{}%
\AgdaBound{r}\AgdaSpace{}%
\AgdaBound{γ}\AgdaSpace{}%
\AgdaBound{q}\<%
\\
\>[4]\AgdaSymbol{...}\AgdaSpace{}%
\AgdaSymbol{|}\AgdaSpace{}%
\AgdaInductiveConstructor{true}\AgdaSpace{}%
\AgdaSymbol{|}\AgdaSpace{}%
\AgdaInductiveConstructor{true}%
\>[24]\AgdaSymbol{|}\AgdaSpace{}%
\AgdaInductiveConstructor{refl}\AgdaSpace{}%
\AgdaSymbol{=}\AgdaSpace{}%
\AgdaFunction{ℰ-close}\AgdaSpace{}%
\AgdaBound{s}\AgdaSpace{}%
\AgdaBound{γ}\AgdaSpace{}%
\AgdaBound{q}\<%
\\
\>[4]\AgdaSymbol{...}\AgdaSpace{}%
\AgdaSymbol{|}\AgdaSpace{}%
\AgdaInductiveConstructor{false}\AgdaSpace{}%
\AgdaSymbol{|}\AgdaSpace{}%
\AgdaInductiveConstructor{false}\AgdaSpace{}%
\AgdaSymbol{|}\AgdaSpace{}%
\AgdaInductiveConstructor{refl}\AgdaSpace{}%
\AgdaSymbol{=}\AgdaSpace{}%
\AgdaFunction{ℰ-close}\AgdaSpace{}%
\AgdaBound{t}\AgdaSpace{}%
\AgdaBound{γ}\AgdaSpace{}%
\AgdaBound{q}\<%
\\
\\[\AgdaEmptyExtraSkip]%
\>[4]\AgdaComment{--\ Reading\ back\ a\ normal\ form\ from\ an\ evaluated\ term}\<%
\\
\>[4]\AgdaComment{--\ preserves\ meaning}\<%
\\
\>[4]\AgdaFunction{⇑-sound}\AgdaSpace{}%
\AgdaSymbol{:}\<%
\\
\>[4][@{}l@{\AgdaIndent{0}}]%
\>[5]\AgdaSymbol{∀}\AgdaSpace{}%
\AgdaSymbol{\{}\AgdaBound{ρ}\AgdaSpace{}%
\AgdaSymbol{:}\AgdaSpace{}%
\AgdaOperator{\AgdaFunction{𝒞⟦}}\AgdaSpace{}%
\AgdaInductiveConstructor{∅}\AgdaSpace{}%
\AgdaOperator{\AgdaFunction{⟧}}\AgdaSymbol{\}}\AgdaSpace{}%
\AgdaSymbol{(}\AgdaBound{a}\AgdaSpace{}%
\AgdaSymbol{:}\AgdaSpace{}%
\AgdaDatatype{Domain}\AgdaSpace{}%
\AgdaGeneralizable{T}\AgdaSymbol{)}\<%
\\
\>[5]\AgdaSymbol{→}\AgdaSpace{}%
\AgdaOperator{\AgdaFunction{𝒟⟦}}\AgdaSpace{}%
\AgdaBound{a}\AgdaSpace{}%
\AgdaOperator{\AgdaFunction{⟧}}\AgdaSpace{}%
\AgdaOperator{\AgdaDatatype{≡}}\AgdaSpace{}%
\AgdaOperator{\AgdaFunction{ℰ⟦}}\AgdaSpace{}%
\AgdaBound{a}\AgdaSpace{}%
\AgdaOperator{\AgdaFunction{⇑}}\AgdaSpace{}%
\AgdaOperator{\AgdaFunction{⟧}}\AgdaSpace{}%
\AgdaBound{ρ}\<%
\\
\>[4]\AgdaFunction{⇑-sound}\AgdaSpace{}%
\AgdaSymbol{\{}\AgdaArgument{ρ}\AgdaSpace{}%
\AgdaSymbol{=}\AgdaSpace{}%
\AgdaBound{ρ}\AgdaSymbol{\}}\AgdaSpace{}%
\AgdaSymbol{(}\AgdaOperator{\AgdaInductiveConstructor{⟨ƛ\AgdaUnderscore{}⟩\AgdaUnderscore{}}}\AgdaSpace{}%
\AgdaSymbol{\{}\AgdaBound{Γ}\AgdaSymbol{\}}\AgdaSpace{}%
\AgdaBound{t}\AgdaSpace{}%
\AgdaBound{γ}\AgdaSymbol{)}\<%
\\
\>[4][@{}l@{\AgdaIndent{0}}]%
\>[6]\AgdaSymbol{=}\AgdaSpace{}%
\AgdaPostulate{fext}\AgdaSpace{}%
\AgdaSymbol{(λ}\AgdaSpace{}%
\AgdaBound{a}\AgdaSpace{}%
\AgdaSymbol{→}\AgdaSpace{}%
\AgdaFunction{ℰ-close}\AgdaSpace{}%
\AgdaBound{t}\AgdaSpace{}%
\AgdaBound{γ}\AgdaSpace{}%
\AgdaSymbol{(}\AgdaBound{ρ}\AgdaSpace{}%
\AgdaOperator{\AgdaFunction{\&}}\AgdaSpace{}%
\AgdaBound{a}\AgdaSymbol{))}\<%
\\
\>[4]\AgdaFunction{⇑-sound}\AgdaSpace{}%
\AgdaInductiveConstructor{true}\AgdaSpace{}%
\AgdaSymbol{=}\AgdaSpace{}%
\AgdaInductiveConstructor{refl}\<%
\\
\>[4]\AgdaFunction{⇑-sound}\AgdaSpace{}%
\AgdaInductiveConstructor{false}\AgdaSpace{}%
\AgdaSymbol{=}\AgdaSpace{}%
\AgdaInductiveConstructor{refl}\<%
\\
\\[\AgdaEmptyExtraSkip]%
\>[2]\AgdaComment{--\ Use\ the\ fact\ that\ reading\ back\ a\ normal\ form\ is\ sound}\<%
\\
\>[2]\AgdaComment{--\ w.r.t.\ denotational\ semantics\ to\ prove\ normalization}\<%
\\
\>[2]\AgdaComment{--\ is\ sound}\<%
\\
\>[2]\AgdaFunction{normalization-sound}\AgdaSpace{}%
\AgdaSymbol{:}\<%
\\
\>[2][@{}l@{\AgdaIndent{0}}]%
\>[4]\AgdaSymbol{∀}\AgdaSpace{}%
\AgdaSymbol{(}\AgdaBound{t}\AgdaSpace{}%
\AgdaSymbol{:}\AgdaSpace{}%
\AgdaInductiveConstructor{∅}\AgdaSpace{}%
\AgdaOperator{\AgdaDatatype{⊢}}\AgdaSpace{}%
\AgdaGeneralizable{T}\AgdaSymbol{)}\AgdaSpace{}%
\AgdaSymbol{(}\AgdaBound{v}\AgdaSpace{}%
\AgdaSymbol{:}\AgdaSpace{}%
\AgdaInductiveConstructor{∅}\AgdaSpace{}%
\AgdaOperator{\AgdaDatatype{⊢}}\AgdaSpace{}%
\AgdaGeneralizable{T}\AgdaSymbol{)}\<%
\\
\>[4]\AgdaSymbol{→}\AgdaSpace{}%
\AgdaBound{t}\AgdaSpace{}%
\AgdaOperator{\AgdaFunction{normalizes-to}}\AgdaSpace{}%
\AgdaBound{v}\<%
\\
\>[4]\AgdaSymbol{→}\AgdaSpace{}%
\AgdaBound{t}\AgdaSpace{}%
\AgdaOperator{\AgdaFunction{ℰ≡}}\AgdaSpace{}%
\AgdaBound{v}\<%
\\
\>[2]\AgdaFunction{normalization-sound}\AgdaSpace{}%
\AgdaBound{t}\AgdaSpace{}%
\AgdaSymbol{\AgdaUnderscore{}}\AgdaSpace{}%
\AgdaSymbol{(\AgdaUnderscore{}}\AgdaSpace{}%
\AgdaOperator{\AgdaInductiveConstructor{,}}\AgdaSpace{}%
\AgdaBound{t⇓}\AgdaSpace{}%
\AgdaOperator{\AgdaInductiveConstructor{,}}\AgdaSpace{}%
\AgdaInductiveConstructor{refl}\AgdaSymbol{)}\AgdaSpace{}%
\AgdaSymbol{\{}\AgdaBound{ρ}\AgdaSymbol{\}}\<%
\\
\>[2][@{}l@{\AgdaIndent{0}}]%
\>[4]\AgdaKeyword{with}\AgdaSpace{}%
\AgdaFunction{normalization}\AgdaSpace{}%
\AgdaBound{t}\<%
\\
\>[2]\AgdaSymbol{...}\AgdaSpace{}%
\AgdaSymbol{|}\AgdaSpace{}%
\AgdaBound{v}\AgdaSpace{}%
\AgdaOperator{\AgdaInductiveConstructor{,}}\AgdaSpace{}%
\AgdaBound{a}\AgdaSpace{}%
\AgdaOperator{\AgdaInductiveConstructor{,}}\AgdaSpace{}%
\AgdaBound{t⇓′}\AgdaSpace{}%
\AgdaOperator{\AgdaInductiveConstructor{,}}\AgdaSpace{}%
\AgdaInductiveConstructor{refl}\<%
\\
\>[2][@{}l@{\AgdaIndent{0}}]%
\>[4]\AgdaKeyword{with}\AgdaSpace{}%
\AgdaFunction{⇓-uniq}\AgdaSpace{}%
\AgdaBound{t⇓}\AgdaSpace{}%
\AgdaBound{t⇓′}\<%
\\
\>[2]\AgdaSymbol{...}\AgdaSpace{}%
\AgdaSymbol{|}\AgdaSpace{}%
\AgdaInductiveConstructor{refl}\AgdaSpace{}%
\AgdaSymbol{=}\<%
\\
\>[2][@{}l@{\AgdaIndent{0}}]%
\>[4]\AgdaOperator{\AgdaFunction{begin}}\<%
\\
\>[4][@{}l@{\AgdaIndent{0}}]%
\>[6]\AgdaOperator{\AgdaFunction{ℰ⟦}}\AgdaSpace{}%
\AgdaBound{t}\AgdaSpace{}%
\AgdaOperator{\AgdaFunction{⟧}}\AgdaSpace{}%
\AgdaBound{ρ}\<%
\\
\>[4]\AgdaFunction{≡⟨}\AgdaSpace{}%
\AgdaFunction{cong}\AgdaSpace{}%
\AgdaOperator{\AgdaFunction{ℰ⟦}}\AgdaSpace{}%
\AgdaBound{t}\AgdaSpace{}%
\AgdaOperator{\AgdaFunction{⟧}}\AgdaSpace{}%
\AgdaSymbol{(}\AgdaFunction{fext-implicit}\AgdaSpace{}%
\AgdaSymbol{(}\AgdaPostulate{fext}\AgdaSpace{}%
\AgdaSymbol{λ}\AgdaSpace{}%
\AgdaSymbol{()))}\AgdaSpace{}%
\AgdaFunction{⟩}\<%
\\
\>[4][@{}l@{\AgdaIndent{0}}]%
\>[5]\AgdaOperator{\AgdaFunction{ℰ⟦}}\AgdaSpace{}%
\AgdaBound{t}\AgdaSpace{}%
\AgdaOperator{\AgdaFunction{⟧}}\AgdaSpace{}%
\AgdaOperator{\AgdaFunction{𝒢⟦}}\AgdaSpace{}%
\AgdaFunction{empty}\AgdaSpace{}%
\AgdaOperator{\AgdaFunction{⟧}}\<%
\\
\>[4]\AgdaFunction{≡⟨}\AgdaSpace{}%
\AgdaFunction{⇓-sound}\AgdaSpace{}%
\AgdaBound{t⇓}\AgdaSpace{}%
\AgdaFunction{⟩}\<%
\\
\>[4][@{}l@{\AgdaIndent{0}}]%
\>[6]\AgdaOperator{\AgdaFunction{𝒟⟦}}\AgdaSpace{}%
\AgdaBound{a}\AgdaSpace{}%
\AgdaOperator{\AgdaFunction{⟧}}\<%
\\
\>[4]\AgdaFunction{≡⟨}\AgdaSpace{}%
\AgdaFunction{⇑-sound}\AgdaSpace{}%
\AgdaBound{a}\AgdaSpace{}%
\AgdaFunction{⟩}\<%
\\
\>[4][@{}l@{\AgdaIndent{0}}]%
\>[6]\AgdaOperator{\AgdaFunction{ℰ⟦}}\AgdaSpace{}%
\AgdaBound{v}\AgdaSpace{}%
\AgdaOperator{\AgdaFunction{⟧}}\AgdaSpace{}%
\AgdaBound{ρ}\<%
\\
\>[4]\AgdaOperator{\AgdaFunction{∎}}\<%
\end{code}
\end{AgdaAlign}

\section{Normalization with full reductions (full proof of $\S$\ref{sec:nbe})}
\label{sec:nbe-full}

Here we provide the full proof presented in $\S$\ref{sec:nbe}, including any
portions that were omitted in the main body of the paper.

\begin{code}[hide]%
\>[0]\AgdaKeyword{module}\AgdaSpace{}%
\AgdaModule{FullNormalization}\AgdaSpace{}%
\AgdaKeyword{where}\<%
\\
\>[0][@{}l@{\AgdaIndent{0}}]%
\>[2]\AgdaKeyword{infix}\AgdaSpace{}%
\AgdaNumber{5}\AgdaSpace{}%
\AgdaOperator{\AgdaInductiveConstructor{⟨ƛ\AgdaUnderscore{}⟩\AgdaUnderscore{}}}\<%
\\
\>[2]\AgdaKeyword{infix}\AgdaSpace{}%
\AgdaNumber{7}\AgdaSpace{}%
\AgdaOperator{\AgdaInductiveConstructor{`\AgdaUnderscore{}}}\<%
\\
\>[2]\AgdaKeyword{infix}\AgdaSpace{}%
\AgdaNumber{4}\AgdaSpace{}%
\AgdaOperator{\AgdaDatatype{\AgdaUnderscore{}·\AgdaUnderscore{}⇓\AgdaUnderscore{}}}\<%
\\
\>[2]\AgdaKeyword{infix}\AgdaSpace{}%
\AgdaNumber{4}\AgdaSpace{}%
\AgdaOperator{\AgdaDatatype{\AgdaUnderscore{}∣\AgdaUnderscore{}⇓\AgdaUnderscore{}}}\<%
\\
\>[2]\AgdaKeyword{infix}\AgdaSpace{}%
\AgdaNumber{4}\AgdaSpace{}%
\AgdaOperator{\AgdaDatatype{\AgdaUnderscore{}∣\AgdaUnderscore{}⇑\AgdaUnderscore{}}}\<%
\\
\>[2]\AgdaKeyword{infix}\AgdaSpace{}%
\AgdaNumber{4}\AgdaSpace{}%
\AgdaOperator{\AgdaDatatype{\AgdaUnderscore{}∣\AgdaUnderscore{}⇑ⁿᵉ\AgdaUnderscore{}}}\<%
\\
\>[2]\AgdaKeyword{infixr}\AgdaSpace{}%
\AgdaNumber{7}\AgdaSpace{}%
\AgdaOperator{\AgdaFunction{\AgdaUnderscore{}⟶\AgdaUnderscore{}}}\<%
\\
\>[2]\AgdaKeyword{infix}\AgdaSpace{}%
\AgdaNumber{4}\AgdaSpace{}%
\AgdaOperator{\AgdaFunction{\AgdaUnderscore{}⊨\AgdaUnderscore{}}}\<%
\\
\>[2]\AgdaKeyword{infix}\AgdaSpace{}%
\AgdaNumber{4}\AgdaSpace{}%
\AgdaOperator{\AgdaFunction{\AgdaUnderscore{}⊨\AgdaUnderscore{}∷\AgdaUnderscore{}}}\<%
\\
\>[2]\AgdaKeyword{infixl}\AgdaSpace{}%
\AgdaNumber{5}\AgdaSpace{}%
\AgdaOperator{\AgdaFunction{\AgdaUnderscore{}++\AgdaUnderscore{}}}\<%
\\
\>[2]\AgdaKeyword{infix}\AgdaSpace{}%
\AgdaNumber{5}\AgdaSpace{}%
\AgdaOperator{\AgdaInductiveConstructor{ƛ\AgdaUnderscore{}}}\<%
\\
\>[2]\AgdaKeyword{infixl}\AgdaSpace{}%
\AgdaNumber{7}\AgdaSpace{}%
\AgdaOperator{\AgdaInductiveConstructor{\AgdaUnderscore{}·\AgdaUnderscore{}}}\<%
\\
\>[2]\AgdaKeyword{infix}\AgdaSpace{}%
\AgdaNumber{4}\AgdaSpace{}%
\AgdaOperator{\AgdaDatatype{\AgdaUnderscore{}⊢\AgdaUnderscore{}∷\AgdaUnderscore{}}}\<%
\\
\>[2]\AgdaKeyword{infix}\AgdaSpace{}%
\AgdaNumber{4}\AgdaSpace{}%
\AgdaOperator{\AgdaDatatype{\AgdaUnderscore{}∷\AgdaUnderscore{}∈\AgdaUnderscore{}}}\<%
\\
\>[2]\AgdaKeyword{infixr}\AgdaSpace{}%
\AgdaNumber{7}\AgdaSpace{}%
\AgdaOperator{\AgdaInductiveConstructor{\AgdaUnderscore{}⇒\AgdaUnderscore{}}}\<%
\\
\>[2]\AgdaKeyword{infixl}\AgdaSpace{}%
\AgdaNumber{5}\AgdaSpace{}%
\AgdaOperator{\AgdaInductiveConstructor{\AgdaUnderscore{}·:\AgdaUnderscore{}}}\<%
\end{code}
\begin{AgdaAlign}

\subsection{STLC (Figure \ref{fig:extrinsic-stlc})}
\begin{code}%
\>[2]\AgdaKeyword{data}\AgdaSpace{}%
\AgdaDatatype{Type}\AgdaSpace{}%
\AgdaSymbol{:}\AgdaSpace{}%
\AgdaPrimitive{Set}\AgdaSpace{}%
\AgdaKeyword{where}\<%
\\
\>[2][@{}l@{\AgdaIndent{0}}]%
\>[4]\AgdaInductiveConstructor{base}\AgdaSpace{}%
\AgdaSymbol{:}\AgdaSpace{}%
\AgdaDatatype{Type}\<%
\\
\>[4]\AgdaOperator{\AgdaInductiveConstructor{\AgdaUnderscore{}⇒\AgdaUnderscore{}}}\AgdaSpace{}%
\AgdaSymbol{:}\AgdaSpace{}%
\AgdaDatatype{Type}\AgdaSpace{}%
\AgdaSymbol{→}\AgdaSpace{}%
\AgdaDatatype{Type}\AgdaSpace{}%
\AgdaSymbol{→}\AgdaSpace{}%
\AgdaDatatype{Type}\<%
\\
\\[\AgdaEmptyExtraSkip]%
\>[2]\AgdaKeyword{variable}\AgdaSpace{}%
\AgdaGeneralizable{S}\AgdaSpace{}%
\AgdaGeneralizable{T}\AgdaSpace{}%
\AgdaSymbol{:}\AgdaSpace{}%
\AgdaDatatype{Type}\<%
\\
\\[\AgdaEmptyExtraSkip]%
\>[2]\AgdaKeyword{data}\AgdaSpace{}%
\AgdaDatatype{Ctx}\AgdaSpace{}%
\AgdaSymbol{:}\AgdaSpace{}%
\AgdaPrimitive{Set}\AgdaSpace{}%
\AgdaKeyword{where}\<%
\\
\>[2][@{}l@{\AgdaIndent{0}}]%
\>[4]\AgdaInductiveConstructor{∅}\AgdaSpace{}%
\AgdaSymbol{:}\AgdaSpace{}%
\AgdaDatatype{Ctx}\<%
\\
\>[4]\AgdaOperator{\AgdaInductiveConstructor{\AgdaUnderscore{}·:\AgdaUnderscore{}}}\AgdaSpace{}%
\AgdaSymbol{:}\AgdaSpace{}%
\AgdaDatatype{Ctx}\AgdaSpace{}%
\AgdaSymbol{→}\AgdaSpace{}%
\AgdaDatatype{Type}\AgdaSpace{}%
\AgdaSymbol{→}\AgdaSpace{}%
\AgdaDatatype{Ctx}\<%
\\
\\[\AgdaEmptyExtraSkip]%
\>[2]\AgdaKeyword{variable}\AgdaSpace{}%
\AgdaGeneralizable{Γ}\AgdaSpace{}%
\AgdaSymbol{:}\AgdaSpace{}%
\AgdaDatatype{Ctx}\<%
\\
\\[\AgdaEmptyExtraSkip]%
\>[2]\AgdaComment{--\ Raw\ terms}\<%
\\
\>[2]\AgdaKeyword{data}\AgdaSpace{}%
\AgdaDatatype{Term}\AgdaSpace{}%
\AgdaSymbol{:}\AgdaSpace{}%
\AgdaPrimitive{Set}\AgdaSpace{}%
\AgdaKeyword{where}\<%
\\
\>[2][@{}l@{\AgdaIndent{0}}]%
\>[4]\AgdaInductiveConstructor{var}\AgdaSpace{}%
\AgdaSymbol{:}\AgdaSpace{}%
\AgdaSymbol{(}\AgdaBound{x}\AgdaSpace{}%
\AgdaSymbol{:}\AgdaSpace{}%
\AgdaDatatype{ℕ}\AgdaSymbol{)}\AgdaSpace{}%
\AgdaSymbol{→}\AgdaSpace{}%
\AgdaDatatype{Term}\<%
\\
\>[4]\AgdaOperator{\AgdaInductiveConstructor{ƛ\AgdaUnderscore{}}}\AgdaSpace{}%
\AgdaSymbol{:}\AgdaSpace{}%
\AgdaDatatype{Term}\AgdaSpace{}%
\AgdaSymbol{→}\AgdaSpace{}%
\AgdaDatatype{Term}\<%
\\
\>[4]\AgdaOperator{\AgdaInductiveConstructor{\AgdaUnderscore{}·\AgdaUnderscore{}}}\AgdaSpace{}%
\AgdaSymbol{:}\AgdaSpace{}%
\AgdaDatatype{Term}\AgdaSpace{}%
\AgdaSymbol{→}\AgdaSpace{}%
\AgdaDatatype{Term}\AgdaSpace{}%
\AgdaSymbol{→}\AgdaSpace{}%
\AgdaDatatype{Term}\<%
\\
\\[\AgdaEmptyExtraSkip]%
\>[2]\AgdaKeyword{variable}\AgdaSpace{}%
\AgdaGeneralizable{x}\AgdaSpace{}%
\AgdaSymbol{:}\AgdaSpace{}%
\AgdaDatatype{ℕ}\<%
\\
\>[2]\AgdaKeyword{variable}\AgdaSpace{}%
\AgdaGeneralizable{r}\AgdaSpace{}%
\AgdaGeneralizable{s}\AgdaSpace{}%
\AgdaGeneralizable{t}\AgdaSpace{}%
\AgdaGeneralizable{u}\AgdaSpace{}%
\AgdaGeneralizable{v}\AgdaSpace{}%
\AgdaSymbol{:}\AgdaSpace{}%
\AgdaDatatype{Term}\<%
\\
\\[\AgdaEmptyExtraSkip]%
\>[2]\AgdaComment{--\ Variable\ lookup}\<%
\\
\>[2]\AgdaKeyword{data}\AgdaSpace{}%
\AgdaOperator{\AgdaDatatype{\AgdaUnderscore{}∷\AgdaUnderscore{}∈\AgdaUnderscore{}}}\AgdaSpace{}%
\AgdaSymbol{:}\AgdaSpace{}%
\AgdaDatatype{ℕ}\AgdaSpace{}%
\AgdaSymbol{→}\AgdaSpace{}%
\AgdaDatatype{Type}\AgdaSpace{}%
\AgdaSymbol{→}\AgdaSpace{}%
\AgdaDatatype{Ctx}\AgdaSpace{}%
\AgdaSymbol{→}\AgdaSpace{}%
\AgdaPrimitive{Set}\AgdaSpace{}%
\AgdaKeyword{where}\<%
\\
\>[2][@{}l@{\AgdaIndent{0}}]%
\>[4]\AgdaInductiveConstructor{here}\AgdaSpace{}%
\AgdaSymbol{:}\AgdaSpace{}%
\AgdaInductiveConstructor{zero}\AgdaSpace{}%
\AgdaOperator{\AgdaDatatype{∷}}\AgdaSpace{}%
\AgdaGeneralizable{T}\AgdaSpace{}%
\AgdaOperator{\AgdaDatatype{∈}}\AgdaSpace{}%
\AgdaGeneralizable{Γ}\AgdaSpace{}%
\AgdaOperator{\AgdaInductiveConstructor{·:}}\AgdaSpace{}%
\AgdaGeneralizable{T}\<%
\\
\>[4]\AgdaInductiveConstructor{there}\AgdaSpace{}%
\AgdaSymbol{:}\AgdaSpace{}%
\AgdaGeneralizable{x}\AgdaSpace{}%
\AgdaOperator{\AgdaDatatype{∷}}\AgdaSpace{}%
\AgdaGeneralizable{T}\AgdaSpace{}%
\AgdaOperator{\AgdaDatatype{∈}}\AgdaSpace{}%
\AgdaGeneralizable{Γ}\AgdaSpace{}%
\AgdaSymbol{→}\AgdaSpace{}%
\AgdaInductiveConstructor{suc}\AgdaSpace{}%
\AgdaGeneralizable{x}\AgdaSpace{}%
\AgdaOperator{\AgdaDatatype{∷}}\AgdaSpace{}%
\AgdaGeneralizable{T}\AgdaSpace{}%
\AgdaOperator{\AgdaDatatype{∈}}\AgdaSpace{}%
\AgdaGeneralizable{Γ}\AgdaSpace{}%
\AgdaOperator{\AgdaInductiveConstructor{·:}}\AgdaSpace{}%
\AgdaGeneralizable{S}\<%
\\
\\[\AgdaEmptyExtraSkip]%
\>[2]\AgdaComment{--\ Typing\ judgement}\<%
\\
\>[2]\AgdaKeyword{data}\AgdaSpace{}%
\AgdaOperator{\AgdaDatatype{\AgdaUnderscore{}⊢\AgdaUnderscore{}∷\AgdaUnderscore{}}}\AgdaSpace{}%
\AgdaSymbol{:}\AgdaSpace{}%
\AgdaDatatype{Ctx}\AgdaSpace{}%
\AgdaSymbol{→}\AgdaSpace{}%
\AgdaDatatype{Term}\AgdaSpace{}%
\AgdaSymbol{→}\AgdaSpace{}%
\AgdaDatatype{Type}\AgdaSpace{}%
\AgdaSymbol{→}\AgdaSpace{}%
\AgdaPrimitive{Set}\AgdaSpace{}%
\AgdaKeyword{where}\<%
\\
\>[2][@{}l@{\AgdaIndent{0}}]%
\>[4]\AgdaInductiveConstructor{⊢var}\AgdaSpace{}%
\AgdaSymbol{:}\AgdaSpace{}%
\AgdaGeneralizable{x}\AgdaSpace{}%
\AgdaOperator{\AgdaDatatype{∷}}\AgdaSpace{}%
\AgdaGeneralizable{T}\AgdaSpace{}%
\AgdaOperator{\AgdaDatatype{∈}}\AgdaSpace{}%
\AgdaGeneralizable{Γ}\AgdaSpace{}%
\AgdaSymbol{→}\AgdaSpace{}%
\AgdaGeneralizable{Γ}\AgdaSpace{}%
\AgdaOperator{\AgdaDatatype{⊢}}\AgdaSpace{}%
\AgdaInductiveConstructor{var}\AgdaSpace{}%
\AgdaGeneralizable{x}\AgdaSpace{}%
\AgdaOperator{\AgdaDatatype{∷}}\AgdaSpace{}%
\AgdaGeneralizable{T}\<%
\\
\>[4]\AgdaInductiveConstructor{⊢abs}\AgdaSpace{}%
\AgdaSymbol{:}\AgdaSpace{}%
\AgdaGeneralizable{Γ}\AgdaSpace{}%
\AgdaOperator{\AgdaInductiveConstructor{·:}}\AgdaSpace{}%
\AgdaGeneralizable{S}\AgdaSpace{}%
\AgdaOperator{\AgdaDatatype{⊢}}\AgdaSpace{}%
\AgdaGeneralizable{t}\AgdaSpace{}%
\AgdaOperator{\AgdaDatatype{∷}}\AgdaSpace{}%
\AgdaGeneralizable{T}\AgdaSpace{}%
\AgdaSymbol{→}\AgdaSpace{}%
\AgdaGeneralizable{Γ}\AgdaSpace{}%
\AgdaOperator{\AgdaDatatype{⊢}}\AgdaSpace{}%
\AgdaOperator{\AgdaInductiveConstructor{ƛ}}\AgdaSpace{}%
\AgdaGeneralizable{t}\AgdaSpace{}%
\AgdaOperator{\AgdaDatatype{∷}}\AgdaSpace{}%
\AgdaGeneralizable{S}\AgdaSpace{}%
\AgdaOperator{\AgdaInductiveConstructor{⇒}}\AgdaSpace{}%
\AgdaGeneralizable{T}\<%
\\
\>[4]\AgdaInductiveConstructor{⊢app}\AgdaSpace{}%
\AgdaSymbol{:}\AgdaSpace{}%
\AgdaGeneralizable{Γ}\AgdaSpace{}%
\AgdaOperator{\AgdaDatatype{⊢}}\AgdaSpace{}%
\AgdaGeneralizable{r}\AgdaSpace{}%
\AgdaOperator{\AgdaDatatype{∷}}\AgdaSpace{}%
\AgdaGeneralizable{S}\AgdaSpace{}%
\AgdaOperator{\AgdaInductiveConstructor{⇒}}\AgdaSpace{}%
\AgdaGeneralizable{T}\AgdaSpace{}%
\AgdaSymbol{→}\AgdaSpace{}%
\AgdaGeneralizable{Γ}\AgdaSpace{}%
\AgdaOperator{\AgdaDatatype{⊢}}\AgdaSpace{}%
\AgdaGeneralizable{s}\AgdaSpace{}%
\AgdaOperator{\AgdaDatatype{∷}}\AgdaSpace{}%
\AgdaGeneralizable{S}\AgdaSpace{}%
\AgdaSymbol{→}\AgdaSpace{}%
\AgdaGeneralizable{Γ}\AgdaSpace{}%
\AgdaOperator{\AgdaDatatype{⊢}}\AgdaSpace{}%
\AgdaGeneralizable{r}\AgdaSpace{}%
\AgdaOperator{\AgdaInductiveConstructor{·}}\AgdaSpace{}%
\AgdaGeneralizable{s}\AgdaSpace{}%
\AgdaOperator{\AgdaDatatype{∷}}\AgdaSpace{}%
\AgdaGeneralizable{T}\<%
\end{code}

\subsection{Natural semantics (Figure \ref{fig:semantics-norm})}
\begin{code}%
\>[2]\AgdaKeyword{mutual}\<%
\\
\>[2][@{}l@{\AgdaIndent{0}}]%
\>[4]\AgdaComment{--\ Environments}\<%
\\
\>[4]\AgdaFunction{Env}\AgdaSpace{}%
\AgdaSymbol{=}\AgdaSpace{}%
\AgdaDatatype{ℕ}\AgdaSpace{}%
\AgdaSymbol{→}\AgdaSpace{}%
\AgdaDatatype{Domain}\<%
\\
\\[\AgdaEmptyExtraSkip]%
\>[4]\AgdaComment{--\ Domain}\<%
\\
\>[4]\AgdaKeyword{data}\AgdaSpace{}%
\AgdaDatatype{Domain}\AgdaSpace{}%
\AgdaSymbol{:}\AgdaSpace{}%
\AgdaPrimitive{Set}\AgdaSpace{}%
\AgdaKeyword{where}\<%
\\
\>[4][@{}l@{\AgdaIndent{0}}]%
\>[6]\AgdaOperator{\AgdaInductiveConstructor{⟨ƛ\AgdaUnderscore{}⟩\AgdaUnderscore{}}}\AgdaSpace{}%
\AgdaSymbol{:}\AgdaSpace{}%
\AgdaDatatype{Term}\AgdaSpace{}%
\AgdaSymbol{→}\AgdaSpace{}%
\AgdaFunction{Env}\AgdaSpace{}%
\AgdaSymbol{→}\AgdaSpace{}%
\AgdaDatatype{Domain}\<%
\\
\>[6]\AgdaOperator{\AgdaInductiveConstructor{`\AgdaUnderscore{}}}\AgdaSpace{}%
\AgdaSymbol{:}\AgdaSpace{}%
\AgdaDatatype{Domainⁿᵉ}\AgdaSpace{}%
\AgdaSymbol{→}\AgdaSpace{}%
\AgdaDatatype{Domain}\<%
\\
\\[\AgdaEmptyExtraSkip]%
\>[4]\AgdaComment{--\ Neutral\ domain}\<%
\\
\>[4]\AgdaKeyword{data}\AgdaSpace{}%
\AgdaDatatype{Domainⁿᵉ}\AgdaSpace{}%
\AgdaSymbol{:}\AgdaSpace{}%
\AgdaPrimitive{Set}\AgdaSpace{}%
\AgdaKeyword{where}\<%
\\
\>[4][@{}l@{\AgdaIndent{0}}]%
\>[6]\AgdaInductiveConstructor{lvl}\AgdaSpace{}%
\AgdaSymbol{:}\AgdaSpace{}%
\AgdaSymbol{(}\AgdaBound{k}\AgdaSpace{}%
\AgdaSymbol{:}\AgdaSpace{}%
\AgdaDatatype{ℕ}\AgdaSymbol{)}\AgdaSpace{}%
\AgdaSymbol{→}\AgdaSpace{}%
\AgdaDatatype{Domainⁿᵉ}\<%
\\
\>[6]\AgdaOperator{\AgdaInductiveConstructor{\AgdaUnderscore{}·\AgdaUnderscore{}}}\AgdaSpace{}%
\AgdaSymbol{:}\AgdaSpace{}%
\AgdaDatatype{Domainⁿᵉ}\AgdaSpace{}%
\AgdaSymbol{→}\AgdaSpace{}%
\AgdaDatatype{Domain}\AgdaSpace{}%
\AgdaSymbol{→}\AgdaSpace{}%
\AgdaDatatype{Domainⁿᵉ}\<%
\\
\\[\AgdaEmptyExtraSkip]%
\>[2]\AgdaKeyword{variable}\AgdaSpace{}%
\AgdaGeneralizable{γ}\AgdaSpace{}%
\AgdaSymbol{:}\AgdaSpace{}%
\AgdaFunction{Env}\<%
\\
\>[2]\AgdaKeyword{variable}\AgdaSpace{}%
\AgdaGeneralizable{a}\AgdaSpace{}%
\AgdaGeneralizable{b}\AgdaSpace{}%
\AgdaGeneralizable{d}\AgdaSpace{}%
\AgdaGeneralizable{f}\AgdaSpace{}%
\AgdaSymbol{:}\AgdaSpace{}%
\AgdaDatatype{Domain}\<%
\\
\>[2]\AgdaKeyword{variable}\AgdaSpace{}%
\AgdaGeneralizable{e}\AgdaSpace{}%
\AgdaSymbol{:}\AgdaSpace{}%
\AgdaDatatype{Domainⁿᵉ}\<%
\\
\>[2]\AgdaKeyword{variable}\AgdaSpace{}%
\AgdaGeneralizable{k}\AgdaSpace{}%
\AgdaSymbol{:}\AgdaSpace{}%
\AgdaDatatype{ℕ}\AgdaSpace{}%
\AgdaComment{--\ de\ Brujin\ level}\<%
\\
\\[\AgdaEmptyExtraSkip]%
\>[2]\AgdaOperator{\AgdaFunction{\AgdaUnderscore{}++\AgdaUnderscore{}}}\AgdaSpace{}%
\AgdaSymbol{:}\AgdaSpace{}%
\AgdaFunction{Env}\AgdaSpace{}%
\AgdaSymbol{→}\AgdaSpace{}%
\AgdaDatatype{Domain}\AgdaSpace{}%
\AgdaSymbol{→}\AgdaSpace{}%
\AgdaFunction{Env}\<%
\\
\>[2]\AgdaSymbol{(\AgdaUnderscore{}}\AgdaSpace{}%
\AgdaOperator{\AgdaFunction{++}}\AgdaSpace{}%
\AgdaBound{a}\AgdaSymbol{)}\AgdaSpace{}%
\AgdaInductiveConstructor{zero}%
\>[19]\AgdaSymbol{=}\AgdaSpace{}%
\AgdaBound{a}\<%
\\
\>[2]\AgdaSymbol{(}\AgdaBound{γ}\AgdaSpace{}%
\AgdaOperator{\AgdaFunction{++}}\AgdaSpace{}%
\AgdaSymbol{\AgdaUnderscore{})}\AgdaSpace{}%
\AgdaSymbol{(}\AgdaInductiveConstructor{suc}\AgdaSpace{}%
\AgdaBound{m}\AgdaSymbol{)}\AgdaSpace{}%
\AgdaSymbol{=}\AgdaSpace{}%
\AgdaBound{γ}\AgdaSpace{}%
\AgdaBound{m}\<%
\\
\\[\AgdaEmptyExtraSkip]%
\>[2]\AgdaComment{--\ Natural\ semantics}\<%
\\
\>[2]\AgdaKeyword{mutual}\<%
\\
\>[2][@{}l@{\AgdaIndent{0}}]%
\>[4]\AgdaKeyword{data}\AgdaSpace{}%
\AgdaOperator{\AgdaDatatype{\AgdaUnderscore{}∣\AgdaUnderscore{}⇓\AgdaUnderscore{}}}\AgdaSpace{}%
\AgdaSymbol{:}\AgdaSpace{}%
\AgdaFunction{Env}\AgdaSpace{}%
\AgdaSymbol{→}\AgdaSpace{}%
\AgdaDatatype{Term}\AgdaSpace{}%
\AgdaSymbol{→}\AgdaSpace{}%
\AgdaDatatype{Domain}\AgdaSpace{}%
\AgdaSymbol{→}\AgdaSpace{}%
\AgdaPrimitive{Set}\AgdaSpace{}%
\AgdaKeyword{where}\<%
\\
\>[4][@{}l@{\AgdaIndent{0}}]%
\>[6]\AgdaInductiveConstructor{evalVar}\AgdaSpace{}%
\AgdaSymbol{:}\AgdaSpace{}%
\AgdaGeneralizable{γ}\AgdaSpace{}%
\AgdaOperator{\AgdaDatatype{∣}}\AgdaSpace{}%
\AgdaInductiveConstructor{var}\AgdaSpace{}%
\AgdaGeneralizable{x}\AgdaSpace{}%
\AgdaOperator{\AgdaDatatype{⇓}}\AgdaSpace{}%
\AgdaGeneralizable{γ}\AgdaSpace{}%
\AgdaGeneralizable{x}\<%
\\
\>[6]\AgdaInductiveConstructor{evalAbs}\AgdaSpace{}%
\AgdaSymbol{:}\AgdaSpace{}%
\AgdaGeneralizable{γ}\AgdaSpace{}%
\AgdaOperator{\AgdaDatatype{∣}}\AgdaSpace{}%
\AgdaOperator{\AgdaInductiveConstructor{ƛ}}\AgdaSpace{}%
\AgdaGeneralizable{t}\AgdaSpace{}%
\AgdaOperator{\AgdaDatatype{⇓}}\AgdaSpace{}%
\AgdaOperator{\AgdaInductiveConstructor{⟨ƛ}}\AgdaSpace{}%
\AgdaGeneralizable{t}\AgdaSpace{}%
\AgdaOperator{\AgdaInductiveConstructor{⟩}}\AgdaSpace{}%
\AgdaGeneralizable{γ}\<%
\\
\>[6]\AgdaInductiveConstructor{evalApp}\AgdaSpace{}%
\AgdaSymbol{:}\AgdaSpace{}%
\AgdaGeneralizable{γ}\AgdaSpace{}%
\AgdaOperator{\AgdaDatatype{∣}}\AgdaSpace{}%
\AgdaGeneralizable{r}\AgdaSpace{}%
\AgdaOperator{\AgdaDatatype{⇓}}\AgdaSpace{}%
\AgdaGeneralizable{f}\AgdaSpace{}%
\AgdaSymbol{→}\AgdaSpace{}%
\AgdaGeneralizable{γ}\AgdaSpace{}%
\AgdaOperator{\AgdaDatatype{∣}}\AgdaSpace{}%
\AgdaGeneralizable{s}\AgdaSpace{}%
\AgdaOperator{\AgdaDatatype{⇓}}\AgdaSpace{}%
\AgdaGeneralizable{a}\AgdaSpace{}%
\AgdaSymbol{→}\AgdaSpace{}%
\AgdaGeneralizable{f}\AgdaSpace{}%
\AgdaOperator{\AgdaDatatype{·}}\AgdaSpace{}%
\AgdaGeneralizable{a}\AgdaSpace{}%
\AgdaOperator{\AgdaDatatype{⇓}}\AgdaSpace{}%
\AgdaGeneralizable{b}\AgdaSpace{}%
\AgdaSymbol{→}\AgdaSpace{}%
\AgdaGeneralizable{γ}\AgdaSpace{}%
\AgdaOperator{\AgdaDatatype{∣}}\AgdaSpace{}%
\AgdaGeneralizable{r}\AgdaSpace{}%
\AgdaOperator{\AgdaInductiveConstructor{·}}\AgdaSpace{}%
\AgdaGeneralizable{s}\AgdaSpace{}%
\AgdaOperator{\AgdaDatatype{⇓}}\AgdaSpace{}%
\AgdaGeneralizable{b}\<%
\\
\\[\AgdaEmptyExtraSkip]%
\>[4]\AgdaKeyword{data}\AgdaSpace{}%
\AgdaOperator{\AgdaDatatype{\AgdaUnderscore{}·\AgdaUnderscore{}⇓\AgdaUnderscore{}}}\AgdaSpace{}%
\AgdaSymbol{:}\AgdaSpace{}%
\AgdaDatatype{Domain}\AgdaSpace{}%
\AgdaSymbol{→}\AgdaSpace{}%
\AgdaDatatype{Domain}\AgdaSpace{}%
\AgdaSymbol{→}\AgdaSpace{}%
\AgdaDatatype{Domain}\AgdaSpace{}%
\AgdaSymbol{→}\AgdaSpace{}%
\AgdaPrimitive{Set}\AgdaSpace{}%
\AgdaKeyword{where}\<%
\\
\>[4][@{}l@{\AgdaIndent{0}}]%
\>[6]\AgdaInductiveConstructor{appClosure}\AgdaSpace{}%
\AgdaSymbol{:}\AgdaSpace{}%
\AgdaGeneralizable{γ}\AgdaSpace{}%
\AgdaOperator{\AgdaFunction{++}}\AgdaSpace{}%
\AgdaGeneralizable{a}\AgdaSpace{}%
\AgdaOperator{\AgdaDatatype{∣}}\AgdaSpace{}%
\AgdaGeneralizable{t}\AgdaSpace{}%
\AgdaOperator{\AgdaDatatype{⇓}}\AgdaSpace{}%
\AgdaGeneralizable{b}\AgdaSpace{}%
\AgdaSymbol{→}\AgdaSpace{}%
\AgdaOperator{\AgdaInductiveConstructor{⟨ƛ}}\AgdaSpace{}%
\AgdaGeneralizable{t}\AgdaSpace{}%
\AgdaOperator{\AgdaInductiveConstructor{⟩}}\AgdaSpace{}%
\AgdaGeneralizable{γ}\AgdaSpace{}%
\AgdaOperator{\AgdaDatatype{·}}\AgdaSpace{}%
\AgdaGeneralizable{a}\AgdaSpace{}%
\AgdaOperator{\AgdaDatatype{⇓}}\AgdaSpace{}%
\AgdaGeneralizable{b}\<%
\\
\>[6]\AgdaInductiveConstructor{appNeutral}\AgdaSpace{}%
\AgdaSymbol{:}\AgdaSpace{}%
\AgdaOperator{\AgdaInductiveConstructor{`}}\AgdaSpace{}%
\AgdaGeneralizable{e}\AgdaSpace{}%
\AgdaOperator{\AgdaDatatype{·}}\AgdaSpace{}%
\AgdaGeneralizable{d}\AgdaSpace{}%
\AgdaOperator{\AgdaDatatype{⇓}}\AgdaSpace{}%
\AgdaOperator{\AgdaInductiveConstructor{`}}\AgdaSpace{}%
\AgdaSymbol{(}\AgdaGeneralizable{e}\AgdaSpace{}%
\AgdaOperator{\AgdaInductiveConstructor{·}}\AgdaSpace{}%
\AgdaGeneralizable{d}\AgdaSymbol{)}\<%
\end{code}

\subsection{Reading a normal form back from an evaluated term (Figure \ref{fig:readback-nf})}

\begin{code}%
\>[2]\AgdaKeyword{variable}\AgdaSpace{}%
\AgdaGeneralizable{n}\AgdaSpace{}%
\AgdaSymbol{:}\AgdaSpace{}%
\AgdaDatatype{ℕ}\AgdaSpace{}%
\AgdaComment{--\ Scope}\<%
\\
\\[\AgdaEmptyExtraSkip]%
\>[2]\AgdaComment{--\ Converting\ a\ de\ Brujin\ level\ to\ a\ de\ Brujin\ index}\<%
\\
\>[2]\AgdaFunction{lvl→idx}\AgdaSpace{}%
\AgdaSymbol{:}\AgdaSpace{}%
\AgdaDatatype{ℕ}\AgdaSpace{}%
\AgdaSymbol{→}\AgdaSpace{}%
\AgdaDatatype{ℕ}\AgdaSpace{}%
\AgdaSymbol{→}\AgdaSpace{}%
\AgdaDatatype{ℕ}\<%
\\
\>[2]\AgdaFunction{lvl→idx}\AgdaSpace{}%
\AgdaBound{k}\AgdaSpace{}%
\AgdaBound{n}\AgdaSpace{}%
\AgdaSymbol{=}\AgdaSpace{}%
\AgdaBound{n}\AgdaSpace{}%
\AgdaOperator{\AgdaPrimitive{-}}\AgdaSpace{}%
\AgdaInductiveConstructor{suc}\AgdaSpace{}%
\AgdaBound{k}\<%
\\
\\[\AgdaEmptyExtraSkip]%
\>[2]\AgdaKeyword{mutual}\<%
\\
\>[2][@{}l@{\AgdaIndent{0}}]%
\>[4]\AgdaComment{--\ Reading\ back\ a\ normal\ term}\<%
\\
\>[4]\AgdaKeyword{data}\AgdaSpace{}%
\AgdaOperator{\AgdaDatatype{\AgdaUnderscore{}∣\AgdaUnderscore{}⇑\AgdaUnderscore{}}}\AgdaSpace{}%
\AgdaSymbol{:}\AgdaSpace{}%
\AgdaDatatype{ℕ}\AgdaSpace{}%
\AgdaSymbol{→}\AgdaSpace{}%
\AgdaDatatype{Domain}\AgdaSpace{}%
\AgdaSymbol{→}\AgdaSpace{}%
\AgdaDatatype{Term}\AgdaSpace{}%
\AgdaSymbol{→}\AgdaSpace{}%
\AgdaPrimitive{Set}\AgdaSpace{}%
\AgdaKeyword{where}\<%
\\
\>[4][@{}l@{\AgdaIndent{0}}]%
\>[6]\AgdaInductiveConstructor{⇑closure}\AgdaSpace{}%
\AgdaSymbol{:}\<%
\\
\>[6][@{}l@{\AgdaIndent{0}}]%
\>[10]\AgdaGeneralizable{γ}\AgdaSpace{}%
\AgdaOperator{\AgdaFunction{++}}\AgdaSpace{}%
\AgdaOperator{\AgdaInductiveConstructor{`}}\AgdaSpace{}%
\AgdaInductiveConstructor{lvl}\AgdaSpace{}%
\AgdaGeneralizable{n}\AgdaSpace{}%
\AgdaOperator{\AgdaDatatype{∣}}\AgdaSpace{}%
\AgdaGeneralizable{t}\AgdaSpace{}%
\AgdaOperator{\AgdaDatatype{⇓}}\AgdaSpace{}%
\AgdaGeneralizable{a}\<%
\\
\>[6][@{}l@{\AgdaIndent{0}}]%
\>[8]\AgdaSymbol{→}\AgdaSpace{}%
\AgdaGeneralizable{n}\AgdaSpace{}%
\AgdaOperator{\AgdaDatatype{∣}}\AgdaSpace{}%
\AgdaGeneralizable{a}\AgdaSpace{}%
\AgdaOperator{\AgdaDatatype{⇑}}\AgdaSpace{}%
\AgdaGeneralizable{v}\<%
\\
\>[8]\AgdaSymbol{→}\AgdaSpace{}%
\AgdaGeneralizable{n}\AgdaSpace{}%
\AgdaOperator{\AgdaDatatype{∣}}\AgdaSpace{}%
\AgdaOperator{\AgdaInductiveConstructor{⟨ƛ}}\AgdaSpace{}%
\AgdaGeneralizable{t}\AgdaSpace{}%
\AgdaOperator{\AgdaInductiveConstructor{⟩}}\AgdaSpace{}%
\AgdaGeneralizable{γ}\AgdaSpace{}%
\AgdaOperator{\AgdaDatatype{⇑}}\AgdaSpace{}%
\AgdaOperator{\AgdaInductiveConstructor{ƛ}}\AgdaSpace{}%
\AgdaGeneralizable{v}\<%
\\
\\[\AgdaEmptyExtraSkip]%
\>[6]\AgdaInductiveConstructor{⇑neutral}\AgdaSpace{}%
\AgdaSymbol{:}\AgdaSpace{}%
\AgdaGeneralizable{n}\AgdaSpace{}%
\AgdaOperator{\AgdaDatatype{∣}}\AgdaSpace{}%
\AgdaGeneralizable{e}\AgdaSpace{}%
\AgdaOperator{\AgdaDatatype{⇑ⁿᵉ}}\AgdaSpace{}%
\AgdaGeneralizable{u}\AgdaSpace{}%
\AgdaSymbol{→}\AgdaSpace{}%
\AgdaGeneralizable{n}\AgdaSpace{}%
\AgdaOperator{\AgdaDatatype{∣}}\AgdaSpace{}%
\AgdaOperator{\AgdaInductiveConstructor{`}}\AgdaSpace{}%
\AgdaGeneralizable{e}\AgdaSpace{}%
\AgdaOperator{\AgdaDatatype{⇑}}\AgdaSpace{}%
\AgdaGeneralizable{u}\<%
\\
\\[\AgdaEmptyExtraSkip]%
\>[4]\AgdaComment{--\ Reading\ back\ a\ neutral\ term}\<%
\\
\>[4]\AgdaKeyword{data}\AgdaSpace{}%
\AgdaOperator{\AgdaDatatype{\AgdaUnderscore{}∣\AgdaUnderscore{}⇑ⁿᵉ\AgdaUnderscore{}}}\AgdaSpace{}%
\AgdaSymbol{:}\AgdaSpace{}%
\AgdaDatatype{ℕ}\AgdaSpace{}%
\AgdaSymbol{→}\AgdaSpace{}%
\AgdaDatatype{Domainⁿᵉ}\AgdaSpace{}%
\AgdaSymbol{→}\AgdaSpace{}%
\AgdaDatatype{Term}\AgdaSpace{}%
\AgdaSymbol{→}\AgdaSpace{}%
\AgdaPrimitive{Set}\AgdaSpace{}%
\AgdaKeyword{where}\<%
\\
\>[4][@{}l@{\AgdaIndent{0}}]%
\>[6]\AgdaInductiveConstructor{⇑lvl}\AgdaSpace{}%
\AgdaSymbol{:}\AgdaSpace{}%
\AgdaGeneralizable{n}\AgdaSpace{}%
\AgdaOperator{\AgdaDatatype{∣}}\AgdaSpace{}%
\AgdaInductiveConstructor{lvl}\AgdaSpace{}%
\AgdaGeneralizable{k}\AgdaSpace{}%
\AgdaOperator{\AgdaDatatype{⇑ⁿᵉ}}\AgdaSpace{}%
\AgdaInductiveConstructor{var}\AgdaSpace{}%
\AgdaSymbol{(}\AgdaFunction{lvl→idx}\AgdaSpace{}%
\AgdaGeneralizable{k}\AgdaSpace{}%
\AgdaGeneralizable{n}\AgdaSymbol{)}\<%
\\
\\[\AgdaEmptyExtraSkip]%
\>[6]\AgdaInductiveConstructor{⇑app}\AgdaSpace{}%
\AgdaSymbol{:}\<%
\\
\>[6][@{}l@{\AgdaIndent{0}}]%
\>[10]\AgdaGeneralizable{n}\AgdaSpace{}%
\AgdaOperator{\AgdaDatatype{∣}}\AgdaSpace{}%
\AgdaGeneralizable{e}\AgdaSpace{}%
\AgdaOperator{\AgdaDatatype{⇑ⁿᵉ}}\AgdaSpace{}%
\AgdaGeneralizable{u}\<%
\\
\>[6][@{}l@{\AgdaIndent{0}}]%
\>[8]\AgdaSymbol{→}\AgdaSpace{}%
\AgdaGeneralizable{n}\AgdaSpace{}%
\AgdaOperator{\AgdaDatatype{∣}}\AgdaSpace{}%
\AgdaGeneralizable{d}\AgdaSpace{}%
\AgdaOperator{\AgdaDatatype{⇑}}\AgdaSpace{}%
\AgdaGeneralizable{v}\<%
\\
\>[8]\AgdaSymbol{→}\AgdaSpace{}%
\AgdaGeneralizable{n}\AgdaSpace{}%
\AgdaOperator{\AgdaDatatype{∣}}\AgdaSpace{}%
\AgdaGeneralizable{e}\AgdaSpace{}%
\AgdaOperator{\AgdaInductiveConstructor{·}}\AgdaSpace{}%
\AgdaGeneralizable{d}\AgdaSpace{}%
\AgdaOperator{\AgdaDatatype{⇑ⁿᵉ}}\AgdaSpace{}%
\AgdaGeneralizable{u}\AgdaSpace{}%
\AgdaOperator{\AgdaInductiveConstructor{·}}\AgdaSpace{}%
\AgdaGeneralizable{v}\<%
\end{code}

\subsection{Asking more of semantic types with candidate spaces (Figure \ref{fig:candidate-space})}
\begin{code}%
\>[2]\AgdaFunction{LogPred}\AgdaSpace{}%
\AgdaSymbol{=}\AgdaSpace{}%
\AgdaDatatype{Domain}\AgdaSpace{}%
\AgdaSymbol{→}\AgdaSpace{}%
\AgdaPrimitive{Set}\<%
\\
\\[\AgdaEmptyExtraSkip]%
\>[2]\AgdaComment{--\ Bottom\ of\ candidate\ space}\<%
\\
\>[2]\AgdaComment{--\ (neutral\ term\ can\ be\ read\ back)}\<%
\\
\>[2]\AgdaFunction{𝔹}\AgdaSpace{}%
\AgdaSymbol{:}\AgdaSpace{}%
\AgdaFunction{LogPred}\<%
\\
\>[2]\AgdaFunction{𝔹}\AgdaSpace{}%
\AgdaSymbol{(}\AgdaOperator{\AgdaInductiveConstructor{`}}\AgdaSpace{}%
\AgdaBound{e}\AgdaSymbol{)}\AgdaSpace{}%
\AgdaSymbol{=}\AgdaSpace{}%
\AgdaSymbol{∀}\AgdaSpace{}%
\AgdaBound{n}\AgdaSpace{}%
\AgdaSymbol{→}\AgdaSpace{}%
\AgdaFunction{∃[}\AgdaSpace{}%
\AgdaBound{u}\AgdaSpace{}%
\AgdaFunction{]}\AgdaSpace{}%
\AgdaBound{n}\AgdaSpace{}%
\AgdaOperator{\AgdaDatatype{∣}}\AgdaSpace{}%
\AgdaBound{e}\AgdaSpace{}%
\AgdaOperator{\AgdaDatatype{⇑ⁿᵉ}}\AgdaSpace{}%
\AgdaBound{u}\<%
\\
\>[2]\AgdaCatchallClause{\AgdaFunction{𝔹}}\AgdaSpace{}%
\AgdaCatchallClause{\AgdaSymbol{\AgdaUnderscore{}}}\AgdaSpace{}%
\AgdaSymbol{=}\AgdaSpace{}%
\AgdaDatatype{⊥}\<%
\\
\\[\AgdaEmptyExtraSkip]%
\>[2]\AgdaComment{--\ Top\ of\ candidate\ space}\<%
\\
\>[2]\AgdaComment{--\ (normal\ term\ can\ be\ read\ back)}\<%
\\
\>[2]\AgdaFunction{𝕋}\AgdaSpace{}%
\AgdaSymbol{:}\AgdaSpace{}%
\AgdaFunction{LogPred}\<%
\\
\>[2]\AgdaFunction{𝕋}\AgdaSpace{}%
\AgdaBound{d}\AgdaSpace{}%
\AgdaSymbol{=}\AgdaSpace{}%
\AgdaSymbol{∀}\AgdaSpace{}%
\AgdaBound{n}\AgdaSpace{}%
\AgdaSymbol{→}\AgdaSpace{}%
\AgdaFunction{∃[}\AgdaSpace{}%
\AgdaBound{v}\AgdaSpace{}%
\AgdaFunction{]}\AgdaSpace{}%
\AgdaBound{n}\AgdaSpace{}%
\AgdaOperator{\AgdaDatatype{∣}}\AgdaSpace{}%
\AgdaBound{d}\AgdaSpace{}%
\AgdaOperator{\AgdaDatatype{⇑}}\AgdaSpace{}%
\AgdaBound{v}\<%
\\
\\[\AgdaEmptyExtraSkip]%
\>[2]\AgdaComment{--\ Bottom\ of\ space\ is\ a\ subset\ of\ top}\<%
\\
\>[2]\AgdaFunction{𝔹⊆𝕋}\AgdaSpace{}%
\AgdaSymbol{:}\AgdaSpace{}%
\AgdaGeneralizable{d}\AgdaSpace{}%
\AgdaOperator{\AgdaFunction{∈}}\AgdaSpace{}%
\AgdaFunction{𝔹}\AgdaSpace{}%
\AgdaSymbol{→}\AgdaSpace{}%
\AgdaGeneralizable{d}\AgdaSpace{}%
\AgdaOperator{\AgdaFunction{∈}}\AgdaSpace{}%
\AgdaFunction{𝕋}\<%
\\
\>[2]\AgdaFunction{𝔹⊆𝕋}\AgdaSpace{}%
\AgdaSymbol{\{}\AgdaOperator{\AgdaInductiveConstructor{`}}\AgdaSpace{}%
\AgdaBound{e}\AgdaSymbol{\}}\AgdaSpace{}%
\AgdaBound{eb}\AgdaSpace{}%
\AgdaBound{n}\<%
\\
\>[2][@{}l@{\AgdaIndent{0}}]%
\>[4]\AgdaKeyword{with}\AgdaSpace{}%
\AgdaBound{eb}\AgdaSpace{}%
\AgdaBound{n}\<%
\\
\>[2]\AgdaSymbol{...}\AgdaSpace{}%
\AgdaSymbol{|}\AgdaSpace{}%
\AgdaBound{u}\AgdaSpace{}%
\AgdaOperator{\AgdaInductiveConstructor{,}}\AgdaSpace{}%
\AgdaBound{e⇑u}\AgdaSpace{}%
\AgdaSymbol{=}\<%
\\
\>[2][@{}l@{\AgdaIndent{0}}]%
\>[4]\AgdaBound{u}\AgdaSpace{}%
\AgdaOperator{\AgdaInductiveConstructor{,}}\AgdaSpace{}%
\AgdaInductiveConstructor{⇑neutral}\AgdaSpace{}%
\AgdaBound{e⇑u}\<%
\end{code}

\subsection{Proof by logical relations}

\begin{code}%
\>[2]\AgdaOperator{\AgdaFunction{\AgdaUnderscore{}⟶\AgdaUnderscore{}}}\AgdaSpace{}%
\AgdaSymbol{:}\AgdaSpace{}%
\AgdaFunction{LogPred}\AgdaSpace{}%
\AgdaSymbol{→}\AgdaSpace{}%
\AgdaFunction{LogPred}\AgdaSpace{}%
\AgdaSymbol{→}\AgdaSpace{}%
\AgdaFunction{LogPred}\<%
\\
\>[2]\AgdaSymbol{(}\AgdaBound{A}\AgdaSpace{}%
\AgdaOperator{\AgdaFunction{⟶}}\AgdaSpace{}%
\AgdaBound{B}\AgdaSymbol{)}\AgdaSpace{}%
\AgdaBound{f}\AgdaSpace{}%
\AgdaSymbol{=}\AgdaSpace{}%
\AgdaSymbol{∀}\AgdaSpace{}%
\AgdaSymbol{\{}\AgdaBound{a}\AgdaSymbol{\}}\AgdaSpace{}%
\AgdaSymbol{→}\AgdaSpace{}%
\AgdaBound{a}\AgdaSpace{}%
\AgdaOperator{\AgdaFunction{∈}}\AgdaSpace{}%
\AgdaBound{A}\AgdaSpace{}%
\AgdaSymbol{→}\AgdaSpace{}%
\AgdaFunction{∃[}\AgdaSpace{}%
\AgdaBound{b}\AgdaSpace{}%
\AgdaFunction{]}\AgdaSpace{}%
\AgdaBound{f}\AgdaSpace{}%
\AgdaOperator{\AgdaDatatype{·}}\AgdaSpace{}%
\AgdaBound{a}\AgdaSpace{}%
\AgdaOperator{\AgdaDatatype{⇓}}\AgdaSpace{}%
\AgdaBound{b}\AgdaSpace{}%
\AgdaOperator{\AgdaFunction{×}}\AgdaSpace{}%
\AgdaBound{b}\AgdaSpace{}%
\AgdaOperator{\AgdaFunction{∈}}\AgdaSpace{}%
\AgdaBound{B}\<%
\\
\\[\AgdaEmptyExtraSkip]%
\>[2]\AgdaOperator{\AgdaFunction{⟦\AgdaUnderscore{}⟧}}\AgdaSpace{}%
\AgdaSymbol{:}\AgdaSpace{}%
\AgdaDatatype{Type}\AgdaSpace{}%
\AgdaSymbol{→}\AgdaSpace{}%
\AgdaFunction{LogPred}\<%
\\
\>[2]\AgdaOperator{\AgdaFunction{⟦}}\AgdaSpace{}%
\AgdaInductiveConstructor{base}\AgdaSpace{}%
\AgdaOperator{\AgdaFunction{⟧}}\AgdaSpace{}%
\AgdaSymbol{=}\AgdaSpace{}%
\AgdaFunction{𝔹}\<%
\\
\>[2]\AgdaOperator{\AgdaFunction{⟦}}\AgdaSpace{}%
\AgdaBound{S}\AgdaSpace{}%
\AgdaOperator{\AgdaInductiveConstructor{⇒}}\AgdaSpace{}%
\AgdaBound{T}\AgdaSpace{}%
\AgdaOperator{\AgdaFunction{⟧}}\AgdaSpace{}%
\AgdaSymbol{=}\AgdaSpace{}%
\AgdaOperator{\AgdaFunction{⟦}}\AgdaSpace{}%
\AgdaBound{S}\AgdaSpace{}%
\AgdaOperator{\AgdaFunction{⟧}}\AgdaSpace{}%
\AgdaOperator{\AgdaFunction{⟶}}\AgdaSpace{}%
\AgdaOperator{\AgdaFunction{⟦}}\AgdaSpace{}%
\AgdaBound{T}\AgdaSpace{}%
\AgdaOperator{\AgdaFunction{⟧}}\<%
\\
\>[2]\AgdaOperator{\AgdaFunction{\AgdaUnderscore{}⊨\AgdaUnderscore{}}}\AgdaSpace{}%
\AgdaSymbol{:}\AgdaSpace{}%
\AgdaDatatype{Ctx}\AgdaSpace{}%
\AgdaSymbol{→}\AgdaSpace{}%
\AgdaFunction{Env}\AgdaSpace{}%
\AgdaSymbol{→}\AgdaSpace{}%
\AgdaPrimitive{Set}\<%
\\
\\[\AgdaEmptyExtraSkip]%
\>[2]\AgdaBound{Γ}\AgdaSpace{}%
\AgdaOperator{\AgdaFunction{⊨}}\AgdaSpace{}%
\AgdaBound{γ}\AgdaSpace{}%
\AgdaSymbol{=}\AgdaSpace{}%
\AgdaSymbol{∀}\AgdaSpace{}%
\AgdaSymbol{\{}\AgdaBound{x}\AgdaSymbol{\}}\AgdaSpace{}%
\AgdaSymbol{\{}\AgdaBound{T}\AgdaSymbol{\}}\AgdaSpace{}%
\AgdaSymbol{→}\AgdaSpace{}%
\AgdaBound{x}\AgdaSpace{}%
\AgdaOperator{\AgdaDatatype{∷}}\AgdaSpace{}%
\AgdaBound{T}\AgdaSpace{}%
\AgdaOperator{\AgdaDatatype{∈}}\AgdaSpace{}%
\AgdaBound{Γ}\AgdaSpace{}%
\AgdaSymbol{→}\AgdaSpace{}%
\AgdaBound{γ}\AgdaSpace{}%
\AgdaBound{x}\AgdaSpace{}%
\AgdaOperator{\AgdaFunction{∈}}\AgdaSpace{}%
\AgdaOperator{\AgdaFunction{⟦}}\AgdaSpace{}%
\AgdaBound{T}\AgdaSpace{}%
\AgdaOperator{\AgdaFunction{⟧}}\<%
\\
\\[\AgdaEmptyExtraSkip]%
\>[2]\AgdaOperator{\AgdaFunction{\AgdaUnderscore{}⊨\AgdaUnderscore{}∷\AgdaUnderscore{}}}\AgdaSpace{}%
\AgdaSymbol{:}\AgdaSpace{}%
\AgdaDatatype{Ctx}\AgdaSpace{}%
\AgdaSymbol{→}\AgdaSpace{}%
\AgdaDatatype{Term}\AgdaSpace{}%
\AgdaSymbol{→}\AgdaSpace{}%
\AgdaDatatype{Type}\AgdaSpace{}%
\AgdaSymbol{→}\AgdaSpace{}%
\AgdaPrimitive{Set}\<%
\\
\>[2]\AgdaBound{Γ}\AgdaSpace{}%
\AgdaOperator{\AgdaFunction{⊨}}\AgdaSpace{}%
\AgdaBound{t}\AgdaSpace{}%
\AgdaOperator{\AgdaFunction{∷}}\AgdaSpace{}%
\AgdaBound{T}\AgdaSpace{}%
\AgdaSymbol{=}\AgdaSpace{}%
\AgdaSymbol{∀}\AgdaSpace{}%
\AgdaSymbol{\{}\AgdaBound{γ}\AgdaSymbol{\}}\AgdaSpace{}%
\AgdaSymbol{→}\AgdaSpace{}%
\AgdaBound{Γ}\AgdaSpace{}%
\AgdaOperator{\AgdaFunction{⊨}}\AgdaSpace{}%
\AgdaBound{γ}\AgdaSpace{}%
\AgdaSymbol{→}\AgdaSpace{}%
\AgdaFunction{∃[}\AgdaSpace{}%
\AgdaBound{a}\AgdaSpace{}%
\AgdaFunction{]}\AgdaSpace{}%
\AgdaBound{γ}\AgdaSpace{}%
\AgdaOperator{\AgdaDatatype{∣}}\AgdaSpace{}%
\AgdaBound{t}\AgdaSpace{}%
\AgdaOperator{\AgdaDatatype{⇓}}\AgdaSpace{}%
\AgdaBound{a}\AgdaSpace{}%
\AgdaOperator{\AgdaFunction{×}}\AgdaSpace{}%
\AgdaBound{a}\AgdaSpace{}%
\AgdaOperator{\AgdaFunction{∈}}\AgdaSpace{}%
\AgdaOperator{\AgdaFunction{⟦}}\AgdaSpace{}%
\AgdaBound{T}\AgdaSpace{}%
\AgdaOperator{\AgdaFunction{⟧}}\<%
\\
\\[\AgdaEmptyExtraSkip]%
\>[2]\AgdaOperator{\AgdaFunction{\AgdaUnderscore{}\textasciicircum{}\AgdaUnderscore{}}}\AgdaSpace{}%
\AgdaSymbol{:}\AgdaSpace{}%
\AgdaGeneralizable{Γ}\AgdaSpace{}%
\AgdaOperator{\AgdaFunction{⊨}}\AgdaSpace{}%
\AgdaGeneralizable{γ}\AgdaSpace{}%
\AgdaSymbol{→}\AgdaSpace{}%
\AgdaGeneralizable{a}\AgdaSpace{}%
\AgdaOperator{\AgdaFunction{∈}}\AgdaSpace{}%
\AgdaOperator{\AgdaFunction{⟦}}\AgdaSpace{}%
\AgdaGeneralizable{S}\AgdaSpace{}%
\AgdaOperator{\AgdaFunction{⟧}}\AgdaSpace{}%
\AgdaSymbol{→}\AgdaSpace{}%
\AgdaGeneralizable{Γ}\AgdaSpace{}%
\AgdaOperator{\AgdaInductiveConstructor{·:}}\AgdaSpace{}%
\AgdaGeneralizable{S}\AgdaSpace{}%
\AgdaOperator{\AgdaFunction{⊨}}\AgdaSpace{}%
\AgdaGeneralizable{γ}\AgdaSpace{}%
\AgdaOperator{\AgdaFunction{++}}\AgdaSpace{}%
\AgdaGeneralizable{a}\<%
\\
\>[2]\AgdaOperator{\AgdaFunction{\AgdaUnderscore{}\textasciicircum{}\AgdaUnderscore{}}}\AgdaSpace{}%
\AgdaSymbol{\AgdaUnderscore{}}\AgdaSpace{}%
\AgdaBound{sa}\AgdaSpace{}%
\AgdaInductiveConstructor{here}\AgdaSpace{}%
\AgdaSymbol{=}\AgdaSpace{}%
\AgdaBound{sa}\<%
\\
\>[2]\AgdaOperator{\AgdaFunction{\AgdaUnderscore{}\textasciicircum{}\AgdaUnderscore{}}}\AgdaSpace{}%
\AgdaBound{⊨γ}\AgdaSpace{}%
\AgdaSymbol{\AgdaUnderscore{}}\AgdaSpace{}%
\AgdaSymbol{(}\AgdaInductiveConstructor{there}\AgdaSpace{}%
\AgdaBound{pf}\AgdaSymbol{)}\AgdaSpace{}%
\AgdaSymbol{=}\AgdaSpace{}%
\AgdaBound{⊨γ}\AgdaSpace{}%
\AgdaBound{pf}\<%
\\
\\[\AgdaEmptyExtraSkip]%
\>[2]\AgdaFunction{fundamental-lemma}\AgdaSpace{}%
\AgdaSymbol{:}\AgdaSpace{}%
\AgdaGeneralizable{Γ}\AgdaSpace{}%
\AgdaOperator{\AgdaDatatype{⊢}}\AgdaSpace{}%
\AgdaGeneralizable{t}\AgdaSpace{}%
\AgdaOperator{\AgdaDatatype{∷}}\AgdaSpace{}%
\AgdaGeneralizable{T}\AgdaSpace{}%
\AgdaSymbol{→}\AgdaSpace{}%
\AgdaGeneralizable{Γ}\AgdaSpace{}%
\AgdaOperator{\AgdaFunction{⊨}}\AgdaSpace{}%
\AgdaGeneralizable{t}\AgdaSpace{}%
\AgdaOperator{\AgdaFunction{∷}}\AgdaSpace{}%
\AgdaGeneralizable{T}\<%
\\
\>[2]\AgdaFunction{fundamental-lemma}\AgdaSpace{}%
\AgdaSymbol{(}\AgdaInductiveConstructor{⊢var}\AgdaSpace{}%
\AgdaSymbol{\{}\AgdaBound{x}\AgdaSymbol{\}}\AgdaSpace{}%
\AgdaBound{pf}\AgdaSymbol{)}\AgdaSpace{}%
\AgdaSymbol{\{}\AgdaBound{γ}\AgdaSymbol{\}}\AgdaSpace{}%
\AgdaBound{⊨γ}\AgdaSpace{}%
\AgdaSymbol{=}\<%
\\
\>[2][@{}l@{\AgdaIndent{0}}]%
\>[4]\AgdaBound{γ}\AgdaSpace{}%
\AgdaBound{x}\AgdaSpace{}%
\AgdaOperator{\AgdaInductiveConstructor{,}}\AgdaSpace{}%
\AgdaInductiveConstructor{evalVar}\AgdaSpace{}%
\AgdaOperator{\AgdaInductiveConstructor{,}}\AgdaSpace{}%
\AgdaBound{⊨γ}\AgdaSpace{}%
\AgdaBound{pf}\<%
\\
\>[2]\AgdaFunction{fundamental-lemma}\AgdaSpace{}%
\AgdaSymbol{\{}\AgdaArgument{t}\AgdaSpace{}%
\AgdaSymbol{=}\AgdaSpace{}%
\AgdaOperator{\AgdaInductiveConstructor{ƛ}}\AgdaSpace{}%
\AgdaBound{t}\AgdaSymbol{\}}\AgdaSpace{}%
\AgdaSymbol{(}\AgdaInductiveConstructor{⊢abs}\AgdaSpace{}%
\AgdaBound{⊢t}\AgdaSymbol{)}\AgdaSpace{}%
\AgdaSymbol{\{}\AgdaBound{γ}\AgdaSymbol{\}}\AgdaSpace{}%
\AgdaBound{⊨γ}\AgdaSpace{}%
\AgdaSymbol{=}\<%
\\
\>[2][@{}l@{\AgdaIndent{0}}]%
\>[4]\AgdaOperator{\AgdaInductiveConstructor{⟨ƛ}}\AgdaSpace{}%
\AgdaBound{t}\AgdaSpace{}%
\AgdaOperator{\AgdaInductiveConstructor{⟩}}\AgdaSpace{}%
\AgdaBound{γ}\AgdaSpace{}%
\AgdaOperator{\AgdaInductiveConstructor{,}}\<%
\\
\>[4]\AgdaInductiveConstructor{evalAbs}\AgdaSpace{}%
\AgdaOperator{\AgdaInductiveConstructor{,}}\<%
\\
\>[4]\AgdaSymbol{λ}%
\>[5034I]\AgdaBound{sa}\AgdaSpace{}%
\AgdaSymbol{→}\<%
\\
\>[.][@{}l@{}]\<[5034I]%
\>[6]\AgdaKeyword{let}\AgdaSpace{}%
\AgdaBound{⊨t}\AgdaSpace{}%
\AgdaSymbol{=}\AgdaSpace{}%
\AgdaFunction{fundamental-lemma}\AgdaSpace{}%
\AgdaBound{⊢t}\AgdaSpace{}%
\AgdaKeyword{in}\<%
\\
\>[6]\AgdaKeyword{let}\AgdaSpace{}%
\AgdaSymbol{(}\AgdaBound{b}\AgdaSpace{}%
\AgdaOperator{\AgdaInductiveConstructor{,}}\AgdaSpace{}%
\AgdaBound{eval-closure}\AgdaSpace{}%
\AgdaOperator{\AgdaInductiveConstructor{,}}\AgdaSpace{}%
\AgdaBound{sb}\AgdaSymbol{)}\AgdaSpace{}%
\AgdaSymbol{=}\AgdaSpace{}%
\AgdaBound{⊨t}\AgdaSpace{}%
\AgdaSymbol{(}\AgdaBound{⊨γ}\AgdaSpace{}%
\AgdaOperator{\AgdaFunction{\textasciicircum{}}}\AgdaSpace{}%
\AgdaBound{sa}\AgdaSymbol{)}\AgdaSpace{}%
\AgdaKeyword{in}\<%
\\
\>[6]\AgdaBound{b}\AgdaSpace{}%
\AgdaOperator{\AgdaInductiveConstructor{,}}\AgdaSpace{}%
\AgdaInductiveConstructor{appClosure}\AgdaSpace{}%
\AgdaBound{eval-closure}\AgdaSpace{}%
\AgdaOperator{\AgdaInductiveConstructor{,}}\AgdaSpace{}%
\AgdaBound{sb}\<%
\\
\>[2]\AgdaFunction{fundamental-lemma}\AgdaSpace{}%
\AgdaSymbol{(}\AgdaInductiveConstructor{⊢app}\AgdaSpace{}%
\AgdaBound{⊢r}\AgdaSpace{}%
\AgdaBound{⊢s}\AgdaSymbol{)}\AgdaSpace{}%
\AgdaBound{⊨γ}\AgdaSpace{}%
\AgdaSymbol{=}\<%
\\
\>[2][@{}l@{\AgdaIndent{0}}]%
\>[4]\AgdaKeyword{let}\AgdaSpace{}%
\AgdaSymbol{(}\AgdaBound{f}\AgdaSpace{}%
\AgdaOperator{\AgdaInductiveConstructor{,}}\AgdaSpace{}%
\AgdaBound{r⇓}\AgdaSpace{}%
\AgdaOperator{\AgdaInductiveConstructor{,}}\AgdaSpace{}%
\AgdaBound{sf}\AgdaSymbol{)}\AgdaSpace{}%
\AgdaSymbol{=}\AgdaSpace{}%
\AgdaFunction{fundamental-lemma}\AgdaSpace{}%
\AgdaBound{⊢r}\AgdaSpace{}%
\AgdaBound{⊨γ}\AgdaSpace{}%
\AgdaKeyword{in}\<%
\\
\>[4]\AgdaKeyword{let}\AgdaSpace{}%
\AgdaSymbol{(}\AgdaBound{a}\AgdaSpace{}%
\AgdaOperator{\AgdaInductiveConstructor{,}}\AgdaSpace{}%
\AgdaBound{s⇓}\AgdaSpace{}%
\AgdaOperator{\AgdaInductiveConstructor{,}}\AgdaSpace{}%
\AgdaBound{sa}\AgdaSymbol{)}\AgdaSpace{}%
\AgdaSymbol{=}\AgdaSpace{}%
\AgdaFunction{fundamental-lemma}\AgdaSpace{}%
\AgdaBound{⊢s}\AgdaSpace{}%
\AgdaBound{⊨γ}\AgdaSpace{}%
\AgdaKeyword{in}\<%
\\
\>[4]\AgdaKeyword{let}\AgdaSpace{}%
\AgdaSymbol{(}\AgdaBound{b}\AgdaSpace{}%
\AgdaOperator{\AgdaInductiveConstructor{,}}\AgdaSpace{}%
\AgdaBound{app-eval}\AgdaSpace{}%
\AgdaOperator{\AgdaInductiveConstructor{,}}\AgdaSpace{}%
\AgdaBound{sb}\AgdaSymbol{)}\AgdaSpace{}%
\AgdaSymbol{=}\AgdaSpace{}%
\AgdaBound{sf}\AgdaSpace{}%
\AgdaBound{sa}\AgdaSpace{}%
\AgdaKeyword{in}\<%
\\
\>[4]\AgdaBound{b}\AgdaSpace{}%
\AgdaOperator{\AgdaInductiveConstructor{,}}\AgdaSpace{}%
\AgdaInductiveConstructor{evalApp}\AgdaSpace{}%
\AgdaBound{r⇓}\AgdaSpace{}%
\AgdaBound{s⇓}\AgdaSpace{}%
\AgdaBound{app-eval}\AgdaSpace{}%
\AgdaOperator{\AgdaInductiveConstructor{,}}\AgdaSpace{}%
\AgdaBound{sb}\<%
\end{code}

\subsection{Semantically typed domain elements inhabit the candidate space}

\begin{code}%
\>[2]\AgdaFunction{𝔹⊆𝕋⟶𝔹}\AgdaSpace{}%
\AgdaSymbol{:}\AgdaSpace{}%
\AgdaOperator{\AgdaInductiveConstructor{`}}\AgdaSpace{}%
\AgdaGeneralizable{e}\AgdaSpace{}%
\AgdaOperator{\AgdaFunction{∈}}\AgdaSpace{}%
\AgdaFunction{𝔹}\AgdaSpace{}%
\AgdaSymbol{→}\AgdaSpace{}%
\AgdaOperator{\AgdaInductiveConstructor{`}}\AgdaSpace{}%
\AgdaGeneralizable{e}\AgdaSpace{}%
\AgdaOperator{\AgdaFunction{∈}}\AgdaSpace{}%
\AgdaFunction{𝕋}\AgdaSpace{}%
\AgdaOperator{\AgdaFunction{⟶}}\AgdaSpace{}%
\AgdaFunction{𝔹}\<%
\\
\>[2]\AgdaFunction{𝔹⊆𝕋⟶𝔹}\AgdaSpace{}%
\AgdaSymbol{\{}\AgdaBound{e}\AgdaSymbol{\}}\AgdaSpace{}%
\AgdaBound{eb}\AgdaSpace{}%
\AgdaSymbol{\{}\AgdaBound{d}\AgdaSymbol{\}}\AgdaSpace{}%
\AgdaBound{dt}\AgdaSpace{}%
\AgdaSymbol{=}\<%
\\
\>[2][@{}l@{\AgdaIndent{0}}]%
\>[4]\AgdaOperator{\AgdaInductiveConstructor{`}}\AgdaSpace{}%
\AgdaSymbol{(}\AgdaBound{e}\AgdaSpace{}%
\AgdaOperator{\AgdaInductiveConstructor{·}}\AgdaSpace{}%
\AgdaBound{d}\AgdaSymbol{)}\AgdaSpace{}%
\AgdaOperator{\AgdaInductiveConstructor{,}}\<%
\\
\>[4]\AgdaInductiveConstructor{appNeutral}\AgdaSpace{}%
\AgdaOperator{\AgdaInductiveConstructor{,}}\<%
\\
\>[4]\AgdaSymbol{λ}%
\>[5120I]\AgdaBound{n}\AgdaSpace{}%
\AgdaSymbol{→}\<%
\\
\>[.][@{}l@{}]\<[5120I]%
\>[6]\AgdaKeyword{let}\AgdaSpace{}%
\AgdaSymbol{(}\AgdaBound{u}\AgdaSpace{}%
\AgdaOperator{\AgdaInductiveConstructor{,}}\AgdaSpace{}%
\AgdaBound{e⇑u}\AgdaSymbol{)}\AgdaSpace{}%
\AgdaSymbol{=}\AgdaSpace{}%
\AgdaBound{eb}\AgdaSpace{}%
\AgdaBound{n}\AgdaSpace{}%
\AgdaKeyword{in}\<%
\\
\>[6]\AgdaKeyword{let}\AgdaSpace{}%
\AgdaSymbol{(}\AgdaBound{v}\AgdaSpace{}%
\AgdaOperator{\AgdaInductiveConstructor{,}}\AgdaSpace{}%
\AgdaBound{d⇑v}\AgdaSymbol{)}\AgdaSpace{}%
\AgdaSymbol{=}\AgdaSpace{}%
\AgdaBound{dt}\AgdaSpace{}%
\AgdaBound{n}\AgdaSpace{}%
\AgdaKeyword{in}\<%
\\
\>[6]\AgdaBound{u}\AgdaSpace{}%
\AgdaOperator{\AgdaInductiveConstructor{·}}\AgdaSpace{}%
\AgdaBound{v}\AgdaSpace{}%
\AgdaOperator{\AgdaInductiveConstructor{,}}\AgdaSpace{}%
\AgdaInductiveConstructor{⇑app}\AgdaSpace{}%
\AgdaBound{e⇑u}\AgdaSpace{}%
\AgdaBound{d⇑v}\<%
\\
\\[\AgdaEmptyExtraSkip]%
\>[2]\AgdaFunction{lvl∈𝔹}\AgdaSpace{}%
\AgdaSymbol{:}\AgdaSpace{}%
\AgdaSymbol{∀}\AgdaSpace{}%
\AgdaBound{k}\AgdaSpace{}%
\AgdaSymbol{→}\AgdaSpace{}%
\AgdaOperator{\AgdaInductiveConstructor{`}}\AgdaSpace{}%
\AgdaInductiveConstructor{lvl}\AgdaSpace{}%
\AgdaBound{k}\AgdaSpace{}%
\AgdaOperator{\AgdaFunction{∈}}\AgdaSpace{}%
\AgdaFunction{𝔹}\<%
\\
\>[2]\AgdaFunction{lvl∈𝔹}\AgdaSpace{}%
\AgdaBound{k}\AgdaSpace{}%
\AgdaBound{n}\AgdaSpace{}%
\AgdaSymbol{=}\AgdaSpace{}%
\AgdaInductiveConstructor{var}\AgdaSpace{}%
\AgdaSymbol{(}\AgdaFunction{lvl→idx}\AgdaSpace{}%
\AgdaBound{k}\AgdaSpace{}%
\AgdaBound{n}\AgdaSymbol{)}\AgdaSpace{}%
\AgdaOperator{\AgdaInductiveConstructor{,}}\AgdaSpace{}%
\AgdaInductiveConstructor{⇑lvl}\<%
\\
\\[\AgdaEmptyExtraSkip]%
\>[2]\AgdaFunction{𝔹⟶𝕋⊆𝕋}\AgdaSpace{}%
\AgdaSymbol{:}\AgdaSpace{}%
\AgdaGeneralizable{d}\AgdaSpace{}%
\AgdaOperator{\AgdaFunction{∈}}\AgdaSpace{}%
\AgdaFunction{𝔹}\AgdaSpace{}%
\AgdaOperator{\AgdaFunction{⟶}}\AgdaSpace{}%
\AgdaFunction{𝕋}\AgdaSpace{}%
\AgdaSymbol{→}\AgdaSpace{}%
\AgdaGeneralizable{d}\AgdaSpace{}%
\AgdaOperator{\AgdaFunction{∈}}\AgdaSpace{}%
\AgdaFunction{𝕋}\<%
\\
\>[2]\AgdaFunction{𝔹⟶𝕋⊆𝕋}\AgdaSpace{}%
\AgdaSymbol{\{}\AgdaOperator{\AgdaInductiveConstructor{⟨ƛ}}\AgdaSpace{}%
\AgdaBound{t}\AgdaSpace{}%
\AgdaOperator{\AgdaInductiveConstructor{⟩}}\AgdaSpace{}%
\AgdaBound{γ}\AgdaSymbol{\}}\AgdaSpace{}%
\AgdaBound{pf}\AgdaSpace{}%
\AgdaBound{n}\<%
\\
\>[2][@{}l@{\AgdaIndent{0}}]%
\>[4]\AgdaKeyword{with}\AgdaSpace{}%
\AgdaBound{pf}\AgdaSpace{}%
\AgdaSymbol{(}\AgdaFunction{lvl∈𝔹}\AgdaSpace{}%
\AgdaBound{n}\AgdaSymbol{)}\<%
\\
\>[2]\AgdaSymbol{...}\AgdaSpace{}%
\AgdaSymbol{|}\AgdaSpace{}%
\AgdaBound{d}\AgdaSpace{}%
\AgdaOperator{\AgdaInductiveConstructor{,}}\AgdaSpace{}%
\AgdaInductiveConstructor{appClosure}\AgdaSpace{}%
\AgdaBound{eval-closure}\AgdaSpace{}%
\AgdaOperator{\AgdaInductiveConstructor{,}}\AgdaSpace{}%
\AgdaBound{dt}\<%
\\
\>[2][@{}l@{\AgdaIndent{0}}]%
\>[4]\AgdaKeyword{with}\AgdaSpace{}%
\AgdaBound{dt}\AgdaSpace{}%
\AgdaBound{n}\<%
\\
\>[2]\AgdaSymbol{...}\AgdaSpace{}%
\AgdaSymbol{|}\AgdaSpace{}%
\AgdaBound{v}\AgdaSpace{}%
\AgdaOperator{\AgdaInductiveConstructor{,}}\AgdaSpace{}%
\AgdaBound{d⇑v}\AgdaSpace{}%
\AgdaSymbol{=}\<%
\\
\>[2][@{}l@{\AgdaIndent{0}}]%
\>[4]\AgdaOperator{\AgdaInductiveConstructor{ƛ}}\AgdaSpace{}%
\AgdaBound{v}\AgdaSpace{}%
\AgdaOperator{\AgdaInductiveConstructor{,}}\AgdaSpace{}%
\AgdaInductiveConstructor{⇑closure}\AgdaSpace{}%
\AgdaBound{eval-closure}\AgdaSpace{}%
\AgdaBound{d⇑v}\<%
\\
\>[2]\AgdaFunction{𝔹⟶𝕋⊆𝕋}\AgdaSpace{}%
\AgdaSymbol{\{}\AgdaOperator{\AgdaInductiveConstructor{`}}\AgdaSpace{}%
\AgdaBound{e}\AgdaSymbol{\}}\AgdaSpace{}%
\AgdaBound{pf}\AgdaSpace{}%
\AgdaBound{n}\<%
\\
\>[2][@{}l@{\AgdaIndent{0}}]%
\>[4]\AgdaKeyword{with}\AgdaSpace{}%
\AgdaBound{pf}\AgdaSpace{}%
\AgdaSymbol{(}\AgdaFunction{lvl∈𝔹}\AgdaSpace{}%
\AgdaBound{n}\AgdaSymbol{)}\<%
\\
\>[2]\AgdaSymbol{...}\AgdaSpace{}%
\AgdaSymbol{|}\AgdaSpace{}%
\AgdaSymbol{\AgdaUnderscore{}}\AgdaSpace{}%
\AgdaOperator{\AgdaInductiveConstructor{,}}\AgdaSpace{}%
\AgdaInductiveConstructor{appNeutral}\AgdaSpace{}%
\AgdaOperator{\AgdaInductiveConstructor{,}}\AgdaSpace{}%
\AgdaBound{et}\<%
\\
\>[2][@{}l@{\AgdaIndent{0}}]%
\>[4]\AgdaKeyword{with}\AgdaSpace{}%
\AgdaBound{et}\AgdaSpace{}%
\AgdaBound{n}\<%
\\
\>[2]\AgdaSymbol{...}\AgdaSpace{}%
\AgdaSymbol{|}\AgdaSpace{}%
\AgdaBound{u}\AgdaSpace{}%
\AgdaOperator{\AgdaInductiveConstructor{·}}\AgdaSpace{}%
\AgdaBound{v}\AgdaSpace{}%
\AgdaOperator{\AgdaInductiveConstructor{,}}\AgdaSpace{}%
\AgdaInductiveConstructor{⇑neutral}\AgdaSpace{}%
\AgdaSymbol{(}\AgdaInductiveConstructor{⇑app}\AgdaSpace{}%
\AgdaBound{e⇑u}\AgdaSpace{}%
\AgdaSymbol{\AgdaUnderscore{})}\AgdaSpace{}%
\AgdaSymbol{=}\<%
\\
\>[2][@{}l@{\AgdaIndent{0}}]%
\>[4]\AgdaBound{u}\AgdaSpace{}%
\AgdaOperator{\AgdaInductiveConstructor{,}}\AgdaSpace{}%
\AgdaInductiveConstructor{⇑neutral}\AgdaSpace{}%
\AgdaBound{e⇑u}\<%
\\
\\[\AgdaEmptyExtraSkip]%
\>[2]\AgdaKeyword{mutual}\<%
\\
\>[2][@{}l@{\AgdaIndent{0}}]%
\>[4]\AgdaOperator{\AgdaFunction{𝕋⟶𝔹⊆⟦\AgdaUnderscore{}⇒\AgdaUnderscore{}⟧}}\AgdaSpace{}%
\AgdaSymbol{:}\AgdaSpace{}%
\AgdaSymbol{∀}\AgdaSpace{}%
\AgdaBound{S}\AgdaSpace{}%
\AgdaBound{T}\AgdaSpace{}%
\AgdaSymbol{→}\AgdaSpace{}%
\AgdaGeneralizable{f}\AgdaSpace{}%
\AgdaOperator{\AgdaFunction{∈}}\AgdaSpace{}%
\AgdaFunction{𝕋}\AgdaSpace{}%
\AgdaOperator{\AgdaFunction{⟶}}\AgdaSpace{}%
\AgdaFunction{𝔹}\AgdaSpace{}%
\AgdaSymbol{→}\AgdaSpace{}%
\AgdaGeneralizable{f}\AgdaSpace{}%
\AgdaOperator{\AgdaFunction{∈}}\AgdaSpace{}%
\AgdaOperator{\AgdaFunction{⟦}}\AgdaSpace{}%
\AgdaBound{S}\AgdaSpace{}%
\AgdaOperator{\AgdaInductiveConstructor{⇒}}\AgdaSpace{}%
\AgdaBound{T}\AgdaSpace{}%
\AgdaOperator{\AgdaFunction{⟧}}\<%
\\
\>[4]\AgdaOperator{\AgdaFunction{𝕋⟶𝔹⊆⟦}}\AgdaSpace{}%
\AgdaBound{S}\AgdaSpace{}%
\AgdaOperator{\AgdaFunction{⇒}}\AgdaSpace{}%
\AgdaBound{T}\AgdaSpace{}%
\AgdaOperator{\AgdaFunction{⟧}}\AgdaSpace{}%
\AgdaBound{pf}\AgdaSpace{}%
\AgdaBound{sa}\<%
\\
\>[4][@{}l@{\AgdaIndent{0}}]%
\>[6]\AgdaKeyword{with}\AgdaSpace{}%
\AgdaOperator{\AgdaFunction{⟦}}\AgdaSpace{}%
\AgdaBound{S}\AgdaSpace{}%
\AgdaOperator{\AgdaFunction{⟧⊆𝕋}}\AgdaSpace{}%
\AgdaBound{sa}\<%
\\
\>[4]\AgdaSymbol{...}\AgdaSpace{}%
\AgdaSymbol{|}\AgdaSpace{}%
\AgdaBound{at}\<%
\\
\>[4][@{}l@{\AgdaIndent{0}}]%
\>[6]\AgdaKeyword{with}\AgdaSpace{}%
\AgdaBound{pf}\AgdaSpace{}%
\AgdaBound{at}\<%
\\
\>[4]\AgdaSymbol{...}\AgdaSpace{}%
\AgdaSymbol{|}\AgdaSpace{}%
\AgdaOperator{\AgdaInductiveConstructor{`}}\AgdaSpace{}%
\AgdaBound{e}\AgdaSpace{}%
\AgdaOperator{\AgdaInductiveConstructor{,}}\AgdaSpace{}%
\AgdaBound{app-eval}\AgdaSpace{}%
\AgdaOperator{\AgdaInductiveConstructor{,}}\AgdaSpace{}%
\AgdaBound{eb}\<%
\\
\>[4][@{}l@{\AgdaIndent{0}}]%
\>[6]\AgdaKeyword{with}\AgdaSpace{}%
\AgdaOperator{\AgdaFunction{𝔹⊆⟦}}\AgdaSpace{}%
\AgdaBound{T}\AgdaSpace{}%
\AgdaOperator{\AgdaFunction{⟧}}\AgdaSpace{}%
\AgdaBound{eb}\<%
\\
\>[4]\AgdaSymbol{...}\AgdaSpace{}%
\AgdaSymbol{|}\AgdaSpace{}%
\AgdaBound{se}\AgdaSpace{}%
\AgdaSymbol{=}\<%
\\
\>[4][@{}l@{\AgdaIndent{0}}]%
\>[6]\AgdaOperator{\AgdaInductiveConstructor{`}}\AgdaSpace{}%
\AgdaBound{e}\AgdaSpace{}%
\AgdaOperator{\AgdaInductiveConstructor{,}}\AgdaSpace{}%
\AgdaBound{app-eval}\AgdaSpace{}%
\AgdaOperator{\AgdaInductiveConstructor{,}}\AgdaSpace{}%
\AgdaBound{se}\<%
\\
\\[\AgdaEmptyExtraSkip]%
\>[4]\AgdaOperator{\AgdaFunction{⟦\AgdaUnderscore{}⇒\AgdaUnderscore{}⟧⊆𝔹⟶𝕋}}\AgdaSpace{}%
\AgdaSymbol{:}\AgdaSpace{}%
\AgdaSymbol{∀}\AgdaSpace{}%
\AgdaBound{S}\AgdaSpace{}%
\AgdaBound{T}\AgdaSpace{}%
\AgdaSymbol{→}\AgdaSpace{}%
\AgdaGeneralizable{f}\AgdaSpace{}%
\AgdaOperator{\AgdaFunction{∈}}\AgdaSpace{}%
\AgdaOperator{\AgdaFunction{⟦}}\AgdaSpace{}%
\AgdaBound{S}\AgdaSpace{}%
\AgdaOperator{\AgdaInductiveConstructor{⇒}}\AgdaSpace{}%
\AgdaBound{T}\AgdaSpace{}%
\AgdaOperator{\AgdaFunction{⟧}}\AgdaSpace{}%
\AgdaSymbol{→}\AgdaSpace{}%
\AgdaGeneralizable{f}\AgdaSpace{}%
\AgdaOperator{\AgdaFunction{∈}}\AgdaSpace{}%
\AgdaFunction{𝔹}\AgdaSpace{}%
\AgdaOperator{\AgdaFunction{⟶}}\AgdaSpace{}%
\AgdaFunction{𝕋}\<%
\\
\>[4]\AgdaOperator{\AgdaFunction{⟦}}%
\>[5295I]\AgdaBound{S}\AgdaSpace{}%
\AgdaOperator{\AgdaFunction{⇒}}\AgdaSpace{}%
\AgdaBound{T}\AgdaSpace{}%
\AgdaOperator{\AgdaFunction{⟧⊆𝔹⟶𝕋}}\AgdaSpace{}%
\AgdaBound{sf}\AgdaSpace{}%
\AgdaSymbol{\{}\AgdaOperator{\AgdaInductiveConstructor{`}}\AgdaSpace{}%
\AgdaBound{e}\AgdaSymbol{\}}\AgdaSpace{}%
\AgdaBound{eb}\<%
\\
\>[.][@{}l@{}]\<[5295I]%
\>[6]\AgdaKeyword{with}\AgdaSpace{}%
\AgdaBound{sf}\AgdaSpace{}%
\AgdaSymbol{(}\AgdaOperator{\AgdaFunction{𝔹⊆⟦}}\AgdaSpace{}%
\AgdaBound{S}\AgdaSpace{}%
\AgdaOperator{\AgdaFunction{⟧}}\AgdaSpace{}%
\AgdaBound{eb}\AgdaSymbol{)}\<%
\\
\>[4]\AgdaSymbol{...}\AgdaSpace{}%
\AgdaSymbol{|}\AgdaSpace{}%
\AgdaBound{d}\AgdaSpace{}%
\AgdaOperator{\AgdaInductiveConstructor{,}}\AgdaSpace{}%
\AgdaBound{app-eval}\AgdaSpace{}%
\AgdaOperator{\AgdaInductiveConstructor{,}}\AgdaSpace{}%
\AgdaBound{sd}\AgdaSpace{}%
\AgdaSymbol{=}\<%
\\
\>[4][@{}l@{\AgdaIndent{0}}]%
\>[6]\AgdaBound{d}\AgdaSpace{}%
\AgdaOperator{\AgdaInductiveConstructor{,}}%
\>[11]\AgdaBound{app-eval}\AgdaSpace{}%
\AgdaOperator{\AgdaInductiveConstructor{,}}\AgdaSpace{}%
\AgdaOperator{\AgdaFunction{⟦}}\AgdaSpace{}%
\AgdaBound{T}\AgdaSpace{}%
\AgdaOperator{\AgdaFunction{⟧⊆𝕋}}\AgdaSpace{}%
\AgdaBound{sd}\<%
\\
\\[\AgdaEmptyExtraSkip]%
\>[4]\AgdaComment{--\ Any\ object\ that\ can\ have\ a\ neutral\ form\ read\ back}\<%
\\
\>[4]\AgdaComment{--\ is\ semantically\ typed}\<%
\\
\>[4]\AgdaOperator{\AgdaFunction{𝔹⊆⟦\AgdaUnderscore{}⟧}}\AgdaSpace{}%
\AgdaSymbol{:}\AgdaSpace{}%
\AgdaSymbol{∀}\AgdaSpace{}%
\AgdaBound{T}\AgdaSpace{}%
\AgdaSymbol{→}\AgdaSpace{}%
\AgdaOperator{\AgdaInductiveConstructor{`}}\AgdaSpace{}%
\AgdaGeneralizable{e}\AgdaSpace{}%
\AgdaOperator{\AgdaFunction{∈}}\AgdaSpace{}%
\AgdaFunction{𝔹}\AgdaSpace{}%
\AgdaSymbol{→}\AgdaSpace{}%
\AgdaOperator{\AgdaInductiveConstructor{`}}\AgdaSpace{}%
\AgdaGeneralizable{e}\AgdaSpace{}%
\AgdaOperator{\AgdaFunction{∈}}\AgdaSpace{}%
\AgdaOperator{\AgdaFunction{⟦}}\AgdaSpace{}%
\AgdaBound{T}\AgdaSpace{}%
\AgdaOperator{\AgdaFunction{⟧}}\<%
\\
\>[4]\AgdaOperator{\AgdaFunction{𝔹⊆⟦}}\AgdaSpace{}%
\AgdaInductiveConstructor{base}\AgdaSpace{}%
\AgdaOperator{\AgdaFunction{⟧}}\AgdaSpace{}%
\AgdaBound{eb}\AgdaSpace{}%
\AgdaSymbol{=}\AgdaSpace{}%
\AgdaBound{eb}\<%
\\
\>[4]\AgdaOperator{\AgdaFunction{𝔹⊆⟦}}\AgdaSpace{}%
\AgdaBound{S}\AgdaSpace{}%
\AgdaOperator{\AgdaInductiveConstructor{⇒}}\AgdaSpace{}%
\AgdaBound{T}\AgdaSpace{}%
\AgdaOperator{\AgdaFunction{⟧}}\AgdaSpace{}%
\AgdaSymbol{=}\AgdaSpace{}%
\AgdaOperator{\AgdaFunction{𝕋⟶𝔹⊆⟦}}\AgdaSpace{}%
\AgdaBound{S}\AgdaSpace{}%
\AgdaOperator{\AgdaFunction{⇒}}\AgdaSpace{}%
\AgdaBound{T}\AgdaSpace{}%
\AgdaOperator{\AgdaFunction{⟧}}\AgdaSpace{}%
\AgdaOperator{\AgdaFunction{∘}}\AgdaSpace{}%
\AgdaFunction{𝔹⊆𝕋⟶𝔹}\<%
\\
\\[\AgdaEmptyExtraSkip]%
\>[4]\AgdaComment{--\ Semantic\ typing\ implies\ a\ normal\ form\ can\ be\ read}\<%
\\
\>[4]\AgdaComment{--\ back}\<%
\\
\>[4]\AgdaOperator{\AgdaFunction{⟦\AgdaUnderscore{}⟧⊆𝕋}}\AgdaSpace{}%
\AgdaSymbol{:}\AgdaSpace{}%
\AgdaSymbol{∀}\AgdaSpace{}%
\AgdaBound{T}\AgdaSpace{}%
\AgdaSymbol{→}\AgdaSpace{}%
\AgdaGeneralizable{d}\AgdaSpace{}%
\AgdaOperator{\AgdaFunction{∈}}\AgdaSpace{}%
\AgdaOperator{\AgdaFunction{⟦}}\AgdaSpace{}%
\AgdaBound{T}\AgdaSpace{}%
\AgdaOperator{\AgdaFunction{⟧}}\AgdaSpace{}%
\AgdaSymbol{→}\AgdaSpace{}%
\AgdaGeneralizable{d}\AgdaSpace{}%
\AgdaOperator{\AgdaFunction{∈}}\AgdaSpace{}%
\AgdaFunction{𝕋}\<%
\\
\>[4]\AgdaOperator{\AgdaFunction{⟦}}\AgdaSpace{}%
\AgdaInductiveConstructor{base}\AgdaSpace{}%
\AgdaOperator{\AgdaFunction{⟧⊆𝕋}}\AgdaSpace{}%
\AgdaSymbol{=}\AgdaSpace{}%
\AgdaFunction{𝔹⊆𝕋}\<%
\\
\>[4]\AgdaOperator{\AgdaFunction{⟦}}\AgdaSpace{}%
\AgdaBound{S}\AgdaSpace{}%
\AgdaOperator{\AgdaInductiveConstructor{⇒}}\AgdaSpace{}%
\AgdaBound{T}\AgdaSpace{}%
\AgdaOperator{\AgdaFunction{⟧⊆𝕋}}\AgdaSpace{}%
\AgdaSymbol{=}\AgdaSpace{}%
\AgdaFunction{𝔹⟶𝕋⊆𝕋}\AgdaSpace{}%
\AgdaOperator{\AgdaFunction{∘}}\AgdaSpace{}%
\AgdaOperator{\AgdaFunction{⟦}}\AgdaSpace{}%
\AgdaBound{S}\AgdaSpace{}%
\AgdaOperator{\AgdaFunction{⇒}}\AgdaSpace{}%
\AgdaBound{T}\AgdaSpace{}%
\AgdaOperator{\AgdaFunction{⟧⊆𝔹⟶𝕋}}\<%
\end{code}

\subsection{Establishing normalization (Figure \ref{fig:nbe})}

\begin{code}%
\>[2]\AgdaComment{--\ Normalization\ of\ STLC}\<%
\\
\>[2]\AgdaFunction{idx→lvl}\AgdaSpace{}%
\AgdaSymbol{:}\AgdaSpace{}%
\AgdaDatatype{ℕ}\AgdaSpace{}%
\AgdaSymbol{→}\AgdaSpace{}%
\AgdaDatatype{ℕ}\AgdaSpace{}%
\AgdaSymbol{→}\AgdaSpace{}%
\AgdaDatatype{ℕ}\<%
\\
\>[2]\AgdaFunction{idx→lvl}\AgdaSpace{}%
\AgdaBound{i}\AgdaSpace{}%
\AgdaBound{n}\AgdaSpace{}%
\AgdaSymbol{=}\AgdaSpace{}%
\AgdaBound{n}\AgdaSpace{}%
\AgdaOperator{\AgdaPrimitive{-}}\AgdaSpace{}%
\AgdaInductiveConstructor{suc}\AgdaSpace{}%
\AgdaBound{i}\<%
\\
\\[\AgdaEmptyExtraSkip]%
\>[2]\AgdaFunction{env}\AgdaSpace{}%
\AgdaSymbol{:}\AgdaSpace{}%
\AgdaDatatype{ℕ}\AgdaSpace{}%
\AgdaSymbol{→}\AgdaSpace{}%
\AgdaFunction{Env}\<%
\\
\>[2]\AgdaFunction{env}\AgdaSpace{}%
\AgdaBound{n}\AgdaSpace{}%
\AgdaBound{i}\AgdaSpace{}%
\AgdaSymbol{=}\AgdaSpace{}%
\AgdaOperator{\AgdaInductiveConstructor{`}}\AgdaSpace{}%
\AgdaInductiveConstructor{lvl}\AgdaSpace{}%
\AgdaSymbol{(}\AgdaFunction{idx→lvl}\AgdaSpace{}%
\AgdaBound{i}\AgdaSpace{}%
\AgdaBound{n}\AgdaSymbol{)}\<%
\\
\\[\AgdaEmptyExtraSkip]%
\>[2]\AgdaOperator{\AgdaFunction{\AgdaUnderscore{}has-normal-form<\AgdaUnderscore{}>\AgdaUnderscore{}}}\AgdaSpace{}%
\AgdaSymbol{:}\AgdaSpace{}%
\AgdaDatatype{Term}\AgdaSpace{}%
\AgdaSymbol{→}\AgdaSpace{}%
\AgdaDatatype{ℕ}\AgdaSpace{}%
\AgdaSymbol{→}\AgdaSpace{}%
\AgdaDatatype{Term}\AgdaSpace{}%
\AgdaSymbol{→}\AgdaSpace{}%
\AgdaPrimitive{Set}\<%
\\
\>[2]\AgdaBound{t}%
\>[5415I]\AgdaOperator{\AgdaFunction{has-normal-form<}}\AgdaSpace{}%
\AgdaBound{n}\AgdaSpace{}%
\AgdaOperator{\AgdaFunction{>}}\AgdaSpace{}%
\AgdaBound{v}\AgdaSpace{}%
\AgdaSymbol{=}\<%
\\
\>[.][@{}l@{}]\<[5415I]%
\>[4]\AgdaFunction{∃[}\AgdaSpace{}%
\AgdaBound{a}\AgdaSpace{}%
\AgdaFunction{]}\AgdaSpace{}%
\AgdaFunction{env}\AgdaSpace{}%
\AgdaBound{n}\AgdaSpace{}%
\AgdaOperator{\AgdaDatatype{∣}}\AgdaSpace{}%
\AgdaBound{t}\AgdaSpace{}%
\AgdaOperator{\AgdaDatatype{⇓}}\AgdaSpace{}%
\AgdaBound{a}\AgdaSpace{}%
\AgdaOperator{\AgdaFunction{×}}\AgdaSpace{}%
\AgdaBound{n}\AgdaSpace{}%
\AgdaOperator{\AgdaDatatype{∣}}\AgdaSpace{}%
\AgdaBound{a}\AgdaSpace{}%
\AgdaOperator{\AgdaDatatype{⇑}}\AgdaSpace{}%
\AgdaBound{v}\<%
\\
\\[\AgdaEmptyExtraSkip]%
\>[2]\AgdaOperator{\AgdaFunction{∣\AgdaUnderscore{}∣}}\AgdaSpace{}%
\AgdaSymbol{:}\AgdaSpace{}%
\AgdaDatatype{Ctx}\AgdaSpace{}%
\AgdaSymbol{→}\AgdaSpace{}%
\AgdaDatatype{ℕ}\<%
\\
\>[2]\AgdaOperator{\AgdaFunction{∣}}\AgdaSpace{}%
\AgdaInductiveConstructor{∅}\AgdaSpace{}%
\AgdaOperator{\AgdaFunction{∣}}\AgdaSpace{}%
\AgdaSymbol{=}\AgdaSpace{}%
\AgdaInductiveConstructor{zero}\<%
\\
\>[2]\AgdaOperator{\AgdaFunction{∣}}\AgdaSpace{}%
\AgdaBound{Γ}\AgdaSpace{}%
\AgdaOperator{\AgdaInductiveConstructor{·:}}\AgdaSpace{}%
\AgdaSymbol{\AgdaUnderscore{}}\AgdaSpace{}%
\AgdaOperator{\AgdaFunction{∣}}\AgdaSpace{}%
\AgdaSymbol{=}\AgdaSpace{}%
\AgdaInductiveConstructor{suc}\AgdaSpace{}%
\AgdaOperator{\AgdaFunction{∣}}\AgdaSpace{}%
\AgdaBound{Γ}\AgdaSpace{}%
\AgdaOperator{\AgdaFunction{∣}}\<%
\\
\\[\AgdaEmptyExtraSkip]%
\>[2]\AgdaOperator{\AgdaFunction{⊨env∣\AgdaUnderscore{}∣}}\AgdaSpace{}%
\AgdaSymbol{:}\AgdaSpace{}%
\AgdaSymbol{∀}\AgdaSpace{}%
\AgdaBound{Γ}\AgdaSpace{}%
\AgdaSymbol{→}\AgdaSpace{}%
\AgdaBound{Γ}\AgdaSpace{}%
\AgdaOperator{\AgdaFunction{⊨}}\AgdaSpace{}%
\AgdaFunction{env}\AgdaSpace{}%
\AgdaOperator{\AgdaFunction{∣}}\AgdaSpace{}%
\AgdaBound{Γ}\AgdaSpace{}%
\AgdaOperator{\AgdaFunction{∣}}\<%
\\
\>[2]\AgdaOperator{\AgdaFunction{⊨env∣\AgdaUnderscore{}∣}}\AgdaSpace{}%
\AgdaBound{Γ}\AgdaSpace{}%
\AgdaSymbol{\{}\AgdaBound{x}\AgdaSymbol{\}}\AgdaSpace{}%
\AgdaSymbol{\{}\AgdaBound{T}\AgdaSymbol{\}}\AgdaSpace{}%
\AgdaSymbol{\AgdaUnderscore{}}\AgdaSpace{}%
\AgdaSymbol{=}\<%
\\
\>[2][@{}l@{\AgdaIndent{0}}]%
\>[4]\AgdaOperator{\AgdaFunction{𝔹⊆⟦}}\AgdaSpace{}%
\AgdaBound{T}\AgdaSpace{}%
\AgdaOperator{\AgdaFunction{⟧}}\AgdaSpace{}%
\AgdaSymbol{(}\AgdaFunction{lvl∈𝔹}\AgdaSpace{}%
\AgdaSymbol{(}\AgdaFunction{idx→lvl}\AgdaSpace{}%
\AgdaBound{x}\AgdaSpace{}%
\AgdaOperator{\AgdaFunction{∣}}\AgdaSpace{}%
\AgdaBound{Γ}\AgdaSpace{}%
\AgdaOperator{\AgdaFunction{∣}}\AgdaSymbol{))}\<%
\\
\\[\AgdaEmptyExtraSkip]%
\>[2]\AgdaFunction{normalization}\AgdaSpace{}%
\AgdaSymbol{:}\AgdaSpace{}%
\AgdaGeneralizable{Γ}\AgdaSpace{}%
\AgdaOperator{\AgdaDatatype{⊢}}\AgdaSpace{}%
\AgdaGeneralizable{t}\AgdaSpace{}%
\AgdaOperator{\AgdaDatatype{∷}}\AgdaSpace{}%
\AgdaGeneralizable{T}\AgdaSpace{}%
\AgdaSymbol{→}\AgdaSpace{}%
\AgdaFunction{∃[}\AgdaSpace{}%
\AgdaBound{v}\AgdaSpace{}%
\AgdaFunction{]}\AgdaSpace{}%
\AgdaGeneralizable{t}\AgdaSpace{}%
\AgdaOperator{\AgdaFunction{has-normal-form<}}\AgdaSpace{}%
\AgdaOperator{\AgdaFunction{∣}}\AgdaSpace{}%
\AgdaGeneralizable{Γ}\AgdaSpace{}%
\AgdaOperator{\AgdaFunction{∣}}\AgdaSpace{}%
\AgdaOperator{\AgdaFunction{>}}\AgdaSpace{}%
\AgdaBound{v}\<%
\\
\>[2]\AgdaFunction{normalization}\AgdaSpace{}%
\AgdaSymbol{\{}\AgdaBound{Γ}\AgdaSymbol{\}}\AgdaSpace{}%
\AgdaSymbol{\{}\AgdaArgument{T}\AgdaSpace{}%
\AgdaSymbol{=}\AgdaSpace{}%
\AgdaBound{T}\AgdaSymbol{\}}\AgdaSpace{}%
\AgdaBound{⊢t}\<%
\\
\>[2][@{}l@{\AgdaIndent{0}}]%
\>[4]\AgdaKeyword{with}\AgdaSpace{}%
\AgdaFunction{fundamental-lemma}\AgdaSpace{}%
\AgdaBound{⊢t}\AgdaSpace{}%
\AgdaOperator{\AgdaFunction{⊨env∣}}\AgdaSpace{}%
\AgdaBound{Γ}\AgdaSpace{}%
\AgdaOperator{\AgdaFunction{∣}}\<%
\\
\>[2]\AgdaSymbol{...}\AgdaSpace{}%
\AgdaSymbol{|}\AgdaSpace{}%
\AgdaBound{a}\AgdaSpace{}%
\AgdaOperator{\AgdaInductiveConstructor{,}}\AgdaSpace{}%
\AgdaBound{t⇓a}\AgdaSpace{}%
\AgdaOperator{\AgdaInductiveConstructor{,}}\AgdaSpace{}%
\AgdaBound{st}\<%
\\
\>[2][@{}l@{\AgdaIndent{0}}]%
\>[4]\AgdaKeyword{with}\AgdaSpace{}%
\AgdaOperator{\AgdaFunction{⟦}}\AgdaSpace{}%
\AgdaBound{T}\AgdaSpace{}%
\AgdaOperator{\AgdaFunction{⟧⊆𝕋}}\AgdaSpace{}%
\AgdaBound{st}\AgdaSpace{}%
\AgdaOperator{\AgdaFunction{∣}}\AgdaSpace{}%
\AgdaBound{Γ}\AgdaSpace{}%
\AgdaOperator{\AgdaFunction{∣}}\<%
\\
\>[2]\AgdaSymbol{...}\AgdaSpace{}%
\AgdaSymbol{|}\AgdaSpace{}%
\AgdaBound{v}\AgdaSpace{}%
\AgdaOperator{\AgdaInductiveConstructor{,}}\AgdaSpace{}%
\AgdaBound{a⇑v}\AgdaSpace{}%
\AgdaSymbol{=}\<%
\\
\>[2][@{}l@{\AgdaIndent{0}}]%
\>[4]\AgdaBound{v}\AgdaSpace{}%
\AgdaOperator{\AgdaInductiveConstructor{,}}\AgdaSpace{}%
\AgdaBound{a}\AgdaSpace{}%
\AgdaOperator{\AgdaInductiveConstructor{,}}\AgdaSpace{}%
\AgdaBound{t⇓a}\AgdaSpace{}%
\AgdaOperator{\AgdaInductiveConstructor{,}}\AgdaSpace{}%
\AgdaBound{a⇑v}\<%
\end{code}

\subsection{Read back produces a normal term}

\begin{code}%
\>[2]\AgdaComment{--\ Normal\ and\ neutral\ terms}\<%
\\
\>[2]\AgdaKeyword{mutual}\<%
\\
\>[2][@{}l@{\AgdaIndent{0}}]%
\>[4]\AgdaKeyword{data}\AgdaSpace{}%
\AgdaDatatype{Normal}\AgdaSpace{}%
\AgdaSymbol{:}\AgdaSpace{}%
\AgdaDatatype{Term}\AgdaSpace{}%
\AgdaSymbol{→}\AgdaSpace{}%
\AgdaPrimitive{Set}\AgdaSpace{}%
\AgdaKeyword{where}\<%
\\
\>[4][@{}l@{\AgdaIndent{0}}]%
\>[6]\AgdaInductiveConstructor{normalAbs}\AgdaSpace{}%
\AgdaSymbol{:}\AgdaSpace{}%
\AgdaDatatype{Normal}\AgdaSpace{}%
\AgdaGeneralizable{t}\AgdaSpace{}%
\AgdaSymbol{→}\AgdaSpace{}%
\AgdaDatatype{Normal}\AgdaSpace{}%
\AgdaSymbol{(}\AgdaOperator{\AgdaInductiveConstructor{ƛ}}\AgdaSpace{}%
\AgdaGeneralizable{t}\AgdaSymbol{)}\<%
\\
\>[6]\AgdaInductiveConstructor{neutral}\AgdaSpace{}%
\AgdaSymbol{:}\AgdaSpace{}%
\AgdaDatatype{Neutral}\AgdaSpace{}%
\AgdaGeneralizable{t}\AgdaSpace{}%
\AgdaSymbol{→}\AgdaSpace{}%
\AgdaDatatype{Normal}\AgdaSpace{}%
\AgdaGeneralizable{t}\<%
\\
\\[\AgdaEmptyExtraSkip]%
\>[4]\AgdaKeyword{data}\AgdaSpace{}%
\AgdaDatatype{Neutral}\AgdaSpace{}%
\AgdaSymbol{:}\AgdaSpace{}%
\AgdaDatatype{Term}\AgdaSpace{}%
\AgdaSymbol{→}\AgdaSpace{}%
\AgdaPrimitive{Set}\AgdaSpace{}%
\AgdaKeyword{where}\<%
\\
\>[4][@{}l@{\AgdaIndent{0}}]%
\>[6]\AgdaInductiveConstructor{neutralVar}\AgdaSpace{}%
\AgdaSymbol{:}\AgdaSpace{}%
\AgdaDatatype{Neutral}\AgdaSpace{}%
\AgdaSymbol{(}\AgdaInductiveConstructor{var}\AgdaSpace{}%
\AgdaGeneralizable{x}\AgdaSymbol{)}\<%
\\
\>[6]\AgdaInductiveConstructor{neutralApp}\AgdaSpace{}%
\AgdaSymbol{:}\AgdaSpace{}%
\AgdaDatatype{Neutral}\AgdaSpace{}%
\AgdaGeneralizable{r}\AgdaSpace{}%
\AgdaSymbol{→}\AgdaSpace{}%
\AgdaDatatype{Normal}\AgdaSpace{}%
\AgdaGeneralizable{s}\AgdaSpace{}%
\AgdaSymbol{→}\AgdaSpace{}%
\AgdaDatatype{Neutral}\AgdaSpace{}%
\AgdaSymbol{(}\AgdaGeneralizable{r}\AgdaSpace{}%
\AgdaOperator{\AgdaInductiveConstructor{·}}\AgdaSpace{}%
\AgdaGeneralizable{s}\AgdaSymbol{)}\<%
\\
\\[\AgdaEmptyExtraSkip]%
\>[2]\AgdaComment{--\ Read\ back\ produces\ a\ normal\ term}\<%
\\
\>[2]\AgdaKeyword{mutual}\<%
\\
\>[2][@{}l@{\AgdaIndent{0}}]%
\>[4]\AgdaFunction{⇑-normal}\AgdaSpace{}%
\AgdaSymbol{:}\AgdaSpace{}%
\AgdaGeneralizable{n}\AgdaSpace{}%
\AgdaOperator{\AgdaDatatype{∣}}\AgdaSpace{}%
\AgdaGeneralizable{a}\AgdaSpace{}%
\AgdaOperator{\AgdaDatatype{⇑}}\AgdaSpace{}%
\AgdaGeneralizable{v}\AgdaSpace{}%
\AgdaSymbol{→}\AgdaSpace{}%
\AgdaDatatype{Normal}\AgdaSpace{}%
\AgdaGeneralizable{v}\<%
\\
\>[4]\AgdaFunction{⇑-normal}\AgdaSpace{}%
\AgdaSymbol{(}\AgdaInductiveConstructor{⇑closure}\AgdaSpace{}%
\AgdaSymbol{\AgdaUnderscore{}}\AgdaSpace{}%
\AgdaBound{a⇑v}\AgdaSymbol{)}\AgdaSpace{}%
\AgdaSymbol{=}\<%
\\
\>[4][@{}l@{\AgdaIndent{0}}]%
\>[6]\AgdaInductiveConstructor{normalAbs}\AgdaSpace{}%
\AgdaSymbol{(}\AgdaFunction{⇑-normal}\AgdaSpace{}%
\AgdaBound{a⇑v}\AgdaSymbol{)}\<%
\\
\>[4]\AgdaFunction{⇑-normal}\AgdaSpace{}%
\AgdaSymbol{(}\AgdaInductiveConstructor{⇑neutral}\AgdaSpace{}%
\AgdaBound{e⇑u}\AgdaSymbol{)}\AgdaSpace{}%
\AgdaSymbol{=}\<%
\\
\>[4][@{}l@{\AgdaIndent{0}}]%
\>[6]\AgdaInductiveConstructor{neutral}\AgdaSpace{}%
\AgdaSymbol{(}\AgdaFunction{⇑ⁿᵉ-neutral}\AgdaSpace{}%
\AgdaBound{e⇑u}\AgdaSymbol{)}\<%
\\
\\[\AgdaEmptyExtraSkip]%
\>[4]\AgdaFunction{⇑ⁿᵉ-neutral}\AgdaSpace{}%
\AgdaSymbol{:}\AgdaSpace{}%
\AgdaGeneralizable{n}\AgdaSpace{}%
\AgdaOperator{\AgdaDatatype{∣}}\AgdaSpace{}%
\AgdaGeneralizable{e}\AgdaSpace{}%
\AgdaOperator{\AgdaDatatype{⇑ⁿᵉ}}\AgdaSpace{}%
\AgdaGeneralizable{u}\AgdaSpace{}%
\AgdaSymbol{→}\AgdaSpace{}%
\AgdaDatatype{Neutral}\AgdaSpace{}%
\AgdaGeneralizable{u}\<%
\\
\>[4]\AgdaFunction{⇑ⁿᵉ-neutral}\AgdaSpace{}%
\AgdaInductiveConstructor{⇑lvl}\AgdaSpace{}%
\AgdaSymbol{=}\AgdaSpace{}%
\AgdaInductiveConstructor{neutralVar}\<%
\\
\>[4]\AgdaFunction{⇑ⁿᵉ-neutral}\AgdaSpace{}%
\AgdaSymbol{(}\AgdaInductiveConstructor{⇑app}\AgdaSpace{}%
\AgdaBound{e⇑u}\AgdaSpace{}%
\AgdaBound{d⇑v}\AgdaSymbol{)}\AgdaSpace{}%
\AgdaSymbol{=}\<%
\\
\>[4][@{}l@{\AgdaIndent{0}}]%
\>[6]\AgdaInductiveConstructor{neutralApp}\AgdaSpace{}%
\AgdaSymbol{(}\AgdaFunction{⇑ⁿᵉ-neutral}\AgdaSpace{}%
\AgdaBound{e⇑u}\AgdaSymbol{)}\AgdaSpace{}%
\AgdaSymbol{(}\AgdaFunction{⇑-normal}\AgdaSpace{}%
\AgdaBound{d⇑v}\AgdaSymbol{)}\<%
\\
\\[\AgdaEmptyExtraSkip]%
\>[2]\AgdaComment{--\ Normal\ form\ is\ truly\ a\ normal\ term}\<%
\\
\>[2]\AgdaFunction{nf-normal}\AgdaSpace{}%
\AgdaSymbol{:}\AgdaSpace{}%
\AgdaGeneralizable{t}\AgdaSpace{}%
\AgdaOperator{\AgdaFunction{has-normal-form<}}\AgdaSpace{}%
\AgdaGeneralizable{n}\AgdaSpace{}%
\AgdaOperator{\AgdaFunction{>}}\AgdaSpace{}%
\AgdaGeneralizable{v}\AgdaSpace{}%
\AgdaSymbol{→}\AgdaSpace{}%
\AgdaDatatype{Normal}\AgdaSpace{}%
\AgdaGeneralizable{v}\<%
\\
\>[2]\AgdaFunction{nf-normal}\AgdaSpace{}%
\AgdaSymbol{(\AgdaUnderscore{}}\AgdaSpace{}%
\AgdaOperator{\AgdaInductiveConstructor{,}}\AgdaSpace{}%
\AgdaSymbol{\AgdaUnderscore{}}\AgdaSpace{}%
\AgdaOperator{\AgdaInductiveConstructor{,}}\AgdaSpace{}%
\AgdaBound{⇑v}\AgdaSymbol{)}\AgdaSpace{}%
\AgdaSymbol{=}\AgdaSpace{}%
\AgdaFunction{⇑-normal}\AgdaSpace{}%
\AgdaBound{⇑v}\<%
\end{code}
\end{AgdaAlign}

\section{Totality of evaluation: extrinsic representation using raw bindings}
\label{sec:extrinsic}

Here we provide a proof of that evaluation of the simply typed lambda calculus
is total using an extrinsic representation of terms and raw variable bindings.

Using raw variable bindings in a mechanized proof by logical relations is
generally not worth the readability it may add as proving substitution lemmas
becomes increasingly cumbersome. As we do not need any such lemmas, our proof
remains nearly as short as before.

In the main body of the paper, we this proof using an intrinsically typed
representation as it simplifies a proof of soundness with respect to a
denotational semantics when we extend the proof to show weak head normalization.
As a result, it is easier to discuss this result in $\S$\ref{sec:correctness}.

\begin{code}[hide]%
\>[0]\AgdaKeyword{module}\AgdaSpace{}%
\AgdaModule{ExtrinsicEvaluationTotality}\AgdaSpace{}%
\AgdaSymbol{(}\AgdaBound{Id}\AgdaSpace{}%
\AgdaSymbol{:}\AgdaSpace{}%
\AgdaPrimitive{Set}\AgdaSymbol{)}\AgdaSpace{}%
\AgdaKeyword{where}\<%
\\
\>[0][@{}l@{\AgdaIndent{0}}]%
\>[2]\AgdaKeyword{infixr}\AgdaSpace{}%
\AgdaNumber{7}\AgdaSpace{}%
\AgdaOperator{\AgdaInductiveConstructor{\AgdaUnderscore{}⇒\AgdaUnderscore{}}}\<%
\\
\>[2]\AgdaKeyword{infixl}\AgdaSpace{}%
\AgdaNumber{5}\AgdaSpace{}%
\AgdaOperator{\AgdaInductiveConstructor{\AgdaUnderscore{},\AgdaUnderscore{}∷\AgdaUnderscore{}}}\<%
\\
\>[2]\AgdaKeyword{infixl}\AgdaSpace{}%
\AgdaNumber{7}\AgdaSpace{}%
\AgdaOperator{\AgdaInductiveConstructor{\AgdaUnderscore{}·\AgdaUnderscore{}}}\<%
\\
\>[2]\AgdaKeyword{infix}\AgdaSpace{}%
\AgdaNumber{5}\AgdaSpace{}%
\AgdaOperator{\AgdaInductiveConstructor{ƛ\AgdaUnderscore{}⟶\AgdaUnderscore{}}}\<%
\\
\>[2]\AgdaKeyword{infix}\AgdaSpace{}%
\AgdaNumber{4}\AgdaSpace{}%
\AgdaOperator{\AgdaDatatype{\AgdaUnderscore{}⊢\AgdaUnderscore{}∷\AgdaUnderscore{}}}\<%
\\
\>[2]\AgdaKeyword{infix}\AgdaSpace{}%
\AgdaNumber{4}\AgdaSpace{}%
\AgdaOperator{\AgdaDatatype{\AgdaUnderscore{}∷\AgdaUnderscore{}∈\AgdaUnderscore{}}}\<%
\\
\>[2]\AgdaKeyword{infix}\AgdaSpace{}%
\AgdaNumber{5}\AgdaSpace{}%
\AgdaOperator{\AgdaInductiveConstructor{⟨ƛ\AgdaUnderscore{}⟶\AgdaUnderscore{}⟩\AgdaUnderscore{}}}\<%
\\
\>[2]\AgdaKeyword{infix}\AgdaSpace{}%
\AgdaNumber{4}\AgdaSpace{}%
\AgdaOperator{\AgdaDatatype{\AgdaUnderscore{}↦\AgdaUnderscore{}∈\AgdaUnderscore{}}}\<%
\\
\>[2]\AgdaKeyword{infix}\AgdaSpace{}%
\AgdaNumber{4}\AgdaSpace{}%
\AgdaOperator{\AgdaDatatype{\AgdaUnderscore{}∣\AgdaUnderscore{}⇓\AgdaUnderscore{}}}\<%
\\
\>[2]\AgdaKeyword{infix}\AgdaSpace{}%
\AgdaNumber{4}\AgdaSpace{}%
\AgdaOperator{\AgdaFunction{\AgdaUnderscore{}⊨\AgdaUnderscore{}}}\<%
\\
\>[2]\AgdaKeyword{infix}\AgdaSpace{}%
\AgdaNumber{4}\AgdaSpace{}%
\AgdaOperator{\AgdaFunction{\AgdaUnderscore{}⊨\AgdaUnderscore{}∷\AgdaUnderscore{}}}\<%
\end{code}
\subsection{Embedding of STLC in Agda}

\begin{AgdaAlign}
\begin{code}%
\>[2]\AgdaKeyword{variable}\AgdaSpace{}%
\AgdaGeneralizable{x}\AgdaSpace{}%
\AgdaGeneralizable{y}\AgdaSpace{}%
\AgdaSymbol{:}\AgdaSpace{}%
\AgdaBound{Id}\<%
\\
\\[\AgdaEmptyExtraSkip]%
\>[2]\AgdaComment{--\ Types}\<%
\\
\>[2]\AgdaKeyword{data}\AgdaSpace{}%
\AgdaDatatype{Type}\AgdaSpace{}%
\AgdaSymbol{:}\AgdaSpace{}%
\AgdaPrimitive{Set}\AgdaSpace{}%
\AgdaKeyword{where}\<%
\\
\>[2][@{}l@{\AgdaIndent{0}}]%
\>[4]\AgdaInductiveConstructor{base}\AgdaSpace{}%
\AgdaSymbol{:}\AgdaSpace{}%
\AgdaDatatype{Type}\<%
\\
\>[4]\AgdaOperator{\AgdaInductiveConstructor{\AgdaUnderscore{}⇒\AgdaUnderscore{}}}\AgdaSpace{}%
\AgdaSymbol{:}\AgdaSpace{}%
\AgdaDatatype{Type}\AgdaSpace{}%
\AgdaSymbol{→}\AgdaSpace{}%
\AgdaDatatype{Type}\AgdaSpace{}%
\AgdaSymbol{→}\AgdaSpace{}%
\AgdaDatatype{Type}\<%
\\
\\[\AgdaEmptyExtraSkip]%
\>[2]\AgdaKeyword{variable}\AgdaSpace{}%
\AgdaGeneralizable{S}\AgdaSpace{}%
\AgdaGeneralizable{T}\AgdaSpace{}%
\AgdaSymbol{:}\AgdaSpace{}%
\AgdaDatatype{Type}\<%
\\
\\[\AgdaEmptyExtraSkip]%
\>[2]\AgdaComment{--\ Contexts}\<%
\\
\>[2]\AgdaKeyword{data}\AgdaSpace{}%
\AgdaDatatype{Ctx}\AgdaSpace{}%
\AgdaSymbol{:}\AgdaSpace{}%
\AgdaPrimitive{Set}\AgdaSpace{}%
\AgdaKeyword{where}\<%
\\
\>[2][@{}l@{\AgdaIndent{0}}]%
\>[4]\AgdaInductiveConstructor{∅}\AgdaSpace{}%
\AgdaSymbol{:}\AgdaSpace{}%
\AgdaDatatype{Ctx}\<%
\\
\>[4]\AgdaOperator{\AgdaInductiveConstructor{\AgdaUnderscore{},\AgdaUnderscore{}∷\AgdaUnderscore{}}}\AgdaSpace{}%
\AgdaSymbol{:}\AgdaSpace{}%
\AgdaDatatype{Ctx}\AgdaSpace{}%
\AgdaSymbol{→}\AgdaSpace{}%
\AgdaBound{Id}\AgdaSpace{}%
\AgdaSymbol{→}\AgdaSpace{}%
\AgdaDatatype{Type}\AgdaSpace{}%
\AgdaSymbol{→}\AgdaSpace{}%
\AgdaDatatype{Ctx}\<%
\\
\\[\AgdaEmptyExtraSkip]%
\>[2]\AgdaKeyword{variable}\AgdaSpace{}%
\AgdaGeneralizable{Γ}\AgdaSpace{}%
\AgdaGeneralizable{Δ}\AgdaSpace{}%
\AgdaSymbol{:}\AgdaSpace{}%
\AgdaDatatype{Ctx}\<%
\\
\\[\AgdaEmptyExtraSkip]%
\>[2]\AgdaComment{--\ Terms}\<%
\\
\>[2]\AgdaKeyword{data}\AgdaSpace{}%
\AgdaDatatype{Term}\AgdaSpace{}%
\AgdaSymbol{:}\AgdaSpace{}%
\AgdaPrimitive{Set}\AgdaSpace{}%
\AgdaKeyword{where}\<%
\\
\>[2][@{}l@{\AgdaIndent{0}}]%
\>[4]\AgdaInductiveConstructor{var}\AgdaSpace{}%
\AgdaSymbol{:}\AgdaSpace{}%
\AgdaBound{Id}\AgdaSpace{}%
\AgdaSymbol{→}\AgdaSpace{}%
\AgdaDatatype{Term}\<%
\\
\>[4]\AgdaOperator{\AgdaInductiveConstructor{ƛ\AgdaUnderscore{}⟶\AgdaUnderscore{}}}\AgdaSpace{}%
\AgdaSymbol{:}\AgdaSpace{}%
\AgdaBound{Id}\AgdaSpace{}%
\AgdaSymbol{→}\AgdaSpace{}%
\AgdaDatatype{Term}\AgdaSpace{}%
\AgdaSymbol{→}\AgdaSpace{}%
\AgdaDatatype{Term}\<%
\\
\>[4]\AgdaOperator{\AgdaInductiveConstructor{\AgdaUnderscore{}·\AgdaUnderscore{}}}\AgdaSpace{}%
\AgdaSymbol{:}\AgdaSpace{}%
\AgdaDatatype{Term}\AgdaSpace{}%
\AgdaSymbol{→}\AgdaSpace{}%
\AgdaDatatype{Term}\AgdaSpace{}%
\AgdaSymbol{→}\AgdaSpace{}%
\AgdaDatatype{Term}\<%
\\
\\[\AgdaEmptyExtraSkip]%
\>[2]\AgdaKeyword{variable}\AgdaSpace{}%
\AgdaGeneralizable{r}\AgdaSpace{}%
\AgdaGeneralizable{s}\AgdaSpace{}%
\AgdaGeneralizable{t}\AgdaSpace{}%
\AgdaSymbol{:}\AgdaSpace{}%
\AgdaDatatype{Term}\<%
\\
\\[\AgdaEmptyExtraSkip]%
\>[2]\AgdaComment{--\ Variable\ lookup\ judgement\ into\ a\ context}\<%
\\
\>[2]\AgdaKeyword{data}\AgdaSpace{}%
\AgdaOperator{\AgdaDatatype{\AgdaUnderscore{}∷\AgdaUnderscore{}∈\AgdaUnderscore{}}}\AgdaSpace{}%
\AgdaSymbol{:}\AgdaSpace{}%
\AgdaBound{Id}\AgdaSpace{}%
\AgdaSymbol{→}\AgdaSpace{}%
\AgdaDatatype{Type}\AgdaSpace{}%
\AgdaSymbol{→}\AgdaSpace{}%
\AgdaDatatype{Ctx}\AgdaSpace{}%
\AgdaSymbol{→}\AgdaSpace{}%
\AgdaPrimitive{Set}\AgdaSpace{}%
\AgdaKeyword{where}\<%
\\
\>[2][@{}l@{\AgdaIndent{0}}]%
\>[4]\AgdaInductiveConstructor{here}\AgdaSpace{}%
\AgdaSymbol{:}\AgdaSpace{}%
\AgdaGeneralizable{x}\AgdaSpace{}%
\AgdaOperator{\AgdaDatatype{∷}}\AgdaSpace{}%
\AgdaGeneralizable{T}\AgdaSpace{}%
\AgdaOperator{\AgdaDatatype{∈}}\AgdaSpace{}%
\AgdaGeneralizable{Γ}\AgdaSpace{}%
\AgdaOperator{\AgdaInductiveConstructor{,}}\AgdaSpace{}%
\AgdaGeneralizable{x}\AgdaSpace{}%
\AgdaOperator{\AgdaInductiveConstructor{∷}}\AgdaSpace{}%
\AgdaGeneralizable{T}\<%
\\
\>[4]\AgdaInductiveConstructor{there}\AgdaSpace{}%
\AgdaSymbol{:}\AgdaSpace{}%
\AgdaGeneralizable{x}\AgdaSpace{}%
\AgdaOperator{\AgdaDatatype{∷}}\AgdaSpace{}%
\AgdaGeneralizable{T}\AgdaSpace{}%
\AgdaOperator{\AgdaDatatype{∈}}\AgdaSpace{}%
\AgdaGeneralizable{Γ}\AgdaSpace{}%
\AgdaSymbol{→}\AgdaSpace{}%
\AgdaOperator{\AgdaFunction{¬}}\AgdaSpace{}%
\AgdaSymbol{(}\AgdaGeneralizable{x}\AgdaSpace{}%
\AgdaOperator{\AgdaDatatype{≡}}\AgdaSpace{}%
\AgdaGeneralizable{y}\AgdaSymbol{)}\AgdaSpace{}%
\AgdaSymbol{→}\AgdaSpace{}%
\AgdaGeneralizable{x}\AgdaSpace{}%
\AgdaOperator{\AgdaDatatype{∷}}\AgdaSpace{}%
\AgdaGeneralizable{T}\AgdaSpace{}%
\AgdaOperator{\AgdaDatatype{∈}}\AgdaSpace{}%
\AgdaGeneralizable{Γ}\AgdaSpace{}%
\AgdaOperator{\AgdaInductiveConstructor{,}}\AgdaSpace{}%
\AgdaGeneralizable{y}\AgdaSpace{}%
\AgdaOperator{\AgdaInductiveConstructor{∷}}\AgdaSpace{}%
\AgdaGeneralizable{S}\<%
\\
\\[\AgdaEmptyExtraSkip]%
\>[2]\AgdaComment{--\ Syntactic\ typing}\<%
\\
\>[2]\AgdaKeyword{data}\AgdaSpace{}%
\AgdaOperator{\AgdaDatatype{\AgdaUnderscore{}⊢\AgdaUnderscore{}∷\AgdaUnderscore{}}}\AgdaSpace{}%
\AgdaSymbol{:}\AgdaSpace{}%
\AgdaDatatype{Ctx}\AgdaSpace{}%
\AgdaSymbol{→}\AgdaSpace{}%
\AgdaDatatype{Term}\AgdaSpace{}%
\AgdaSymbol{→}\AgdaSpace{}%
\AgdaDatatype{Type}\AgdaSpace{}%
\AgdaSymbol{→}\AgdaSpace{}%
\AgdaPrimitive{Set}\AgdaSpace{}%
\AgdaKeyword{where}\<%
\\
\>[2][@{}l@{\AgdaIndent{0}}]%
\>[4]\AgdaInductiveConstructor{⊢var}\AgdaSpace{}%
\AgdaSymbol{:}\AgdaSpace{}%
\AgdaGeneralizable{x}\AgdaSpace{}%
\AgdaOperator{\AgdaDatatype{∷}}\AgdaSpace{}%
\AgdaGeneralizable{T}\AgdaSpace{}%
\AgdaOperator{\AgdaDatatype{∈}}\AgdaSpace{}%
\AgdaGeneralizable{Γ}\AgdaSpace{}%
\AgdaSymbol{→}\AgdaSpace{}%
\AgdaGeneralizable{Γ}\AgdaSpace{}%
\AgdaOperator{\AgdaDatatype{⊢}}\AgdaSpace{}%
\AgdaInductiveConstructor{var}\AgdaSpace{}%
\AgdaGeneralizable{x}\AgdaSpace{}%
\AgdaOperator{\AgdaDatatype{∷}}\AgdaSpace{}%
\AgdaGeneralizable{T}\<%
\\
\\[\AgdaEmptyExtraSkip]%
\>[4]\AgdaInductiveConstructor{⊢abs}\AgdaSpace{}%
\AgdaSymbol{:}\AgdaSpace{}%
\AgdaGeneralizable{Γ}\AgdaSpace{}%
\AgdaOperator{\AgdaInductiveConstructor{,}}\AgdaSpace{}%
\AgdaGeneralizable{x}\AgdaSpace{}%
\AgdaOperator{\AgdaInductiveConstructor{∷}}\AgdaSpace{}%
\AgdaGeneralizable{S}\AgdaSpace{}%
\AgdaOperator{\AgdaDatatype{⊢}}\AgdaSpace{}%
\AgdaGeneralizable{t}\AgdaSpace{}%
\AgdaOperator{\AgdaDatatype{∷}}\AgdaSpace{}%
\AgdaGeneralizable{T}\AgdaSpace{}%
\AgdaSymbol{→}\AgdaSpace{}%
\AgdaGeneralizable{Γ}\AgdaSpace{}%
\AgdaOperator{\AgdaDatatype{⊢}}\AgdaSpace{}%
\AgdaOperator{\AgdaInductiveConstructor{ƛ}}\AgdaSpace{}%
\AgdaGeneralizable{x}\AgdaSpace{}%
\AgdaOperator{\AgdaInductiveConstructor{⟶}}\AgdaSpace{}%
\AgdaGeneralizable{t}\AgdaSpace{}%
\AgdaOperator{\AgdaDatatype{∷}}\AgdaSpace{}%
\AgdaGeneralizable{S}\AgdaSpace{}%
\AgdaOperator{\AgdaInductiveConstructor{⇒}}\AgdaSpace{}%
\AgdaGeneralizable{T}\<%
\\
\\[\AgdaEmptyExtraSkip]%
\>[4]\AgdaInductiveConstructor{⊢app}\AgdaSpace{}%
\AgdaSymbol{:}\AgdaSpace{}%
\AgdaGeneralizable{Γ}\AgdaSpace{}%
\AgdaOperator{\AgdaDatatype{⊢}}\AgdaSpace{}%
\AgdaGeneralizable{r}\AgdaSpace{}%
\AgdaOperator{\AgdaDatatype{∷}}\AgdaSpace{}%
\AgdaGeneralizable{S}\AgdaSpace{}%
\AgdaOperator{\AgdaInductiveConstructor{⇒}}\AgdaSpace{}%
\AgdaGeneralizable{T}\AgdaSpace{}%
\AgdaSymbol{→}\AgdaSpace{}%
\AgdaGeneralizable{Γ}\AgdaSpace{}%
\AgdaOperator{\AgdaDatatype{⊢}}\AgdaSpace{}%
\AgdaGeneralizable{s}\AgdaSpace{}%
\AgdaOperator{\AgdaDatatype{∷}}\AgdaSpace{}%
\AgdaGeneralizable{S}\AgdaSpace{}%
\AgdaSymbol{→}\AgdaSpace{}%
\AgdaGeneralizable{Γ}\AgdaSpace{}%
\AgdaOperator{\AgdaDatatype{⊢}}\AgdaSpace{}%
\AgdaGeneralizable{r}\AgdaSpace{}%
\AgdaOperator{\AgdaInductiveConstructor{·}}\AgdaSpace{}%
\AgdaGeneralizable{s}\AgdaSpace{}%
\AgdaOperator{\AgdaDatatype{∷}}\AgdaSpace{}%
\AgdaGeneralizable{T}\<%
\end{code}

\subsection{Natural semantics}

\begin{code}%
\>[2]\AgdaComment{--\ Environments\ +\ domains}\<%
\\
\>[2]\AgdaKeyword{mutual}\<%
\\
\>[2][@{}l@{\AgdaIndent{0}}]%
\>[4]\AgdaKeyword{data}\AgdaSpace{}%
\AgdaDatatype{Env}\AgdaSpace{}%
\AgdaSymbol{:}\AgdaSpace{}%
\AgdaPrimitive{Set}\AgdaSpace{}%
\AgdaKeyword{where}\<%
\\
\>[4][@{}l@{\AgdaIndent{0}}]%
\>[6]\AgdaInductiveConstructor{∅}\AgdaSpace{}%
\AgdaSymbol{:}\AgdaSpace{}%
\AgdaDatatype{Env}\<%
\\
\>[6]\AgdaOperator{\AgdaInductiveConstructor{\AgdaUnderscore{},\AgdaUnderscore{}↦\AgdaUnderscore{}}}\AgdaSpace{}%
\AgdaSymbol{:}\AgdaSpace{}%
\AgdaDatatype{Env}\AgdaSpace{}%
\AgdaSymbol{→}\AgdaSpace{}%
\AgdaBound{Id}\AgdaSpace{}%
\AgdaSymbol{→}\AgdaSpace{}%
\AgdaDatatype{Domain}\AgdaSpace{}%
\AgdaSymbol{→}\AgdaSpace{}%
\AgdaDatatype{Env}\<%
\\
\\[\AgdaEmptyExtraSkip]%
\>[4]\AgdaKeyword{data}\AgdaSpace{}%
\AgdaDatatype{Domain}\AgdaSpace{}%
\AgdaSymbol{:}\AgdaSpace{}%
\AgdaPrimitive{Set}\AgdaSpace{}%
\AgdaKeyword{where}\<%
\\
\>[4][@{}l@{\AgdaIndent{0}}]%
\>[6]\AgdaOperator{\AgdaInductiveConstructor{⟨ƛ\AgdaUnderscore{}⟶\AgdaUnderscore{}⟩\AgdaUnderscore{}}}\AgdaSpace{}%
\AgdaSymbol{:}\AgdaSpace{}%
\AgdaBound{Id}\AgdaSpace{}%
\AgdaSymbol{→}\AgdaSpace{}%
\AgdaDatatype{Term}\AgdaSpace{}%
\AgdaSymbol{→}\AgdaSpace{}%
\AgdaDatatype{Env}\AgdaSpace{}%
\AgdaSymbol{→}\AgdaSpace{}%
\AgdaDatatype{Domain}\<%
\\
\\[\AgdaEmptyExtraSkip]%
\>[2]\AgdaKeyword{variable}\AgdaSpace{}%
\AgdaGeneralizable{γ}\AgdaSpace{}%
\AgdaGeneralizable{δ}\AgdaSpace{}%
\AgdaSymbol{:}\AgdaSpace{}%
\AgdaDatatype{Env}\<%
\\
\>[2]\AgdaKeyword{variable}\AgdaSpace{}%
\AgdaGeneralizable{a}\AgdaSpace{}%
\AgdaGeneralizable{b}\AgdaSpace{}%
\AgdaGeneralizable{f}\AgdaSpace{}%
\AgdaSymbol{:}\AgdaSpace{}%
\AgdaDatatype{Domain}\<%
\\
\\[\AgdaEmptyExtraSkip]%
\>[2]\AgdaComment{--\ Variable\ lookup\ judgement\ into\ an\ environment}\<%
\\
\>[2]\AgdaKeyword{data}\AgdaSpace{}%
\AgdaOperator{\AgdaDatatype{\AgdaUnderscore{}↦\AgdaUnderscore{}∈\AgdaUnderscore{}}}\AgdaSpace{}%
\AgdaSymbol{:}\AgdaSpace{}%
\AgdaBound{Id}\AgdaSpace{}%
\AgdaSymbol{→}\AgdaSpace{}%
\AgdaDatatype{Domain}\AgdaSpace{}%
\AgdaSymbol{→}\AgdaSpace{}%
\AgdaDatatype{Env}\AgdaSpace{}%
\AgdaSymbol{→}\AgdaSpace{}%
\AgdaPrimitive{Set}\AgdaSpace{}%
\AgdaKeyword{where}\<%
\\
\>[2][@{}l@{\AgdaIndent{0}}]%
\>[4]\AgdaInductiveConstructor{here}\AgdaSpace{}%
\AgdaSymbol{:}\AgdaSpace{}%
\AgdaGeneralizable{x}\AgdaSpace{}%
\AgdaOperator{\AgdaDatatype{↦}}\AgdaSpace{}%
\AgdaGeneralizable{a}\AgdaSpace{}%
\AgdaOperator{\AgdaDatatype{∈}}\AgdaSpace{}%
\AgdaGeneralizable{γ}\AgdaSpace{}%
\AgdaOperator{\AgdaInductiveConstructor{,}}\AgdaSpace{}%
\AgdaGeneralizable{x}\AgdaSpace{}%
\AgdaOperator{\AgdaInductiveConstructor{↦}}\AgdaSpace{}%
\AgdaGeneralizable{a}\<%
\\
\>[4]\AgdaInductiveConstructor{there}\AgdaSpace{}%
\AgdaSymbol{:}\AgdaSpace{}%
\AgdaGeneralizable{x}\AgdaSpace{}%
\AgdaOperator{\AgdaDatatype{↦}}\AgdaSpace{}%
\AgdaGeneralizable{a}\AgdaSpace{}%
\AgdaOperator{\AgdaDatatype{∈}}\AgdaSpace{}%
\AgdaGeneralizable{γ}\AgdaSpace{}%
\AgdaSymbol{→}\AgdaSpace{}%
\AgdaOperator{\AgdaFunction{¬}}\AgdaSpace{}%
\AgdaSymbol{(}\AgdaGeneralizable{x}\AgdaSpace{}%
\AgdaOperator{\AgdaDatatype{≡}}\AgdaSpace{}%
\AgdaGeneralizable{y}\AgdaSymbol{)}\AgdaSpace{}%
\AgdaSymbol{→}\AgdaSpace{}%
\AgdaGeneralizable{x}\AgdaSpace{}%
\AgdaOperator{\AgdaDatatype{↦}}\AgdaSpace{}%
\AgdaGeneralizable{a}\AgdaSpace{}%
\AgdaOperator{\AgdaDatatype{∈}}\AgdaSpace{}%
\AgdaGeneralizable{γ}\AgdaSpace{}%
\AgdaOperator{\AgdaInductiveConstructor{,}}\AgdaSpace{}%
\AgdaGeneralizable{y}\AgdaSpace{}%
\AgdaOperator{\AgdaInductiveConstructor{↦}}\AgdaSpace{}%
\AgdaGeneralizable{b}\<%
\\
\\[\AgdaEmptyExtraSkip]%
\>[2]\AgdaComment{--\ Natural\ semantics}\<%
\\
\>[2]\AgdaKeyword{data}\AgdaSpace{}%
\AgdaOperator{\AgdaDatatype{\AgdaUnderscore{}∣\AgdaUnderscore{}⇓\AgdaUnderscore{}}}\AgdaSpace{}%
\AgdaSymbol{:}\AgdaSpace{}%
\AgdaDatatype{Env}\AgdaSpace{}%
\AgdaSymbol{→}\AgdaSpace{}%
\AgdaDatatype{Term}\AgdaSpace{}%
\AgdaSymbol{→}\AgdaSpace{}%
\AgdaDatatype{Domain}\AgdaSpace{}%
\AgdaSymbol{→}\AgdaSpace{}%
\AgdaPrimitive{Set}\AgdaSpace{}%
\AgdaKeyword{where}\<%
\\
\>[2][@{}l@{\AgdaIndent{0}}]%
\>[4]\AgdaInductiveConstructor{evalVar}\AgdaSpace{}%
\AgdaSymbol{:}\AgdaSpace{}%
\AgdaGeneralizable{x}\AgdaSpace{}%
\AgdaOperator{\AgdaDatatype{↦}}\AgdaSpace{}%
\AgdaGeneralizable{a}\AgdaSpace{}%
\AgdaOperator{\AgdaDatatype{∈}}\AgdaSpace{}%
\AgdaGeneralizable{γ}\AgdaSpace{}%
\AgdaSymbol{→}\AgdaSpace{}%
\AgdaGeneralizable{γ}\AgdaSpace{}%
\AgdaOperator{\AgdaDatatype{∣}}\AgdaSpace{}%
\AgdaInductiveConstructor{var}\AgdaSpace{}%
\AgdaGeneralizable{x}\AgdaSpace{}%
\AgdaOperator{\AgdaDatatype{⇓}}\AgdaSpace{}%
\AgdaGeneralizable{a}\<%
\\
\\[\AgdaEmptyExtraSkip]%
\>[4]\AgdaInductiveConstructor{evalAbs}\AgdaSpace{}%
\AgdaSymbol{:}\AgdaSpace{}%
\AgdaGeneralizable{γ}\AgdaSpace{}%
\AgdaOperator{\AgdaDatatype{∣}}\AgdaSpace{}%
\AgdaOperator{\AgdaInductiveConstructor{ƛ}}\AgdaSpace{}%
\AgdaGeneralizable{x}\AgdaSpace{}%
\AgdaOperator{\AgdaInductiveConstructor{⟶}}\AgdaSpace{}%
\AgdaGeneralizable{t}\AgdaSpace{}%
\AgdaOperator{\AgdaDatatype{⇓}}\AgdaSpace{}%
\AgdaOperator{\AgdaInductiveConstructor{⟨ƛ}}\AgdaSpace{}%
\AgdaGeneralizable{x}\AgdaSpace{}%
\AgdaOperator{\AgdaInductiveConstructor{⟶}}\AgdaSpace{}%
\AgdaGeneralizable{t}\AgdaSpace{}%
\AgdaOperator{\AgdaInductiveConstructor{⟩}}\AgdaSpace{}%
\AgdaGeneralizable{γ}\<%
\\
\\[\AgdaEmptyExtraSkip]%
\>[4]\AgdaInductiveConstructor{evalApp}\AgdaSpace{}%
\AgdaSymbol{:}\AgdaSpace{}%
\AgdaGeneralizable{γ}\AgdaSpace{}%
\AgdaOperator{\AgdaDatatype{∣}}\AgdaSpace{}%
\AgdaGeneralizable{r}\AgdaSpace{}%
\AgdaOperator{\AgdaDatatype{⇓}}\AgdaSpace{}%
\AgdaOperator{\AgdaInductiveConstructor{⟨ƛ}}\AgdaSpace{}%
\AgdaGeneralizable{x}\AgdaSpace{}%
\AgdaOperator{\AgdaInductiveConstructor{⟶}}\AgdaSpace{}%
\AgdaGeneralizable{t}\AgdaSpace{}%
\AgdaOperator{\AgdaInductiveConstructor{⟩}}\AgdaSpace{}%
\AgdaGeneralizable{δ}\<%
\\
\>[4][@{}l@{\AgdaIndent{0}}]%
\>[9]\AgdaSymbol{→}\AgdaSpace{}%
\AgdaGeneralizable{γ}\AgdaSpace{}%
\AgdaOperator{\AgdaDatatype{∣}}\AgdaSpace{}%
\AgdaGeneralizable{s}\AgdaSpace{}%
\AgdaOperator{\AgdaDatatype{⇓}}\AgdaSpace{}%
\AgdaGeneralizable{a}\<%
\\
\>[9]\AgdaSymbol{→}\AgdaSpace{}%
\AgdaGeneralizable{δ}\AgdaSpace{}%
\AgdaOperator{\AgdaInductiveConstructor{,}}\AgdaSpace{}%
\AgdaGeneralizable{x}\AgdaSpace{}%
\AgdaOperator{\AgdaInductiveConstructor{↦}}\AgdaSpace{}%
\AgdaGeneralizable{a}\AgdaSpace{}%
\AgdaOperator{\AgdaDatatype{∣}}\AgdaSpace{}%
\AgdaGeneralizable{t}\AgdaSpace{}%
\AgdaOperator{\AgdaDatatype{⇓}}\AgdaSpace{}%
\AgdaGeneralizable{b}\<%
\\
\>[9]\AgdaSymbol{→}\AgdaSpace{}%
\AgdaGeneralizable{γ}\AgdaSpace{}%
\AgdaOperator{\AgdaDatatype{∣}}\AgdaSpace{}%
\AgdaGeneralizable{r}\AgdaSpace{}%
\AgdaOperator{\AgdaInductiveConstructor{·}}\AgdaSpace{}%
\AgdaGeneralizable{s}\AgdaSpace{}%
\AgdaOperator{\AgdaDatatype{⇓}}\AgdaSpace{}%
\AgdaGeneralizable{b}\<%
\end{code}

\subsection{Proof by logical relations}

\begin{code}%
\>[2]\AgdaComment{--\ Semantic\ Types}\<%
\\
\>[2]\AgdaOperator{\AgdaFunction{⟦\AgdaUnderscore{}⟧}}\AgdaSpace{}%
\AgdaSymbol{:}\AgdaSpace{}%
\AgdaDatatype{Type}\AgdaSpace{}%
\AgdaSymbol{→}\AgdaSpace{}%
\AgdaSymbol{(}\AgdaDatatype{Domain}\AgdaSpace{}%
\AgdaSymbol{→}\AgdaSpace{}%
\AgdaPrimitive{Set}\AgdaSymbol{)}\<%
\\
\>[2]\AgdaOperator{\AgdaFunction{⟦}}\AgdaSpace{}%
\AgdaInductiveConstructor{base}\AgdaSpace{}%
\AgdaOperator{\AgdaFunction{⟧}}\AgdaSpace{}%
\AgdaSymbol{\AgdaUnderscore{}}\AgdaSpace{}%
\AgdaSymbol{=}\AgdaSpace{}%
\AgdaDatatype{⊥}\<%
\\
\>[2]\AgdaOperator{\AgdaFunction{⟦}}%
\>[5975I]\AgdaBound{S}\AgdaSpace{}%
\AgdaOperator{\AgdaInductiveConstructor{⇒}}\AgdaSpace{}%
\AgdaBound{T}\AgdaSpace{}%
\AgdaOperator{\AgdaFunction{⟧}}\AgdaSpace{}%
\AgdaSymbol{(}\AgdaOperator{\AgdaInductiveConstructor{⟨ƛ}}\AgdaSpace{}%
\AgdaBound{x}\AgdaSpace{}%
\AgdaOperator{\AgdaInductiveConstructor{⟶}}\AgdaSpace{}%
\AgdaBound{t}\AgdaSpace{}%
\AgdaOperator{\AgdaInductiveConstructor{⟩}}\AgdaSpace{}%
\AgdaBound{δ}\AgdaSymbol{)}\AgdaSpace{}%
\AgdaSymbol{=}\<%
\\
\>[.][@{}l@{}]\<[5975I]%
\>[4]\AgdaSymbol{∀}\AgdaSpace{}%
\AgdaSymbol{\{}\AgdaBound{a}\AgdaSpace{}%
\AgdaSymbol{:}\AgdaSpace{}%
\AgdaDatatype{Domain}\AgdaSymbol{\}}\<%
\\
\>[4]\AgdaSymbol{→}\AgdaSpace{}%
\AgdaBound{a}\AgdaSpace{}%
\AgdaOperator{\AgdaFunction{∈}}\AgdaSpace{}%
\AgdaOperator{\AgdaFunction{⟦}}\AgdaSpace{}%
\AgdaBound{S}\AgdaSpace{}%
\AgdaOperator{\AgdaFunction{⟧}}\<%
\\
\>[4]\AgdaSymbol{→}\AgdaSpace{}%
\AgdaFunction{∃[}\AgdaSpace{}%
\AgdaBound{b}\AgdaSpace{}%
\AgdaFunction{]}\AgdaSpace{}%
\AgdaBound{δ}\AgdaSpace{}%
\AgdaOperator{\AgdaInductiveConstructor{,}}\AgdaSpace{}%
\AgdaBound{x}\AgdaSpace{}%
\AgdaOperator{\AgdaInductiveConstructor{↦}}\AgdaSpace{}%
\AgdaBound{a}\AgdaSpace{}%
\AgdaOperator{\AgdaDatatype{∣}}\AgdaSpace{}%
\AgdaBound{t}\AgdaSpace{}%
\AgdaOperator{\AgdaDatatype{⇓}}\AgdaSpace{}%
\AgdaBound{b}\AgdaSpace{}%
\AgdaOperator{\AgdaFunction{×}}\AgdaSpace{}%
\AgdaBound{b}\AgdaSpace{}%
\AgdaOperator{\AgdaFunction{∈}}\AgdaSpace{}%
\AgdaOperator{\AgdaFunction{⟦}}\AgdaSpace{}%
\AgdaBound{T}\AgdaSpace{}%
\AgdaOperator{\AgdaFunction{⟧}}\<%
\\
\\[\AgdaEmptyExtraSkip]%
\>[2]\AgdaComment{--\ Semantic\ typing\ for\ environments}\<%
\\
\>[2]\AgdaOperator{\AgdaFunction{\AgdaUnderscore{}⊨\AgdaUnderscore{}}}\AgdaSpace{}%
\AgdaSymbol{:}\AgdaSpace{}%
\AgdaDatatype{Ctx}\AgdaSpace{}%
\AgdaSymbol{→}\AgdaSpace{}%
\AgdaDatatype{Env}\AgdaSpace{}%
\AgdaSymbol{→}\AgdaSpace{}%
\AgdaPrimitive{Set}\<%
\\
\>[2]\AgdaBound{Γ}\AgdaSpace{}%
\AgdaOperator{\AgdaFunction{⊨}}\AgdaSpace{}%
\AgdaBound{γ}\AgdaSpace{}%
\AgdaSymbol{=}%
\>[6021I]\AgdaSymbol{∀}\AgdaSpace{}%
\AgdaSymbol{\{}\AgdaBound{x}\AgdaSpace{}%
\AgdaSymbol{:}\AgdaSpace{}%
\AgdaBound{Id}\AgdaSymbol{\}}\AgdaSpace{}%
\AgdaSymbol{\{}\AgdaBound{T}\AgdaSpace{}%
\AgdaSymbol{:}\AgdaSpace{}%
\AgdaDatatype{Type}\AgdaSymbol{\}}\<%
\\
\>[.][@{}l@{}]\<[6021I]%
\>[10]\AgdaSymbol{→}\AgdaSpace{}%
\AgdaBound{x}\AgdaSpace{}%
\AgdaOperator{\AgdaDatatype{∷}}\AgdaSpace{}%
\AgdaBound{T}\AgdaSpace{}%
\AgdaOperator{\AgdaDatatype{∈}}\AgdaSpace{}%
\AgdaBound{Γ}\<%
\\
\>[10]\AgdaSymbol{→}\AgdaSpace{}%
\AgdaFunction{∃[}\AgdaSpace{}%
\AgdaBound{a}\AgdaSpace{}%
\AgdaFunction{]}\AgdaSpace{}%
\AgdaSymbol{(}\AgdaBound{x}\AgdaSpace{}%
\AgdaOperator{\AgdaDatatype{↦}}\AgdaSpace{}%
\AgdaBound{a}\AgdaSpace{}%
\AgdaOperator{\AgdaDatatype{∈}}\AgdaSpace{}%
\AgdaBound{γ}\AgdaSymbol{)}\AgdaSpace{}%
\AgdaOperator{\AgdaFunction{×}}\AgdaSpace{}%
\AgdaBound{a}\AgdaSpace{}%
\AgdaOperator{\AgdaFunction{∈}}\AgdaSpace{}%
\AgdaOperator{\AgdaFunction{⟦}}\AgdaSpace{}%
\AgdaBound{T}\AgdaSpace{}%
\AgdaOperator{\AgdaFunction{⟧}}\<%
\\
\\[\AgdaEmptyExtraSkip]%
\>[2]\AgdaComment{--\ Extending\ semantically\ typed\ environments}\<%
\\
\>[2]\AgdaOperator{\AgdaFunction{\AgdaUnderscore{}\textasciicircum{}\AgdaUnderscore{}}}\AgdaSpace{}%
\AgdaSymbol{:}\AgdaSpace{}%
\AgdaGeneralizable{Γ}\AgdaSpace{}%
\AgdaOperator{\AgdaFunction{⊨}}\AgdaSpace{}%
\AgdaGeneralizable{γ}\AgdaSpace{}%
\AgdaSymbol{→}\AgdaSpace{}%
\AgdaGeneralizable{a}\AgdaSpace{}%
\AgdaOperator{\AgdaFunction{∈}}\AgdaSpace{}%
\AgdaOperator{\AgdaFunction{⟦}}\AgdaSpace{}%
\AgdaGeneralizable{T}\AgdaSpace{}%
\AgdaOperator{\AgdaFunction{⟧}}\AgdaSpace{}%
\AgdaSymbol{→}\AgdaSpace{}%
\AgdaGeneralizable{Γ}\AgdaSpace{}%
\AgdaOperator{\AgdaInductiveConstructor{,}}\AgdaSpace{}%
\AgdaGeneralizable{x}\AgdaSpace{}%
\AgdaOperator{\AgdaInductiveConstructor{∷}}\AgdaSpace{}%
\AgdaGeneralizable{T}\AgdaSpace{}%
\AgdaOperator{\AgdaFunction{⊨}}\AgdaSpace{}%
\AgdaGeneralizable{γ}\AgdaSpace{}%
\AgdaOperator{\AgdaInductiveConstructor{,}}\AgdaSpace{}%
\AgdaGeneralizable{x}\AgdaSpace{}%
\AgdaOperator{\AgdaInductiveConstructor{↦}}\AgdaSpace{}%
\AgdaGeneralizable{a}\<%
\\
\>[2]\AgdaOperator{\AgdaFunction{\AgdaUnderscore{}\textasciicircum{}\AgdaUnderscore{}}}\AgdaSpace{}%
\AgdaSymbol{\{}\AgdaArgument{a}\AgdaSpace{}%
\AgdaSymbol{=}\AgdaSpace{}%
\AgdaBound{a}\AgdaSymbol{\}}\AgdaSpace{}%
\AgdaSymbol{\{}\AgdaArgument{x}\AgdaSpace{}%
\AgdaSymbol{=}\AgdaSpace{}%
\AgdaBound{x}\AgdaSymbol{\}}\AgdaSpace{}%
\AgdaSymbol{\AgdaUnderscore{}}\AgdaSpace{}%
\AgdaBound{sa}\AgdaSpace{}%
\AgdaInductiveConstructor{here}\AgdaSpace{}%
\AgdaSymbol{=}\AgdaSpace{}%
\AgdaBound{a}\AgdaSpace{}%
\AgdaOperator{\AgdaInductiveConstructor{,}}\AgdaSpace{}%
\AgdaInductiveConstructor{here}\AgdaSpace{}%
\AgdaOperator{\AgdaInductiveConstructor{,}}\AgdaSpace{}%
\AgdaBound{sa}\<%
\\
\>[2]\AgdaOperator{\AgdaFunction{\AgdaUnderscore{}\textasciicircum{}\AgdaUnderscore{}}}\AgdaSpace{}%
\AgdaBound{⊨γ}\AgdaSpace{}%
\AgdaSymbol{\AgdaUnderscore{}}\AgdaSpace{}%
\AgdaSymbol{(}\AgdaInductiveConstructor{there}\AgdaSpace{}%
\AgdaBound{x∈Γ}\AgdaSpace{}%
\AgdaBound{x≢y}\AgdaSymbol{)}\AgdaSpace{}%
\AgdaSymbol{=}\<%
\\
\>[2][@{}l@{\AgdaIndent{0}}]%
\>[4]\AgdaKeyword{let}\AgdaSpace{}%
\AgdaSymbol{(}\AgdaBound{b}\AgdaSpace{}%
\AgdaOperator{\AgdaInductiveConstructor{,}}\AgdaSpace{}%
\AgdaBound{x∈γ}\AgdaSpace{}%
\AgdaOperator{\AgdaInductiveConstructor{,}}\AgdaSpace{}%
\AgdaBound{sb}\AgdaSymbol{)}\AgdaSpace{}%
\AgdaSymbol{=}\AgdaSpace{}%
\AgdaBound{⊨γ}\AgdaSpace{}%
\AgdaBound{x∈Γ}\AgdaSpace{}%
\AgdaKeyword{in}\<%
\\
\>[4]\AgdaBound{b}\AgdaSpace{}%
\AgdaOperator{\AgdaInductiveConstructor{,}}\AgdaSpace{}%
\AgdaInductiveConstructor{there}\AgdaSpace{}%
\AgdaBound{x∈γ}\AgdaSpace{}%
\AgdaBound{x≢y}\AgdaSpace{}%
\AgdaOperator{\AgdaInductiveConstructor{,}}\AgdaSpace{}%
\AgdaBound{sb}\<%
\\
\\[\AgdaEmptyExtraSkip]%
\>[2]\AgdaComment{--\ Semantic\ typing\ for\ terms}\<%
\\
\>[2]\AgdaOperator{\AgdaFunction{\AgdaUnderscore{}⊨\AgdaUnderscore{}∷\AgdaUnderscore{}}}\AgdaSpace{}%
\AgdaSymbol{:}\AgdaSpace{}%
\AgdaDatatype{Ctx}\AgdaSpace{}%
\AgdaSymbol{→}\AgdaSpace{}%
\AgdaDatatype{Term}\AgdaSpace{}%
\AgdaSymbol{→}\AgdaSpace{}%
\AgdaDatatype{Type}\AgdaSpace{}%
\AgdaSymbol{→}\AgdaSpace{}%
\AgdaPrimitive{Set}\<%
\\
\>[2]\AgdaBound{Γ}\AgdaSpace{}%
\AgdaOperator{\AgdaFunction{⊨}}\AgdaSpace{}%
\AgdaBound{t}\AgdaSpace{}%
\AgdaOperator{\AgdaFunction{∷}}\AgdaSpace{}%
\AgdaBound{T}\AgdaSpace{}%
\AgdaSymbol{=}%
\>[6118I]\AgdaSymbol{∀}\AgdaSpace{}%
\AgdaSymbol{\{}\AgdaBound{γ}\AgdaSpace{}%
\AgdaSymbol{:}\AgdaSpace{}%
\AgdaDatatype{Env}\AgdaSymbol{\}}\<%
\\
\>[.][@{}l@{}]\<[6118I]%
\>[14]\AgdaSymbol{→}\AgdaSpace{}%
\AgdaBound{Γ}\AgdaSpace{}%
\AgdaOperator{\AgdaFunction{⊨}}\AgdaSpace{}%
\AgdaBound{γ}\<%
\\
\>[14]\AgdaSymbol{→}\AgdaSpace{}%
\AgdaFunction{∃[}\AgdaSpace{}%
\AgdaBound{a}\AgdaSpace{}%
\AgdaFunction{]}\AgdaSpace{}%
\AgdaBound{γ}\AgdaSpace{}%
\AgdaOperator{\AgdaDatatype{∣}}\AgdaSpace{}%
\AgdaBound{t}\AgdaSpace{}%
\AgdaOperator{\AgdaDatatype{⇓}}\AgdaSpace{}%
\AgdaBound{a}\AgdaSpace{}%
\AgdaOperator{\AgdaFunction{×}}\AgdaSpace{}%
\AgdaBound{a}\AgdaSpace{}%
\AgdaOperator{\AgdaFunction{∈}}\AgdaSpace{}%
\AgdaOperator{\AgdaFunction{⟦}}\AgdaSpace{}%
\AgdaBound{T}\AgdaSpace{}%
\AgdaOperator{\AgdaFunction{⟧}}\<%
\\
\\[\AgdaEmptyExtraSkip]%
\>[2]\AgdaComment{--\ Well-typed\ terms\ are\ semantically\ typed}\<%
\\
\>[2]\AgdaFunction{fundamental-lemma}\AgdaSpace{}%
\AgdaSymbol{:}\AgdaSpace{}%
\AgdaGeneralizable{Γ}\AgdaSpace{}%
\AgdaOperator{\AgdaDatatype{⊢}}\AgdaSpace{}%
\AgdaGeneralizable{t}\AgdaSpace{}%
\AgdaOperator{\AgdaDatatype{∷}}\AgdaSpace{}%
\AgdaGeneralizable{T}\AgdaSpace{}%
\AgdaSymbol{→}\AgdaSpace{}%
\AgdaGeneralizable{Γ}\AgdaSpace{}%
\AgdaOperator{\AgdaFunction{⊨}}\AgdaSpace{}%
\AgdaGeneralizable{t}\AgdaSpace{}%
\AgdaOperator{\AgdaFunction{∷}}\AgdaSpace{}%
\AgdaGeneralizable{T}\<%
\\
\>[2]\AgdaFunction{fundamental-lemma}\AgdaSpace{}%
\AgdaSymbol{(}\AgdaInductiveConstructor{⊢var}\AgdaSpace{}%
\AgdaBound{x∈Γ}\AgdaSymbol{)}\AgdaSpace{}%
\AgdaBound{⊨γ}\AgdaSpace{}%
\AgdaSymbol{=}\<%
\\
\>[2][@{}l@{\AgdaIndent{0}}]%
\>[4]\AgdaKeyword{let}\AgdaSpace{}%
\AgdaSymbol{(}\AgdaBound{a}\AgdaSpace{}%
\AgdaOperator{\AgdaInductiveConstructor{,}}\AgdaSpace{}%
\AgdaBound{x∈γ}\AgdaSpace{}%
\AgdaOperator{\AgdaInductiveConstructor{,}}\AgdaSpace{}%
\AgdaBound{sa}\AgdaSymbol{)}\AgdaSpace{}%
\AgdaSymbol{=}\AgdaSpace{}%
\AgdaBound{⊨γ}\AgdaSpace{}%
\AgdaBound{x∈Γ}\AgdaSpace{}%
\AgdaKeyword{in}\<%
\\
\>[4]\AgdaBound{a}\AgdaSpace{}%
\AgdaOperator{\AgdaInductiveConstructor{,}}\AgdaSpace{}%
\AgdaInductiveConstructor{evalVar}\AgdaSpace{}%
\AgdaBound{x∈γ}\AgdaSpace{}%
\AgdaOperator{\AgdaInductiveConstructor{,}}\AgdaSpace{}%
\AgdaBound{sa}\<%
\\
\>[2]\AgdaFunction{fundamental-lemma}\AgdaSpace{}%
\AgdaSymbol{\{}\AgdaArgument{t}\AgdaSpace{}%
\AgdaSymbol{=}\AgdaSpace{}%
\AgdaOperator{\AgdaInductiveConstructor{ƛ}}\AgdaSpace{}%
\AgdaBound{x}\AgdaSpace{}%
\AgdaOperator{\AgdaInductiveConstructor{⟶}}\AgdaSpace{}%
\AgdaBound{t}\AgdaSymbol{\}}\AgdaSpace{}%
\AgdaSymbol{(}\AgdaInductiveConstructor{⊢abs}\AgdaSpace{}%
\AgdaBound{⊢t}\AgdaSymbol{)}\AgdaSpace{}%
\AgdaSymbol{\{}\AgdaBound{γ}\AgdaSymbol{\}}\AgdaSpace{}%
\AgdaBound{⊨γ}\AgdaSpace{}%
\AgdaSymbol{=}\<%
\\
\>[2][@{}l@{\AgdaIndent{0}}]%
\>[4]\AgdaOperator{\AgdaInductiveConstructor{⟨ƛ}}\AgdaSpace{}%
\AgdaBound{x}\AgdaSpace{}%
\AgdaOperator{\AgdaInductiveConstructor{⟶}}\AgdaSpace{}%
\AgdaBound{t}\AgdaSpace{}%
\AgdaOperator{\AgdaInductiveConstructor{⟩}}\AgdaSpace{}%
\AgdaBound{γ}\AgdaSpace{}%
\AgdaOperator{\AgdaInductiveConstructor{,}}\<%
\\
\>[4]\AgdaInductiveConstructor{evalAbs}\AgdaSpace{}%
\AgdaOperator{\AgdaInductiveConstructor{,}}\<%
\\
\>[4]\AgdaSymbol{λ}\AgdaSpace{}%
\AgdaBound{sa}\AgdaSpace{}%
\AgdaSymbol{→}\AgdaSpace{}%
\AgdaFunction{fundamental-lemma}\AgdaSpace{}%
\AgdaBound{⊢t}\AgdaSpace{}%
\AgdaSymbol{(}\AgdaBound{⊨γ}\AgdaSpace{}%
\AgdaOperator{\AgdaFunction{\textasciicircum{}}}\AgdaSpace{}%
\AgdaBound{sa}\AgdaSymbol{)}\<%
\\
\>[2]\AgdaFunction{fundamental-lemma}\AgdaSpace{}%
\AgdaSymbol{(}\AgdaInductiveConstructor{⊢app}\AgdaSpace{}%
\AgdaBound{⊢r}\AgdaSpace{}%
\AgdaBound{⊢s}\AgdaSymbol{)}\AgdaSpace{}%
\AgdaBound{⊨γ}\<%
\\
\>[2][@{}l@{\AgdaIndent{0}}]%
\>[4]\AgdaKeyword{with}\AgdaSpace{}%
\AgdaFunction{fundamental-lemma}\AgdaSpace{}%
\AgdaBound{⊢r}\AgdaSpace{}%
\AgdaBound{⊨γ}\<%
\\
\>[2]\AgdaSymbol{...}\AgdaSpace{}%
\AgdaSymbol{|}\AgdaSpace{}%
\AgdaOperator{\AgdaInductiveConstructor{⟨ƛ}}\AgdaSpace{}%
\AgdaBound{x}\AgdaSpace{}%
\AgdaOperator{\AgdaInductiveConstructor{⟶}}\AgdaSpace{}%
\AgdaBound{t}\AgdaSpace{}%
\AgdaOperator{\AgdaInductiveConstructor{⟩}}\AgdaSpace{}%
\AgdaBound{δ}\AgdaSpace{}%
\AgdaOperator{\AgdaInductiveConstructor{,}}\AgdaSpace{}%
\AgdaBound{r⇓}\AgdaSpace{}%
\AgdaOperator{\AgdaInductiveConstructor{,}}\AgdaSpace{}%
\AgdaBound{sf}\AgdaSpace{}%
\AgdaSymbol{=}\<%
\\
\>[2][@{}l@{\AgdaIndent{0}}]%
\>[4]\AgdaKeyword{let}\AgdaSpace{}%
\AgdaSymbol{(}\AgdaBound{a}\AgdaSpace{}%
\AgdaOperator{\AgdaInductiveConstructor{,}}\AgdaSpace{}%
\AgdaBound{s⇓}\AgdaSpace{}%
\AgdaOperator{\AgdaInductiveConstructor{,}}\AgdaSpace{}%
\AgdaBound{sa}\AgdaSymbol{)}\AgdaSpace{}%
\AgdaSymbol{=}\AgdaSpace{}%
\AgdaFunction{fundamental-lemma}\AgdaSpace{}%
\AgdaBound{⊢s}\AgdaSpace{}%
\AgdaBound{⊨γ}\AgdaSpace{}%
\AgdaKeyword{in}\<%
\\
\>[4]\AgdaKeyword{let}\AgdaSpace{}%
\AgdaSymbol{(}\AgdaBound{b}\AgdaSpace{}%
\AgdaOperator{\AgdaInductiveConstructor{,}}\AgdaSpace{}%
\AgdaBound{f⇓}\AgdaSpace{}%
\AgdaOperator{\AgdaInductiveConstructor{,}}\AgdaSpace{}%
\AgdaBound{sb}\AgdaSymbol{)}\AgdaSpace{}%
\AgdaSymbol{=}\AgdaSpace{}%
\AgdaBound{sf}\AgdaSpace{}%
\AgdaBound{sa}\AgdaSpace{}%
\AgdaKeyword{in}\<%
\\
\>[4]\AgdaBound{b}\AgdaSpace{}%
\AgdaOperator{\AgdaInductiveConstructor{,}}\AgdaSpace{}%
\AgdaInductiveConstructor{evalApp}\AgdaSpace{}%
\AgdaBound{r⇓}\AgdaSpace{}%
\AgdaBound{s⇓}\AgdaSpace{}%
\AgdaBound{f⇓}\AgdaSpace{}%
\AgdaOperator{\AgdaInductiveConstructor{,}}\AgdaSpace{}%
\AgdaBound{sb}\<%
\\
\\[\AgdaEmptyExtraSkip]%
\>[2]\AgdaComment{--\ Arbitrary\ environment\ as\ environments\ are\ no\ longer}\<%
\\
\>[2]\AgdaComment{--\ scoped\ by\ a\ context\ in\ extrinsic\ representation}\<%
\\
\>[2]\AgdaKeyword{variable}\AgdaSpace{}%
\AgdaGeneralizable{empty}\AgdaSpace{}%
\AgdaSymbol{:}\AgdaSpace{}%
\AgdaDatatype{Env}\<%
\\
\\[\AgdaEmptyExtraSkip]%
\>[2]\AgdaComment{--\ Totality\ of\ evaluation\ for\ well-typed\ terms\ in}\<%
\\
\>[2]\AgdaComment{--\ well-typed\ environments}\<%
\\
\>[2]\AgdaFunction{⇓-total}\AgdaSpace{}%
\AgdaSymbol{:}\AgdaSpace{}%
\AgdaInductiveConstructor{∅}\AgdaSpace{}%
\AgdaOperator{\AgdaDatatype{⊢}}\AgdaSpace{}%
\AgdaGeneralizable{t}\AgdaSpace{}%
\AgdaOperator{\AgdaDatatype{∷}}\AgdaSpace{}%
\AgdaGeneralizable{T}\AgdaSpace{}%
\AgdaSymbol{→}\AgdaSpace{}%
\AgdaFunction{∃[}\AgdaSpace{}%
\AgdaBound{a}\AgdaSpace{}%
\AgdaFunction{]}\AgdaSpace{}%
\AgdaGeneralizable{empty}\AgdaSpace{}%
\AgdaOperator{\AgdaDatatype{∣}}\AgdaSpace{}%
\AgdaGeneralizable{t}\AgdaSpace{}%
\AgdaOperator{\AgdaDatatype{⇓}}\AgdaSpace{}%
\AgdaBound{a}\<%
\\
\>[2]\AgdaFunction{⇓-total}\AgdaSpace{}%
\AgdaBound{⊢t}\AgdaSpace{}%
\AgdaSymbol{=}\<%
\\
\>[2][@{}l@{\AgdaIndent{0}}]%
\>[4]\AgdaKeyword{let}\AgdaSpace{}%
\AgdaSymbol{(}\AgdaBound{a}\AgdaSpace{}%
\AgdaOperator{\AgdaInductiveConstructor{,}}\AgdaSpace{}%
\AgdaBound{t⇓a}\AgdaSpace{}%
\AgdaOperator{\AgdaInductiveConstructor{,}}\AgdaSpace{}%
\AgdaSymbol{\AgdaUnderscore{})}\AgdaSpace{}%
\AgdaSymbol{=}\AgdaSpace{}%
\AgdaFunction{fundamental-lemma}\AgdaSpace{}%
\AgdaBound{⊢t}\AgdaSpace{}%
\AgdaSymbol{(λ}\AgdaSpace{}%
\AgdaSymbol{())}\AgdaSpace{}%
\AgdaKeyword{in}\<%
\\
\>[4]\AgdaBound{a}\AgdaSpace{}%
\AgdaOperator{\AgdaInductiveConstructor{,}}\AgdaSpace{}%
\AgdaBound{t⇓a}\<%
\end{code}
\end{AgdaAlign}

\end{document}